\documentclass[a4paper, 11pt, doubleside]{Thesis} 

\graphicspath{{Figures/}} 

\usepackage[square, numbers, comma, sort&compress]{natbib} 

\usepackage{verbatim}

\usepackage{vector}

\usepackage{acronym}

\usepackage{mathrsfs}

\usepackage{psfrag}
\usepackage{pstool}
\usepackage{upgreek}
\usepackage{multirow}

\hypersetup{urlcolor=blue, colorlinks=false}

\begin{document}
\frontmatter	  

\acrodef{ACTD}{Amplitude Corrected Time-Domain}
\acrodef{BBH}{binary black hole}
\acrodef{BNS}{binary neutron star}
\acrodef{NSBH}{neutron star-black hole binary}
\acrodef{CBC}{compact binary coalescence}
\acrodef{DAG}{direct acrylic graph}
\acrodef{DFT}{discrete Fourier transform}
\acrodef{EFE}{Einstein Field Equations}
\acrodef{EOB}{effective one-body}
\acrodef{ET}{Einstein Telescope}
\acrodef{FAR}{false alarm rate}
\acrodef{FD}{frequency domain}
\acrodef{FBLO}{first beyond leading order}
\acrodef{FLSO}{frequency of last stable orbit}
\acrodef{FT}{Fourier transform}
\acrodef{FWF}{full waveform}
\acrodef{HIPE}{Hierarchical Inspiral Pipeline Executable}
\acrodef{IFAR}{inverse false alarm rate}
\acrodef{IFO}{interferometer}
\acrodef{IFT}{inverse Fourier transform}
\acrodef{IMBBH}{intermediate mass binary black hole}
\acrodef{IMR}{inspiral-merger-ringdown}
\acrodef{ISCO}{innermost stable circular orbit}
\acrodef{LIGO}{Laser Interferometic Gravitational-Wave Observatory}
\acrodef{LHS}{left hand side}
\acrodef{LSC}{LIGO Scientific Collaboration}
\acrodef{PN}{post-Newtonian}
\acrodef{PSD}{power spectral density}
\acrodef{RWF}{restricted waveform}
\acrodef{S4}{LIGO's 4th science run}
\acrodef{S5}{LIGO's 5th science run}
\acrodef{S51YR}{S5 first year}
\acrodef{SNR}{signal-to-noise ratio}
\acrodef{SPA}{stationary-phase approximation}
\acrodef{RHS}{right hand side}
\acrodef{TD}{time domain}
\acrodef{TT3}{Taylor-T3}
\acrodef{VSR1}{Virgo Science Run $1$}

\def\Msun{\ensuremath{M_{\odot}}}
\def\be{\begin{equation}}
\def\ee{\end{equation}}

\def\BNSul{\ensuremath{1.4 \times 10^{-2}}}
\def\BHNSul{\ensuremath{3.6 \times 10^{-3}}}
\def\BBHul{\ensuremath{7.3 \times 10^{-4}}}
\def\SBHNSul{\ensuremath{4.4 \times 10^{-3}}}
\def\SBBHul{\ensuremath{9.0 \times 10^{-4}}}

\def\gsim{\mathrel{
 \rlap{\raise 0.511ex \hbox{$>$}}{\lower 0.511ex
 \hbox{$\sim$}}}}

\def\lsim{\mathrel{
 \rlap{\raise 0.511ex \hbox{$<$}}{\lower 0.511ex
 \hbox{$\sim$}}}}

\title{On the use of higher order waveforms\\
in the search for gravitational waves \\
emitted by compact binary coalescences}
\authors{David J. A. McKechan}            
           
\addresses  {\groupname\\\deptname\\\univname}
\date       {\today}
\subject    {}

\maketitle

\setstretch{1.3}

\fancyhead{}  
\rhead{\thepage}  
\lhead{}  

\pagestyle{fancy}

\mycopyright{}
\clearpage

\newpage
\thispagestyle{empty}
\mbox{}
\newpage

\pagestyle{empty}  

\null\vfill

\textit{`I ran a \acs{LIGO} search for gravitational waves and all I got was
a lousy chapter in my thesis...'}

\begin{flushright}
The Author.
\end{flushright}

\vfill\vfill\vfill\vfill\vfill\vfill\null
\cleardoublepage  

\Declaration{

\addtocontents{toc}{}  

\begin{itemize} 
\item \textsc{Declaration:}\\ This work has not previously been accepted in 
substance for any degree and is not concurrently submitted in candidature 
for any degree.
\\
\\
Signed: $_{..................................................................}$
(candidate)\ \ Date: $_{...................}$ 
\\
 
\item \textsc{Statement 1:}\\ This thesis is being submitted in partial 
fulfillment of the requirements for the degree of Doctor of Philosophy (PhD).
\\
\\
Signed: $_{..................................................................}$
(candidate)\ \ Date: $_{...................}$ 
\\

\item \textsc{Statement 2:}\\ This thesis is the result of my own independent
work/investigation, except where otherwise stated. Other sources are 
acknowledged by explicit references.  
\\
\\
Signed: $_{..................................................................}$
(candidate)\ \ Date: $_{...................}$ 
\\
 
\item \textsc{Statement 3}\\ I hereby give consent for my thesis, if accepted,
to be available for photocopying and for inter-library loan, and for the
title and summary to be made available to outside organisations.
\\
\\
Signed: $_{..................................................................}$
(candidate)\ \ Date: $_{...................}$ 
\\
%
\end{itemize}

}
\cleardoublepage  

\addtotoc{Summary of Thesis}  
\abstract{
\addtocontents{toc}{}  
\begin{singlespace} 
This thesis concerns the use,
in gravitational wave data analysis, 
of higher order waveform models of the gravitational radiation
emitted by compact binary coalescences.
We begin with an introductory chapter that includes
an overview of the theory of general relativity,
gravitational radiation and ground-based interferometric gravitational wave
detectors. 
We then discuss, in Chapter 2, the
gravitational waves emitted by compact binary coalescences,
with an explanation of higher order waveforms and how they differ from
leading order waveforms; we also introduce the post-Newtonian formalism.
In Chapter~\ref{howchap} the method and results of a gravitational
wave search for low mass compact binary 
coalescences
using a subset of \acl{S5} data are presented and in
the subsequent chapter we
examine how one could use higher order waveforms in such analyses.
We follow the
development of a \textit{new} search algorithm that
incorporates higher order waveforms with promising results for
detection efficiency and parameter estimation.
In Chapter~\ref{winchap}, a new
method of windowing time-domain waveforms that offers benefit to
gravitational wave searches is presented. The final chapter covers the
development of a game designed 
as an outreach project
to raise public awareness and understanding of 
the search for gravitational waves.
\end{singlespace}
}

\cleardoublepage  

\setstretch{1.3}  

\acknowledgements{
\addtocontents{toc}{}  
I must begin by thanking my supervisor, Bangalore Sathyaprakash, for
all of his considered advice, patience and directorship that has been of
great benefit to me
over the past three and a half years here at Cardiff. It has truly been a
pleasure to work with Sathya and I am still unsure which is greater: his
passion for gravitational waves and physics, or his knowledge.
Both attributes are inspiring and I wish that there had been more time to hear 
Sathya talk with great enthusiasm about his favourite subjects.

I gratefully acknowledge those that have directly contributed to the
projects presented in this thesis. I would first like to thank Chris Van 
Den Broeck, whom I regarded as my higher harmonics mentor; 
and Craig Robinson, who was always at hand to help me
squash bugs in various analysis codes (not always mine).
Both Chris and Craig were able to fill in the gaps of my
mathematical knowledge and explain `what Sathya meant' whenever I was
confused (as I often was).
I am grateful to Stephen Fairhurst, whom I have
considered as my vice-supervisor,
for his invaluable advice and insight, particularly regarding the
12-to-18 analysis. I also thank the 12-to-18 monthers who made it very easy for
me to `lead' the project and who all did such a great job:
Collin Capano, Ian Harry, Chris, Lucia Santamaria, Michele Vallisneri, 
Sukanta Bose, Andrew Lundgren and Duncan Brown. Oh, and thanks for putting up
with `Dictator Dave'.

I have thoroughly enjoyed working with all the members of 
the Cardiff Gravitational Physics group, past and present. 
Thomas Cokelaer, Anand Sengupta, Jack Yu, Deepak Baskaran, Patrick Sutton,
Leonid Grishchuk,
Laura Nuttall, Thomas Dent, Duncan MacLeod, John Veitch, James Clark, 
Gerald Davies, Devanka Pathak, 
Alexander Dietz,
Gareth Jones, Wen Zhao, Ioannis Kamaretsos,
Mark Edwards, Valeriu Predoi, Chris, Craig, Ian, Steve and Sathya: 
thank you all for being easy to work with,
willing to help whenever asked and for the interesting 
conversations (of varying sobriety) we have had over the years.
Leo - thanks for never letting me believe I had any talent at football.
I also acknowledge Louise Winter, Nicola Hunt, 
Philip Treadgold and all other staff in the school office
for their friendly administrative support.

I wish to thank everyone in the CBC group for all their help.
I am sure to forget a few names here, for which I apologise, but I would like
to mention a few in particular who have helped me on various
occasions: Alan Weinstein,
Fredrique Marion, Gabriella Gonzales, Drew Keppel, Jessica Clayton, 
Ruslan Vaulin, Patrick Brady, Gianluca Guidi, Scott Koranda, 
Jolien Creighton, Nickolaus Fotopoulos, Marie Ann Bizourd, John Whelan, 
and Adam Mercer. 

I am also grateful to the LSC, for permitting the use of LIGO S4 data,
and to the STFC for my studentship (PPA/S/S/2006/4330).

On a personal level I thank my friends from home: Steve, Adam, Sarah,
Jonny, Soph, Chris, Steve, Rob, Emily and Will. They may not realise it, but
the times that we have shared during the holidays have just about 
preserved my sanity (particularly Christmas 2009).
I also thank my family for all their support
over the years, Nan, Wendy, Nigel; Laura and Paul for their 
helpful advice,
even whilst busy looking after my baby niece Ella.

I am indebted to my beautiful girlfriend Amanda; for her unending love and 
support, especially over the past six months where I have done little but
write this thesis, moan about writing this thesis, go running 
and complete job applications. 

Finally, I dedicate this work to my parents who have given me everything.

}
\cleardoublepage  

\copapers{
\addtocontents{toc}{}  
Sections of this thesis include collaborative work published in
the following papers:
\begin{enumerate}

\item Chapter~\ref{howchap}: \textit{`Search for Gravitational Waves from 
Low Mass Compact Binary Coalescence} [sic] \textit{in 186 Days of LIGO's fifth 
Science Run'}~\cite{Abbott:2009qj}.

The author analysed one month of data and was the corresponding author
for the publication. The author also contributed to tuning the pipeline,
the development of ihope and the ihope results page.

\textsc{Disclaimer}: Although this chapter presents results previously 
published, the views expressed in the chapter are the author's own and
are in no way endorsed by the \ac{LSC}.

\item Chapter~\ref{winchap}: \textit{`A tapering window for time-domain
templates and simulated signals in the detection of gravitational
waves from coalescing compact binaries.'}~\cite{mrs:2009}.

The author developed the implementation of the new window and conducted
the majority of the corresponding studies.

\item Chapter~\ref{hunter}: `\textit{Black Hole Hunter: The game that lets
YOU search for gravitational waves.'}~\cite{cdf:bhh}.

The author developed the code that produced  the audio and image files
used in the game.
\end{enumerate}
}
\cleardoublepage  

\pagestyle{fancy}

\lhead{\emph{Contents}}  
\tableofcontents  

\lhead{\emph{List of figures}}  
\listoffigures

\lhead{\emph{List of tables}}  
\listoftables  

\clearpage  

\lhead{\emph{List of acronyms}}
\begin{spacing}{1.3}{
    \setlength{\parskip}{1pt}
    \chapter*{List of acronyms}
    \addtotoc{List of acronyms}
    \begin{tabbing}
      \hspace{3.5cm} \= \\
      ACTD \> Amplitude Corrected Time-Domain.\\
BBH \> Binary black hole.\\
BNS \> Binary neutron star.\\
NSBH \> Neutron star-black hole binary.\\
CBC \> Compact binary coalescence.\\
DAG \> Direct acrylic graph.\\
DFT \> Discrete Fourier transform.\\
EFE \> Einstein Field Equations.\\
EOB \> Effective one-body.\\
ET \> Einstein Telescope.\\
FAR \> False alarm rate.\\
FD \> Frequency domain.\\
FBLO \> First beyond leading order.\\
FLSO \> Frequency of last stable orbit.\\
FT \> Fourier transform.\\
FWF \> Full waveform.\\
HIPE \> Hierarchical Inspiral Pipeline Executable.\\
IFAR \> Inverse false alarm rate.\\
IFO \> Interferometer.\\
IFT \> Inverse Fourier transform.\\
IMBBH \> Intermediate mass binary black hole.\\
IMR \> Inspiral-merger-ringdown.\\
ISCO \> Innermost stable circular orbit.\\
LIGO \> Laser Interferometic Gravitational-Wave Observatory.\\
LHS \> Left hand side.\\
LSC \> LIGO Scientific Collaboration.\\
PN \> Post-Newtonian.\\
PSD \> Power spectral density.\\
RWF \> Restricted waveform.\\
S4 \> LIGO's 4th science run.\\
S5 \> LIGO's 5th science run.\\
S51YR \> S5 first year.\\
SNR \> Signal-to-noise ratio.\\
SPA \> Stationary-phase approximation.\\
RHS \> Right hand side.\\
TD \> Time domain.\\
TT3 \> Taylor-T3.\\
VSR1 \> Virgo Science Run $1$.\\

    \end{tabbing}
    \cleardoublepage
}\end{spacing}

\lhead{\emph{List of constants and variables}}
\begin{spacing}{1.3}{
    \setlength{\parskip}{1pt}
    \chapter*{List of constants and variables}
    \addtotoc{List of constants and variables}
    \begin{tabbing}
      \hspace{3.5cm} \= \\
      $c$ \> Speed of light in a vacuum. \\
$G$ \> Gravitational constant.\\
$\Msun$ \> Mass of the Sun.\\
\\
$\eta$ \> Symmetric mass ratio. (Also Minkowski metric in Chapter 1.)\\
$F_{+,\times}$ \> Detector response functions.\\
${\cal M}$ \> Chirp mass.\\
$\rho$ \> Signal-to-noise ratio (SNR).\\
$S_n(f)$ \> Power spectral density.\\

    \end{tabbing}
    \cleardoublepage
}\end{spacing}

\lhead{\emph{Conventions}}  
\begin{spacing}{1.3}{
    \setlength{\parskip}{1pt}
    \chapter*{Conventions}
    \addtotoc{Conventions}
\begin{enumerate}
\item Unless stated otherwise,
Greek indices run from $0,\ldots,3$ and Latin indices from
$1,\ldots,3$.

\item The mass of the Sun is $\Msun=1.99\times10^{30}\, \rm kg$.

\item The speed of light is $c=3\times10^{8}\, {\rm ms^{-1}}$.

\item Unless stated otherwise,
natural units will be used where the gravitational
constant and the speed of light are unity, i.e., $G=c=1$.

\item Three component spatial vectors will be represented as bold and 
non-italic text, e.g.,
\begin{equation*}
\textup{\textbf{F}} = m \textup{\textbf{a}}\, .
\end{equation*}

\item An overhead dot represents the derivative with respect to time, e.g.,
\begin{equation*}
\textup{\textbf{F}} 
  = m \textup{\textbf{a}}
  = m \frac{d\textup{\textbf{v}}}{dt} = m \dot{\textup{\textbf{v}}}
  = m \frac{d^2\textup{\textbf{x}}}{{dt}^2} = m \ddot{\textup{\textbf{x}}}\, .
\end{equation*}

\item The derivative is represented by a comma, e.g.,
\begin{equation*}
A^{\nu}_{\ ,\mu} = \frac{\partial A^{\nu}}{\partial x^\mu}\, .
\end{equation*}
and the covariant derivative is represented by a semi-colon, e.g.,
\begin{equation*}
A^{\nu}_{\ ;\mu} = 
  \frac{\partial A^{\nu}}{\partial x^\mu}
  + \Gamma^{\nu}_{\, \mu\alpha}A^\alpha\, .
\end{equation*}
where $\Gamma^{\nu}_{\mu\alpha}$ is the Christoffel symbol.

\item The Einstein summation convention is used,
\be
a_ix^i = \sum_i a_ix^i\, .
\ee

\item The expectation value is represented by an overline,
\be
\overline{f(x)} = \int f(x) P(x) dx\, ,
\ee
where $P(x)$ is the probability density function.

\item The abbreviation for the \ac{FWF} refers to any \ac{PN} waveform
that is greater than 0\ac{PN} in amplitude. For instance a waveform that is
0.5\ac{PN} in amplitude and 2\ac{PN} in phase may still be referred to as 
the
\ac{FWF}. Where the amplitude order is specified it is written in brackets, 
e.g., a waveform at 2\ac{PN} in amplitude will be written as 
\ac{FWF} (2\ac{PN}). N.B.: this does not necessarily
indicate the phase order.
\item The \acs{LIGO} interferometers are denoted by their location, 
H for Hanford and L for Livingston; and by their arm length, 1 for $4\, \rm km$
and $2$ for $2\, \rm km$.
The initial \acs{LIGO} configuration 
consisted of two interferometers at Hanford, H1 and H2 and a single
interferometer at Livingston, L1.
Data collected by all three detectors were called triple time or
H1H2L1 time, alternatively there were the double times, H1L1 and H2L1, when
only one of the Hanford interferometers was operating
\footnote{H1H2 double time data are not considered.}. Triggers that are 
coincident between the sites may be labelled in the same 
manner, e.g., a trigger
coincident between H1 and L1 during triple time could be 
described as an H1L1 trigger in H1H2L1 time.
\end{enumerate}
    \cleardoublepage
}\end{spacing}

\newpage

\dedicatory{For Mum and Dad.}
\clearpage

\addtocontents{toc}{\vspace{2em}}  

\mainmatter	  
\pagestyle{fancy}  


\onehalfspacing

\chapter*{\textit{Prologue}}
\addtotoc{Prologue}
\textit{
Nicolaus Copernicus was the first to develop a thorough and detailed
Heliocentric theory of the Universe, with the Sun at the centre and,
perhaps more importantly, the Earth in orbit around the Sun. 
Observation and appreciation of celestial mechanics was the first step
on the path towards understanding gravity.
\\
\\
Just 200 years 
later Kepler had developed his laws of planetary motion and 
Newton, in turn, his universal law of gravitation. In a short time gravity had 
developed from insignificance to a simple inverse-square law, 
explaining the motion of all the stars and
all the planets, and why objects fall to the Earth...nearly.
\\
\\
Another 200 years later, Einstein completed his theory of 
general relativity. With the advent of relativity, 
gravitational field information, like everything else, was bound to 
the universal speed limit of light. Thereafter, any
theory of gravity obeying the principles of special relativity,
was obliged to permit gravitational waves.
\\
\\
Today, approaching the centenary of general relativity, we are on the
cusp of direct gravitational wave detection that will open a new window
from which to view the Universe, illuminating our understanding.}

\chapter{Introduction}
\label{Intro}
\lhead{Chapter~\ref{Intro} \emph{Introduction}}
\rule{15.7cm}{0.05cm}

We begin with an introduction to relativity that gently introduces the
fundamental concepts and an idea of curved spacetime before quickly progressing
on to the theory of gravitational waves by understanding how they propagate
and interact with free particles. The chapter concludes with 
an overview of
ground-based interferometric detection of gravitational waves with an
introduction to the \acs{LIGO} detectors and their operation.

\section{A very brief course in relativity}
\subsection{The Principle of relativity}
The \emph{principle of relativity} is the simple requirement that
the \emph{the laws of physics are the same in every inertial frame}. A
passenger inside a train moving at a constant velocity can perform no
experiment to determine the \emph{absolute} speed of the train, measuring
the same physical constants etc., as his or her 
companion waiting at rest on the station platform. 
Under Newtonian physics their frames of reference are related by a
\emph{Galilean transformation}, which applies to the spatial dimensions
with both observers measuring the same \emph{absolute time}. However, 
Galilean transformations do not work when applied to light
emitted from objects moving relative to one another.
When doubts of the existence of a luminiferous aether arose,
it became clear that Galilean transformations were not 
entirely consistent with the principle of relativity.

Einstein abandoned the concept of absolute time. He introduced a second
postulate to the principle of relativity, that \emph{the speed of light is the
same in all inertial reference frames regardless of their relative motion}.
In fact if Maxwell's equations, which reveal the nature of light as
electromagnetic radiation, are the same in all inertial frames,
then the
second postulate is implied by the principle of relativity regardless.
Einstein had
developed his theory of \emph{special relativity}, where the coordinates of
two inertial frames are related by a \emph{Lorentz transformation}, 
which applies to the three spatial coordinates \emph{and} the time-coordinate.

From this simple construct, all the popular wonders of special relativity
arise: time-dilation, length-contraction and mass-energy equivalence.
However, special relativity does not account for non-inertial frames of 
reference, i.e., it can not be applied in an accelerating
frame\footnote{Accelerating frames can be studied in special relativity by 
using an \emph{instantaneous rest frame}.}. Furthermore, Newton's law
of gravity is not consistent with special relativity.

The \emph{general} principle of relativity requires that \emph{the laws of
physics are the same in all reference frames} - both inertial and
non-inertial - and forms the basis of Einstein's theory of
\emph{general relativity}, a theory of gravity that is consistent with
special relativity.

\subsection{Tidal forces and the curvature of spacetime}
An astonishing coincidence of nature is the equivalence of gravitational 
and inertial mass, i.e., the property of matter that determines 
the force an object experiences
due to gravity is the same property that determines its 
resistance to an applied 
external force. Einstein realised that a person at rest on 
the Earth's surface, where the gravitational acceleration is 
$\textup{\textbf{g}}_E$, is
indistinguishable from another person, inside a spaceship accelerating
at $\textup{\textbf{a}}=\textup{\textbf{g}}_E$, 
far away from any gravitational field. 
Moreover, a person in 
free-fall, over a \emph{short-period of time}, is equivalent to another 
in a spaceship, also far away from any gravitational field, but undergoing 
no acceleration. Thus the equivalence principle is defined:
\emph{In a freely falling laboratory, in a small region of spacetime, the
laws of physics are those of special relativity}. 

Consider a pair of identical sky-diving twins, who have jumped
simultaneously from a plane using doors on either side of an
aircraft and who are now in free-fall. Initially, they are at the
same distance from the centre of the Earth, but separated by a short 
horizontal distance.
As each twin is falling on a path that extends radially from the centre
of the Earth, they will gradually drift towards each other. Had they jumped
one after the other, so that they were separated by a short vertical distance,
the first twin to jump would undergo a slightly stronger acceleration and their
vertical separation would gradually increase. The effect on the twins'
horizontal or vertical separation is tidal acceleration; due to a
non-uniform gravitational field which gives the \emph{tidal force}.

Einstein concluded that an object in free fall is \emph{not}
subjected to a gravitational force, i.e., although the sky divers' horizontal
separation decreases, there is no horizontal force acting upon them. Rather,
spacetime is curved due to the Earth's mass and energy - the sky divers are
instead
following separate \emph{geodesic} paths,
the `straight lines' of a curved space.

\subsection{The geometry of spacetime and the Einstein Field Equations}
In relativity, the geometry of spacetime is defined as a 
pseudo-Riemannian manifold.
In special relativity the interval, ${ds}^2$,  
between two events on 
the spacetime manifold is given by the Minkowski metric, $\eta$, where
\be
{ds}^2 = \eta_{\mu\nu} dx^\mu dx^\nu\, ,
\ee
\be
x^\mu = (t, x, y, z)\, ,
\ee
and
\be
\eta_{\mu\nu} = 
\begin{pmatrix}
 -1 & 0 & 0 & 0 \\
  0 & 1 & 0 & 0 \\
  0 & 0 & 1 & 0 \\
  0 & 0 & 0 & 1 \\
\end{pmatrix}\, .
\ee
If the Minkowski metric is that of flat spacetime geometry,
then in general relativity the interval between two events in spacetime is
defined by a general metric, $g$,
\be
{ds}^2 = g_{\mu\nu} dx^\mu dx^\nu\, .
\ee

The metric $g$ contains the information about the curvature of spacetime. Our
sky diving twins are experiencing a tidal force, their horizontal
separation is decreasing. Under gravitational free-fall, both are
following geodesic paths that were
initially parallel to each other but are converging due to the curvature of
spacetime. The curvature is quantified by the Riemann tensor, 
$R^\mu_{\ \nu\rho\sigma}$. The Riemann tensor is defined entirely by the
spacetime metric and its first and second derivatives; it is equal to zero
in a flat spacetime.

Einstein linked the curvature of spacetime to the energy-momentum tensor,
$T_{\mu\nu}$, which contains the momentum and energy 
densities and their fluxes in
a region of spacetime (see \ref{sec:tmunu}), in the form of ten
second order PDEs known as the \ac{EFE}\footnote{Rather than sixteen
equations due to the symmetry of the metric tensor and $R_{\mu\nu}$.},
\be
\label{eq:efe}
G_{\mu\nu} = R_{\mu\nu} - \frac{1}{2}g_{\mu\nu}R = \kappa T_{\mu\nu}\, ,
\ee
where
\be
R_{\mu\nu} = R^\alpha_{\ \mu\alpha\nu}\, ,
\ee
\be
R = R^\mu_{\ \mu}\, ,
\ee
and
\be
\kappa = \frac{8\pi G}{c^2}.
\ee
The \ac{EFE} have the important properties that:
\begin{itemize}
\item energy and momentum are conserved, 
\be
  T^{\mu\nu}_{\ \ ;\mu} = 0\, ;
\ee
\item Newtonian gravity is recovered in the correct limits, i.e., where
$v\ll1$ and the internal stresses are small;
\item they are tensor equations and are manifestly
\emph{invariant} under coordinate transformations!
\end{itemize}

\section{The weak field approximation}
The \ac{EFE} are difficult, if not impossible, to solve in all but the most
simple of situations. One approach is that of the weak field approximation
where the
spacetime metric is expressed simply as Minkowski spacetime plus a small
perturbation,
\be
\label{eq:gmunu}
g_{\mu\nu} = \eta_{\mu\nu} + h_{\mu\nu}\, ,
\ee
where
\be
|h_{\mu\nu}| \ll 1\, .
\ee
As the perturbation $h_{\mu\nu}$ and its derivatives are very small, one can 
retain only their first order terms, i.e., terms \emph{linear} in
$h_{\mu\nu}$ and $\partial h_{\mu\nu}$. In doing so, the Riemann tensor takes
the simple form,
\be
\label{eq:rch}
R_{\mu\nu\rho\sigma} = 
  \frac{1}{2}
  \left(
    h_{\mu\sigma,\nu\rho} + h_{\nu\rho,\mu\sigma}
  - h_{\mu\rho,\nu\sigma} - h_{\nu\sigma,\mu\rho}
  \right)\, .
\ee

Thus we will obtain the \emph{linearised} \ac{EFE} by
substituting (\ref{eq:gmunu}) and (\ref{eq:rch}) in
(\ref{eq:efe}). Before doing so we should recall that in general relativity
we are free to make
any coordinate transformation that we wish. Interestingly, it can be shown
that under a \emph{small} coordinate transformation the metric can remain 
unchanged. Given that
\be
g^\prime_{\mu\nu} = \frac{\partial x^\rho}{\partial x^{\prime\mu}}
  \frac{\partial x^\sigma}{\partial x^{\prime\nu}} g_{\rho\sigma}\, ,
\ee
and
\be
x^{\prime\mu} = x^\mu + \epsilon^\alpha\ , 
\ee
the metric (\ref{eq:gmunu}) will transform as
\begin{subequations}
\begin{align}
g^\prime_{\mu\nu} & = \frac{\partial x^\rho}{\partial x^{\prime\mu}}
  \frac{\partial x^\sigma}{\partial x^{\prime\nu}}
  \left[\eta_{\rho\sigma} + h_{\rho\sigma}\right]\, ,\\
  & = 
  \eta_{\mu\nu} + 
  \frac{\partial x^\rho}{\partial x^{\prime\mu}}
  \frac{\partial x^\sigma}{\partial x^{\prime\nu}}
  h_{\rho\sigma}\, ,\\
  & = 
  \eta_{\mu\nu} + h_{\mu\nu} - \epsilon_{\mu,\nu} - \epsilon_{\nu,\mu}\, .
\end{align}
\end{subequations}
Hence the coordinate transformation
simply re-defines the metric perturbation,
$h_{\mu\nu}\rightarrow h^\text{(new)}_{\mu\nu}$. Provided the weak field
condition is still met, $|h^\text{(new)}_{\mu\nu}|\ll1$, 
one can make any 
coordinate transformation; such changes are known as gauge transformations.
The freedom to choose any gauge allows us to greatly
simplify the \ac{EFE}. 

The \emph{trace-reverse}
of the perturbation $h_{\mu\nu}$ is defined as
\be
\bar{h}_{\mu\nu} = h_{\mu\nu} 
  - \frac{1}{2}\eta_{\mu\nu}h^\alpha_{\ \alpha}\, .
\ee
If we make use of the trace-reverse of $h_{\mu\nu}$ 
and choose the Lorentz gauge condition,
\be
\label{eq:gauge1}
\partial^\mu \bar{h}_{\mu\nu} = 0\, ,
\ee
we find the linearised \ac{EFE} can be written
elegantly as
\be
\label{eq:lefe}
\square\bar{h}_{\mu\nu} = -2\kappa T_{\mu\nu}\, .
\ee

\section{Gravitational waves}
\subsection{Vacua solutions to the linearised \acs{EFE}}
In vacua, (\ref{eq:lefe}) reduces to 
\be
\square\bar{h}_{\mu\nu} = 0\, ,
\ee
which is a wave equation with solutions that are superpositions of plane
waves of the form
\be
\bar{h}_{\mu\nu} = A_{\mu\nu}\exp\left(ik_\alpha x^\alpha\right)\, ,
\ee
where the equality
\be
A_{\mu\nu}k^\nu = 0\, ,
\ee
must always be true to satisfy (\ref{eq:gauge1}).

Let us pause for reflection here, we now understand that the 
perturbation of the spacetime metric, $h$, i.e., the gravitational field,
propagates through empty spacetime as a \emph{gravitational wave}. 

The gravitational wave vector, $k^\alpha$, where the wave is of frequency,
$\omega$, may be written as
\be
k^\alpha = \left(\omega, \textup{\textbf{k}}\right)\, .
\ee
The magnitude of $k^\alpha$ is
\be
k^2 = -\omega^2 + {\textup{\textbf{k}}}^2\, .
\ee
The EFE imply that $k^\alpha$ is null, i.e., ${|k|}^2=0$. Therefore,
\be
\omega = |\textup{\textbf{k}}|\, .
\ee
Recall that the general wave-vector $k=\omega/v$, therefore $v=1=c$ and thus
gravitational waves propagate at the speed of light.
Furthermore, in satisfying the Lorentz gauge condition,
we conclude that the amplitude matrix, 
$A_{\mu\nu}$, is orthogonal to the wave vector and, therefore, gravitational
waves are \emph{transverse}.

\subsection{The transverse-traceless gauge}
Before we imposed the gauge conditions (\ref{eq:gauge1}), the
linear \ac{EFE} consisted of ten equations, afterwards there were six.
The linearised \ac{EFE} are further reduced to just two equations
with the \emph{additional} choice of gauge conditions
\begin{subequations}
\begin{align}
\bar{h}_{0\mu} &= 0\, ,\\
\label{eq:traceless}
\bar{h}^\alpha_{\ \alpha} &= 0\, ,
\end{align}
\end{subequations}
known as the \emph{transverse-traceless} gauge conditions.
From here on we shall indicate the transverse-traceless gauge with the 
superscript $TT$ and make use of the fact that under these gauge conditions
$\bar{h}_{\mu\nu}^{TT} = h_{\mu\nu}^{TT}$.

With the conditions that $h^{TT}_{\mu\nu}$ is symmetric and traceless
the \ac{EFE} reduce to just two components. A gravitational
wave propagating
in the $z$-direction takes the form
\be
h_{\mu\nu}^{TT} = 
  \begin{pmatrix}
    0 & 0 & 0 & 0 \\
    0 & h_+ & h_\times & 0 \\
    0 & h_\times & -h_+ & 0 \\
    0 & 0 & 0 & 0 \\
  \end{pmatrix}
\, ,
\ee
where
\be
h_+ = A^{TT}_{xx} \exp\left(ik_\alpha x^\alpha\right)
\ee
and
\be
h_\times = A^{TT}_{xy} \exp\left(ik_\alpha x^\alpha\right)\, .
\ee
The two degrees of freedom, $h_+$ and $h_\times$, are known as the \emph{plus} 
($+$) 
and \emph{cross} ($\times$)
polarisations respectively. A gravitational wave in this
gauge could consist of either polarisation alone or a combination of the two.

One can now write the \emph{time-dependent} weak field metric as
\be
\label{eq:wfmh}
g_{\mu\nu} = 
  \begin{pmatrix}
    -1 & 0 & 0 & 0 \\
    0 & 1 + h_+(t) & h_\times(t) & 0 \\
    0 & h_\times(t) & 1 - h_+(t) & 0 \\
    0 & 0 & 0 & 1 \\
  \end{pmatrix}
\, .
\ee

Throughout this chapter we shall continue to consider gravitational waves
propagating in the $z$-direction with respect to our chosen coordinates.

\subsection{Effect of gravitational waves on a free particle}
\begin{figure}
\centering
\psfragfig{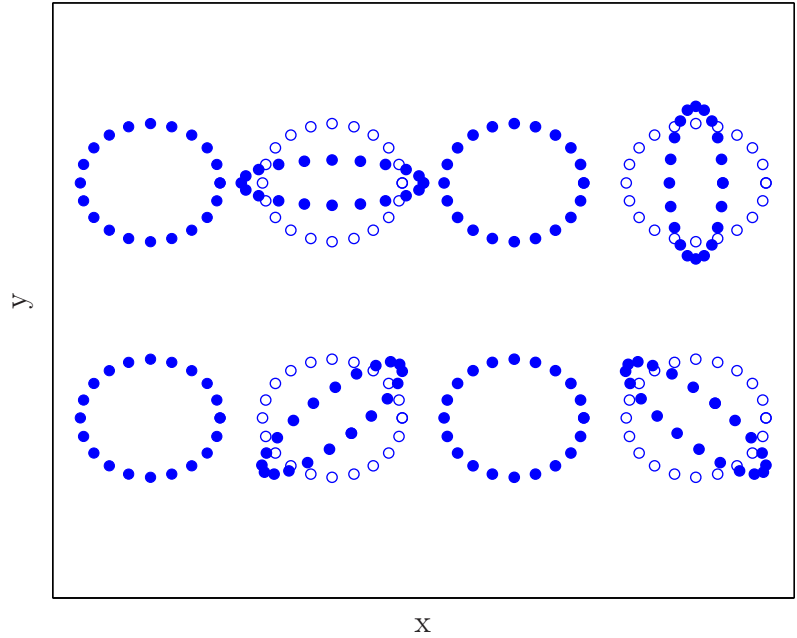}
\caption[Gravitational wave polarisations]
{The displacement of a ring of test 
particles due to a $+$ polarised 
gravitational wave (top) and a $\times$
polarised gravitational wave (bottom). 
The two polarisations 
are related by a $45^\circ$ rotation. From left to right we see the
$0,\pi/2,\pi,3\pi/2$ phases of the gravitational wave, respectively.
The empty circles represent the 
original separation of the particles.}
\label{fig:polars}
\end{figure}

The motion of a test particle\footnote{
A small particle of negligible mass free from any external forces.}
initially at rest in our
chosen coordinates is given by the geodesic equation. It can be shown that
in the $TT$ gauge, the effect of a passing gravitational wave will not
change the particle's four-velocity, i.e., it will remain at rest. Thus in
our coordinates, particles do not move due to a passing gravitational wave.
However, the \emph{proper distance}, $L_x$, between a particle at the origin
and another at $x=L_0$ is given by
\be
L_x = \int_0^{L_0} \sqrt{g_{xx}{dx}^2}\, ,
\ee
which is \emph{time-dependent} when a gravitational wave
passes,
e.g., if the wave is 
propagating
in the $z$-direction, as given in (\ref{eq:wfmh}), we have
\be
L_x(t) = \int_0^{L_0} \sqrt{\left(1+h_+(t)\right){dx}^2}\, .
\ee
Hence the effect of a passing gravitational wave can be seen by observing
the change in proper distance between two test particles. 
Figure~\ref{fig:polars} shows the effect of a passing gravitational wave
on a ring of particles for both polarisations. The particles experience
a time-dependent tidal force. 
One can quantify the effect of a passing gravitational wave by comparing 
the original separation of the particles with the new separation.

Returning
to the case of a particle at the origin and another at $x=L_0$ we can 
calculate the change in length, $\Delta L_x$, due to the metric perturbation
\be
\Delta L_x 
  = L_x(t) - L_0
  = \int_0^{L_0} \sqrt{1+h_+(t)} dx - L_0\, .       
\ee
Using the binomial expansion of the square root and keeping only first order
terms, we rewrite the instantaneous separation as
\be
\label{eq:deltaL}
\Delta L_x 
  = \int_0^{L_0} \left(1+\frac{1}{2}h_+(t)\right) dx - L_0
  = \frac{1}{2}L_0 h_+(t)\, . 
\ee
The fact that the separation of the particles, due to the
effect of a passing gravitational wave, is proportional to the original 
separation, $L_0$, is of great importance when considering a gravitational
wave detector. With that in mind we can rewrite (\ref{eq:deltaL}) as
\be
h_+ = 2 \frac{\Delta L_x}{L_0}\, ,
\ee
where we refer to $h_+$ as the \emph{gravitational wave strain}.

\subsection{Sources of gravitational waves}
\label{sec:sources}
The generation of gravitational waves is understood by finding a general
solution to (\ref{eq:lefe}) and will be discussed in 
appropriate detail in Chapter~\ref{cbc}, where we will pay close 
attention to gravitational waves radiated by \acp{CBC}.

The solution
reveals that gravitational waves are \emph{quadrupolar} in nature and
are generated when a mass accelerates
in a non-spherically symmetric manner, e.g., an inwards-spiralling
binary system (inspiral) or a spinning non-axisymmetric neutron 
star.
Other potential sources of gravitational waves include supernovae,
progenitors of gamma ray bursts, flaring magnetars, pulsars glitches
and
a stochastic background composed of many overlapping signals from the
distant Universe as well as primordial gravitational waves generated in the 
early Universe.

In principle, one could generate gravitational waves in the laboratory,
for instance by rotating a dumb-bell which will have similar characteristics
to a binary. However, even if the impracticalities of detection are neglected,
the gravitational wave strain from such sources
would be far too small ever to be measured~\cite{Saulson:1994}.

\subsection{Indirect evidence of gravitational waves}
\label{sec:hulse}
Observations of binary systems consisting of at least one pulsar 
provide conclusive evidence of the emission of
gravitational radiation in accordance with general relativity. The most famous
of these is PSR B1913+16,
consisting of one pulsar with a companion neutron star. 
The pulsar allows for accurate
measurements of the motion of the two objects, in particular the timing
of the orbital period. General relativity predicts that the system will emit 
rotational energy of the system
as gravitational radiation, causing the orbital separation and period
to decrease. 
Observations of the binary system over nearly 40 years have shown that
the evolution of the orbital period has matched that predicted by 
general relativity to remarkable accuracy. Hulse and Taylor,
who first observed the system,
were duly awarded the 1993 Nobel prize in Physics for their 
discovery which, for the first time, enabled general relativity
to be tested in the strong field dissipative regime~\cite{Hulse:1994}.

The gravitational waves emitted by PSR B1913+16 cannot currently be detected 
directly as they are very small in amplitude and are also of the wrong
frequency to be detected by ground based detectors.
As the binary evolves and the separation between the stars
decreases, the gravitational radiation will increase in 
frequency and amplitude, but 
is not likely to be to be detectable for another three-hundred million years
when the components will coalesce.

\subsection{Direct detection of gravitational waves}
To date gravitational waves have not been detected directly. Efforts began
in the 1960s with resonant bar detectors, the sensitivity of which
has now been surpassed by ground-based interferometric detectors~\cite{
Fafone:2006uy,LIGO-E950018-02-E}, which we 
will discuss below.
In the future, we can look forward to space-based
detectors~\cite{Shaddock:2009za,Corbin:2005ny}
that are free from some of the noise sources that inhibit ground-based
experiments.
Another possibility is the use of accurate pulsar timing 
arrays~\cite{Hobbs:2009yy,Verbiest:2009av}, that could measure fluctuations,
due to a passing gravitational wave,
in the timings of a known set of millisecond pulsars.
 
\section{Interferometric gravitational wave detectors}
The concept of an interferometric detector is simple. Suppose we have
an \ac{IFO} with arms of length $L_0$,
such that a beam splitter sends half the light from
a monochromatic laser along an arm aligned with the $x$-axis and half 
along an arm aligned with the $y$-axis. 
The two beams will be reflected by the end-mirrors
at coordinates $x=L_0$ and $y=L_0$, respectively, before being
superposed upon returning to the beam splitter. If the mirrors are suspended
such that they are freely falling, i.e., 
free from all external forces other than `gravity', they will
behave with respect to the origin in the same manner as
the test particles shown in Figure~\ref{fig:polars}. When a gravitational
wave passes, the separation between the mirrors and the beam splitter will
vary, which can be measured.

We quantify the light travel time along each arm of the interferometer 
using the null interval. For the $x$-axis we have
\be
{ds}^2 = 0 = -{dt}^2 + (1 + h_+){dx}^2\, .
\ee
The time, $\tau_{x1}$, 
of light travel along the $x$-axis from the beam 
splitter to the mirror is, therefore,
\be
\label{eq:tout}
\int_0^{\tau_{x1}} dt
  \approx \int_0^{L_0} \left(1 + \frac{1}{2}h_+\right)dx
  = L_0 + \Delta L\, ,
\ee
where $\Delta L$ is given by (\ref{eq:deltaL}). The return time, $\tau_{x2}$,
is found by swapping the limits
of integration in (\ref{eq:tout}) and noting that the velocity is now in the
negative $x$-direction (or, more simply, multiplying by $2$),
which gives a total light travel time of
\be
\tau_{x} = 
  2L_0 + 2\Delta L\, .
\ee
Similarly, for the arm aligned with the $y$-axis we have a travel time
\be
\tau_{y} = 
  2L_0 - 2\Delta L\, .
\ee
In the absence of a gravitational wave ($h_+=0$ and $\Delta L=0$),
the difference in the travel times between the two arms is
$\Delta\tau=0$. However, in the presence of
a gravitational wave\footnote{Assuming $h_+$ is  
constant for the period of the round trip.}, the difference is
\be
\Delta\tau = 4\Delta L\, .
\ee
Alternatively, written as the phase-shift of the laser light returning to the
beam splitter:
\be
\label{eq:dphi}
\Delta\phi(t) = 4\Delta L(t) \frac{2\pi}{\lambda}
= \frac{4\pi}{\lambda} L_0 h_+(t)\, ,
\ee
where $\lambda$ is the wavelength of the laser. Thus
the passing of a gravitational wave may be observed 
by measuring the phase shift between the light
beams when they are superposed at the beam splitter.

\subsection{Sensitivity}
Supposing the minimum phase difference one can measure is $10^{-9}$; using 
laser light of wavelength $500\, \rm nm$ and an \ac{IFO} of $4\, \rm km$ 
in length,
we find the minimum gravitational wave strain measurable to be $\sim10^{-20}$.
To reach a minimum strain of $\sim10^{-22}$, the \ac{IFO} would need to be 
one hundred times longer. 
However, an effective extension in the arm length can be 
achieved by using \emph{Fabry-Perot} cavities that \emph{fold}
the light, i.e., reflect the light
up and down the arm multiple times before it is superposed at the beam 
splitter. 

In Section~\ref{sec:interest} we will estimate the gravitational wave
strain that is measurable on Earth, due to gravitational 
radiation emitted by \acp{CBC} in the nearby Universe and see that it
is greater than $\sim10^{-22}$.

\subsection{Antenna response functions}
\begin{figure}
\centering
\psfragfig{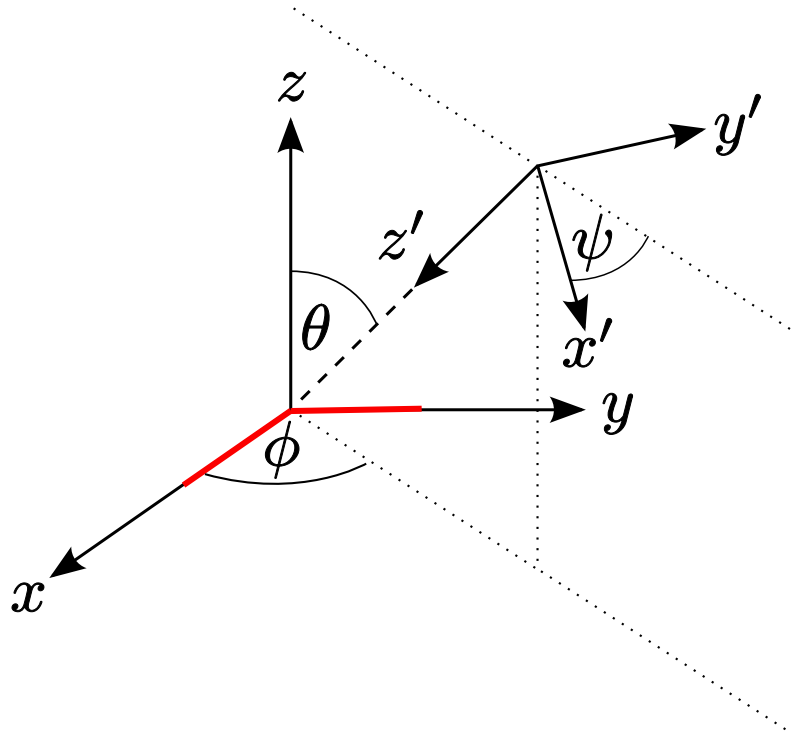}
\caption[The sky angles]
{The sky angles $\theta$, $\phi$ and $\psi$,
between an \acs{IFO} (located at the origin of the unprimed
coordinates and aligned with the $x$-$y$ axes) and a gravitational wave
propagating in the $z^\prime$ direction in the $TT$ gauge.}
\label{fig:skyangle}
\end{figure}

Thus far we have considered a gravitational wave
travelling in the $z$-direction with the detector arms aligned with the $x$- 
and
$y$-axes. In general, the gravitational wave strain in a detector will
be a linear combination of each polarisation multiplied by the \emph{antennae 
response} functions, $F_+$ and $F_\times$, such that
\be
\label{eq:sindet}
h(t) = F_+h_+(t) + F_\times h_\times(t)\, .
\ee
The antenna response functions depend upon the orientation of the source
with respect to the detector, namely the three sky angles $\theta$, $\phi$
and $\psi$ (see Figure~\ref{fig:skyangle}):
\be
F_+(\theta,\phi,\psi) = 
  \frac{1}{2}\cos2\psi\left(1+\cos^2\theta\right)\cos2\phi
  -\sin2\psi\cos\theta\sin2\phi\, ,
\ee
\be
F_\times(\theta,\phi,\psi) = 
  \frac{1}{2}\sin2\psi\left(1+\cos^2\theta\right)\cos2\phi
  -\cos2\psi\cos\theta\sin2\phi\, .
\ee
The angles $\theta$ and $\phi$ give the location of the source,
where $\theta$~$+$~$\pi$ is the angle between the detector's zenith and the
propagation direction of the gravitational wave, $z^\prime$, and
$\phi$ is the azimuth angle between the detector's $x$-axis and the
projection of $z^\prime$ in the $x$-$y$ plane. 
Finally, $\psi$ is the \emph{polarisation} angle, which is the 
angle between the detector's zenith projected on the sky and $x^\prime$.

\subsection{Sources of noise}
A noise source in an \ac{IFO} detector is any process other than
a passing gravitational wave that causes a change in the measured
phase offset (\ref{eq:dphi}). There are four main sources of noise:

\paragraph*{Seismic noise}
Mechanical vibrations of the mirrors (the test masses), 
will occur due to seismic activity that could be caused by anything from an 
earthquake, to the wind or a passing train.
Seismic noise is typically of a low frequency and is the dominant source of
noise below $40\, \rm Hz$~\cite{Saulson:1994}. The seismic noise
may be reduced by isolating
the test masses using suspension systems,
but becomes technically challenging, if not impossible, below $\sim1\,\rm Hz$.

\paragraph*{Thermal noise}
The test masses and their suspension systems will vibrate due to their thermal
energy. The strain\footnote{In units of $1/\sqrt{\rm Hz}$.} 
induced in a detector due to thermal vibrations decreases 
linearly with the natural logarithm
of the frequency and dominates the noise budget
between $40$-$200\, \rm Hz$~\cite{Saulson:1994}.
Ideally, the resonant frequency of the
detector materials will be outside the frequency range of interest 
(the gravitational wave frequency) and will have a high Q-value. 
Thermal noise can also be reduced by designing a cryogenic
detector, e.g., LCGT~\cite{Uchiyama:2004vr}, although detectors
typically operate at ambient temperature. 

\paragraph*{Shot noise}
The number of photons returning 
from each arm of an \ac{IFO} is Poisson distributed with a mean value,
$N$, and standard deviation, $\sqrt{N}$. 
Fluctuations in the number of photons
limits the minimum possible $\Delta\phi$
that can be measured as it appears identical
to a fluctuation in phase, since the phase is estimated by measuring 
output power.  
It
follows that the shot noise is inversely proportional to $\sqrt{N}$, or 
the square root of the 
input power of the laser~\cite{Saulson:1994}.
To reduce the shot
noise to acceptable levels, light that exits the beam splitter is 
recycled, by use of a mirror that returns light that exits the beam splitter
in the direction of the input laser. 
In due course the power builds up in the detector such that
the laser is simply balancing the light losses due to 
imperfections in the mirrors and diffraction losses as well as the light
that exits towards the photodiode. 
The shot noise increases with the square root of the laser 
frequency~\cite{Saulson:1994}.

\paragraph*{Radiation pressure}
Each photon will impart twice its momentum on the test masses upon 
reflection. This radiation pressure will vary with the intensity of the 
photons and, although shot noise can be reduced by increasing laser power,
conversely the intensity fluctuations \emph{increase}
with laser power. Hence a trade-off occurs between improvements in shot
noise and the radiation pressure noise. This trade-off, however, is not a
concern for initial detectors where the laser power is not large enough
for the radiation pressure noise to exceed other low frequency noise
sources, such
as seismic and thermal.

\section{Operation of \acs{LIGO}}
The initial operation of the \acf{LIGO}
consisted of three interferometric
detectors at two sites: Hanford, WA and Livingston, LA. Each site 
had a $4\, \rm km$ \ac{IFO}, but there was a second $2\, \rm km$
\ac{IFO} at Hanford co-aligned with the $4\, \rm km$ detector. 
Indeed, the $4\, \rm km$ \ac{IFO}s are still operating as part
of Enhanced \ac{LIGO}~\cite{Smith:2009bx}.
Here we shall consider one of the $4\, \rm km$ detectors.

\begin{figure}
\centering
\psfragfig{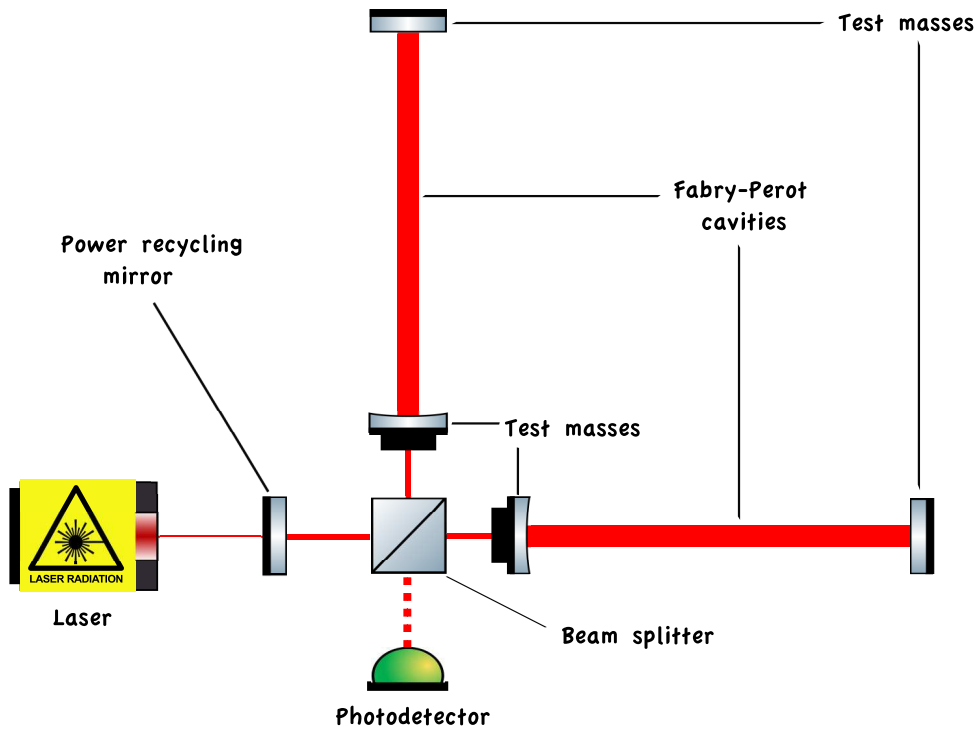}
\caption[Schematic of \acs{LIGO} optical layout (Simplified)]
{A simplified schematic of the \acs{LIGO} optical layout (not to scale). The
Fabry-Perot cavities are $4 \rm km$ long.}
\label{fig:optics}
\end{figure}

\begin{figure}
\centering
\psfragfig{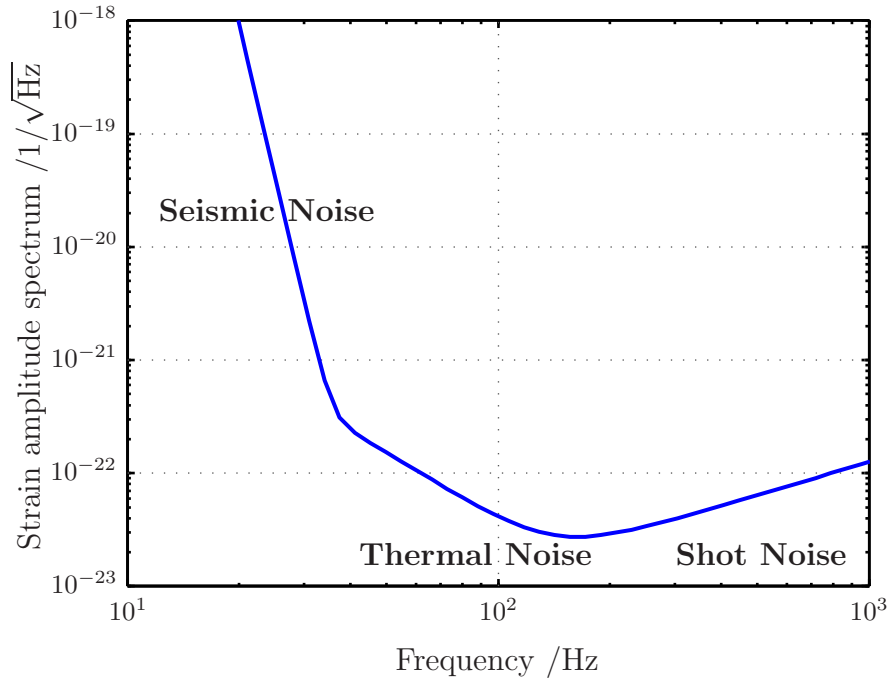}
\caption[\acs{LIGO} design noise budget]
{The design noise spectrum for \ac{LIGO} $4 \rm km$ detectors 
shown as the strain in units of
$1/\sqrt{ \rm Hz}$. 
Seismic noise dominates the lower frequency range, whilst shot
noise dominates the high frequency range. The frequency range of between 
$40-200 \rm Hz$ is dominated by thermal noise.}
\label{fig:designpsd}
\end{figure}

Figure~\ref{fig:optics}\footnote{Figure~\ref{fig:optics}
was produced using svg files
originally created by Alexander Franzen.}
 shows a simplified layout of the \ac{LIGO} 
optics including the Fabry-Perot cavities and power recycling mirror
that were discussed above. 
The Fabry-Perot cavities increase the \ac{LIGO} optical path
length by a factor of approximately $100$. Thus sensitivities of $10^{-22}$
can be achieved, as can be seen in Figure~\ref{fig:designpsd}, 
which shows the design strain amplitude spectrum, i.e., the total noise,
of the \ac{LIGO}
design~\cite{LIGO-E950018-02-E}.

\subsection{Feedback control system - data calibration}
When collecting data, the \ac{LIGO} detector is configured
such that the superposition of the light from each arm gives approximately
null output at the photodiode. So that the
detector can collect data continuously, it is
kept in stable operation by use of a feedback
system. The signal output at the photodiode is returned back into 
the Fabry-Perot cavities as a control strain that maintains 
the null superposition. When the feedback control system
is operating correctly the detector is said to be
in `lock'.
Use of the feedback control means that the output of the 
\ac{LIGO} detector is not a gravitational wave strain, $h(f)$, but an 
error signal, $q(f)$, from which the gravitational
wave strain is obtained using the calibration equation
\be
h(f) = R(f)q(f)\, ,
\ee
where $R(f)$ depends upon various 
quantities, e.g., the recorded strain, the control strain, feedback gains
etc.~\cite{Siemens:2004pr}.

\section{Concluding remarks}
We have covered the basics behind the theory of gravitational waves and 
their detection. In the following chapters we will learn in more detail
the nature of gravitational waves emitted by \acp{CBC} and how a search
for gravitational waves using LIGO data is performed. 
During its fifth science run, 
LIGO collected data of unprecedented sensitivity and
bandwidth. The results of a search for gravitational
waves from low mass \acp{CBC} in a subset of
\ac{S5} data are presented in Chapter~\ref{howchap}.


\chapter{Gravitational waves radiated from binary systems} 
\label{cbc}
\lhead{Chapter~\ref{cbc}
\emph{Gravitational waves radiated from binary systems}}
\rule{15.7cm}{0.05cm}
In this chapter we study the nature of gravitational waves
radiated by \acfp{CBC}, i.e., 
binary systems consisting of neutron stars or black holes that
lose energy via gravitational wave emission, until the objects 
eventually merge.
We consider \emph{compact}
objects, rather than say main-sequence stars, as they can be treated as point
particles. Specifically, they need to be compact enough so that their
surfaces are not touching when their orbital frequency is in the range
of interest for detection.

We begin by finding a general
solution to the linearised \ac{EFE} before proceeding to
the dynamics of a binary system and discussing the \acf{PN} formalism 
used to characterise the waveforms emitted by such objects. 

N.B.: in this chapter we will closely follow the derivations of 
Maggiore~\cite{maggiore}.

\section{The general solution to the linearised \acs{EFE}}
The linearised \ac{EFE} can be solved by the method of Green's function,
where the solution will depend upon the appropriate choice of boundary 
conditions. We recall the 
Lorentz gauge condition (\ref{eq:gauge1}), that energy
and momentum are conserved and choose the boundary condition that there is
no-incoming radiation, i.e., the system that we are studying is isolated
from all other bodies in the Universe. Under such conditions we use
the \emph{retarded} Green's function to solve (\ref{eq:lefe}). 
Since we are interested in the solutions at a distance 
$r\sim\infty$, i.e., in the
far zone where the weak field equations are valid, we can write the
solution as it is when transformed into the $TT$ gauge via the
Lambda tensor, $\Lambda_{ij,kl}(\hat{\textup{\textbf{n}}})$, 
(see Appendix~\ref{sec:lambda}), giving
\be
\label{eq:efesol}
h^{TT}_{ij}(t,\textup{\textbf{x}})
  = \Lambda_{ij,kl}(\hat{\textup{\textbf{n}}})
    \frac{\kappa}{4\pi}
    \int 
      \frac{d^3x^\prime}
           {\left|\textup{\textbf{x}}-\textup{\textbf{x}}^\prime\right|} 
  T^{kl}\left(t - \left|\textup{\textbf{x}}-\textup{\textbf{x}}^\prime\right|,
              \textup{\textbf{x}}^\prime\right)\, ,
\ee
where $\hat{\textup{\textbf{n}}}$ is the unit vector in the direction
to the observer from the source,
the primed coordinates represent that of the source and the unprimed
coordinates are of the observer in the far zone.
N.B.: in the $TT$ gauge, $h^{TT}_{0\mu}=0$ and therefore we
only need to use the spatial indices.

\subsection{Low-velocity expansion}
\label{sec:lvexp}
Let us consider a system whose motion, induced by gravity, consists
of non-relativistic velocities, 
$v\ll1$.
The frequency of the emitted gravitational waves, $\omega$,
will be of the same order as the frequency of the source,
$\omega_s$, which is proportional to $v$,
\be
\omega \sim \omega_s \sim \dfrac{v}{a}\, ,
\ee
where $a$ is the size of the source. In this low-velocity limit
we note that the wavelength of the
emitted gravitational waves will be much longer than
$a$. When we consider solutions to 
(\ref{eq:efesol}) at distances $D\gg a$, we may expand
\be
\left|\textup{\textbf{x}}-\textup{\textbf{x}}^\prime\right|
   = D - \textup{\textbf{x}}^\prime\cdot\hat{\textup{\textbf{n}}} 
   + \ldots\, ,
\ee
but keep only the leading term in the denominator. Hence at large distances
(\ref{eq:efesol}) is simplified to
\be
\label{eq:efesol2}
h^{TT}_{ij}(t,\textup{\textbf{x}}) 
  = \Lambda_{ij,kl}(\hat{\textup{\textbf{n}}})
  \frac{\kappa}{4\pi D}\int d^3x^\prime 
  \ T^{kl}\left(t - D + \textup{\textbf{x}}^\prime
    \cdot\hat{\textup{\textbf{n}}},\textup{\textbf{x}}^\prime\right)\, ,
\ee
As $\textup{\textbf{x}}^\prime\cdot\hat{\textup{\textbf{n}}}\ll D$,
we can Taylor expand (\ref{eq:efesol2}),
\be
\label{eq:lvexp}
h^{TT}_{ij}(t,\textup{\textbf{x}}) = 
  \Lambda_{ij,kl}(\hat{\textup{\textbf{n}}})
  \frac{\kappa}{4\pi D}
  \times
  \left[
    S^{kl} + n_m\dot{S}^{kl,m} +
    \frac{1}{2}n_mn_p\ddot{S}^{kl,mp} + \ldots
  \right]\, ,
\ee
where $S^{kl}$ are the moments of $T^{ij}$
and are related
to the moments, $M$, of the energy density, $T^{00}$, as 
\be
\label{eq:sij}
S^{ij} = \frac{1}{2}\ddot{M}^{ij}\, ,
\ee
(see Appendix~\ref{sec:momident}).
The metric perturbation may also be expressed
as a multipole expansion, in which case $S^{kl}$ is proportional to
the second time 
derivative of the \emph{quadrupole} moment, which we define as
\begin{subequations}\begin{align}
Q^{ij} & = M^{ij} - \frac{1}{3}\delta^{ij}M_{kk} \\
& = \int d^3x 
  \rho(t,\textup{\textbf{x}})\left(x^ix^j 
                                      - \frac{1}{3}r^2\delta^{ij}\right)\, ,
\end{align}\end{subequations}
where $\rho=T^{00}$, which in the low-velocity expansion is dominated
by the rest mass of the binary.  
It is interesting to note that as the quadrupole moment is the 
leading order term there exists no monopole or dipole 
gravitational radiation.

The moments of the energy 
density and the linear moments
are discussed in more detail in Appendix~\ref{AppendixB}, and will be used in
Section~\ref{sec:howf}. N.B.: in the ${TT}$ gauge $Q^{ij} = M^{ij}$.

\subsection{Quadrupole radiation}
\label{sec:quadrad}
Physically the absence of monopole and dipole gravitational radiation
are typically 
understood as the conservation of energy and angular momentum 
respectively, which is the correct explanation in linearised theory
(see Appendix~\ref{sec:momident}), but
is not true in general. Indeed it is clear that if we wish to detect
gravitational waves we require energy to be emitted so that it can
cause tidal forces to be imposed upon our detector. However, it is generally 
true that monopole and dipole gravitational radiation do not exist. 
The correct explanation, given in, 
e.g., \cite{maggiore}, is that the graviton has helicity $\pm2$ and therefore
cannot have a total angular momentum
of $0$ or $1$ that would correspond to the monopole and dipole,
respectively.

We can now understand the nature of sources of gravitational waves.
The gravitational quadrupole is a measure of the distribution of mass that is
non-zero for an asymmetric system. Additionally, for
radiation to be emitted, the quadrupole moment of the system must have
a non-zero second time derivative, i.e., it must be accelerating. 
Some types of astrophysical sources that are expected to emit such gravitational
radiation were briefly described in Section~\ref{sec:sources}, including
\acp{CBC}.

\subsection{Calculating the polarisations}
The contraction of the quadrupole moment with the lambda tensor yields
the quadrupole gravitational radiation as
\be
\label{eq:hquad}
h^{TT}_{ij} = \frac{\kappa}{2\pi D}\ddot{M}^{TT}_{ij}\, ,
\ee
where
\be
\ddot{M}^{TT}_{ij} = \ddot{Q}^{TT}_{ij} = \Lambda_{ij,kl}\ddot{Q}_{kl}\, .
\ee

However, we would like to relate (\ref{eq:hquad}) to the $+$ and $\times$
polarisations observed by a detector. It can be shown
(e.g.~\cite{maggiore}) that when the propagation direction
$\hat{\textup{\textbf{n}}}$ is in the $z$ direction, the polarisations are
simply
\begin{subequations}
\label{eq:hquad2}
\begin{align}
h_+ & = \frac{\kappa}{2\pi D}\left(\ddot{M}_{11} - \ddot{M}_{22}\right)\, ,\\
h_\times & = \frac{\kappa}{4\pi D} \ddot{M}_{12}\, .
\end{align}\end{subequations}
The general solution for an observer in any direction
depends upon all six moments, $M_{ij}$, and 
two angles, $i$ and $\phi$, that relate the source
frame to the propagation frame. The former is the 
inclination angle between the $z$-axis of the source
frame and the direction of propagation. 
The latter is the simply the phase offset, i.e., 
the angle of rotation of the binary with respect to the $y$-axis.

\section{A binary system}
Let us now turn our attention to the gravitational wave polarisations
emitted from a binary system. We assume the binary consists of compact objects
of mass $m_1$ and $m_2$, that are moving in a circular orbit
with a separation distance $a$
 in the $x$-$y$
plane. 
We model the evolution and gravitational wave emission of the binary assuming
adiabatic circular motion using Newtonian orbital mechanics and the
lowest order \ac{PN} corrections that give the 
energy loss due to the gravitational radiation. Higher order
corrections will be introduced in Section~\ref{sec:howf}.

Switching to the centre-of-mass frame the binary may be represented
by a single body of reduced mass
\be
\mu = \frac{m_1m_2}{m_1+m_2}\, ,
\ee
that moves in an effective potential and 
whose evolution is described with the following relative coordinates
\begin{subequations}\begin{align}
x_0(t) & = a\cos\left(\omega_s t\right)\, ,\\
y_0(t) & = a\sin\left(\omega_s t\right)\, ,\\
z_0(t) & = 0\, ,
\end{align}\end{subequations}
that give its position relative to the centre of mass of the two bodies
(see Figure~\ref{fig:eob}). This single body approach allows us to obtain
simple expressions for the mass moments (see Appendix~\ref{sec:eobsimp}),
namely,
\be
\label{eq:mquad}
M_{ij}(t) = \mu x_{i0}(t)x_{j0}(t)\, .
\ee
We therefore find
\begin{subequations}
\label{eq:moments}
\begin{align}
M_{11} & = \mu a^2 \frac{1 + \cos\left(2\omega_s t\right)}{2}\, ,\\
M_{22} & = \mu a^2 \frac{1 - \cos\left(2\omega_s t\right)}{2}\, ,\\
M_{12} & = \mu a^2 \frac{\sin\left(2\omega_s t\right)}{2}\, .
\end{align}
\end{subequations}
We can see from (\ref{eq:moments}) that the frequency of the 
gravitational waves emitted from a binary system are
\emph{twice} the orbital frequency. In qualitative
terms this can be understood by the symmetry of the system; if the objects
are of equal mass then the binary has the same configuration twice
every orbit. 
\begin{figure}
\hspace{-0.3cm}
\begin{minipage}[h]{7.5cm}
\centering
\psfragfig{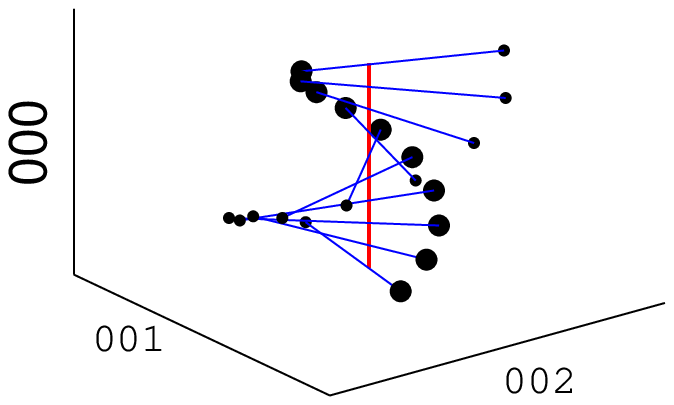}
\end{minipage}
\hspace{0.15cm}
\begin{minipage}[h]{7.5cm}
\centering
\psfragfig{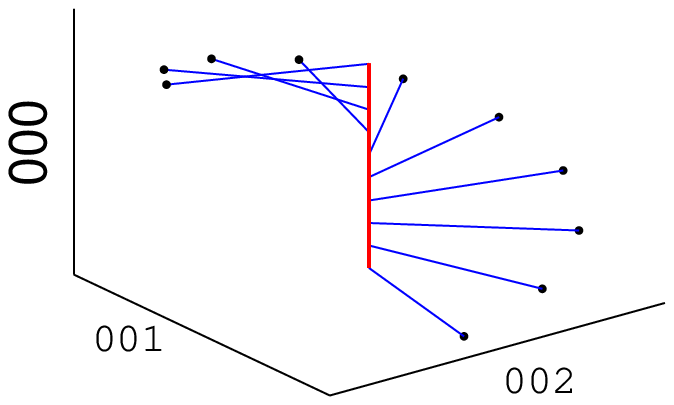}
\end{minipage}
\caption[Cartoon of the centre-of-mass binary dynamics]
{The left plot 
shows a binary system of component masses $m_1=2m_2$ orbiting their
centre of mass, indicated by the vertical red line. The centre-of-mass,
single-body
representation of the same system is shown on the right. The single body
is orbiting the same point as on the left with a reduced mass, $\mu$,
attracted to a `ghost' mass $m_1+m_2$.}
\label{fig:eob}
\end{figure}

Finally by calculating the second time
derivatives of (\ref{eq:moments}) and using the general solution for
the polarisations as opposed
to (\ref{eq:hquad2}), as shown in, e.g, \cite{maggiore}, we find the observed
gravitational wave polarisations to be
\begin{subequations}\label{eq:hquad3}\begin{align}
\label{eq:hquad3p}
h_+ & = \frac{\kappa}{4\pi D}a^2\mu\omega_s^2\frac{\left(1+\cos^2i\right)}{2}
        \cos2\omega_st\, ,\\
h_\times & = \frac{\kappa}{4\pi D}a^2\mu\omega_s^2 \cos i\sin2\omega_st\, .
\end{align}\end{subequations}
In this case the general solution does not depend upon the
angle $\phi$, which is a rotation around the $z$-axis, equivalent to a
time-shift and can instead be represented by choosing a value 
$t^\prime = t + t_0$. As stated above, the inclination angle, $i$, is that
between the $z$ axis of the source and the direction of propagation towards
the observer. Hence if $i=0$,
we see the binary `face-on' and both 
polarisations are of equal amplitude, in which case the gravitational waves
are said to be circularly polarised. However, if the binary is `edge-on', 
$i=\pi/2$, then the gravitational waves are linearly polarised and only consist
of the $+$ polarisation. 
This can be qualitatively understood from
Figure~\ref{fig:polars}. Observing an edge-on binary, and recalling that
gravitational waves are transverse we would only need one dimension to
describe the motion of the binary. On the other hand a face-on binary
requires two dimensions to describe its motion. N.B.: for inclination
angles between $0$ and $\pi/2$, there will be unequal contributions from
the $+$ and $\times$ polarisations; such gravitational waves are said to be
elliptically polarised.

One further point of significance, seen in (\ref{eq:hquad3}), is 
that the gravitational wave amplitude depends upon the frequency 
and amplitude squared, which is the same order as the square of the 
source velocities, i.e., $v^2\sim a^2\omega_s^2$.

\subsection{Energy emission}
We expect a priori the emission of gravitational waves to take energy away
from binary systems. The loss of energy causes the orbital separation to
decrease and the bodies \emph{inspiral} towards each
other\footnote{Indeed, this 
very process has been observed~\cite{Hulse:1994}
(see section~\ref{sec:hulse}).}.

The energy carried by gravitational waves is found by calculating the 
energy-momentum tensor due to the gravitational wave
itself, from which the gravitational wave flux in a
given direction can be found. Integrating the flux over a sphere gives the
total luminosity, $\mathscr{L}$, of the emitted gravitational waves.
This results in the energy balance equation
\be
\label{eq:ebal}
\frac{dE}{dt}  
  = -\mathscr{L} 
  = -\frac{1}{5} \left<\dddot{Q}^{TT}_{ij} \dddot{Q}^{TTij}\right>\, ,
\ee
where $E$ is the total energy of the binary and 
the brackets indicate that we are averaging over several
wavelengths\footnote{A detailed derivation is given in~\cite{maggiore},
a more accessible derivation can be found in~\cite{hobson}.}.
Thus to calculate the energy loss of a binary we take the third time 
derivatives
of (\ref{eq:moments})
\begin{subequations}
\label{eq:dmoments}
\begin{align}
\dddot{M}_{11} & = - 4\mu a^2 \omega_s^3 \sin\left(2\omega_s t\right)\, ,\\
\dddot{M}_{22} & = - \dddot{M}_{11}\, ,\\
\dddot{M}_{12} & = 4\mu a^2 \omega_s^3 \cos\left(2\omega_s t\right)\, .
\end{align}
\end{subequations}
It can be shown that the time dependent parts average
out and the energy loss is 
\be
\label{eq:dedt}
\frac{dE}{dt} = - \frac{32}{5} \mu^2 a^4 \omega_s^6\, .
\ee

\subsection{Evolution of the binary}
Under the assumption that the orbit is adiabatic we can use Kepler's
equations to understand the dynamics of the source. Kepler's third law states
that the square of the orbital period is proportional to the cube of the
semi-major axis, which gives the relation between the frequency of the 
source and the separation
\be
\label{eq:kepler3}
\omega^2_s = \frac{M}{a^3}\, ,
\ee
where $M=m_1+m_2$ is the total mass.
This simple relation shows us that
as the binary inspirals the orbital frequency increases.
We can, therefore, conclude
that the gravitational wave frequency and amplitude increase as the system 
evolves.
We can then determine the evolution of the binary system by substituting
(\ref{eq:kepler3}) into (\ref{eq:dedt}),
giving
\be
\frac{dE}{dt} = - \frac{32}{5} \frac{\mu^2 M^3}{a^5}\, .
\ee
The total energy of a binary system in the Newtonian limit is simply
\be
E = \frac{1}{2}\mu v^2 - \frac{\mu M}{a} = -\frac{\mu M}{2a}\, ,
\ee
from which we can obtain $dE/da$ and subsequently
\be
\frac{da}{dt} = - \frac{64}{5} \frac{\mu M^2}{a^3}\, ,
\ee
which we integrate to find the evolution of the binary separation
\be
\label{eq:aoft}
a(t) = 
  {\left(\frac{256}{5} \mu M^2\right)}^\frac{1}{4}
  {\left(t_c - t\right)}^\frac{1}{4}\, ,
\ee
where $t_c$ is the coalescence time ($a=0$). The evolution of the orbital
frequency is found simply by substituting (\ref{eq:kepler3}) into
(\ref{eq:aoft}) which yields
\be
\omega_s(t) =
  {\left(\frac{256}{5}\right)}^{-\frac{3}{8}} 
  {\cal M}^{-\frac{5}{8}}
  {\left(t_c - t\right)}^{-\frac{3}{8}}\, .
\ee
where we define the \emph{chirp mass}
\be
\label{eq:chirp}
{\cal M} = \eta^\frac{3}{5}M\, ,
\ee
and the \emph{symmetric mass ratio}
\be
\eta = \frac{\mu}{M}\, .
\ee
Finally we define the orbital phase of the binary
\begin{subequations}
\begin{align}
\varphi(t) &= \int \omega_s(t) dt\, ,\\
\varphi(t) &=
  -\frac{8}{5}{\left(\frac{256}{5}\right)}^{-\frac{3}{8}} 
  {\cal M}^{-\frac{5}{8}}
  {\left(t_c - t\right)}^{\frac{5}{8}}\, .
\end{align}
\end{subequations}

We now have all that is required to understand the evolution of the 
gravitational waves radiated from a binary system. It is useful to express
the polarisations in terms of their amplitude and phase
evolution: 
\begin{subequations}
\begin{align}
\label{eq:atcosphi}
h_+(t) & = A(t) \frac{\left(1+\cos^2i\right)}{2}\cos( 2\varphi(t) ) \\
h_\times(t) & = A(t)\cos i\sin( 2\varphi(t) )\, ,
\end{align}
\end{subequations}
where 
\be
\label{eq:amplitude}
A(t) = \frac{\kappa\mu M}{4\pi Da(t)}\, .
\ee
N.B.: the polarisations are $\pi/2$ out of phase and hence they
may also be referred to as the two ``phases'' of the
gravitational wave.

\begin{figure}
\hspace{-0.3cm}
\begin{minipage}[h]{7.5cm}
\centering
\psfragfig{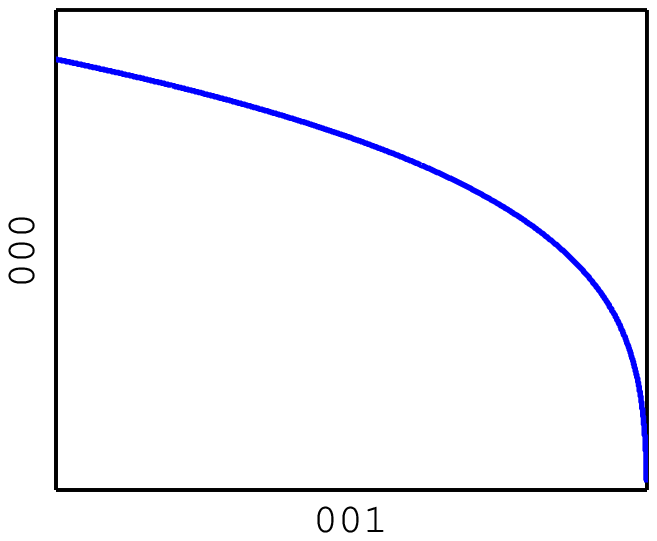}
\end{minipage}
\hspace{0.15cm}
\begin{minipage}[h]{7.5cm}
\centering
\psfragfig{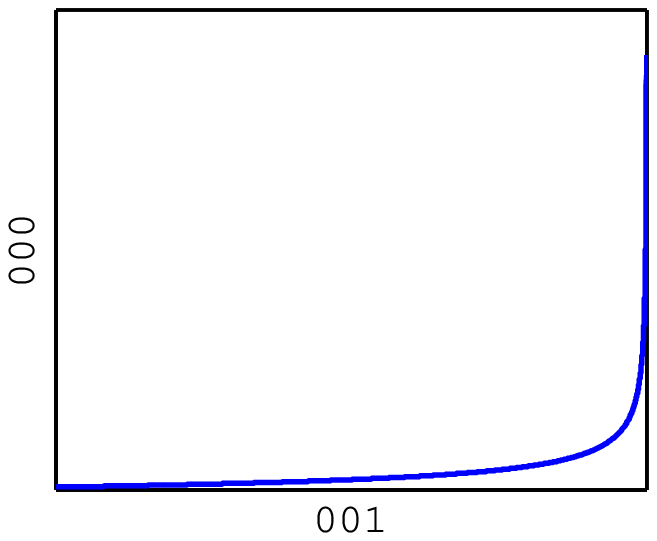}
\end{minipage}
\vspace{0.5cm}
\hspace{-0.3cm}
\begin{minipage}[h]{7.5cm}
\centering
\psfragfig{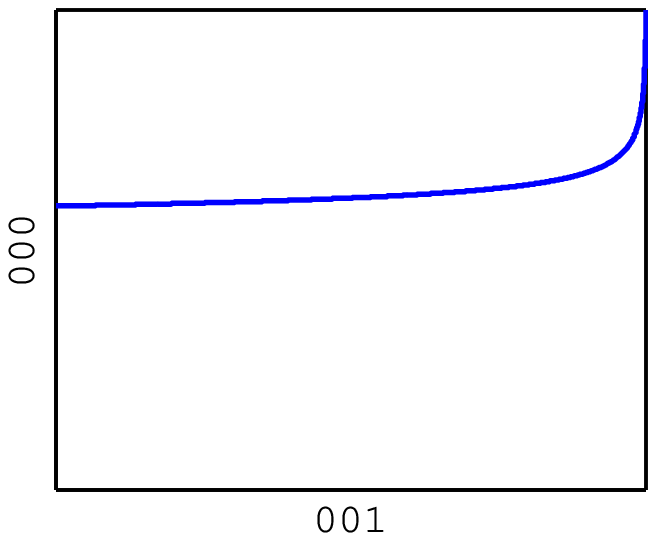}
\end{minipage}
\hspace{0.15cm}
\begin{minipage}[h]{7.5cm}
\centering
\psfragfig{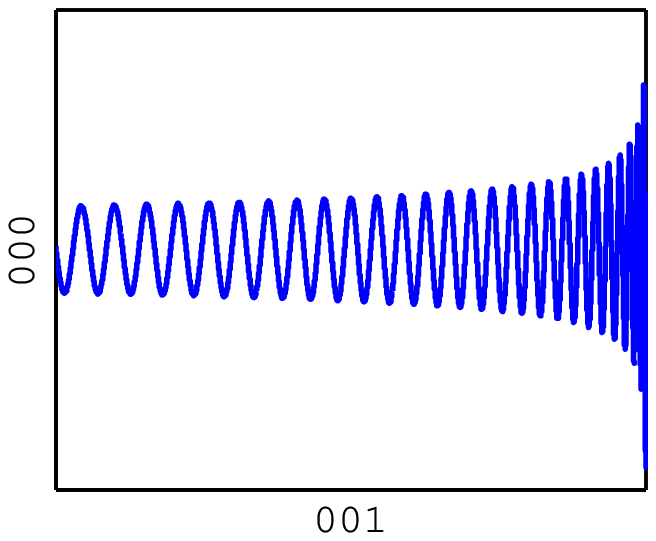}
\end{minipage}
\caption[Evolution of a binary system and the emitted gravitational waves]
{Evolution of the orbital separation (top left), the orbital frequency (top
right), the gravitational wave amplitude (bottom left) and the $+$ polarisation
(bottom right) for a given system shown in arbitrary units.}
\label{fig:evolve}
\end{figure}

\begin{figure}
\centering
\psfragfig{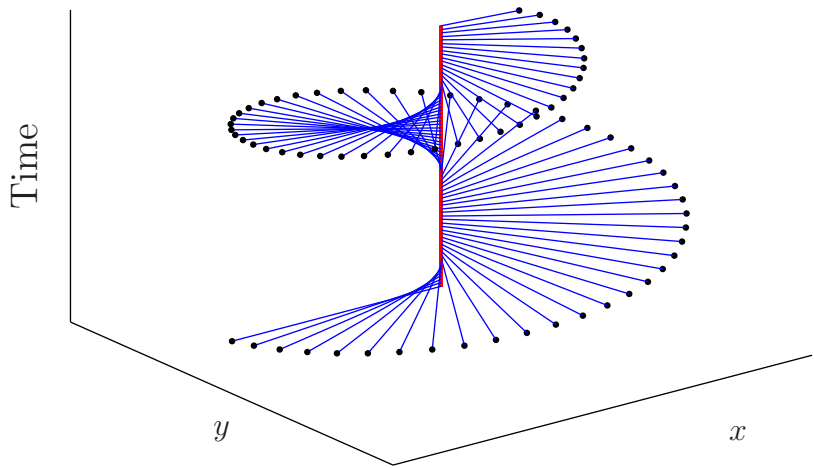}
\caption[Cartoon of a single-body inspiral]
{Evolution of the
centre-of-mass, single-body 
representation of a binary inspiral with a single-body,
sampled at fixed time intervals.}
\label{fig:eobplunge}
\end{figure}

Figure~\ref{fig:evolve} shows qualitatively the evolution of the orbital 
separation, the source frequency, 
the amplitude of the gravitational wave and its
$+$ polarisation. The amplitude and frequency increase as the waveform
evolves, giving it a `chirp` characteristic that depends upon the
chirp mass (\ref{eq:chirp}).
Figure~\ref{fig:eobplunge}
shows a cartoon evolution of the single-body representation in centre-of-mass 
frame.

\subsection{Inspiral waveforms}
The evolution of the binary has thus far been derived
assuming adiabatic circular motion, which is only valid until the binary
reaches the \ac{ISCO}. This period in the binary evolution is known as
the `inspiral' stage  after which the above equations cannot be used to
describe the system.
The \ac{ISCO} of the Schwarschild metric occurs at a
distance of three times the Schwarschild
radius ($6M$). Therefore, inspiral waveforms are usually evolved until the
separation reaches that value.
Inspiral waveform models are often evolved via the 
gravitational wave frequency and are terminated at the corresponding
\ac{FLSO}, which is easily calculated from (\ref{eq:kepler3}).  

As the binary approaches merger, the two objects begin to plunge towards
each other before forming a single black 
hole\footnote{Recall that we are considering compact objects where the
minimum mass system would consist of two neutron stars.} 
that settles into equilibrium by emitting
gravitational waves in what is known as the `ringdown' 
stage~\cite{Leaver:1985ax,Brandt:1994ee}.
Inspiral-merger-ringdown \acused{IMR}(\ac{IMR}) waveforms that include
the merger and ringdown phase 
can be calculated by matching the inspiral stage to the merger and
ringdown stages computed using numerical relativity
simulations of merger dynamics~\cite{Buonanno:2007pf,
Damour:2007yf}. In this thesis, we will consider inspiral-only waveforms and
not \ac{IMR} waveforms, with the exception of Chapter~\ref{winchap}.

\section{Why are gravitational waves from CBCs of interest?}
\label{sec:interest}
In Chapter~\ref{Intro} we learned that ground-based interferometric 
detectors can reach sensitivities of $10^{-22}$ in the frequency range
of around $100$-$1000\,\rm Hz$ (see Figure~\ref{fig:designpsd}). 
We now have everything we need to estimate the amplitude and frequency
of gravitational waves radiated by \acsp{CBC}.
Let us consider a \ac{NSBH}, of component masses $1.4\Msun$ and $10\Msun$,
close to its \ac{ISCO} ($a=8M$), observed
at a distance of $100\,\rm Mpc$ (The Virgo 
galaxy cluster is a mere $18\,\rm Mpc$ from the Milky Way). The amplitude is
given by (\ref{eq:amplitude}), which, after substituting
$a=8M$, we multiply by $G/c^2$ 
to convert from natural units to standard units giving
\be
h \sim 10^{-21}\, .
\ee
We then use (\ref{eq:kepler3}) to determine the frequency. Recalling that
the gravitational wave frequency is twice that of the source, we find
\be
f \sim 250\,\rm Hz\, ,
\ee
Thus the gravitational waves are of the required frequency and amplitude 
to be observable by LIGO!

As the binary approaches the merger stage its frequency 
sweeps across LIGO's sensitive band, reaching its \ac{FLSO} of 
$f\sim400\,\rm Hz$.
Binary neutron star (\acs{BNS})
\acused{BNS}
systems are lower in mass and their frequency evolution sweeps 
across the entire sensitive band with an \ac{FLSO} of $f\sim2000\,\rm Hz$,
whereas \ac{BBH} systems reach their \ac{FLSO}\footnote{
The \ac{FLSO} is explicitly defined in Section~\ref{sec:massreach}.}
in the most sensitive part
of the detector's band - at about $220\,\rm Hz$ for a $(10,10)\Msun$ binary. 
Thus all of these systems are ideal detection
candidates for LIGO.

\section{Higher order waveforms}
\label{sec:howf}
When calculating the gravitational wave polarisations we 
chose\footnote{It was not really a choice as we were working in linearised
theory.}
to keep only
the leading order term in the expansion of $T_{kl}$. We will begin our 
journey into the use of higher order 
waveforms by considering the polarisations in 
linearised theory that include the \ac{FBLO} term. However,
as we shall, see that is merely the tip of the iceberg.

\subsection{First beyond leading order linearised polarisations}
It can be shown, e.g.,~\cite{maggiore}, using (\ref{eq:momident}),
(\ref{eq:momident2}) and (\ref{eq:momident3}), that the
second term in the expansion of $T_{kl}$ (\ref{eq:lvexp}) is
\be
\label{eq:foblo}
\dot{S}^{ij,k} = \frac{1}{6}\dddot{M}^{ijk} 
  + \frac{1}{3}\left(\ddot{P}^{i,jk} + \ddot{P}^{j,ik} - 2\ddot{P}^{k,ij}
  \right)\, . 
\ee
Typically the two terms on the \acs{RHS} of (\ref{eq:foblo}) are
separated into the
moments of the energy density and the momentum density (see
Section~\ref{sec:mos}),
respectively, where the former corresponds to the mass 
octopole moment and the latter the
current quadrupole 
moment\footnote{The quadrupole of the angular momentum density.}.
However,
in the approximation for non-relativistic (low-velocity)
particles it is straightforward to compute $\dot{S}^{ij,k}$ directly. From 
(\ref{eq:momidentS2}) we can write
\be
S^{kl,m} = \mu\dot{x}^k\dot{x}^lx^m\, ,
\ee
and its time derivative
\be
\label{eq:dotsklm}
\dot{S}^{kl,m} = \mu\left[\left(\ddot{x}^k\dot{x}^l + \dot{x}^k\ddot{x}^l
                          \right)x^m + \dot{x}^k\dot{x}^l\dot{x}^m\right]\, .
\ee
We can already note two interesting things about the mass octopole and
current quadrupole
gravitational radiation. Firstly, the \ac{FBLO} term in (\ref{eq:lvexp})
depends upon the direction to the observer from the source, specifically
$n^m$, such that if the observer is orthogonal to the orbital plane then
$\hat{n}\cdot\textup{\textbf{x}}=0$ and, hence, the FBLO term disappears.
Thus we see that there must be motion of the binary components in 
the direction
of the observer (i.e., the inclination angle must be non-zero), 
or there will be no gravitational radiation of this order towards the
observer.
Secondly, we can see that both terms in (\ref{eq:dotsklm}) will have
a factor ${(a\omega_s)}^3$ in the amplitude, an extra factor of $a\omega_s$
compared with the leading order term. 
Recall that this is the same order as the of the velocities of the source
that are small compared to unity.
Therefore, an increase in the order of velocity 
leads to a smaller amplitude of the radiation.

For a binary system with an inclination angle, $i$,
between the $z$-axis of the
source coordinates and the rotational axis
 of the binary, the equations of motion are
\begin{align}
x_0(t) & = a\cos\left(\omega_s t\right)\, ,\\
y_0(t) & = a\cos{i}\sin\left(\omega_s t\right)\, ,\\
z_0(t) & = a\sin{i}\sin\left(\omega_s t\right)\, .
\end{align}
For a gravitational wave propagating along $z$, the polarisations are found
by calculating the FBLO term in (\ref{eq:lvexp}),
\be
{\left( h^{TT}_{ij}\right)}_\text{FBLO}
  = \dfrac{\kappa}{4\pi D}\dot{S}^{TT}_{ij,3}\, ,
\ee
which gives
\begin{subequations}
\label{eq:fbloterms}
\begin{align}
{\left( h^{TT}_+\right)}_\text{FBLO}
  & = \dfrac{\kappa}{4\pi D}\dfrac{1}{8} \mu a^3 \omega_s^3 \sin i
    \left[\left(\cos^2i-3\right)\cos\left(\omega_st\right)
    - 3\left(1+\cos^2i\right)\cos\left(3\omega_st\right)\right]\, ,\\
{\left( h^{TT}_\times\right)}_\text{FBLO}
  & = \dfrac{\kappa}{4\pi D}\dfrac{1}{8} \mu a^3 \omega_s^3 \sin\left(2i\right)
    \left[\sin\left(\omega_st\right)
    - 3\sin\left(3\omega_st\right)\right]\, .
\end{align}
\end{subequations}
Interestingly the FBLO gravitational radiation introduces a
first and third harmonic of the orbital frequency.


\section{Post-Newtonian formalism}
Thus far we have described the nature of gravitational waves, in particular
those radiated from \acp{CBC}, using linearised theory. The leading
order term in the gravitational radiation corresponded to the
mass quadrupole moment
and higher order terms could be calculated as required via the
Taylor expansion (\ref{eq:lvexp}). 
However, the gravity of the source itself and the 
effects of energy-momentum emission on the orbital dynamics, 
which produce corrections to the leading order term, were not taken into
account.
Hence, in linearised theory, without these corrections,
we cannot \emph{correctly} calculate the terms
beyond leading order, i.e., $\mathcal{O}(v^3)$, 
including (\ref{eq:fbloterms}). 

The \acf{PN} formalism is an iterative, perturbative approach to 
solving the \ac{EFE}, that gives an expansion in terms of $(v^2/c^2)$.
Hence for the rest of this chapter we 
shall drop the natural units to keep to the tradition of the PN formalism.
PN theory can be used to provide highly accurate waveform models 
of the \emph{expected} 
gravitational radiation emitted by CBCs. In gravitational wave
data analysis it is very important~\cite{Cutler:1992tc,Cutler:1994}
to have accurate models of the phase
evolution when using the matched filter (see Chapter~\ref{howchap}).

The \ac{PN} expansion for binary systems is typically used to
calculate the energy of the binary and the luminosity, both to 
high order, e.g., $(v/c)^7$~\cite{Blanchet:2001ax}.
The phase evolution of the binary
may then be constructed by a variety of different methods using the 
energy balance equation.

\subsection{Basic overview}
The PN formalism is a complex subject; here a very basic overview
of the process is given (see, e.g.,~\cite{maggiore}).
\begin{itemize}
  \item
    The spacetime metric is written again as flat spacetime plus a
    perturbation $\textup{h}_{\mu\nu}$, where $\textup{h}_{\mu\nu}$ may contain
    non-linear terms, i.e.,
\be
\textup{h}_{\mu\nu} = \Sigma_{n=1}^\infty G_n\textup{h}^n_{\mu\nu}\, .
\ee
  \item
    In the Lorentz gauge, the EFE are written as
    \be
      \label{eq:relax}
      \square\textup{h}_{\mu\nu} = +\dfrac{16\pi G}{c^2} \tau_{\mu\nu}\, ,
    \ee
    where $\tau_{\mu\nu}$ consists of the energy-momentum tensor
    \emph{and} highly non-linear terms of the perturbation.
  \item
    As before, in linearised theory, (\ref{eq:relax}) can be integrated using
    the retarded
    Green's function. However, the result has the perturbation on both sides
    of the equation, for which an analytical solution cannot be found.
  \item
    Outside the source the energy-momentum tensor is zero. Writing 
    $\textup{h}_{\mu\nu}$ as an expansion in powers of $G$
    one can match terms of
    the same order on the \acs{LHS} and \acs{RHS}. The process is iterative:
    first $\textup{h}_{\mu\nu}$
    is found to order $G$, and recycled in to the solution to find
    the term of order $G^2$ etc.
  \item
    The solution outside the source is then written as a multipole expansion
    that depends upon two sets of moments, $I$ and $J$, which are unknown.
  \item
    To determine $I$ and $J$, one must use the above iterative process inside
    the source. In this case $\textup{h}_{\mu\nu}$ and $\tau_{\mu\nu}$ are
    expanded in terms of $(1/c)$. As before terms of the same order are matched
    in an iterative process.
  \item
    By re-expanding the solution outside the source in powers
    of $(v/c)$ the moments $I$ and $J$ can be matched with the solution inside
    the source, which then yields the gravitational wave polarisations.
\end{itemize}

\subsection{PN order}
The results from the PN approximation differ from linearised theory, but 
share the same characteristics. 
The leading order term in the
amplitude is of order $v^2/c^2$ and the frequency of the gravitational
wave is twice the orbital frequency. The next term 
introduces a first and third harmonic and its amplitude is of order
$v^3/c^3$. 

The leading order term is denoted 0PN in order, whereas the
FBLO term is 0.5PN in order. The next highest term
has an amplitude of order $v^4/c^4$ and is denoted 1PN in order, etc.
The PN notation is used to described other quantities, e.g., flux,
acceleration, etc., where the leading, 0PN, term is of a general order 
${(v/c)^n}$. The FBLO, 0.5PN, term is then of order ${(v/c)}^{n+1}$, etc.

\section{PN phase approximants}
Once the PN expressions for the binary's energy and luminosity are determined,
the gravitational wave phase may be calculated using the energy-balance 
equation (\ref{eq:ebal}). Defining the flux as
\be
F(v) = -M\dfrac{dE(v)}{dt}\, ,
\ee
the energy balance equation may be written as
\be
\label{eq:ebal2}
- \dfrac{dE}{dt} = - \dfrac{dE}{dv}\dfrac{dv}{dt} = \dfrac{F}{M}\, .
\ee
Using Kepler's laws we see that the velocity of the source, $v$, is
related to the source frequency as
\be
\label{eq:vel}
v = (M\omega_s)^{\frac{1}{3}}\, .
\ee
Therefore, (\ref{eq:ebal}) and (\ref{eq:vel}) 
lead to two non-linear, ordinary differential equations
\be
\label{eq:ode1}
\dfrac{dv}{dt} = - \dfrac{1}{M}\dfrac{F}{dE/dv}\, ,
\ee
and
\be
\label{eq:ode2}
\dfrac{d\varphi}{dt} = \dfrac{v^3}{M}\, .
\ee
The flux, $F(v)$, is calculated by the \ac{PN} method~\cite{Blanchet:2004ek}.
Ideally we wish to find $v(t)$, by integrating (\ref{eq:ode1}), and then
$\varphi(t)$ by integrating (\ref{eq:ode2}). However, the \acs{RHS}
of (\ref{eq:ode1}) consists of a fraction where both numerator and
denominator are polynomial functions of $v$. There are three popular
ways in which one can find $\varphi(t)$, known as the Taylor-T1, Taylor-T2 and 
Taylor-T3 approximants~\cite{Damour:2000zb}\footnote{There is also a 
Taylor-T4 approximant among others~\cite{Damour:2000zb}.}. 
The Taylor-T1 approximant is 
found by simply integrating (\ref{eq:ode1})
numerically to find $v(t)$. To find the Taylor-T2 approximant one expresses 
${F}/\left({dE/dv}\right)$ 
as an infnite series in $v$, truncating at the appropriate
order before integrating. Finally, the Taylor-T3 approximant is found by
using the infinite series of the Taylor-T2 approximant and inverting it to
find $v(t)$.

\begin{figure}
  \centering
  \psfragfig{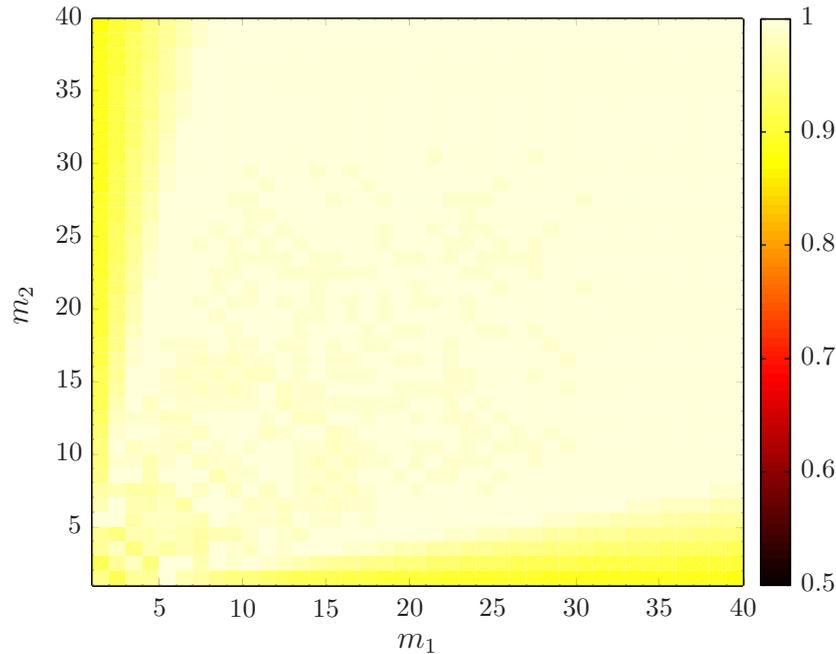}
  \caption[Stability of the TT3 approximant at 2PN]
  {The ratio of the termination frequency to the FLSO for the TT3 approximant
   at 2PN. The ratio
   is typically above 0.9, which is adequate; the frequency is increasing
   dramatically as the binary 
   approaches ISCO. The phase evolution will therefore
   be terminated at a time very close to that at which it would reach FLSO.}
  \label{fig:stable}
\end{figure}

We know from the evolution of
the binary that the phase should be monotonically increasing. 
When generating a waveform model for data analysis, 
the above approximants are considered invalid if the condition
\be
\dfrac{df}{dt} > 0\, ,
\ee
is violated,
at which point the evolution of the waveform should be terminated.
The stability of the each of the Taylor approximants will vary with the
parameters of the waveform and with the PN order at which the phase is
determined. The \acl{TT3} approximant is found to be particularly stable at 
2PN as shown in Figure~\ref{fig:stable}, where the ratio of termination 
frequency to the \ac{FLSO} is plotted for a range of binaries, 
characterised by their component masses.

In later chapters we shall use the \acf{TT3} approximant at 2PN 
(see Appendix~\ref{sec:tt3phase})
and also its FD analog, the \acf{SPA}. 



\section{Restricted and full PN waveforms}
The gravitational wave polarisations have been solved to 
3PN order~\cite{Blanchet:2008je} and the gravitational wave phase has been
solved to 3.5PN order~\cite{Blanchet:2001ax}. The polarisations may be
expressed as
\be
h_{+,\times} = 
  \frac{2G\mu x}{c^2R}\left \{  
    H^{\left (0 \right )}_{+,\times} +
    x^\frac{1}{2}H^{\left (\frac{1}{2} \right )}_{+,\times} +
    xH^{\left (1 \right )}_{+,\times} +
    x^\frac{3}{2}H^{\left (\frac{3}{2} \right )}_{+,\times} +
    x^2H^{\left (2 \right )}_{+,\times} +
    x^\frac{5}{2}H^{\left (\frac{5}{2} \right )}_{+,\times} +
    x^3H^{\left (3 \right )}_{+,\times}
 \right \}\, ,
\ee
where
\be
x=(v^2/c^2)\, ,
\ee
is the PN parameter and the quantities $H$ are the polarisations at each PN
order, e.g.,
\be
H^{(0)}_+ = (1 + \cos^2i)\cos(2\varphi)\, ,
\ee
is the 0PN term for the $+$ polarisation.

To date, gravitational wave searches (e.g.,~\cite{LIGOS3S4Tuning,LIGOS3S4all,
Collaboration:2009tt,Allen:2005fk,Abbott:2009qj})
have used only the leading order amplitude term that consists of the
dominant harmonic at twice the orbital frequency, i.e., the 
$H^{\left (0 \right )}_{+,\times}$ term. However, the phase is used
to a higher order (as it must be).
The waveforms are, therefore, 0PN in amplitude and, say,
2PN in phase. Such waveforms are known as \emph{\acfp{RWF}}. The 
\emph{\acf{FWF}}, on the other hand, retains the higher order amplitude 
terms and contains many interesting features.

We have already seen in
linearised theory that the FBLO term introduces a first and third harmonic
of the orbital frequency. Below we will discuss in more detail the
differences between the \ac{RWF} and the \ac{FWF}, including the higher
order terms that contain other harmonics of the orbital frequency and
\emph{amplitude corrections} to the existing harmonics.

\subsection{Harmonics and amplitude corrections}
Table~\ref{tab:harmonics} shows how the higher order amplitude terms
contribute to the polarisations. As we know, the FBLO term introduces
a first and third harmonic of the orbital frequency.
The 1PN amplitude term consists of a correction to the,
dominant, second harmonic and
a fourth harmonic. Each of the remaining higher order terms contain corrections
to existing harmonics and introduce a new harmonic. 
\begin{table}
\begin{center}
\begin{tabular}{|c|c|c|c|c|c|c|c|c|c|}
\hline
\multicolumn{10}{|c|}
{An arbitrary binary (both polarisations)}\\
\hline
\hline
PN order & $\mathcal{O}$ &
  $\varphi$ & $2\varphi$ & $3\varphi$ & $4\varphi$ & $5\varphi$ & $6\varphi$
  & $7\varphi$ & $8\varphi$ \\
\hline 
\hline 
  0   & $v^2/c^2$ 
      & $\cdot$ & $\bullet$ & $\cdot$ & $\cdot$ & $\cdot$ & $\cdot$ & $\cdot$ 
        & $\cdot$ \\
  0.5 & $v^3/c^3$ 
      & $\bullet$ & $\cdot$ & $\bullet$ & $\cdot$ & $\cdot$ & $\cdot$ & $\cdot$ 
        & $\cdot$ \\
  1   & $v^4/c^4$ 
      & $\cdot$ & $\bullet$ & $\cdot$ & $\bullet$ & $\cdot$ & $\cdot$ & $\cdot$ 
        & $\cdot$ \\
  1.5 & $v^5/c^5$ 
      & $\bullet$ & $\bullet$ & $\bullet$ & $\cdot$ & $\bullet$ & $\cdot$ 
        & $\cdot$ & $\cdot$ \\
  2   & $v^6/c^6$ 
      & $\bullet$ & $\bullet$ & $\bullet$ & $\bullet$ & $\cdot$ & $\bullet$ 
        & $\cdot$ & $\cdot$  \\
  2.5 & $v^7/c^7$ 
      & $\bullet$ & $\bullet$ & $\bullet$ & $\bullet$ & $\bullet$ & $\cdot$ 
        & $\bullet$ & $\cdot$ \\
  3   & $v^8/c^8$ 
      & $\bullet$ & $\bullet$ & $\bullet$ & $\bullet$ & $\bullet$ & $\bullet$
        & $\cdot$ & $\bullet$ \\
\hline
\end{tabular} 
\caption[Harmonics and amplitude corrections]
{The harmonics/amplitude corrections present in each of the
PN amplitude terms.
N.B.: this table is for the general case and
some of the above contributions may be zero for particular binary systems
and/or source orientations.}
\label{tab:harmonics}
\end{center}
\end{table}

The $+$ and $\times$ polarisations, of course, have different 
coefficients and they are generally out of phase by $\pi/2$,
e.g.,
\begin{subequations}
\label{eq:H0p5}
\begin{align}
H^{(0.5)}_+ & = 
  -\Delta \sin i\left[
    \left(\dfrac{5}{8} + \dfrac{1}{8}\cos^2i\right)\cos\varphi
    -\left(\dfrac{9}{8} + \dfrac{9}{8}\cos^2i\right)\cos\left(3\varphi\right)
  \right]\, ,\\
\label{eq:H0p5x}
H^{(0.5)}_\times & = 
  -\Delta \sin i\cos i\left[-\dfrac{3}{4}\sin\varphi 
    + \dfrac{9}{4}\sin\left(3\varphi\right)
  \right]\, ,
\end{align}
\end{subequations}
where
\be
\label{eq:delta}
\Delta = \dfrac{m_1 - m_2}{m_1 + m_2}\, .
\ee
However, some of the higher order terms are `mixed', i.e., they
have amplitude corrections at both phases, e.g., $H^{(2)}_+$
contains \emph{apparent} 
amplitude corrections of the first and third harmonic of the 
$H^{(0.5)}_\times$ term, (\ref{eq:H0p5x}).

The polarisations up to 2PN are listed in Appendix~\ref{sec:hphc}.

N.B.: as expected
the 0.5PN polarisations (\ref{eq:H0p5}) differ from the FBLO term in 
linearised theory (\ref{eq:fbloterms}). 

\subsection{Dependence on inclination angle and mass difference}
The polarisations $h_+$ and $h_\times$ describe the gravitational wave
propagating in the direction of the observer. We saw in linearised theory,
(\ref{eq:fbloterms}),
that the first and third harmonic only propagate towards the observer
if the binary is inclined with respect to the propagation direction, 
i.e., if the inclination angle is non-zero. The
result is the same in the PN approximation, as can be seen in
(\ref{eq:H0p5}). In fact, none of the higher order terms contribute to the
polarisations in the direction of the observer if the binary is
face-on, except for the amplitude corrections to the second harmonic of the
orbital phase, as summarised in Table~\ref{tab:faceon}.
\begin{table}
\begin{center}
\begin{tabular}{|c|c|c|c|c|c|c|c|c|c|}
\hline
\multicolumn{10}{|c|}{Binary observed face-on ($i=0$)}\\
\hline
\hline
PN order & $\mathcal{O}$ &
  $\varphi$ & $2\varphi$ & $3\varphi$ & $4\varphi$ & $5\varphi$ & $6\varphi$
  & $7\varphi$ & $8\varphi$ \\
\hline 
\hline 
  0   & $v^2/c^2$ 
      & $\cdot$ & $\bullet$ & $\cdot$ & $\cdot$ & $\cdot$ & $\cdot$ & $\cdot$ 
        & $\cdot$ \\
  0.5 & $v^3/c^3$ 
      & $\cdot$ & $\cdot$ & $\cdot$ & $\cdot$ & $\cdot$ & $\cdot$ & $\cdot$ 
        & $\cdot$ \\
  1   & $v^4/c^4$ 
      & $\cdot$ & $\bullet$ & $\cdot$ & $\cdot$ & $\cdot$ & $\cdot$ & $\cdot$ 
        & $\cdot$ \\
  1.5 & $v^5/c^5$ 
      & $\cdot$ & $\bullet$ & $\cdot$ & $\cdot$ & $\cdot$ & $\cdot$ & $\cdot$ 
        & $\cdot$ \\
  2   & $v^6/c^6$ 
      & $\cdot$ & $\bullet$ & $\cdot$ & $\cdot$ & $\cdot$ & $\cdot$ & $\cdot$
        & $\cdot$ \\
  2.5 & $v^7/c^7$ 
      & $\cdot$ & $\bullet$ & $\cdot$ & $\cdot$ & $\cdot$ & $\cdot$ & $\cdot$
      & $\cdot$ \\
  3   & $v^8/c^8$ 
      & $\cdot$ & $\bullet$ & $\cdot$ & $\cdot$ & $\cdot$ & $\cdot$ & $\cdot$ 
      & $\cdot$ \\
\hline
\end{tabular} 
\caption[Harmonics and amplitude corrections (face-on)]
{The harmonics/amplitude corrections present in each of the
PN amplitude terms, for a binary observed face-on.}
\label{tab:faceon}
\end{center}
\end{table}
However, if the binary is observed `edge-on' ($i=90^\circ$) then the contributions
are the same as given in Table~\ref{tab:harmonics}, except that
the gravitational wave is linearly
polarised and only consists of the $+$ polarisation.

A result of the PN expansion is that the odd harmonics also depend upon
the mass difference, $\Delta$, such that if the binary components are
of equal mass the
odd harmonics at all orders vanish, see, e.g., (\ref{eq:H0p5}). This may be
understood qualitatively by returning to the argument as to why the 
gravitational wave frequency is twice that of the orbital frequency. We
argued that the binary returns to its start position twice every orbit, due
to the symmetry of the system.
However, the system is less symmetric when the masses are
unequal and so one might expect odd harmonics in that case.
Table~\ref{tab:equal} summarises the contributions to the
polarisations of an equal mass binary.
\begin{table}
\begin{center}
\begin{tabular}{|c|c|c|c|c|c|c|c|c|c|}
\hline
\multicolumn{10}{|c|}
{An equal mass binary ($i\ne0$)}\\
\hline
\hline
PN order & $\mathcal{O}$ &
  $\varphi$ & $2\varphi$ & $3\varphi$ & $4\varphi$ & $5\varphi$ & $6\varphi$
  & $7\varphi$ & $8\varphi$ \\
\hline 
\hline 
  0   & $v^2/c^2$ 
      & $\cdot$ & $\bullet$ & $\cdot$ & $\cdot$ & $\cdot$ & $\cdot$ & $\cdot$ 
        & $\cdot$ \\
  0.5 & $v^3/c^3$ 
      & $\cdot$ & $\cdot$ & $\cdot$ & $\cdot$ & $\cdot$ & $\cdot$ & $\cdot$ 
        & $\cdot$ \\
  1   & $v^4/c^4$ 
      & $\cdot$ & $\bullet$ & $\cdot$ & $\bullet$ & $\cdot$ & $\cdot$ & $\cdot$ 
        & $\cdot$ \\
  1.5 & $v^5/c^5$ 
      & $\cdot$ & $\bullet$ & $\cdot$ & $\cdot$ & $\cdot$ & $\cdot$ 
        & $\cdot$ & $\cdot$ \\
  2   & $v^6/c^6$ 
      & $\cdot$ & $\bullet$ & $\cdot$ & $\bullet$ & $\cdot$ & $\bullet$ 
        & $\cdot$ & $\cdot$  \\
  2.5 & $v^7/c^7$ 
      & $\cdot$ & $\bullet$ & $\cdot$ & $\bullet$ & $\cdot$ & $\cdot$ 
        & $\cdot$ & $\cdot$ \\
  3   & $v^8/c^8$ 
      & $\cdot$ & $\bullet$ & $\cdot$ & $\bullet$ & $\cdot$ & $\bullet$
        & $\cdot$ & $\bullet$ \\
\hline
\end{tabular} 
\caption[Harmonics and amplitude corrections (equal mass binary)]
{The harmonics/amplitude corrections present in each of the
PN amplitude terms, for equal mass binary systems.}
\label{tab:equal}
\end{center}
\end{table}

\subsection{Influence of the amplitude corrections on the 
structure of the waveform}
\label{sec:tddiffs}
\begin{figure}
  \centering
  \psfragfig{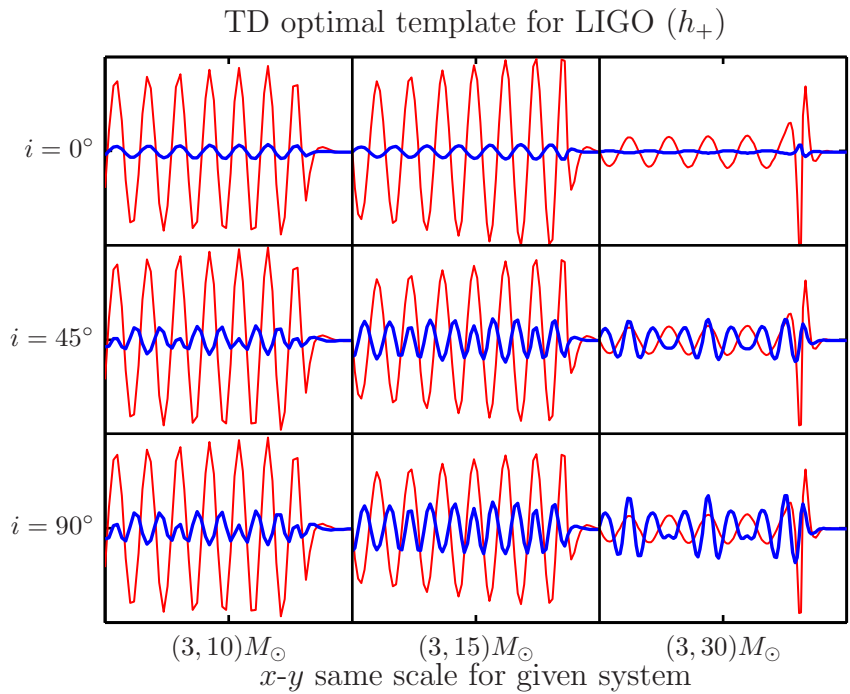}
  \caption[Influence of the amplitude corrections on the structure of the
           waveform]
  {The RWF is plotted in the background and the \emph{difference} between
  the RWF and the FWF (2PN) is plotted in the foreground (bold).
  The waveforms are as observed
   by LIGO for a variety of inclination angles and mass ratios.}
  \label{fig:tddiffs}
\end{figure}

Although the higher order terms are much smaller in amplitude they can
lead to considerable differences between the RWF and the FWF
in the \ac{TD}.
Figure~\ref{fig:tddiffs} shows the difference for a variety of systems
as observed by LIGO, where the FWF is at 2PN in amplitude. As expected
the differences are greater for non-zero inclination angles and larger
mass ratios. 

Considering that FWF should be regarded a closer representation
of nature's gravitational waves, we have a clear motivation for investigating
the use of higher order waveforms in the search for 
gravitational waves. We will discuss this further in Chapter~\ref{ampchap},
where the spectra of the RWF and FWF are compared in Section~\ref{sec:pe}.

\section{Predicted rates of observable \acsp{CBC}}
Here we briefly outline recent work by Abadie et al.~\cite{:2010cf}, 
who produced a summary
of the expected rates of \acp{CBC} observable by current and future
ground-based interferometric detectors. The detection 
rates were predicted using
various sources of information
including observations of GRBs and radio binary pulsars,
the results of previous gravitational-wave searches, and galaxy
catalogs that provide population information of the local Universe.
The predicted rates, of course,
vary for the different types of binaries, i.e., the 
component masses. Here we shall be interested in \ac{BNS}, \ac{BBH}, and
\ac{NSBH}\footnote{These binary systems are precisely defined in
Chapter~\ref{howchap}.}.

There are large uncertainties in the predicted rates due to small statistics
of the 
observations, unknown parameters in astrophysical models etc. The rates
are, therefore, given with quite a large range between the 
lower and upper bounds. Table~\ref{tab:mergers} gives the rates of 
coalescences, whereas Table~\ref{tab:rates} quotes the subsequent
detection rates, predicted
for initial and advanced  LIGO-Virgo~\cite{Acernese:2008}
detector networks, 
giving the lower and upper bounds and a `realistic' 
estimate~\cite{:2010cf}.
\begin{table}
\centering
\begin{tabular}{|c|c|c|c|}
\multicolumn{4}{c}{Rate of mergers}\\
\hline
Source & $N_\text{lower}$ /${\rm L}_{10}^{-1}{\rm yr}^{-1}$ &  
  $N_\text{realistic}$ /${\rm L}_{10}^{-1}{\rm yr}^{-1}$  &  
  $N_\text{upper}$ /${\rm L}_{10}^{-1}{\rm yr}^{-1}$ \\
\hline
\hline
 BNS  & $6\times10^{-7}$ & $6\times10^{-5}$ & $6\times10^{-4}$ \\
 NSBH & $3\times10^{-8}$ & $2\times10^{-6}$ & $6\times10^{-5}$ \\
 BBH  & $6\times10^{-9}$ & $2\times10^{-7}$ & $2\times10^{-5}$ \\
\hline
\end{tabular}
\caption{The number of \acsp{CBC} in the Universe per year per
blue-light luminosity, measured in 
${\rm L}_{10}$, where the Milky Way has a blue-light luminosity of 
$\sim1.7{\ \rm L}_{10}$~\cite{Kalogera:2000dz}.}
\label{tab:mergers}
\end{table}

\begin{table}
\centering
\begin{tabular}{|c|c|c|c|c|}
\multicolumn{5}{c}{Rate of detections}\\
\hline
Detector network &
Source & $N_\text{lower}$ /${\rm yr}^{-1}$ &  
  $N_\text{realistic}$ /${\rm yr}^{-1}$  &  
  $N_\text{upper}$ /${\rm yr}^{-1}$ \\
\hline
\hline
 & BNS  & $2\times10^{-4}$ & $0.02$  & $0.2$  \\
Initial LIGO-Virgo & NSBH & $7\times10^{-5}$ & $0.004$ & $0.1$  \\
 & BBH & $2\times10^{-4}$ & $0.007$ & $0.5$  \\
\hline
\hline
 & BNS  & $0.4$ & $40$  & $400$  \\
Advanced LIGO-Virgo & NSBH & $0.2$ & $10$ & $300$  \\
 & BBH & $0.4$ & $20$ & $1000$  \\
\hline
\end{tabular}
\caption[Predicted rates of \acsp{CBC} observable by LIGO]
{The number of \acsp{CBC} observable by a network of LIGO-Virgo 
detectors per year for
the initial and Advanced detector networks~\cite{:2010cf}.} 
\label{tab:rates}
\end{table}
The detection rates correspond to a \ac{SNR} of 8 
in each detector of the network,
assuming the signal is a \ac{RWF}. In practice, differences between the 
true signal and the \ac{RWF}, such as sub-dominant harmonics, could lead
to a loss in \ac{SNR} (and, hence, detection rates)
if the search relies on \ac{RWF} models.

A network of initial LIGO and Virgo
detectors could be expected to detect inspirals from coalescing \ac{BNS}
systems 
at a rate of one every five years (optimistic rate) to one every
five-thousand years 
(pessimistic rate). However, with a network 
of advanced detectors we can be confident of making the first 
direct gravitational wave detections within ten years. 
Advanced detectors should herald a new era by opening the
gravitational window for observational astronomy. 
Observations with advanced detectors should answer many questions in 
relativistic astrophysics such as direct evidence for the existence of 
black holes, strong field tests of general relativity, 
black hole no-hair theorem, progenitors of gamma-ray bursts, precursors of
magnetar flares, etc. It may be the gravitational window that will one day
reveal what happened shortly after the big bang by detecting primordial
gravitational waves.


\chapter{How to search for gravitational waves from \aclp{CBC}} 
\label{howchap}
\lhead{Chapter~\ref{howchap}.
\emph{How to search for gravitational waves from \acsp{CBC}}} 
\rule{15.7cm}{0.05cm}

In essence, the search for gravitational waves from \acp{CBC} is a 
simple affair.
The expected
waveforms are 
accurately modelled and once gravitational wave strain data are available
(which is of course a huge task for experimenters) a correlation integral is
performed
over a set of signal templates that cover the parameter space
of the search. In practice, however, data analysis 
pipelines become quite complicated once all the considerations of a
running a search,
such as data reduction, coincidence analysis, 
background estimation, detection efficiency,
and dealing with non-stationary noise etc., are taken into 
account. 

We begin with an introduction to the concept of signal processing
before presenting a detailed derivation of the matched filter, 
closely following that of Wainstein and Zubakov~\cite{wainstein:1962}.
The latter parts of the chapter include the results of a search for 
gravitational waves from low mass \acp{CBC} in 186 days of \acf{S5}
data, beginning with an overview of the data analysis pipeline,
which makes use of the matched filter, followed by close attention 
to the analysis of 20 days of data.

\section{Signal processing and filters}
Signal processing refers to the act of performing \emph{useful}
mathematical operations upon a continuous or discrete time series. 
There are many useful applications 
of signal processing, e.g., radar that was developed
during the first half of the 20th century and famously used
by the RAF to win the Battle of Britain in World War II. 

Here, we are interested in processing
\emph{input data} that may consist of only noise or \emph{both}
signal and noise. 
A \emph{filter} will perform operations on the input data, producing
\emph{output data}. When the signal is present in the input data, the 
output data will ideally consist of the transformed signal, i.e., 
 \emph{the filter
extracts the signal from noise}.
In practice, the output data will consist partly of the transformed noise and
partly the transformed signal. Thus we wish to use a filter that
maximises the \ac{SNR}. 

Where the expected signal is known, as it is in the search for
gravitational waves from \acp{CBC}, the filter that provides the largest
\ac{SNR} is the \emph{matched filter}.

\section{The matched filter}
\label{sec:match}
Before we derive the matched filter it is useful to understand the 
following:
\begin{enumerate}
\item The matched filter is a \emph{linear} filter, i.e., the output data 
result from
linear operations of the filter on the input data. The output, 
$y(t)$, of a filter, ${\cal K}$, acting on the input, $x(t)$,
takes the form
\begin{subequations}
\begin{align}
y(t) & = {\cal K}x(t)\, ,\\
y(t) & = \int_{-\infty}^\infty k(\tau)x(t-\tau)d\tau\, ,
\end{align}
\end{subequations}
where $k(\tau)$ is the \emph{impulse response} function of the filter, i.e., 
the response of the filter to a unit impulse (the delta function), 
\be
k(\tau) = \int_{-\infty}^\infty k(\tau-t)\delta(t)dt\, .
\ee
\item The \emph{transfer function}, $K(f)$, of the filter
is the \ac{FT} of the impulse response function.
\item The matched filter is only the optimum filter 
where the noise is a stationary and 
normal random process, i.e., a stationary random process obeying a
Gaussian distribution. It is often convenient to use
the \ac{PSD} of the noise, $S_n(f)$, defined as the \ac{FT} of the 
auto-correlation
function, $R_n(\tau)$,
\be
\label{eq:psd}
S_n(f) = \int_{-\infty}^\infty R_n(\tau)e^{-2\pi i f \tau}d\tau\, ,
\ee
where
\be
R_n(\tau) = \overline{n(t)n(t-\tau)}\, .
\ee
The noise is said to be stationary if the auto-correlation function
depends upon only the value of the time offset, $\tau$, and not the time, $t$.
At $\tau=0$, the auto-correlation function reduces to the mean-square value
of the noise,
\be
R_n(\tau) = R_n(0) = \overline{n^2}\, .
\ee
\end{enumerate}

\subsection{Derivation of the matched filter}
Let us apply a linear filter, ${\cal K}$, to some data, $x(t)$,
that gives the output data
\be
y(t) = {\cal K} x(t)\, .
\ee 
If the data is a linear combination of noise, $n(t)$, and a known signal,
$m(t)$, i.e.,
\be
x(t) = n(t) + m(t)\, ,
\ee
the output of the filter is simply,
\begin{subequations}\begin{align}
y(t) & = {\cal K} n(t) + {\cal K} m(t)\, ,\\
     & = \nu(t) + \mu(t)\, ,
\end{align}\end{subequations}
where $\nu(t)$ and $\mu(t)$ are the filtered values of the noise and 
the signal
respectively,
\be
\nu(t) = \int_{-\infty}^\infty k(t^\prime) n(t - t^\prime) dt^\prime\, ,
\ee
and
\be
\label{eq:sigfilter1}
\mu(t) = \int_{-\infty}^\infty k(t^\prime) m(t - t^\prime) dt^\prime\, .
\ee

Under the assumption that the noise is Gaussian and stationary with a mean 
value of zero ($\overline{n} = 0$), we will find it easier to work with the
\ac{PSD} using the relation (\ref{eq:psd}).
The mean square of the output of the filter with the noise $\nu(t)$ is then
\be
\overline{\nu^2} = \int_{-\infty}^\infty {\left|K(f)\right|}^2 S_n(f) df\, ,
\ee
where $K(f)$ is the transfer function. Applying the convolution 
theorem to (\ref{eq:sigfilter1}) allows us to write the filtered signal
as the \ac{IFT} of the filtered value in the \ac{FD},
\be
\mu(t) = \int_{-\infty}^{\infty} e^{i2\pi ft}K(f) \widetilde{m}(f) df\, .
\ee

The \ac{SNR}, $\rho^2(t_0)$, is then defined as
\be
\label{eq:snrdef}
\rho^2(t_0) =
  \frac{\mu^2(t_0)}{\overline{\nu^2}}
  =  
  \frac{{\left | \int_{-\infty}^\infty e^{i2\pi ft_0} K(f) 
          \widetilde{m}(f) df \right |}^2}
       {\int_{-\infty}^\infty {\left|K(f)\right|}^2 S_n(f) df  }\, .
\ee
The filter, $K(f)$, is a \emph{matched filter} if it is the
best at extracting the signal from the noise, i.e., it must maximise
the SNR.

Multiplying $\mu(t_0)$ by $\sqrt{S_n(f)/S_n(f)}$ and using the 
\emph{Cauchy-Schwarz} inequality we have
\be
{\left | \int_{-\infty}^\infty e^{i2\pi ft_0} 
  K(f)\widetilde{m}(f) df \right |}^2
\le
  \int_{-\infty}^\infty {\left|K(f)\right|}^2 S_n(f) df \int_{-\infty}^\infty
  \frac{ {\left|\widetilde{m}(f)\right|}^2}{ S_n(f) } df
\ee
or
\be
\label{eq:csineq}
{\left | \int_{-\infty}^\infty e^{i2\pi ft_0}
  K(f) \widetilde{m}(f) df \right |}^2
\le
  \ \overline{\nu^2}
  \ \int_{-\infty}^\infty 
    \frac{ {\left|\widetilde{m}(f)\right|}^2}{ S_n(f) } df\, .
\ee
If we now divide both sides of (\ref{eq:csineq}) by $\overline{\nu^2}$,
we can rewrite the \ac{SNR} as
\be
\label{eq:maxineq}
\rho^2(t_0) \le
  \ \int_{-\infty}^\infty 
    \frac{ {\left|\widetilde{m}(f)\right|}^2}{ S_n(f) }df\, ,
\ee
where the RHS is the expected value.
Comparing (\ref{eq:snrdef}) and (\ref{eq:maxineq}) we can see that
$\rho^2$ is maximised with the filter that has the transfer function
\be
K(f) = \gamma e^{-i2\pi ft_0}\frac{ \widetilde{m}^*(f) }{ S_n(f) }\, ,
\ee
where $\gamma$ is an arbitrary constant. Thus the matched filter is defined.


\subsection{Application of the matched filter}
To understand how the filter is applied in a search for gravitational waves
from \acp{CBC}, we must consider the signal as seen in the detector, i.e.,
(\ref{eq:sindet}). Here we will consider a \ac{RWF} signal. We recall
(\ref{eq:sindet}):
\begin{equation*}
h(t) = F_+ h_+ + F_\times h_\times\, . 
\end{equation*}
Suppose
\begin{subequations}
\begin{align}
h_+ = h_0\cos\left(2\varphi(t)\right)\, ,\\
h_\times = h_\frac{\pi}{2}\sin\left(2\varphi(t)\right)\, .
\end{align}
\end{subequations}
The resulting expression can be simplified as
\be
\label{eq:sigindet2}
h(t) = A(t) \cos\left(2\varphi(t) - \Phi_0\right)\, , 
\ee
where
\begin{subequations}\begin{align}
A(t) & = {\left[F_+^2 h_0^2 + F_\times^2h_\frac{\pi}{2}^2\right]}^{1/2}\, ,\\
\cos\Phi_0 & = \frac{F_+ h_0}{A}\, ,\\
\sin\Phi_0 & = \frac{F_\times h_\frac{\pi}{2}}{A}\, .\\ 
\end{align}\end{subequations}
The angle $\Phi_0$ in (\ref{eq:sigindet2}) contains the information about 
the two polarisations and depends upon the sky position and the inclination 
of the source relative to the detector. These angles cannot be known 
\emph{a priori} and therefore must be maximised over.

Now that the matched filter is derived, we can search data, $x$, with
a \emph{template} of the expected signal, $h$, by defining 
the following \emph{inner product} 
as the output of
matched filtering $x$ with $h$. Choosing $\gamma=1$ we have
\be
\label{eq:inp}
\left<x,h\right> := 
 4\text{Re} \int_{0}^{\infty} e^{-i2\pi ft_0} 
    \frac{\widetilde{h}^*(f) \widetilde{x}(f)}
         {S_n^1(f)} df\, ,
\ee
where we have used the \emph{one-sided} \ac{PSD}, $S_n^1(f)$\footnote{
The \acs{PSD}, as
defined before, is an even function, i.e., $S_n(f)=S_n(-f)$.
The one sided PSD uses only the positive frequencies and introduces a 
factor of $2$.}. 

\paragraph{Signal-to-noise ratio}
The \ac{SNR} is given by normalising the matched filter so 
that the recovered signal can be scaled by its amplitude in the noise,
\be
\rho = \frac{\left<x,h\right>}{{\left<h,h\right>}^\frac{1}{2}}\ .
\ee

If the template is normalised, such that the inner product
with itself\footnote{
The square root of the
inner product of two normalised quantities is known as the 
\emph{overlap}.} is equal to unity, i.e.,
\be
\left<\bar{h},\bar{h}\right> = 1\, ,
\ee
where
\be
\bar{h} = \frac{h}{{\left<h,h\right>}^\frac{1}{2}}\, ,
\ee 
then the SNR may be written as
\be
\rho = \left<x,\bar{h}\right>\, .
\ee
The expected value of the SNR, in the presence of a signal
that exactly matches the template, is given by
${\left<h,h\right>}^\frac{1}{2}$.

If we consider a two-phase template of the form (\ref{eq:sigindet2}), we 
can define the following two phases as
\begin{subequations}
\begin{align}
h_c = A(t)\cos2\varphi(t)\, ,\\
h_s = A(t)\sin2\varphi(t)\, .
\end{align}
\end{subequations}
If we then filter the data we have~\cite{Sathyaprakash:1991mt},
\be
\label{eq:filmax1}
\left<x,\bar{h}\right> = \left<x,\bar{h}_c\right>\cos\Phi_0 
+ \left<x,\bar{h}_s\right>\sin\Phi_0\, ,\ee
which we can rewrite as,
\be
\label{eq:filmax2}
\left<x,\bar{h}\right> = {\left[{\left<x,\bar{h}_c\right>}^2 
+ {\left<x,\bar{h}_s\right>}^2\right]}^{1/2} \cos\left(\Phi_0 
- \alpha\right)\, ,
\ee
where
\be
\alpha = \tan^{-1} \frac{\left<x,\bar{h}_s\right>}{
\left<x,\bar{h}_c\right>}\, .
\ee
We cannot know the angle, $\Phi_0-\alpha$, a priori, but can assume it has
a uniform distribution between $0$ and $2\pi$.
It is clear that the maximum value of (\ref{eq:filmax2}) will occur when
\be
\Phi_0 = \alpha\, .
\ee
Therefore, we can write the maximum output of the \emph{two-phase}
matched filter as
\be
\label{eq:maxfilter}
\left<x,\bar{h}\right> = {\left[{\left<x,\bar{h}_c\right>}^2 
+ {\left<x,\bar{h}_s\right>}^2\right]}^{1/2}\, .
\ee

\section{The \acs{LIGO} search pipeline}
\label{sec:pipeline}
The pipeline described here is similar to that used in several \ac{LIGO}
searches~\cite{LIGOS3S4Tuning,LIGOS3S4all,
Collaboration:2009tt,Allen:2005fk} and also~\cite{Abbott:2009qj}, 
for which the results are presented later in this chapter. Each stage of the 
pipeline will be described in detail, but we begin with a basic overview:
\begin{enumerate}
\item A template bank is generated covering the parameters of the search.

\item The data is matched filtered with each template generating 
\emph{first-stage single-detector triggers}.

\item The first stage single detector triggers from the two \ac{LIGO} sites
are compared to see if coincident events exist, producing a list of
\emph{first-stage coincident triggers}.

\item The data is matched filtered using only the templates
associated with first-stage coincident triggers. The new triggers are subjected 
to signal-based vetoes, producing a list of
\emph{second-stage single-detector triggers}.

\item The second stage single-detector triggers are checked for 
coincidence between the \ac{LIGO} sites, producing a list of 
\textit{second-stage coincident triggers}.

\item The second-stage coincident triggers are ranked according to their
\ac{FAR} when compared with background trials.
\end{enumerate}

\subsection{Generating a template bank}
\label{bankgen}
A signal model of $n$ parameters will form a manifold of $n$ dimensions 
on which templates are placed discretely to construct a template bank. 
If spin and higher harmonics are neglected and the sky angles
are maximised over as in (\ref{eq:maxfilter}), then the templates can be
placed on a two-dimensional manifold corresponding to the component masses 
of the binary.

The discreteness of the template bank will cause a loss in \ac{SNR} for
signals whose parameters do not exactly match any of the templates in the
bank. This loss can be limited by setting a threshold known as the 
minimum match, $M_{min}$, of the bank, e.g, $M_{min}=0.95$
(recall the maximum
overlap is unity).
The match, $M$,
between
two nearby templates, $h(\lambda^\mu)$ and 
$h(\lambda^\mu+\Delta\lambda^\mu)$, where $\lambda^\mu$ are
the intrinsic parameters\footnote{The Greek indices run
from $1,\ldots,n$, where $n$ is the number of parameters.} (e.g.,
the component masses as opposed to the sky location), is given by
\be
M = 
  \left<h(\lambda^\mu), h(\lambda^\mu + \Delta\lambda^\mu)\right>\, ,
\ee
which can be Taylor expanded:
\be
M = \left<h(\lambda^\mu),h(\lambda^\mu)\right>
  + \frac{\partial M}{\partial\lambda^\mu}\Delta\lambda^\mu
  + \frac{1}{2}
  \frac{\partial^2M}{\partial\lambda^\mu\partial\lambda^\nu}
  \Delta\lambda^\mu\Delta\lambda^\nu + \ldots\, .
\ee
The first term in the expansion is equal to unity by definition; the second
term will be neglected as it will tend to zero around the maxima of $M$
at $\Delta\lambda^\mu=0$; terms beyond the second derivative will be 
discarded as they are negligible. The resulting expression for the match
becomes
\be
M = 1 + \frac{1}{2}
  \frac{\partial^2M}{\partial\lambda^\mu\partial\lambda^\nu}
  \Delta\lambda^\mu\Delta\lambda^\nu\, .
\ee

If we define the metric tensor of the template manifold as~\cite{Owen:1995tm}
\be
g_{\mu\nu} = -
  \frac{1}{2}
  \frac{\partial^2M}{\partial\lambda^\mu\partial\lambda^\nu}
  \Delta\lambda^\mu\Delta\lambda^\nu\, ,
\ee
we can rewrite the \emph{mismatch} between two nearby templates in 
terms of the metric tensor:
\be
1-M = g_{\mu\nu}\Delta\lambda^\mu\Delta\lambda^\nu\, .
\ee
Thus templates are then placed such that the \emph{maximum}
distance between one template and another in the direction of each parameter,
$x^\mu$, is
\be
{\Delta\lambda^{\mu}}^2 = \frac{2(1 - M)}{g_{\mu\mu}}\, .
\ee
Therefore, a signal that is of the same family as the templates, but
without exactly matching parameters,
would suffer a loss in SNR of no greater than $5\%$ for a minimum match
of 0.95. N.B.: In practice, placing templates using the spacing in the
direction of single parameters will leave some areas of the parameter
space uncovered and, therefore, the actual placement algorithm may use
a smaller spacing~\cite{hexabank}.

The optimum template placement is obtained  using a hexagonal template
placement
algorithm~\cite{hexabank} in the $\left(\tau_0,\tau_3\right)$ parameter space,
where $\tau_0$ and $\tau_3$ are the chirp 
times\footnote{The duration of the signal evolution from the initial to 
the final frequency.}
of the 0 and 1.5\ac{PN}
contributions to the phase. 
The chirp time parameters are used because
their metric is approximately flat, as opposed to that of
 the component masses ($m_1$ and $m_2$). Therefore,
the metric distance between templates
can be considered constant across the entire parameter space, reducing the 
computational cost of template placement.

\subsection{First stage analysis}
The data from each interferometer are matched filtered independently 
over the \emph{entire}
template bank resulting in a \ac{SNR} time series for each template. 
A `trigger' is generated when the \ac{SNR} time series exceeds a given
threshold, $\rho_*$, which is a tunable parameter. 
A low SNR threshold will produce a large number of triggers, i.e., have
a high false alarm probability.
On the other hand a high SNR threshold will reduce the sensitivity 
of the search. Therefore, the threshold is typically set low enough so
that the search remains as sensitive as possible, whilst still being 
computationally manageable.
Given a large trigger rate, where many triggers may be associated with a 
single template at adjacent values in the SNR
time series, the data is reduced by clustering
over the duration of the
template. For each template, the trigger with the largest value of the \ac{SNR} 
time series within that time window is recorded, whilst the others
are discarded. 

Furthermore, a single noise transient (or a signal!) will cause many 
\emph{different}
templates to register triggers at the same time.
Therefore, the triggers are further reduced by 
clustering those that are adjacent in the template bank. 
A three dimensional metric is generated,
($\tau_0,\tau_3,t$), that is used to cluster 
the triggers over time as well as the template
bank parameters.
Starting with a seed trigger on the metric, an error ellipsoid of constant 
metric distance, $\epsilon_f$, is constructed.
Further error ellipsoids are then
generated for all the surrounding triggers 
within a time window, $\pm T_t$,
of the seed trigger\footnote{The time 
window $T_t$ is simply twice the maximum value that an error ellipsoid can
extend in the $t$ direction, i.e., error ellipsoids are not drawn
for triggers so far away in time that they cannot be clustered.}.
Any trigger with an error ellipsoid that overlaps
with the seed trigger's ellipsoid, is clustered with the seed trigger.
This process is repeated for each trigger 
within the original
cluster until no further triggers can be added, at which point 
the trigger with the greatest \ac{SNR} in the cluster is saved whilst all the
others are discarded. 

\subsection{First stage coincidence}
Due to a considerable amount of environmental background noise, a trigger 
cannot be considered as a gravitational wave detection candidate 
unless it is observed in coincidence by detectors at different 
locations. Therefore, we require triple or double coincidence between the 
two \ac{LIGO}
sites, i.e., an H1H2L1 trigger in all three detectors or an H1L1/H2L1
trigger. However, since H1 is twice as sensitive as H2
and colocated, triggers that are 
\emph{only}
found in H2, when H1 is operating normally, are rejected. This will be 
discussed in more detail in section~\ref{coinctwo}. Triggers can also
be found in H1H2 coincidence, but are not analysed
(see Section~\ref{sec:backest}).

The coincidence algorithm~\cite{Robinson:2008} is similar to the first stage
data reduction algorithm.
First the
triggers from each interferometer are time ordered. Then an error ellipsoid
is defined around the first
trigger in the list. The size of the error ellipsoid depends upon the
template's location on the metric, but cannot be greater than a tunable
parameter, $\epsilon_t$, known as \emph{e-thinca}.
Further metric-dependent 
error ellipsoids of the same maximum size 
are then defined for all the triggers from
each of the interferometers within a time window $\pm T_c$. In this
case the time window is set in the same manner as $T_t$, but also accounts
for the light-travel time between the \ac{LIGO} sites (i.e., there could be
a time delay of up to the distance between the sites divided by the speed 
of light).
If any of the
additional triggers'
ellipsoids overlap with the original trigger then they are recorded together
as a coincident trigger.
This process is repeated for all the remaining triggers in the list.
Triggers that are not found to be in coincidence are discarded.
The final list may contain coincident triggers that are duplicated, i.e., an
H1L1 trigger that also exists as part of an H1H2L1 trigger,
in which case the H1L1 trigger is removed from the list.

\subsection{Template bank reduction}
The second stage of the analysis introduces signal-based vetoes and
consistency checks. The checks are potentially computationally expensive
and would considerably increase the latency of the pipeline 
if used when the entire
template bank is matched filtered. Instead, the template bank is reduced
to a subset known as the `trigbank'. The trigbank consists of all the templates
that were part of a coincident trigger at the end of the first stage. This
process can dramatically reduce the number of templates used to analyse a
segment of data. For example, the template bank in Figure~\ref{fig:samplebank}
was reduced from 7477 templates to 1851.  

\subsection{Second stage analysis with signal vetoes}
The second stage analysis is similar to the first stage, but
uses the
trigbank for matched filtering rather than the template bank
and introduces two signal-based vetoes, namely $\chi^2$ and $r^2$.

\paragraph*{The $\chi^2$ veto}
For a given trigger, the $\chi^2$ discriminator measures the consistency 
of power distribution between the data and the template. The template, 
$h$, is divided into $n$ bins 
that provide equal contribution to the expected \ac{SNR},
\be
\left<h_k,h_k\right> 
  = \frac{\left<h,h\right>}{n}\, ,
\ee
for all values of the index $k=1,\ldots,n$.
The $\chi^2$ statistic computes the SNR for each of the bins and
compares with the expected value; taking into account the power 
distribution from both phases of the filter (\ref{eq:maxfilter}), it is
defined as
\be
\label{eq:chisq}
\chi^2 = \sum_{k=1}^{n}\left[
  \biggl({\left<x,h_{0k}\right>} 
    - \frac{\left<h_0,h_0\right>}{n}\biggr)^2
  + 
  \biggl(\left<x,h_{\frac{\pi}{2}k}\right> 
    - \frac{\left<h_\frac{\pi}{2},h_\frac{\pi}{2}\right>}{n}
      \biggr)^2
  \right]\, .
\ee
It is clear that if the data and template match exactly, 
the $\chi^2$ value is zero by definition. More realistically, if
the data consists of Gaussian noise, plus a signal exactly matching the
template, the function (\ref{eq:chisq}) 
follows a classic $\chi^2$ distribution with
$2n-2$ degrees of freedom~\cite{Allen:2004}.
The $\chi^2$ veto is useful because transient sources of noise are 
very unlikely to have the same power
distribution as the template and will therefore have large values of $\chi^2$.

Before setting the threshold, a few things must be taken into consideration.
Firstly, real detector noise is not Gaussian and there will be more excess 
power than expected from the noise. Additionally the template 
and signal parameters are unlikely to match exactly because of the 
discreteness of the bank and the models used to generate templates
will not be exact matches of nature's gravitational wave signals.
Consequently, a genuine signal with a large \ac{SNR} can
be expected to have a large $\chi^2$. Therefore, the $\chi^2$ 
veto is weighted by the \ac{SNR}, defining a new quantity
\be
\xi^2 = \frac{\chi^2}{n + \delta\rho^2}\, .
\ee
Triggers are vetoed when
\be
\xi^2 > \xi^2_*\, ,
\ee
where $\xi_*$, $\delta$ and the number of bins, $n$, are tunable parameters.

A combination of the $\chi^2$ value and the \ac{SNR}, called the 
\emph{effective \ac{SNR}}, 
$\rho_\text{eff}$, is used to rank triggers at the second stage of the
pipeline. The effective \ac{SNR} weights the \ac{SNR} of a trigger by its
$\chi^2$ value and is defined as
\be
\rho^2_\text{eff} = 
  \frac{\rho^2}
       {
          { \left[
          \left(\frac{\chi^2}{2n-2}\right) 
          \left(1 + \frac{\rho^2}{m}\right) 
          \right]}^{1/2}
       }\, ,
\ee
where $m$ is a tunable parameter. The effective \ac{SNR} reduces the 
ranking of triggers with high values of $\chi^2$, which are more likely 
to originate from noise glitches than a signal.

\paragraph*{The $r^2$ veto}
An additional quantity, $r^2$, is defined by renormalising $\chi^2$ 
such that it has an expectation value of $\sim2$, 
\be
r^2 = \frac{\chi^2}{n}\, .
\ee
For a given trigger, 
the veto is constructed by measuring the $r^2$ value in a time
window that precedes the time of the trigger\footnote{Recall that
inspiral-only waveforms model up to the coalescence time and we cannot
know the expected $\chi^2$ in the time following the trigger.}
A trigger will 
be vetoed if the $r^2$ value exceeds a threshold, $r_*^2$,
for a duration
$\Delta t > \Delta t_*$,
where $r^2_*$ and
$\Delta t_*$ are tunable parameters. 

In practice, two $r^2$ thresholds are set; one that is constant for low SNR 
triggers ($\rho < 10$), 
and another that increases linearly with SNR to account for 
the fact that these triggers will have a larger $\chi^2$.

\subsection{Second stage coincidence with signal consistency checks}
\label{coinctwo}
At the second stage coincidence analysis,
two further checks are made between the consistency of what is seen 
in the co-aligned and co-located detectors, H1 and H2.
The effective distance
cut compares the amplitude of a 
trigger recorded in both detectors, whereas
the amplitude consistency check
rejects triggers that were seen in only one detector that \emph{should}
have been seen in both.

\paragraph*{The effective distance cut}
The effective distance, $D_\text{eff}$, is the distance attributed to a 
trigger under the assumption that it is directly overhead the detector and
optimally orientated or, in other words, it is the furthest distance
(up to Gaussian fluctuations)
at which
a source could have produced a trigger of a given \ac{SNR}. The effective 
distance is, therefore, independent of detector sensitivity and a gravitational
wave detected in H1 and H2, in principle, should have the same effective
distance, 
\be
\label{eq:effdist}
D_\text{eff} = \frac{\left<h_{1\,\rm Mpc},h_{1\, \rm Mpc}\right>}
                    {\left<x_,h_{1\, \rm Mpc}\right>}\, ,
\ee
where the template $h_{1\, \rm Mpc}$ was generated at a distance of 
$1\, \rm Mpc$ so that the effective distance has units of ${\rm Mpc}$.

The effective distance cut sets a threshold on the allowed
difference between the effective distance of triggers measured 
coincidently between H1 and H2, defined as
\be
\kappa = \frac{2\left|D_\text{eff,H1} - D_\text{eff,H2}\right|}
{D_\text{eff,H1} + D_\text{eff,H2}}\, .
\ee
The cut will be applied when $\kappa$ is greater than a tunable 
parameter, $\kappa_*$.

\paragraph*{The amplitude consistency check}
The effective distance cut can also be applied when a trigger is present 
in only one of two co-aligned detectors.
The range, $R$, (also known as the horizon distance) of a detector for a given 
template, $h$, is defined as the distance at which an optimally
orientated source (that exactly matches the template) has an expectation
value of the SNR equal to 8, i.e., 
\be
\label{eq:range}
R = \frac{{\left<h, h\right>}^\frac{1}{2}}{8}\, .
\ee
The range of the detector and the effective distance of a trigger are 
related by the \ac{SNR}, allowing the effective distance cut
to be rewritten in terms of the ranges. In the absence of 
a trigger in H2, the maximum expected \ac{SNR} in H1 is then defined as
\be
\label{eq:ampcons}
\rho_\text{max H1} = \frac{R_\text{H1}}{R_\text{H2}}
\frac{\left(2 + \kappa_*\right)}{\left(2 - \kappa_*\right)}\rho_*\, . 
\ee
Thus any triggers present in H1 only, when H2 is operating, 
will be discarded if $\rho > \rho_\text{max H1}$.

\subsection{Data quality vetoes}
\label{sec:dq}
The behaviour of the \ac{LIGO} detectors varies due to environmental factors
that affect the quality of the data, e.g., periods of
seismic activity may cause
a high rate of triggers. When the data are analysed, as many of the 
known environmental factors as possible must be taken in to account and
periods of corrupt data may be vetoed, i.e., removed from the analysis. 
There are four categories of vetoes, for which the analysis
requires a list
of times when they are active.
The vetoes are typically identified by studying \emph{auxiliary channels},
i.e., channels that monitor the state of the detector.

The vetoes are categorised in the following order:
\begin{itemize}
\item{ Category 1: The data is known to be severely corrupted, or even 
missing.}
\item{ Category 2: An auxiliary channel exhibits anomalous behaviour
and a known coupling between the
channel and the gravitational wave strain channel exists.}
\item{ Category 3: An auxiliary channel exhibits anomalous behaviour,
but a less well established coupling between the
channel and the gravitational wave strain channel exists.}
\item{Category 4: An auxiliary channel exhibits anomalous behaviour, but there
is little knowledge of the coupling between channels, although a 
correlation is known to exist.}
\end{itemize}

When running an analysis, the pipeline is usually run first with no
vetoes applied, then repeated with category 1 vetoes, then category 2 etc. 
The information obtained from each
run may be useful for characterising the detectors\footnote{This does not
affect the need for a blind analysis (see Section~\ref{sec:blind}).}.
The remaining data after application of category 1 and 2 vetoes are 
usually
considered good enough to search for gravitational wave candidates. However, 
often category 3 vetoes are also applied. Category 4 vetoes may
later be used to scrutinise any potential gravitational wave candidates.

\subsection{Background estimation}
\label{sec:backest}
To estimate the background the pipeline is run multiple times 
using time-slide data, i.e.,
the data of the two LIGO sites are time shifted by a time greater than the
light-travel time between the detectors. Therefore, any
coincident triggers that 
occur in the time shifted analyses cannot be from a gravitational wave signal
and indicate the background rate. 
The time-shifted data are known
as the background data whereas the non-time-shifted data are known as the
foreground or zero-lag data.
Typically 
one-hundred time slides are performed
and the number of coincident triggers of a given \emph{ranking} present 
in the foreground
are compared with the average number of coincident triggers of
equivalent or higher ranking present in the background.

\subsection{Detection statistic - \acf{FAR}}
\label{sec:detstat}
The detection statistic compares the zero-lag data with the average of the
time-shifted data. At the first stage of the analysis, a simple approach was 
to rank triggers by their effective \ac{SNR}. However, higher mass waveforms
have fewer gravitational wave cycles in the detectors' sensitive band
and the signal based vetoes are not as effective. Thus the rate of background
triggers is expected to be higher. If the loudest foreground trigger is
a \ac{BNS} template it could be hidden due to equally loud high mass
\ac{BBH} template triggers in the background.

When computing a detection statistic, one can
divide up the parameter space into different mass regions. For each of the
foreground triggers a \ac{FAR} can be defined by comparing with the number
of equally loud background triggers in that region of the parameter space.
The \ac{FAR} then allows triggers
from different regions of the parameter space to be compared and ranked
together.
One must also consider that different types of coincidences (H1L1 or H1H2L1)
will have different background rates and should also be compared independently. 

When the different categories from each observation time are recombined to
give the final detection statistic,
the FAR of each trigger needs to be renormalised by the number of trials
(i.e., the number of categories), such that the expected FAR of the loudest
trigger is $1/T$ where $T$ is the observation time.

As H1 and H2 are co-located, their noise is correlated
and the time-shift method cannot be used to
measure the background. Therefore, a FAR cannot be calculated for
H1H2 triggers and they are not included in the final trigger ranking or
the upper limit calculation. H1H2 triggers in H1H2 time may be looked at
in case a gold-plated detection candidate exists, but as it is not known how
to estimate the background
it would be difficult to attribute a level of significance to them.

\subsection{Upper limits}
\label{sec:ulexp}
Once the search is completed an upper limit on the rate of \acp{CBC} can
be calculated for the nearby Universe.
The procedure for calculating upper limits
is described in detail
in~\cite{Fairhurst:2007qj,loudestGWDAW03,Biswas:2007ni} and requires the 
following: the sensitivity of the search, the loudest event and the
background probability.

\paragraph*{Brief description of the upper limit calculation}
For a given rate, $R$, of \acp{CBC},  
the probability of obtaining no triggers `louder'
than a given \ac{FAR}, $x$, due to the background \emph{or} a signal,
is defined as
\be
P\left(x|B,R,T\right) = P_B(x) e^{-RC_L(x)T}\, ,
\ee
where $B$ is the background rate,
$P_B(x)$ is the probability of obtaining no background triggers louder than
$x$, $T$ is the duration of the search and $C_L(x)$ is the
sensitivity of the
search, defined as the cumulative luminosity to which the search can see
a trigger of ranking $x$\footnote{When
using the \ac{FAR} as a detection statistic, a lower value
is louder, e.g., a one-false-alarm-per-year event is louder
than a two-false-alarm-per-year event.}. 
Given that no triggers were louder than the
loudest event, one can define a posterior rate distribution based on the 
\ac{FAR} of the loudest event, $x_m$:
\be
p\left(R|x_m,T,B\right) \propto p(R)\left[
\frac{1 + \Lambda R C_L(x_m)T}{1 + \Lambda}\right] e^{-RC_L(x_m)T}\, ,
\ee
where $p(R)$ is the prior probability distribution on the rates, usually the
result of the previous search, and $\Lambda$ is the likelihood that the
loudest event is due to a gravitational wave as opposed to a background event,
which depends upon the background and sensitivity distributions:
\be
\Lambda = \frac{C^\prime_L(x_m)}{\left|P^\prime_B(x_m)\right|}
\frac{P_B(x_m)}{C_L(x_m)}\, ,
\ee
where $C^\prime_L(x_m) = dC_L/dx$, etc. One can then compute the rate upper
limit, $R_*$, for a given confidence level, $\alpha$,
\be
\alpha = \int_0^{R_*} p\left(R|x_m,T,B\right)\, .
\ee

\paragraph*{The search sensitivity}
In describing the upper limit calculation above, the search sensitivity,
$C_L$, was introduced, which is the \emph{cumulative luminosity}: 
the blue-light luminosity, measured 
in units of ${\rm L}_{10}$\footnote{$\rm L_{10}$ is $10^{10}$ times the
blue solar luminosity (the Milky Way contains 
$\sim1.7\ {\rm L}_{10}$~\cite{Kalogera:2000dz}).}
, of all the local galaxies that may contain \acp{CBC} 
to which the search is sensitive to. To calculate $C_L$ one must know the 
efficiency of the search as a function of distance and chirp mass, 
$\epsilon(D_\text{eff},M_c)$ and
the luminosity of the local Universe, also as a function of distance
and chirp mass, $L(D_\text{eff}, M_c)$. The cumulative luminosity is then 
defined as
\be
C_L = \int \epsilon(D_\text{eff},M_c) L(D_\text{eff},M_c) 
  dD_\text{eff} dM_c\, .
\ee
The blue-light luminosity is used as it is assumed that the rate of \acp{CBC}
is proportional to the star formation rate, which is in turn proportional to 
the blue-light luminosity~\cite{LIGOS3S4Galaxies}.

The efficiency function is calculated by adding simulated
signals (injections)
to the data and evaluating the fraction of detected signals, louder
than $x_m$, for a given set of parameters. The luminosity function
is calculated by multiplying the efficiency of signal
recovery for the search as a function of distance by the physical
luminosity as a function of distance and integrating their
product over distance.

\paragraph*{Uncertainties in calculating the rate upper limit}
There are a number of uncertainties which affect the upper limit calculation,
including Monte Carlo statistics, detector calibration, distances and 
luminosities of galaxies listed in the galaxy catalog~\cite{LIGOS3S4Galaxies}
and differences between the templates used to evaluate the efficiency of the 
search and the true waveforms of nature. All of these uncertainties
may be marginalised over when computing the posterior rate distribution
~\cite{Fairhurst:2007qj}.

\section{The \acs{S5} low mass \acs{CBC} search}
\label{sec:blind}
The fifth science run of \ac{LIGO} 
began in November 2005 and concluded in September
2007, with all three detectors operating at design sensitivity.
A search for gravitational waves from low mass \acp{CBC} 
was performed on the data, with
the analysis divided into three epochs. The \ac{S51YR} search consisted of
data
collected between November 4th, 2005 and November 14th, 
2006~\cite{Collaboration:2009tt}. 
Towards the end of \ac{S5}, as of May 18th 2007, the Virgo detector collected
\ac{VSR1} data in coincidence with \ac{LIGO}. The analysis pipeline of the 
joint search using both \ac{LIGO} and Virgo data required significant 
changes from that used in the \ac{S51YR} search,
thus defining the third epoch~\cite{s5vsr1}. 
The so-called `12-to-18 month' search, described in this chapter, 
used the 186 days of \ac{S5} data recorded after the \ac{S51YR} search
concluded,
but before \ac{VSR1} began~\cite{Abbott:2009qj}. 
In total there were $\sim0.3\, \rm yr$ of
data analysed as opposed to $\sim0.7\,\rm yr$ in the \ac{S51YR} search.

Unlike the \ac{S51YR} search that analysed all of the data in one instance
of the pipeline, the 12-to-18 search analysed each
`month'\footnote{Four weeks of data.} 
of data independently. The detector behaviour varied over the course
of the search, 
hence, analysing the data monthly allowed foreground triggers to be compared
with background triggers that better reflected the behaviour at the time
of the candidates. The results of `month 1'\footnote{Month 1 was not the 
first month of the search, but the second. Sometimes, as in this case,
physicists count from zero.}
are described in detail in this section, along with the final results of the
complete search.

\paragraph*{Blind analysis and search tuning}
In order to avoid any biases that may be introduced by the data analysts, all
tunable parameters, such as the \ac{SNR} threshold, minimum match, the metric
distance used for clustering etc. must be chosen before the 
foreground data is analysed.
This process
prevents the data analysts tuning the search on the basis of a trigger found
in the foreground and is known as a blind analysis. 
However, roughly ten percent 
of the data is marked as `playground' data, which are analysed 
at zero-lag
to check
that the pipeline performs as expected, produces reasonable results and that
the data quality procedures are adequate. 
Alongside the playground data, the analysts are able to look at time-shifted
data, as any coincident triggers cannot be true signals. The time shifted
data can be used to check background rates and the playground analysis
can be compared with these. The tuning of the signal based vetoes is
achieved by performing the analysis with simulated signals added to the data,
known as `injection runs'.

\begin{figure}
\centering
\includegraphics[width=14cm]{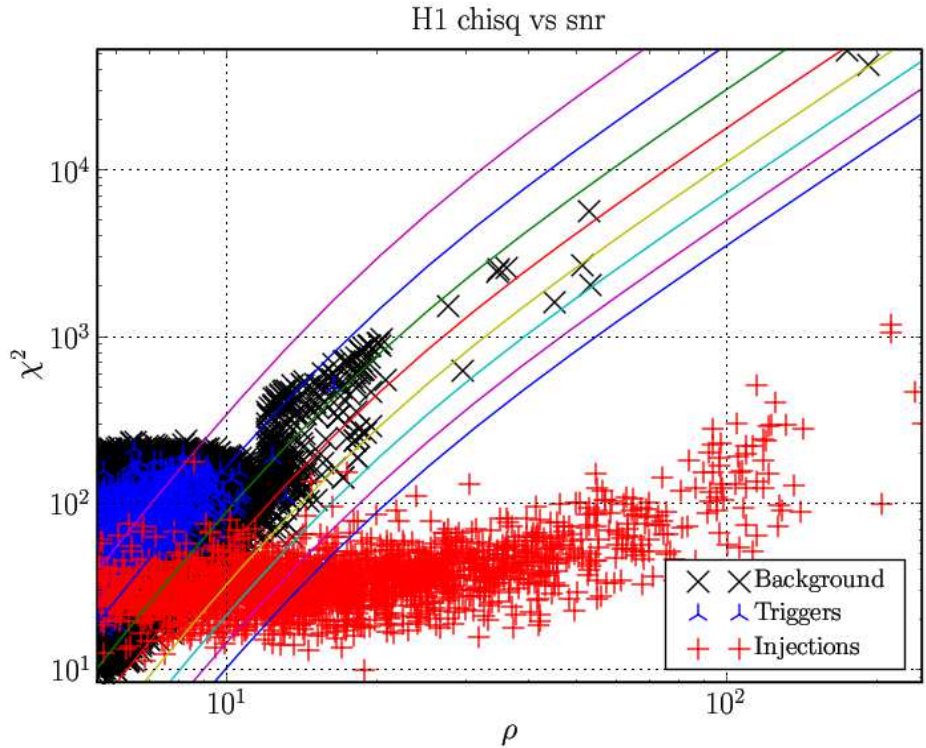}
\caption[Separation of injections from background, \acs{SNR} vs.~$\chi^2$]
{This plot of $\chi^2$ vs.~SNR shows how the effective SNR can be used to 
separate software injections 
from background
triggers in H1. The coloured lines show contours of constant
effective SNR. The sharp cutoff in the background triggers reflects the fact
that there are two $r^2$ thresholds.}
\label{fig:samplechisq}
\end{figure}

As the pipeline used for the 12-to-18 search was nearly identical to
the first year search, the playground analysis used the parameters as
tuned for the first year search. There were no anomalies in the playground
analysis or injection runs and therefore the tuned 
parameters were not altered. 
Figure~\ref{fig:samplechisq} shows the separation of the
software injections from the background using the $\chi^2$ discriminator.
The figure was made after
the analysis was un-blinded and so also includes the foreground triggers, which
are consistent with the background.
Table~\ref{tab:tune} lists a selection of the tuned parameters.
\begin{table}
\centering
\begin{tabular}{|c|c|c|}
\hline
Parameter & Symbol  & Value \\
\hline
\hline
Lower cut-off frequency & $f_l$  & $40.0\, \rm Hz$ \\
\ac{SNR} threshold & $\rho_*$ & 5.5 \\
Minimum match & $M$ & 0.97 \\
Effective distance threshold & $\kappa_*$ & 0.6 \\
Single IFO error ellipsoid & $\epsilon_f$ & 0.06 \\
e-thinca& $\epsilon_t$ & 0.5 \\
$\chi^2$ veto threshold & $\xi^2_*$ & 10 \\
number of $\chi^2$ bins & $n$ & 16 \\
$r^2$ threshold & $r^2_*$ & 16 \\
\hline
\end{tabular}
\caption[\acs{S5} 12-to-18 tuned parameters]{A selection of the tuned 
parameters used in the 12-to-18 search.}
\label{tab:tune}
\end{table}

\subsection{Month 1: Data information and first stage analysis}
Month 1 of the 12-to-18 search began on December 12th, 2006 and finished
on January 9th, 2007 (849974770-852393970 GPS time). The quantity of 
data analysed, before and after the application of the vetoes, is
listed in Table~\ref{tab:12to18data}.

\begin{figure}
\centering
\includegraphics[width=14cm]{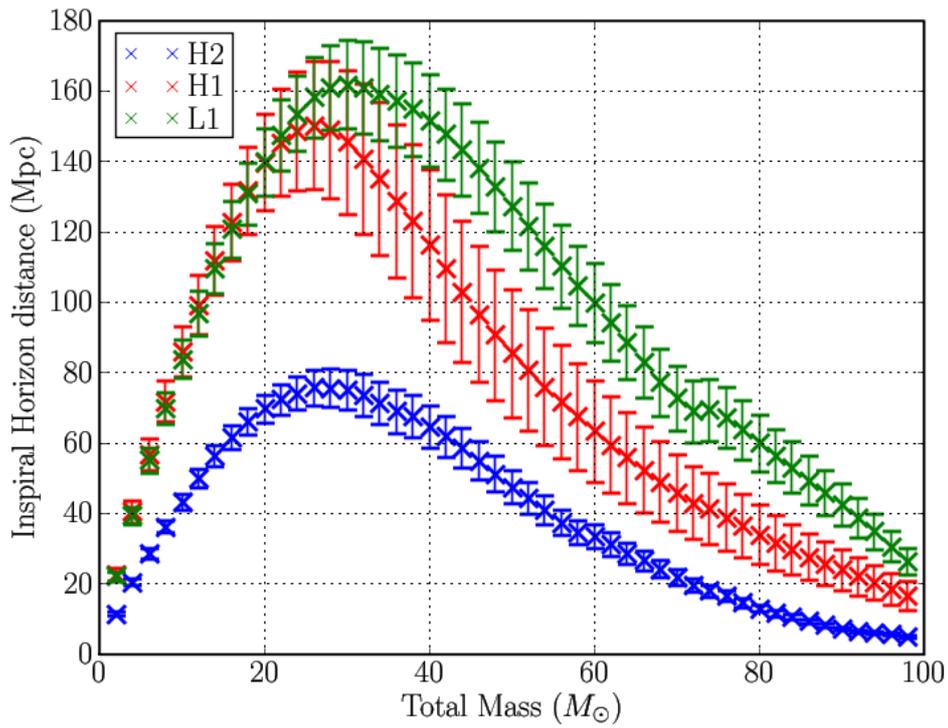}
\caption[Month 1 range vs.~mass]
{The inspiral-range (\ref{eq:range}) vs.~mass for equal mass systems
for each of the interferometers 
averaged over the course of month 1.} 
\label{fig:rangevmass}
\end{figure}

\begin{figure}
\centering
\includegraphics[width=14cm]{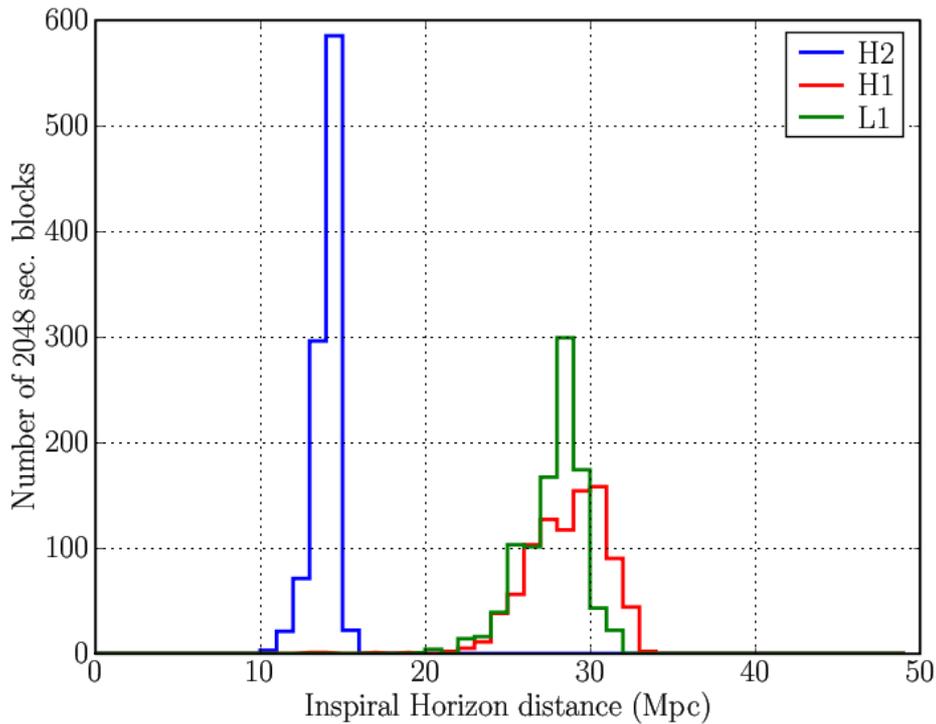}
\caption[Month 1 \acs{BNS} range histogram]
{Histograms of the inspiral-range of a BNS 
system in each detector for month 1.}
\label{fig:rangehist}
\end{figure}

\begin{figure}
\centering
\includegraphics[width=14cm]{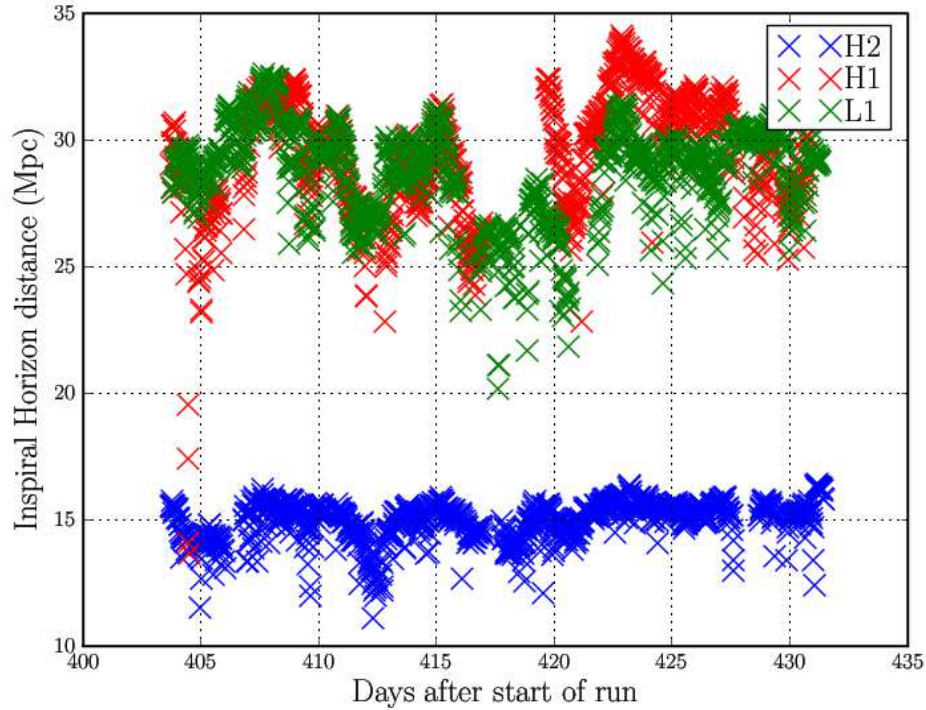}
\caption[Month 1 BNS range vs.~time]
{The inspiral-range of a BNS 
system plotted for each segment of data for month 1 of the 12-to-18 
month search. 
N.B.: days after start of run refers to the start of \acs{S5}.}
\label{fig:rangeplot}
\end{figure}

The data were divided into segments of length $2048\, \rm s$ for analysis. Each 
segment had a different \ac{PSD}, according to the varying
detector behaviour and the noise environment at the time the data was
recorded.
Thus the sensitivity of the search varied for each segment and
can be expressed as the range (\ref{eq:range}), e.g., of a \ac{BNS} system. 
Figures~\ref{fig:rangevmass},~\ref{fig:rangehist} and~\ref{fig:rangeplot} 
indicate the sensitivity of the detectors during month 1. 
\begin{table}
\centering
\begin{tabular}{|c|c|c|c|c|c|}
\hline
Interferometer & Science Segments (days) & After Cat 1 &
 Cat 1,2 & Cat 1,2,3 & Cat 1,2,3,4 \\
\hline
\hline
H1 & 19.8171 & 19.8156 & 19.1730 & 18.8534 & 14.2322 \\
H2 & 21.5143 & 21.5125 & 19.8991 & 18.0748 & 13.4888 \\
L1 & 21.0350 & 21.0254 & 20.8062 & 19.3211 & 18.8299 \\
\hline
\end{tabular}
\caption[\acs{S5} second year month 1 data information]
{The \acs{LIGO} data recorded during month 1 of the 12-to-18 search. The
duration is shown in days before and after data quality vetoes have been 
applied.} 
\label{tab:12to18data}
\end{table}

The data were initially sampled at $16384\, \rm Hz$, but were
reduced by down-sampling to $4096\, \rm Hz$
for analysis. Frequencies below $30\, \rm Hz$ are limited by the `seismic wall'
of \ac{LIGO}'s noise curve and are high pass-filtered during this process.
 
The data segments were chosen to overlap by $256\, \rm s$, 
allowing the first and
last $64\, \rm s$ of each segment to be discarded when matched filtering.
Hence all of the data can be searched, without any corruption occurring 
due to the edge effects of wrapping the \ac{SNR} time series.

The data were analysed in different categories according to which detectors
were operating, denoted triple time (H1H2L1)
when all three detectors are operating
and double time when only one of the Hanford detectors is operating
(H1L1 and H2L1). These
times were redefined after application of each of the data quality vetoes.

\paragraph*{The \acs{PSD}} The \ac{PSD} was calculated for each segment of 
data by dividing it into fifteen overlapping smaller segments of 
$256\, \rm s$ and
taking the \ac{FT} of each of these. The \ac{PSD} was then given by the median 
of each frequency bin. 

\paragraph*{Template bank}
\label{sec:tmpltbank}
The template bank was constructed as described in Section~\ref{bankgen}. It 
consisted of non-spinning RWFs at a phase order of 2PN.
The templates were generated in the \ac{FD} using the \ac{SPA}
with a total mass range of between 
$2$-$35\Msun$ and a minimum component mass of $1\Msun$. The minimal match 
due to the discreteness of the bank was $0.97$. 
\begin{figure}
\centering
\includegraphics[width=14cm]{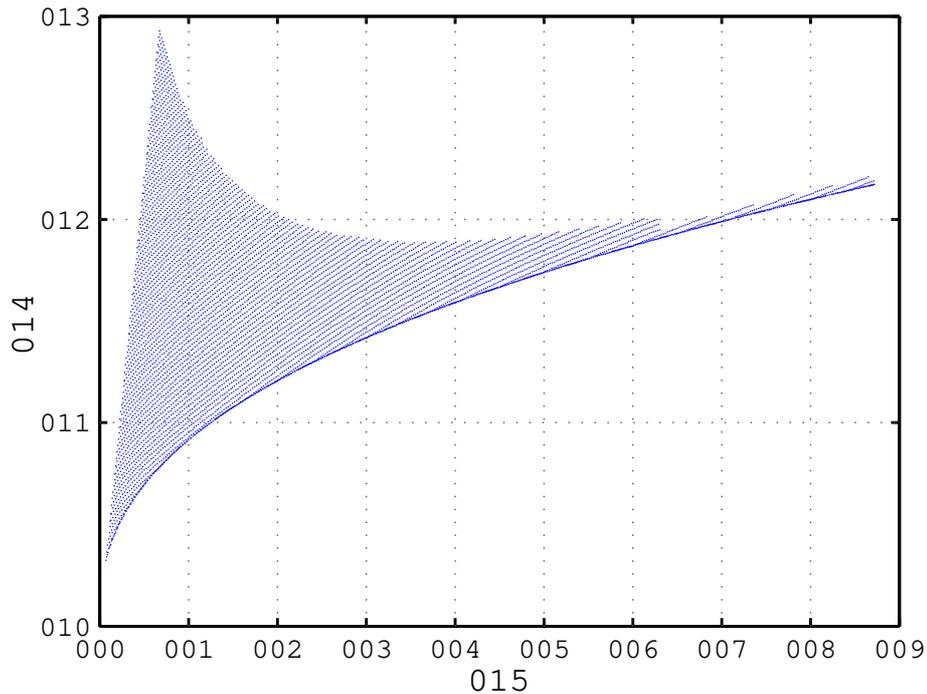}
\caption[Month 1 sample L1 template bank]
{The template bank generated for a $2048\, \rm s$ L1 segment starting
at 852351639 GPS time. There were 7477 templates in this bank.}
\label{fig:samplebank}
\end{figure}
The template bank placement depends upon the \ac{PSD} and therefore varied
for each data segment.
Figure~\ref{fig:samplebank} shows a 
template bank generated for a sample L1 data segment of month 1.

\begin{figure}
\centering
\includegraphics[width=14cm]{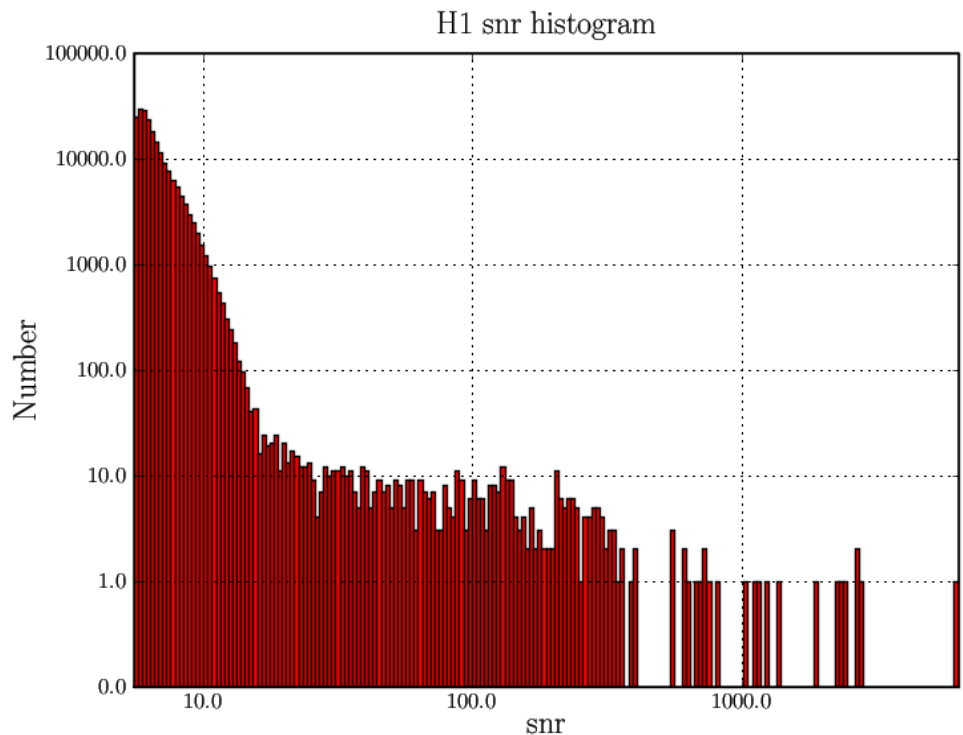}
\caption[Month 1 \#triggers vs.~\acs{SNR}]
{A histogram of the number of triggers vs.~\acs{SNR} for H1.}
\label{fig:h1firstsnr}
\end{figure}

\begin{figure} 
\centering
\includegraphics[width=14cm]{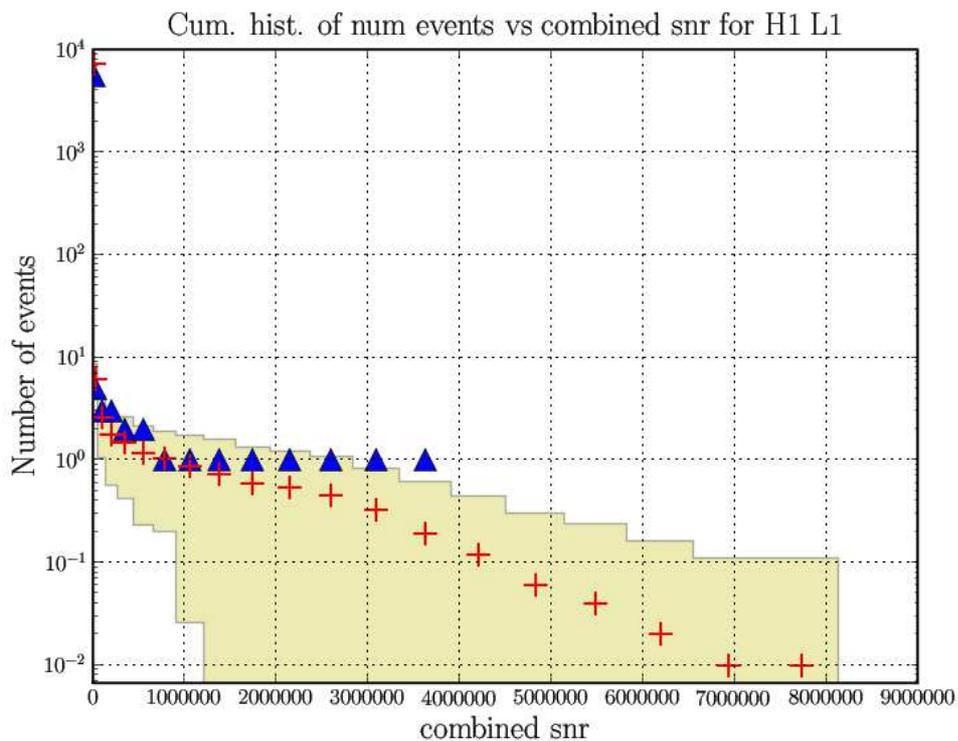}
\caption[Month 1 first stage triggers]
{Histogram comparing foreground and background triggers coincident in
H1 and L1 after the first stage analysis in triple time.
The blue triangles show the
foreground triggers, whereas the red crosses show the background triggers
(with their one sigma errors shown as the yellow area. The combined
SNR is the sum of squares of the individual SNRs in H1 and L1.}
\label{fig:firstthinca}
\end{figure}

\paragraph*{First stage triggers}
Figure~\ref{fig:h1firstsnr} shows the number of triggers in H1 plotted 
against \ac{SNR} after application of first stage trigger clustering. 
Figure~\ref{fig:firstthinca} shows the number
of foreground triggers coincident in H1 and L1 in triple time 
compared to the average number in the time-slides, 
plotted with their one sigma error. 

\begin{figure} 
\centering
\includegraphics[width=14cm]{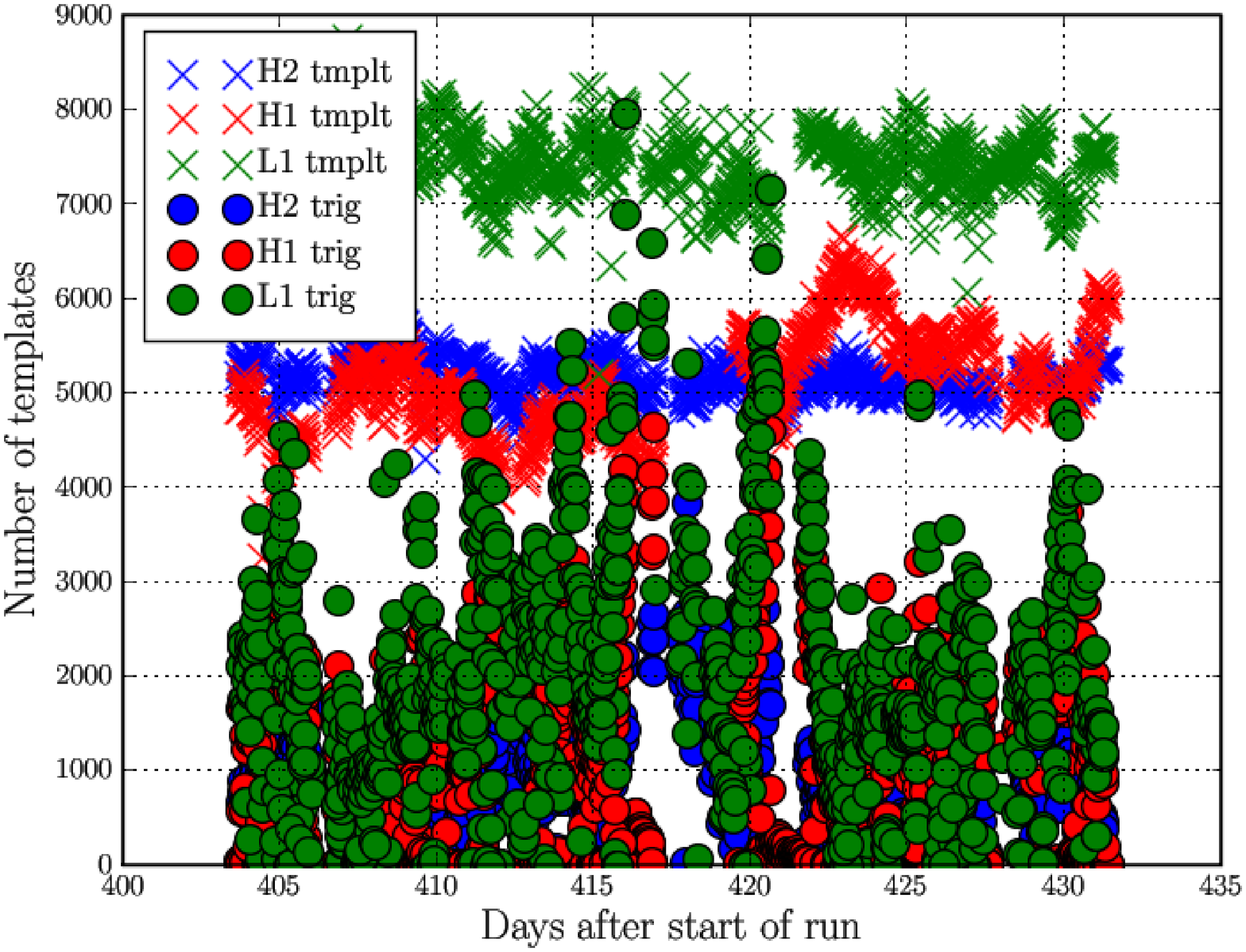}
\caption[Template bank reduction]
{Template bank size compared to trigbank size for the data segments
in month 1.}
\label{fig:numtemplates}
\end{figure}

\paragraph*{Trigbank}
Figure~\ref{fig:numtemplates} shows the variation in template bank size
compared to the trigbank size
for month 1. In several instances the number of
templates in the trigbank are of the same value as the original template bank,
indicating poor quality data as the trigger rates at first stage must have
been large to produce so many coincident events.

\subsection{Month 1: Second stage analysis and loudest triggers}

\paragraph*{The amplitude consistency check revisited}
The 12-to-18 analysis originally produced loud foreground triggers coincident 
in H2 and L1 at times when H1 was operating normally, thus they had passed
the amplitude consistency check, (\ref{eq:ampcons}), between H1 and H2.
However, as H1 was typically twice as sensitive as H2, the maximum \ac{SNR}
for a trigger to be present in H2, but not H1, 
$\rho_\text{max H2}$, was only just above threshold in H2. Using the horizon
distances of each segment it was shown that $\rho_\text{max H2} > 
\left(\rho_* = 5.5\right),$ for just $12.9\%$, of the triple time during
month 1. Further analysis showed that $\rho_\text{max H2} > 5.7$ for 
$3.7\%$ of the time and $\rho_\text{max H2} > 6.0$ for 
$0.2\%$\footnote{These times were calculated before the application of
data quality vetoes. This means that the true percentages would differ.}.
Hence we see that for an H2 trigger to pass the consistency check, it can only
just be above the threshold. As H1 was operating normally it is intuitive to
believe that the H2 triggers were due to the background and happened to be
quiet enough to pass the consistency check whilst having similar
parameters to an L1 trigger, or in other words,
they were not due to a gravitational wave!

Furthermore,
the percentages of times when an H2L1 trigger could occur in triple time
varied for each of the time-slides (due to the L1 vetoes) producing poor 
background estimation and in some cases 
potentially
elevating the ranking of an H2L1 trigger. 
The 12-to-18 analysis was rerun, but with a new cut that rejected
\emph{all}
H2L1 triggers in triple time. This decision was made after the analysis
was un-blinded, as it was considered to be changing a mistake with the
original analysis rather than re-tuning the search. 
Hence in triple time, only H1H2L1 and H1L1 coincident 
triggers are considered.

\paragraph*{The loudest triggers}
As stated in Section~\ref{sec:detstat} the \ac{FAR} allows foreground triggers
of different mass categories and IFO combinations
to be directly compared. In the 12-to-18 search
templates were categorised by their chirp mass into three
ranges defined by the chirp mass of 
\emph{equal} mass systems of a total mass between
2-8, 8-17 and 17-35$\Msun$. When calculating the
detection statistic for triple time data, triple coincidence triggers
are separated from double coincidence triggers, i.e., 
H1H2L1 triggers do not contribute to the background of H1L1 triggers.
The final ranking
statistic used was the \ac{IFAR} in units of ${\rm yr}$. The loudest 
trigger of the month had an \ac{IFAR} of $0.16\, {\rm yr}$,
meaning that a trigger
equally as loud can be expected due to background in every 0.16 years of 
data\footnote{Equivalent livetime.}. 
A summary of the loudest triggers in month 1 is listed in
Table~\ref{tab:m1trig}.
\begin{table}
\centering
\begin{tabular}{|c|c|c|c|c|c|}
\hline
Rank & \acs{IFO} Time & Coincident Type & \acs{IFAR} /yr  \\
\hline
\hline
1 & H1H2L1 & H1H2L1 & 0.16  \\
2 & H1H2L1 & H1H2L1 & 0.10 \\
3 & H1L1 & H1L1 & 0.08 \\ 
4 & H1H2L1 & H1L1 & 0.03 \\
5 & H1H2L1 & H1H2L1 & 0.03 \\
\hline
\end{tabular}
\caption[Month 1 loudest triggers]
  {The 5 loudest triggers for all mass categories 
  ranked by their \acs{IFAR} for month 1
  of the 12-to-18 search. The coincident type refers to whether the trigger
  was coincident in all three detectors (H1H2L1) or in just two (H1L1/H2L1)}
\label{tab:m1trig}
\end{table}

\begin{figure} 
\centering
\includegraphics[width=14cm]{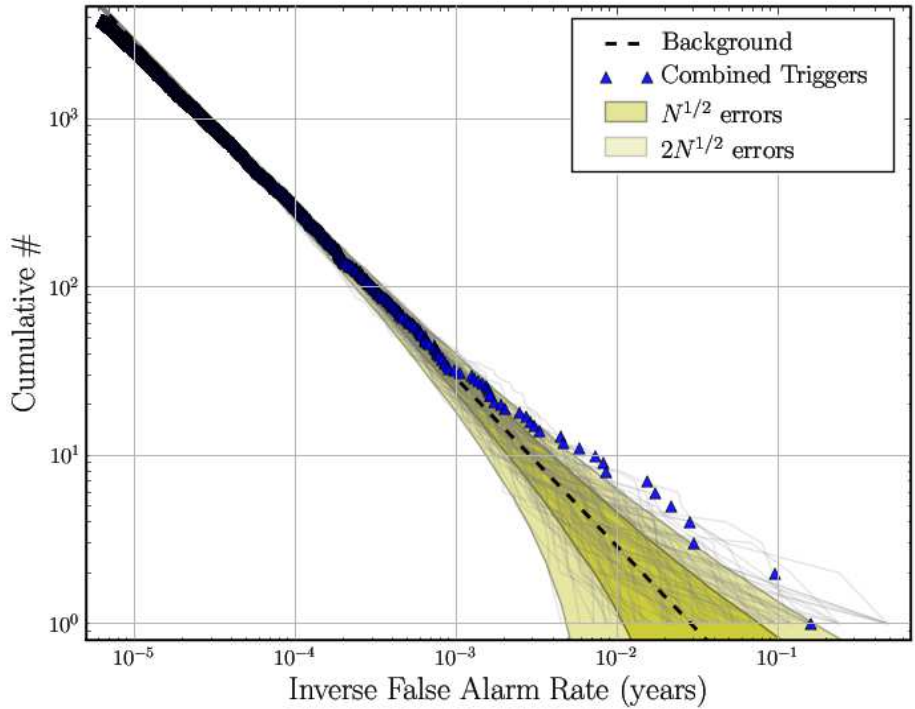}
\caption[Month 1 loudest triggers]
{The loudest triggers of month 1 analysed after application of
data quality vetoes 1,2 and 3.}
\label{fig:ifar}
\end{figure}

Figure~\ref{fig:ifar} shows the cumulative number of foreground triggers in 
H1H2L1 time against \ac{IFAR} for month 1. The triggers are marked as blue
triangles. The dashed black line is the expected background plotted with one
and two sigma error regions. The expected background is simply the \ac{IFAR}
normalised to one year, i.e., in one year we would expect one event in the
background with an \ac{IFAR} of 1. After application of category vetoes 1-3
there were 10.5 days of H1H2L1 data in month 1. Therefore one
would expect the loudest background event to have an \ac{IFAR} of 
$10.5/365.25 \sim 0.03\, {\rm yr}$.
The background events from the time slides
are also plotted as grey lines.
The loudest event was above the expected background for month 1, although
not significantly enough so to be of any interest; it is within the 2-sigma 
background errors and quieter than several of the background trials.
Moreover it was also the loudest event
of the entire 12-to-18 search in which there were $0.21{\rm yr}$ of H1H2L1
data\footnote{Therefore the \ac{IFAR} of the loudest expected background event
is $0.21$ years.}, which places the loudest trigger slightly below the
expected background of the complete search and within one sigma.

\section{Upper limits}
\label{sec:12to18res}
The 12-to-18 search calculated rate upper limits for coalescing 
binaries consisting of neutron stars 
and/or black holes assuming \ac{BNS} systems of
$m_1$ $=$ $m_2$ $=$ $\left(1.35 \pm 0.04\right)\Msun$; \ac{BBH} systems 
consist of $m_1$ $=$ $m_2$ $=$ $\left(5 \pm 1\right)\Msun$; and \ac{NSBH}
systems consist of $m_1$ $=$ $\left(5 \pm 1\right)\Msun$ and
$m_2$ $=$ $\left(1.35 \pm 0.04\right)\Msun$.
For \acp{BBH} 
the upper limits were also calculated as a function of the total
mass of the binary and, for \ac{NSBH} binaries, as a function of
the black hole mass. The effects due to the spin of the sources
visible to \ac{LIGO} are expected to be negligible for \ac{BNS}
waveforms~\cite{ATNF:psrcat,Apostolatos:1994}, and limited for black holes.
The main results of the search were therefore presented assuming non-spinning
sources, however, the upper limits were also calculated for spinning black
holes, assuming their spin is uniformly distributed between zero and 
a maximal value of $m^{2}$, in accordance with theoretical limitations.

The posterior rate distributions were calculated for each month of the 
12-to-18 search separately using a uniform prior. These results were then
combined to produce final posterior rate distributions, 
using the \ac{S51YR} search results as
 the prior, from which the $90\%$ confidence rates were 
calculated. As described in section~\ref{sec:ulexp}, the sensitivity of the
search is measured using the cumulative luminosity, hence the rate upper 
limits are quoted in units of ${\rm yr}^{-1} {\rm L}_{10}^{-1}$. 
The upper limits were calculated on the data after the
application of category vetoes 1-3.

\subsection{12-to-18 search results}
\begin{figure}
  \begin{minipage}[h]{16cm}
    \centering
    \psfragfig{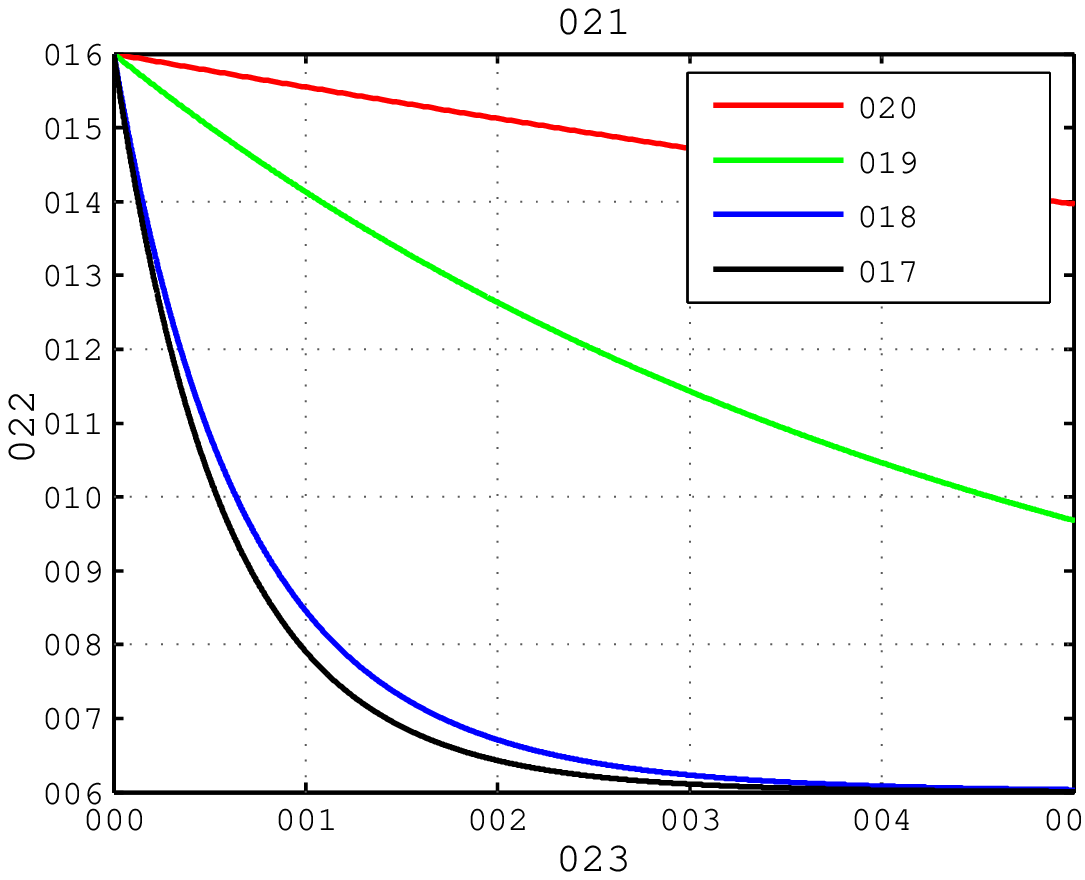}
  \end{minipage}
  \vspace{1cm}
  \begin{minipage}[h]{16cm}
    \centering
    \psfragfig{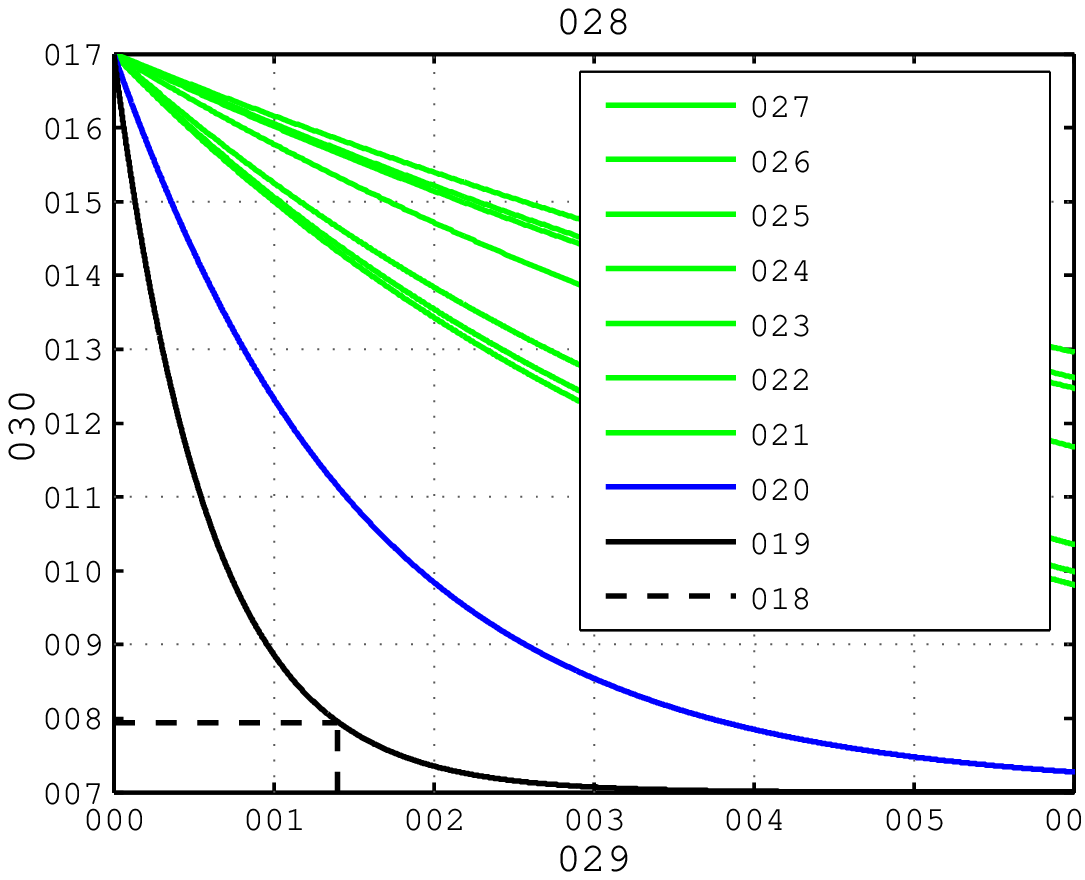}
  \end{minipage}
  \caption[Month 1 \acs{BNS} posterior]
  {The BNS posterior distribution of the rates of coalescing 
  \acs{BNS} systems, neglecting spin, for month 1 (top) and for the entire
  12-to-18 search (bottom). In the top figure we see that H1H2L1 data
  constrains the rates better than double time data. The bottom plot shows
  the contributions of each of the months (green - 
  listed in the legend in the order that 
  they appear from top to bottom on the plot).
   We see that month 4 was the `best'
  and that month 1 was the `worst'.
   The S5YR1 result is shown in blue and the complete S5
   result is shown in black.}
\label{fig:post}
\end{figure}

Figure~\ref{fig:post} shows the posterior rate distributions of
non-spinning \ac{BNS} systems for month 1 (top) and for the complete
12-to-18 search including the S51YR prior (bottom).
The month 1 distributions show that observations using H1L1 and H2L1 data 
constrain the rates far less than those using triple time data, as we 
would expect given the much larger
duration of triple time compared to double time in
the search. The 90\% upper limit on the rates are obtained by normalising
the posterior distributions and integrating to $0.9$. However, 
Figure~\ref{fig:post} shows the non-normalised distributions so that each
curve can be compared qualitatively.
In the bottom plot of Figure~\ref{fig:post}, each month is listed in the
legend in the order that it appears from top to bottom, or rather in the
order of least constraining to most constraining.
We see that month 1 was in fact
the `worst' month of the search, due to poorer data quality. The latter months
are the most constraining on the rates as they consisted of the best 
quality \ac{LIGO} data of S5 (prior to VSR1).
It is interesting to see that although the
S51YR result is far better than any of the months individually, the combined
upper limit is considerably improved with the additional 12-to-18 data.

\begin{figure}
  \begin{minipage}[h]{16cm}
    \centering
    \psfragfig{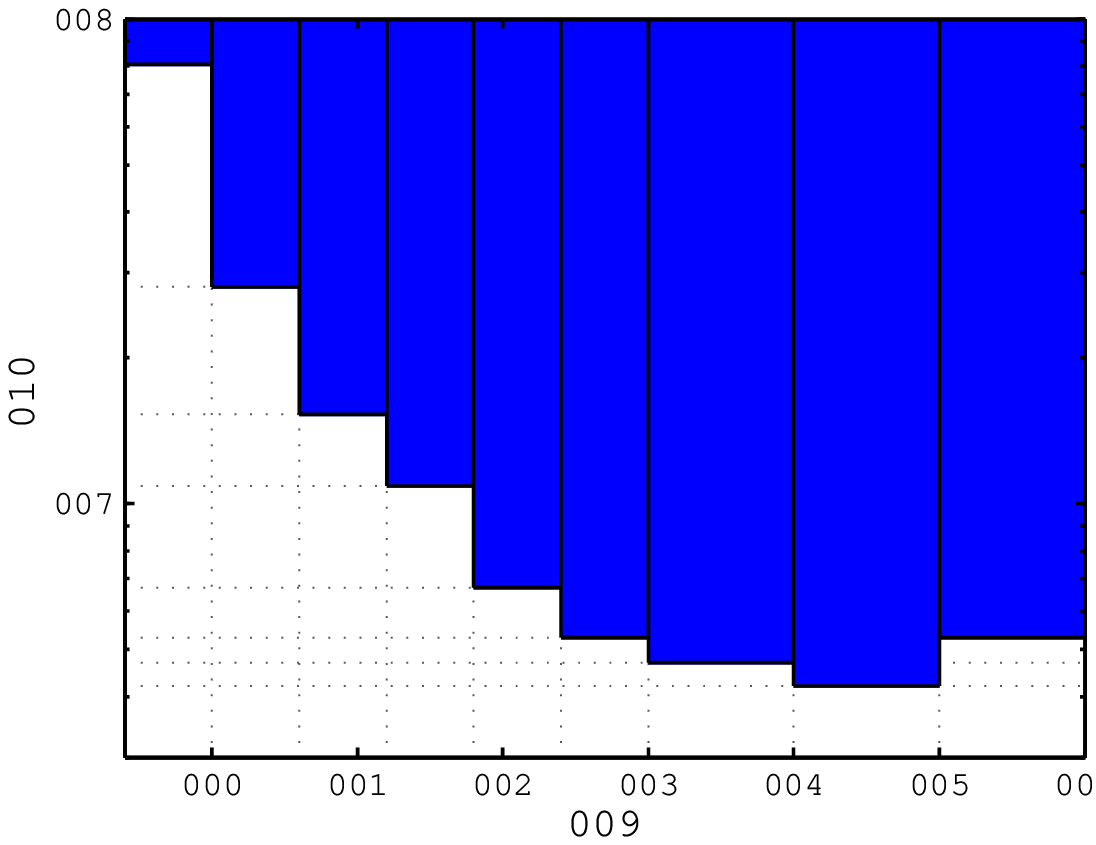}
  \end{minipage}
  \vspace{1cm}
  \begin{minipage}[h]{16cm}
    \centering
    \psfragfig{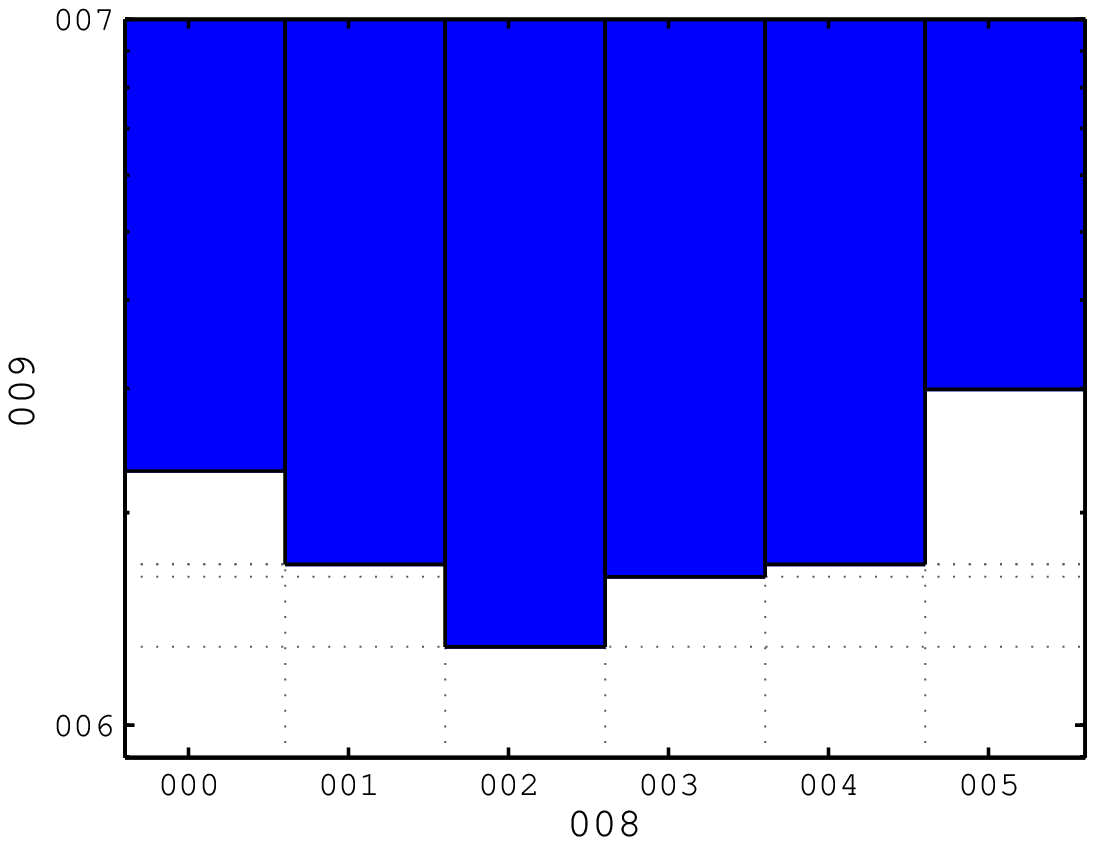}
  \end{minipage}
  \caption[12-to-18 90\% upper limits vs.~mass]{
  Top: The 90\% upper limit for non-spinning BBH systems
  vs.~total mass.\\
  Bottom: The 90\% upper limit for non-spinning 
  NSBH systems vs.~black hole 
  component mass assuming a neutron star mass of $1.35\Msun$}
\label{fig:ulmass}
\end{figure}

Table~\ref{tab:times} shows the quantity of data for each of the IFO
times and Table~\ref{tab:uls} shows the marginalised 90\% rate upper limits, 
the range (averaged over the time of the search) and the
cumulative luminosity to which the search was 
sensitive above the loudest event for
times when all three \ac{LIGO} detectors were operational.  The first
set of upper limits are those obtained for binaries with non-spinning
components. 
\begin{table}
\center
\begin{tabular}{|c|c|c|c|}
\hline
IFO combination & H1H2L1 & H1L1 & H2L1 \\
\hline
Observation time /$\,\rm yr$ & 0.21 & 0.02 & 0.01 \\  
\hline
\end{tabular}
\caption[12-to-18 observation time]
{The observation time for each of the IFO combinations. The vast
majority of the data was triple time.}
\label{tab:times}
\end{table}
\begin{table}
\center
\begin{tabular}{| c | c | c | c |}
\hline
Component masses /$\, \Msun$ & (1.35,1.35) & (5.0,5.0) & (1.35,5.0) \\
\hline
\hline

$D_{\rm horizon}$ /$\, {\rm Mpc}$ 
& $\sim 30$ & $\sim 100$ & $\sim 60$ \\
\hline
Cumulative luminosity /$\, {{\rm L}_{10}}$ & 490 & 11000 & 2100 \\
\hline
Upper limit (non-spinning) /$\, {{\rm yr}^{-1} {\rm L}_{10}^{-1}}$
& \BNSul & \BBHul & \BHNSul \\
\hline
Upper Limit (spinning) /$\,{{\rm yr}^{-1}} {\rm L}_{10}^{-1}$
& -- & \SBBHul & \SBHNSul \\
\hline
\end{tabular}
\caption[12-to-18 upper limits]
{Overview of results from \ac{BNS}, \ac{BBH} and \ac{NSBH}
searches for the 12-to-18 search~\cite{Abbott:2009qj}, including the
S51YR results.
$D_\text{horizon}$
 is the horizon distance~\ref{eq:range} averaged over the time of the search.
The cumulative luminosity is the luminosity to which the search was
sensitive above the loudest event for times when all three LIGO detectors
were operational and is quoted to two significant figures.
However, the upper limits are the combined results for all three IFO
times.}
\label{tab:uls}
\end{table}
Finally, as the rates for systems containing black holes
vary considerably depending on the mass choice,
Figure~\ref{fig:ulmass} shows the marginalized 90\% rate upper limits
as a function of mass for \ac{BBH} (top) and \ac{NSBH} systems (bottom). 
In the former case, the $90\%$ upper limits on the rates are plotted against
the total mass of the system, whereas for the latter the neutron 
star mass is assumed to be $1.35\Msun$ and the $90\%$ upper limits
are plotted against the black hole component mass.
The mass dependent upper limits were
calculated using only H1H2L1 data since the relatively small amount
of H1L1 and H2L1 data made it difficult to evaluate the cumulative luminosity
in the individual mass bins. Therefore, there is a slight reduction in 
the estimate of the sensitivity when calculating these upper limits and
they will be slightly larger as a result. 

\subsection{12-to-18 upper limits summary}
The upper limits were approximately a factor of three lower than those of
the \ac{S51YR} search. The result
is a significant improvement and was obtained using approximately two
thirds as much data. Such an improvement was possible partly
due to improved detector sensitivity, measured as an increase in the range, and
partly due to improvements in data quality and 
stationarity. Moreover, by analysing the data
in separate months, many of the loudest events were significantly quieter than
the loudest event of the \ac{S51YR} search,
thus increasing the cumulative luminosity of 
the search. The astrophysical estimates for \ac{CBC} rates have been discussed
in Chapter~\ref{cbc}.
The results of the 12-to-18 search are $1$-$2$ orders of magnitude
above the optimistic rates. 

A key factor in the improved
upper limits of the 12-to-18 search is the larger
cumulative luminosities in comparison with the \ac{S51YR} year search, which
had, e.g., a cumulative luminosity of $250\, \rm L_{10}$ for BNS 
systems~\cite{Collaboration:2009tt}.
The difference is a little surprising as both searches
quote a \ac{BNS} horizon distance of~$\sim30\, \rm Mpc$. It is therefore a
useful exercise to verify that the results of the 12-to-18 search
are consistent with what we would expect given the duration of the search,
the range and the loudest events.

We will first estimate the cumulative luminosity using the horizon distance.
We note that the horizon distances given
in Table~\ref{tab:uls} are quoted to one significant figure. In fact
many of the months of the search had a horizon distance of~$\sim33\, \rm Mpc$. 
For a given month, we use that slightly larger range to
approximate a distance, $D_c$, up to which the 
search was sensitive using the \ac{SNR} of the loudest event, $\rho_m$,
\be
D_c = \frac{8}{\rho_m}D_\text{horizon}\, .
\ee
The typical SNR of the loudest event in each month
was $\sim6.5$, thus we find $D_c\sim40\, \rm Mpc$. Using equations (4) and (5) 
of~\cite{:2010cf} we find the cumulative luminosity for \ac{BNS} systems
to be~$\sim460\, \rm L_{10}$, 
which is a reasonable match with the result quoted
in Table~\ref{tab:uls}.

Where the loudest trigger is demonstrably
due to background, the signal likelihood,
$\Lambda$, is equal to zero and the 
calculation of the $90\%$ upper limit on the rates is simplified to
\be
\label{eq:ulpred}
\alpha_{90} = \dfrac{2.303}{C_LT}\, .
\ee
The 12-to-18 search analysed $\sim0.3\, {\rm yr}$ data, 
hence we approximate the 
rate upper limit for \ac{BNS} systems to be
$\sim1.7\times10^{-2}{\rm yr}^{-1}{\rm L}^{-1}_{10}$, which 
is consistent with the results of the search. Using the same reasoning, we
estimate the \ac{BBH} and \ac{NSBH} upper limits to be
$\sim6\times10^{-4}{\rm yr}^{-1}{\rm L}^{-1}_{10}$ 
and
$\sim3\times10^{-3}{\rm yr}^{-1}{\rm L}^{-1}_{10}$ 
respectively.

\section{Search automation: ihope}
The 12-to-18 search used an automated pipeline called ihope
- \emph{``I hope it works''}.
 
\paragraph*{ihope}
The analysis pipeline (see Section~\ref{sec:pipeline})
was run with an executable
called the Heirachical-Inspiral-Pipeline-Exectuable
(\acs{HIPE})~\cite{LAL}. 
\acs{HIPE} will run a single instance of the pipeline when
provided with the GPS start and end time of the search, 
a list of data segments, a cache file containing information of the location
of the data files, a list of times for each category veto and an input file
containing the tuned parameters of the search. However, to 
run a
complete analysis,
\acs{HIPE} must be run many times to generate playground results,
and for all of the
injection runs required for tuning and calculating the search efficiency.

ihope was designed to automate the entire process, enabling a search to be run
just by providing the GPS times and the input options. ihope is under constant
development, but at the time of the 12-to-18 search it did the following:
\begin{enumerate}
\item Downloaded a list of GPS containing information regarding when 
data category vetoes should be applied from a provided server.
\item Generated lists of data segments to be analysed.
\item Set up all the required instances of \acs{HIPE}.
\item Set up instances of other executables to produce tuning and result plots.
\item Created a \ac{DAG} file that allows all of the 
data analysis
jobs to be run in 
parallel using Condor\footnote{A management program for scheduling and 
managing distributive computing tasks.}~\cite{beowulfbook-condor}. 
\end{enumerate}
Automation of the pipeline allowed the analysis to be broken into months with
the confidence that each month was run in the same way, without human error.
Dividing the search into months meant that foreground triggers were compared
to background triggers that better reflected the behaviour of the
interferometers at the time, as opposed to the \ac{S51YR} search
where the entire year
of the search was used to estimate the background. Indeed, the behaviour of the
detectors did vary over the search, which is why the analysis of 
month 4 constrained the upper
limit more than that of month 1 (see Figure~\ref{fig:post}).

\begin{figure}
\centering
\includegraphics[width=14cm]{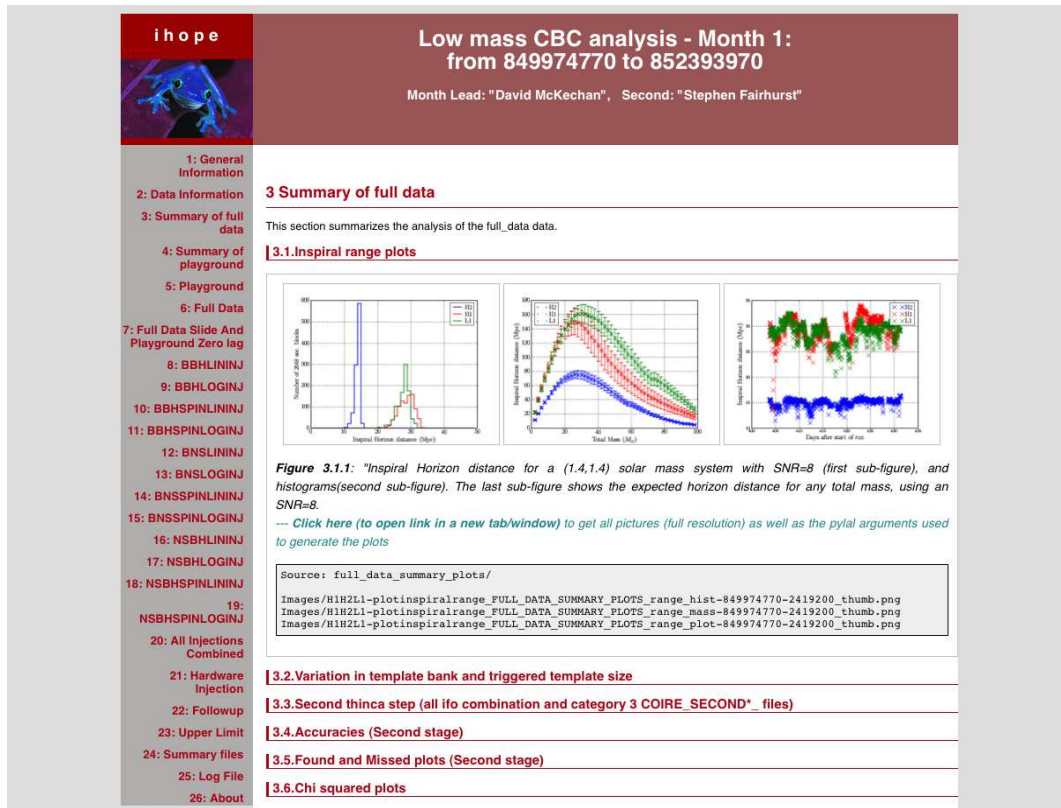}
\caption[ihope results page]{The automated ihope results page for month
1.}
\label{fig:ihopepage}
\end{figure}

\paragraph*{ihope results page}
In order to collate all of the results 
ihope generates
an automated web page
that catalogs all of the relevant information about a run and all of the
tuning and results plots\footnote{At the time the \ac{IFAR} detection
statistic was not included in ihope and the final results
available on the web page were ranked by the effective \ac{SNR}.}.
ihope was first run with playground and injections only to check the tuning of
the parameters. The analysis group then used the web pages to decide whether
the analysis should be un-blinded. Figure~\ref{fig:ihopepage} shows the
ihope results page for month 1. On the left there are links for all 
relevant information, including the injection runs. This page was made 
after the analysis was un-blinded and, therefore, includes the `Full Data' 
result plots.

\section{Concluding remarks
 - is this the best way to search for gravitational waves?}
The title of this chapter, `How to search for gravitational waves from
\acp{CBC}', may lead the reader to believe that he/she is in
possession of an authoritative instruction manual. 
Yet it cannot be claimed that the search method presented here is optimal.
For instance, the core of the search lies with the matched filter,
but that is derived under the assumption that the noise is stationary and 
Gaussian, which is simply not the case with real detector data. How much
of an impact does that have on the search?
We also see the use of binning the template bank into different mass regions
that are then treated as equally likely when
compared with their own backgrounds.
Does that really account for the likelihoods of detection of 
different templates or the various combinations of \ac{IFO} times?
Furthermore, the templates do not include the effects of spin 
or higher order amplitude terms, how does that affect detection efficiencies? 
What are the limitations of the search pipeline presented here and what
can be done to improve it?

\subsection{Gaussian data} Although it is known that real detector
data are non-Gaussian (see Section~\ref{sec:dq} for instance), the
effects have not been quantified before. Recently, 
Robinson et al. have compared a week of \ac{LIGO}
data
(month 4 of the 12-to-18 search)
post category
4 vetoes 
with Gaussian data by running the pipeline on both~\cite{rbs:gaussian}.
The results promise to be interesting and show that although in
many stages of the pipeline, e.g., template reduction, first stage
triggers etc., the \ac{LIGO} data is clearly far from Gaussian
in behaviour, the pipeline performs reasonably well in comparison.
This suggests that if data quality methods, e.g., vetoes and
detector characterisation, are highly robust the non-Gaussian aspect of
real data may not be too much of a hindrance in the search
for gravitational waves. However, month 4 consisted of some of the best
data of S5. 

\subsection{Likelihood}
In the first joint LIGO-Virgo search, the differences between the
detectors meant that for particular combinations of IFO times, some signals
were much more likely to be detected than others. E.g., at the time
Virgo was much more sensitive to BNS systems
than it was to NSBH or BBH systems and this
had to be taken into account when formulating a detection statistic.
For each IFO time, coincident type and mass bin an `efficiency factor'
was calculated that was then compared with the background rate. The final
detection statistic was given as the
`Likelihood' of a trigger based on its background
rate and efficiency factor~\cite{s5vsr1}. The statistic performed better than
\ac{IFAR} and a method of this kind will likely form the basis of future
gravitational wave searches.

\subsection{Template families}
The search we have described 
in this chapter
used \ac{RWF} templates that do not
include spin or higher order amplitude terms, both of which can
have an effect on the detection efficiency and parameter 
estimation. The use of higher order waveforms is discussed
in the following chapter. Upper limits were calculated for spinning
black holes and are not significantly larger than for the non-spinning case,
so it is not clear how much an improvement can be gained by incorporating
spin. 
However, spin has been included in LIGO data analysis
previously~\cite{S3_BCVSpin}
and there are several studies on the inclusion
of spin and its benefits~\cite{Buonanno:2005pt,Buonanno:2004yd,Pan:2003qt}.


\chapter{Higher order waveforms in data analysis}
\label{ampchap}
\lhead{Chapter~\ref{ampchap}
\emph{Higher order waveforms in data analysis}}
\rule{15.7cm}{0.05cm}

In this chapter, we will
study the use of the \ac{FWF} in gravitational wave data analysis.
In Chapter~\ref{howchap} we saw that waveform models may be used as both
injections and templates in the search for gravitational waves from
\acp{CBC}.
The use of the \ac{FWF} for
injections presents no complications and,
indeed, it has been shown that using the \ac{RWF} for injections, 
rather than \ac{FWF}, can significantly overestimate
the \ac{SNR}~\cite{VanDenBroeck:2006qi},
which could arguably lead to artificially
lower upper limits on the rate of \acp{CBC}\footnote{If nature's
gravitational waves are better represented by the FWF then one
would overestimate the search efficiency and consequently the 
cumulative luminosity would also be
overestimated, hence reducing the upper limit.}.
On the other hand, the use of the \ac{FWF} for templates 
when matched filtering is not straightforward.
One can no longer use the matched filter as presented
in (\ref{eq:maxfilter}), since
the maximisation is derived for templates of the form
(\ref{eq:sigindet2}). 

We begin with a brief overview of 
the motivations behind using \ac{FWF} templates in gravitational
wave searches, whilst the rest of the chapter presents in detail
the development and
results of a matched filtering
algorithm that uses \ac{FWF} templates of $0.5$\ac{PN} in 
amplitude.

N.B.: throughout this chapter we shall drop the convention that Latin
indices run over $1,\ldots,3$.

\section{Motivations}
\label{sec:fwfmotiv}
\subsection{Mass reach}
\label{sec:massreach}
When considering inspiral-only waveforms, the mass reach of a detector, 
in terms of the total mass of the binary, 
\emph{may} be determined by the \ac{FLSO} and the detector's 
lower cut-off frequency. For example, \ac{LIGO} is dominated
by seismic noise below $40\,\rm Hz$~\cite{LIGO-E950018-02-E}
and therefore is not considered sensitive
to binary systems with an \ac{FLSO} below that frequency. 
Using the \ac{FLSO}, the theoretical mass reach of \ac{LIGO},
is $\sim100\Msun$; such a system has an \ac{FLSO}
of $\sim43\, \rm Hz$. 
However, when the binary reaches its \ac{ISCO}, the 
higher harmonics contain power at frequencies greater than 
the \ac{FLSO}, albeit at lower amplitudes. 
Nevertheless, including higher harmonics can still be
significant, particularly for advanced detectors. The \ac{FLSO}
scales linearly with the \ac{PN} order, $k$, of the waveform,
\be
f_{LSO} = \left(k + 1\right)f_0(M_T)\, ,
\ee 
where $f_0$ is the \ac{FLSO} of the dominant harmonic. It can be shown
that the detector's mass reach scales in the same manner.
Thus if waveforms of $0.5$\ac{PN} in amplitude
are considered, \ac{LIGO}'s mass reach extends to $\sim150\Msun$; at 
$3$\ac{PN} it theoretically extends to $400\Msun$
\footnote{N.B.: we are considering inspiral-only
waveforms with which it would not be appropriate to study \ac{CBC} systems
of such high mass, as the inspiral stage
would contain only a few cycles in the \ac{LIGO}'s sensitive band;
essentially the template may look like a glitch.
To study high mass systems, \ac{IMR} waveforms, that include the merger and
ringdown of the \ac{CBC}, should be used. Indeed, a recent study indicates
that \ac{IMR} waveforms should be used in data analysis for systems
as low as $12\Msun$~\cite{Buonanno:2009zt}. However, the motivation
that including higher
harmonics extends a detector's mass reach 
also applies to \ac{IMR} waveforms.}.

\begin{figure}
\centering
\psfragfig{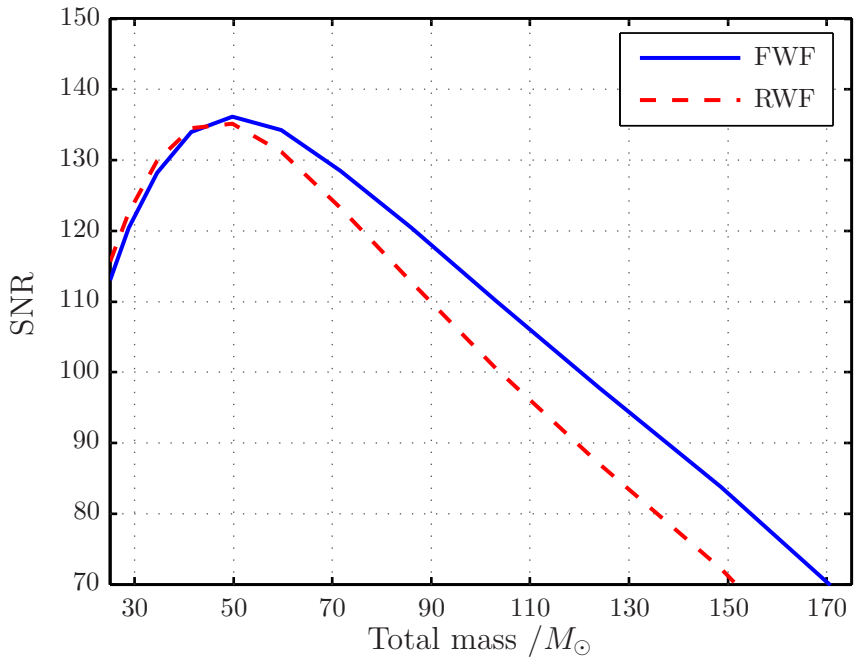}
\caption[Advanced \acs{LIGO} \acs{SNR} vs.~mass]
{\acs{SNR} vs.~mass for both the \acs{RWF} and \ac{FWF} in Advanced \ac{LIGO}
The signals are overhead the detector
at a distance of $100\, \rm Mpc$, with a constant mass ratio of $1:4$
and inclination angle of $45^\circ$.}
\label{fig:adligosnr}
\end{figure}

The expectation value of the \ac{SNR} for a signal in stationary Gaussian 
noise, where the signal and template match exactly, may be calculated as  
\begin{equation}
\label{eq:snrest}
\left<h,h\right> = 4 
\int_{f_{L}}^{f_{ny}}\frac{{|\widetilde{h}(f)|}^2}{S_n(f)} df\, ,
\end{equation}
where 
$f_{ny}$ is the nyquist frequency
and $f_{L}$ is the lower cut-off frequency chosen, such that the contribution
to the \ac{SNR} from frequencies $f<f_{L}$ would be negligible.
Figure~\ref{fig:adligosnr} shows $\left<h,h\right>$, calculated using
the Advanced \ac{LIGO}
\ac{PSD}~\cite{Owen:1998dk}, 
assuming a lower cut-off frequency of $20\, \rm Hz$,
plotted against total mass for both the 
\ac{RWF} and the \ac{FWF} (2\ac{PN}).
The \ac{SNR}s of the two waveforms
agree until $\sim40\Msun$, but thereafter the contribution of
the higher harmonics leads to a larger
\ac{SNR} for the \ac{FWF}. In this example the SNRs are well
above any realistic detection threshold, but, because the SNR scales linearly 
with effective distance, one can chose any value of $\rho_*$ to see
how the mass reach is extended: here, at an SNR of 70,
the mass reach of Advanced LIGO is extended 
from $\sim150\Msun$ to $\sim170\Msun$.
The expected
SNR for the FWF is slightly less than that of the RWF at lighter
values of total mass. This effect is due to the contributions from the 
different amplitude orders in the FWF interfering with each other. As the
total mass increases, the RWF has less power in the detector's sensitive
frequency band. Hence, the
higher harmonics in the FWF lead to a greater SNR.

\subsection{Parameter estimation}
\label{sec:pe}
\begin{figure}
\begin{minipage}[h]{16cm}
\centering
\psfragfig{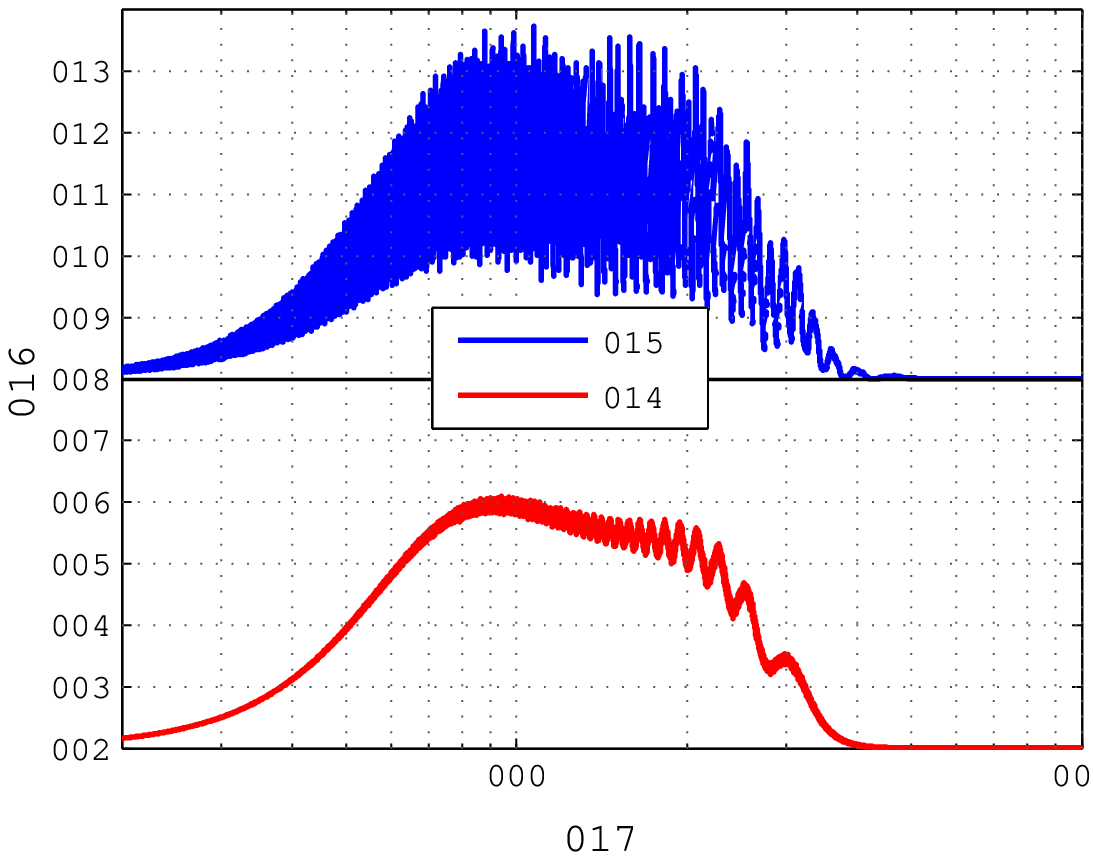}
\end{minipage}
\vspace{1cm}
\begin{minipage}[h]{16cm}
\centering
\psfragfig{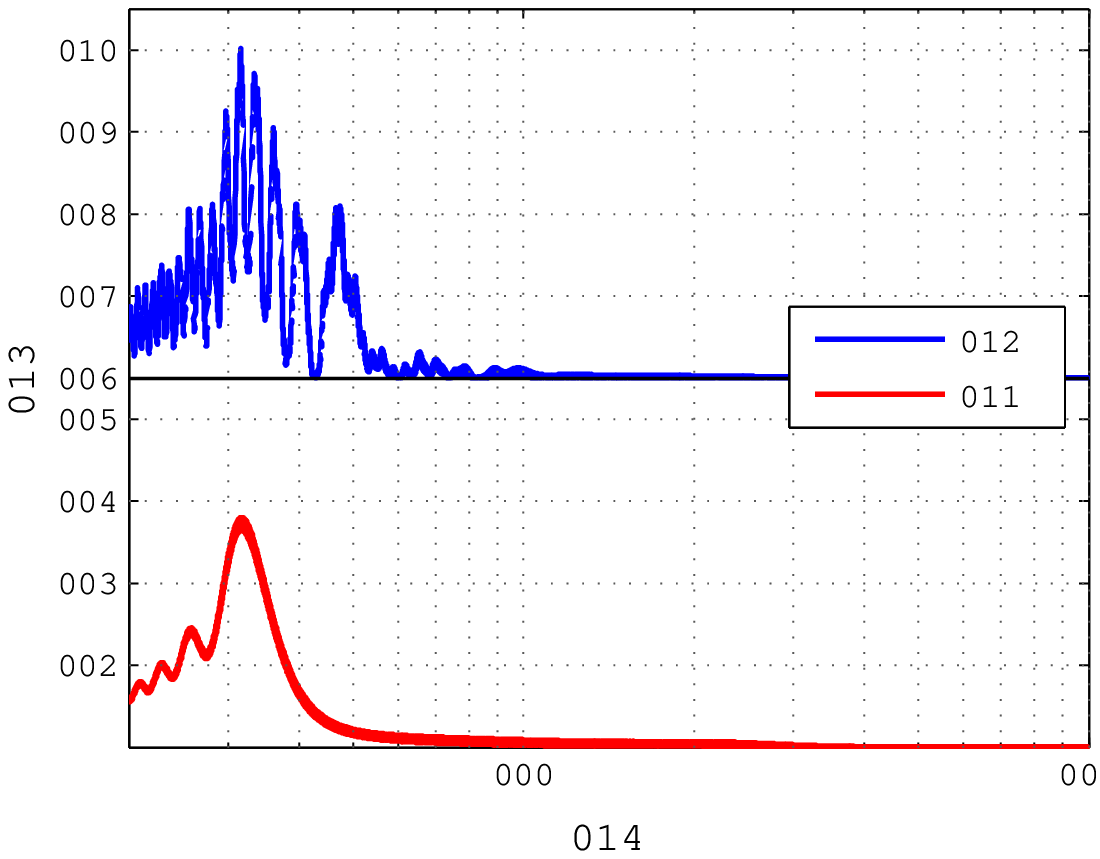}
\end{minipage}
\caption[Observed Spectra]
{A comparison of the observed spectra for \ac{CBC} waveforms overhead an
Advanced \ac{LIGO} detector, with component masses (1,10)$\Msun$ [top] and
(10,100)$\Msun$ [bottom]. The sources are at a distance of $100\, \rm Mpc$ and
have an inclination angle of $45^\circ$.}
\label{fig:spectra}
\end{figure}

When performing a gravitational wave search, one has a family of templates
defined by a set of parameters $\mu_i$. In the case of detection,
the signal will have parameters $\widetilde{\mu_i}$,
which will differ
from the measured parameters, $\hat{\mu_i}$. The measurement error
is caused by differences between the
templates and nature's gravitational waves and the discreteness of the
template bank. Moreover, the presence of noise will cause a measurement
error even if the signal exactly matches one of the templates. 
There have been several studies that compare the ability to recover the 
intrinsic and extrinsic parameters of
\ac{CBC} when using \ac{FWF} templates as opposed to 
\ac{RWF}~\cite{Sintes:1999cg,Sintes:1999ch,Moore:1999zw,Hellings:2002si,
VanDenBroeck:2006ar}. 
The usual approach to estimate the uncertainty in the measured parameters 
is to use the covariance matrix formalism, first applied in this context
by Finn and Chernoff~\cite{FinnChernoff:1993}. At large \ac{SNR}s, the 
measurement errors 
follow a multivariate Gaussian probability
distribution that depends upon the Fisher information matrix,
$\Gamma_{ij}$, which
is the inverse of the covariance matrix, $C_{ij}$. In this formalism, 
the root-mean-square error in the measurement of a 
parameter $\mu_i$ is given by,
\be
\label{eq:pe}
\Delta\mu_i = \sqrt{{\tilde{\mu}_i - \hat{\mu}_i}^2} 
 = \sqrt{C_{ii}} = \sqrt{\Gamma_{ii}^{-1}}\, .
\ee 
As we saw in section~\ref{bankgen}, the 
Fisher information matrix is calculated from the inner products
(\ref{eq:inp}) of the derivatives of the
waveform, $h$, with respect to the parameters.
Hence the Fisher information matrix, and therefore the parameter estimation,
will depend upon the spectra of the waveform and the detector PSD. 
It is useful to plot the
`observed 
spectrum'\footnote{The \ac{SNR} contribution per logarithmic frequency bin
for a given PSD.}~\cite{VanDenBroeck:2006ar}, 
$\mathscr{P}(f)$, which is defined as
\be
\mathscr{P}(f) = 
  \frac{ f{|\widetilde{h}(f)|}^2  }{ S_h(f) }\, ,
\ee
and bears a direct relation to the way that
a waveform is seen by a detector, dependent on the sensitivity and the
waveform itself. Figure~\ref{fig:spectra} shows the observed spectra for the
\ac{RWF} and the \ac{FWF} (2\ac{PN}), 
overhead  Advanced \ac{LIGO} for two different choices of
total mass.
In both cases it is clear that the spectra of the \ac{FWF} contains
more structure, which is due to the interaction of the different harmonics.
This structure leads one to expect an improvement in parameter estimation
under the covariance matrix formalism when using the \ac{FWF}.

Van Den Broeck and Sengupta
calculated measurement errors using (\ref{eq:pe}) for various
intrinsic and extrinsic parameters using the \ac{SPA} \ac{FWF} at
2.5\ac{PN} in amplitude and phase, with promising
results~\cite{VanDenBroeck:2006ar}.
E.g., they found that in Advanced \ac{LIGO} the error in
time-of-coalescence (arrival-time) may reduce by a factor of five
compared to the
\ac{RWF} at low masses and by a much larger factor at high masses. 
Furthermore, the individual
component masses of the binary are expected to be found with errors as low
as a few percent in Advanced \ac{LIGO}, as opposed to being poorly 
determined by the \ac{RWF}.

In the lower plot of
Figure~\ref{fig:spectra}, the \ac{FWF} contains significant power at
frequencies beyond the \ac{FLSO} of the \ac{RWF} ($40\, \rm Hz$), 
demonstrating how
the mass reach of a detector may be extended with the \ac{FWF}.

\section{Developing a filter}
The motivations for using \ac{FWF} templates in the search
for gravitational waves are strong, but thus far the practicalities of 
implementing such a filter have not been addressed.
An algorithm is required that allows one to search
for gravitational waves, as demonstrated in chapter~\ref{howchap}, but
using \ac{FWF} templates. As before, 
the templates will need to be correctly normalised
and maximised over the sky angles etc.   
As we shall see, use of the \ac{FWF} complicates matters somewhat.
For that reason, we will only consider templates of 
0.5\ac{PN} in amplitude, where there are additional harmonics
but no amplitude corrections.

\subsection{Constructing the 0.5PN templates}
The 0.5\ac{PN} waveform, as seen in a detector with response functions
$F_+$ and $F_\times$, will take the form:
\be
\label{eq:tmphoft}
h(t) = 
  \frac{2G\mu x}{c^2R}\left\{F_+\left( H^0_+ + x^{1/2}H^{0.5}_+\right)
  + F_\times\left( H^0_\times + x^{1/2}H^{0.5}_\times\right)\right\}\, ,
\ee
where
\begin{subequations}\begin{align}
  H^{0}_{+} & = a_{+ 2}\cos2\varphi(t)\, ,\\
  H^{0}_{\times} & = a_{\times 2}\sin2\varphi(t)\, ,  \\
  H^{0.5}_{+} & = a_{+ 1}\cos\varphi(t) + a_{+ 3}\cos3\varphi(t)\, , \\
  H^{0.5}_{\times} & 
    = a_{\times 1}\sin\varphi(t)) + a_{\times 3}\sin3\varphi(t)\, ,
\end{align}\end{subequations}
and, recalling that $\Delta$ is a measure of the mass difference 
(\ref{eq:delta}),
\begin{subequations}
\label{eq:amplitudes}
\begin{align}
  a_{+2} & = \left( 1 + \cos^2i\right)\, ,\\
  a_{\times2} & = 2 \cos i\, ,\\
  a_{+1} & = -\Delta \sin i \left( \tfrac{5}{8} 
    + \tfrac{1}{8}\cos^2i \right)\, ,\\
  a_{\times1} & = \Delta \tfrac{3}{4} \sin i\cos i\, ,\\
  a_{+3} & = \Delta \sin i \left( \tfrac{9}{8} 
      + \tfrac{9}{8}\cos^2i\right)\, ,\\
  a_{\times3} & = -\Delta\tfrac{9}{4} \sin i\cos i\, .
\end{align}\end{subequations}
We can simplify (\ref{eq:tmphoft}) with the following relations,
\be
\label{eq:amplitudes2}
A_k := {\left( F^2_+a^2_{+ k} + F^2_\times a^2_{\times k}\right)}^{1/2}
\ee
and
\be
k\psi_k := 
  \tan^{-1}\left( \frac{F_\times a_{\times k}}{F_+a_{+ k}}\right )\, ,
\ee
where $k=1,2,3$ and represents the first three harmonics of the orbital phase.
Let us define
\be
V_{0} = \frac{2G\mu x}{c^2R}\, ,
\ee
and now write
\be
\begin{split}
h(t) & = V_0\biggl[ A_2\left(\right. \cos2\psi_2\cos2\varphi(t) 
  + \sin2\psi_2\sin2\varphi(t) \left.\right)\\
  &\quad + x^{1/2} A_1\left(\right.
    \cos\psi_1\cos\varphi(t) + \sin\psi_1\sin\varphi(t) \left.\right)\\
  &\quad + x^{1/2} A_3\left(\right. 
    \cos3\psi_3\cos3\varphi(t) + \sin3\psi_3\sin3\varphi(t) \left.\right)
  \biggr]\, .
\end{split}
\ee
After using the double angle formulae we have
\be
\label{eq:tmplt}
h(t) = \sum_{k=1}^3 A_k V_{k} \cos k(\varphi(t) - \psi_k) = 
\sum_{k=1}^3 \mathcal{H}_k\, ,
\ee
where $V_{2} = V_{0}$ and  $V_{1} = V_{3} = x^{1/2} V_{0}$. On the right
hand side of (\ref{eq:tmplt}) the template is simply written as the sum of
three terms, ${\mathcal H}_k$, representing the first, second and third harmonic.

We will also find it useful to define the following:
\begin{subequations}\begin{align}
\label{eq:plusamplitudes}
h_{+1} = h_1 & = V_1\, a_{+1}\, \cos\varphi(t)\, ,\\
h_{+2} = h_2 & = V_2\, a_{+2}\, \cos2\varphi(t)\, ,\\
h_{+3} = h_3 & = V_3\, a_{+3}\, \cos3\varphi(t)\, ,\\ 
h_{\times1} = h_4 & = V_1\, a_{\times1}\, \sin\varphi(t)\, ,\\
h_{\times2} = h_5 & = V_2\, a_{\times2}\, \sin2\varphi(t)\, ,\\
h_{\times3} = h_6 & = V_3\, a_{\times3}\, \sin3\varphi(t)\, . 
\end{align}\end{subequations}

The waveform 
(\ref{eq:tmplt}) will be used as a matched filter. It should be
immediately noted that there are three phase offset angles $\psi_{1,2,3}$.
These angles depend upon the sky
position and orientation of the source independently
of one another
and will need to be
maximised over - it is not simply the case that $\psi_2 = 2\psi_1$ etc.

\subsection{Orthonormalisation}
At first glance, one may assume that the three terms of (\ref{eq:tmplt}) could
be matched filtered separately, maximising the three phase offset angles as in
(\ref{eq:maxfilter}), before recombining by taking the sum of squares
of the \ac{SNR}s, a process that neglects any correlation
between the harmonics.

To examine the above idea we will look at one 
polarisation, i.e., $h_{1,2,3}$. Recalling 
that each template must be normalised
such that its overlap is unity, a priori one
might expect that to good accuracy
\be
\label{eq:north}
\left<\bar{h},\bar{h}\right> = \frac{\left<h_1,h_1\right>
+ \left<h_2,h_2\right>
+ \left<h_3,h_3\right>}
{\left<h,h\right>} = 1\, ,
\ee
where the numerator is the output of matched filtering each of 
$h_{1,2,3}$ separately 
and the denominator is the normalisation factor, which
by definition is
\be
\label{eq:north2}
\sigma_h := \left<h,h\right> = 
\left<h_1,h_1\right>
+ \left<h_2,h_2\right>
+ \left<h_3,h_3\right>
+ 2\left<h_1,h_2\right>
+ 2\left<h_1,h_3\right>
+ 2\left<h_2,h_3\right>\, ,
\ee
Let us define the \emph{diagonal} and \emph{cross} terms
\be
\sigma_\perp = \left<h_1,h_1\right>
+ \left<h_2,h_2\right>
+ \left<h_3,h_3\right>\, ,
\ee
and
\be
\sigma_c = 2\left<h_1,h_2\right>
+ 2\left<h_1,h_3\right>
+ 2\left<h_2,h_3\right>\, ,
\ee
respectively, that add to give
\be
\sigma_h = \sigma_\perp + \sigma_c\, .
\ee
Thus  for (\ref{eq:north}) to hold, the cross terms, 
$\sigma_c$, should sum to zero for any choice of
waveform parameters and sky location. 

\begin{figure}
\centering
\psfragfig{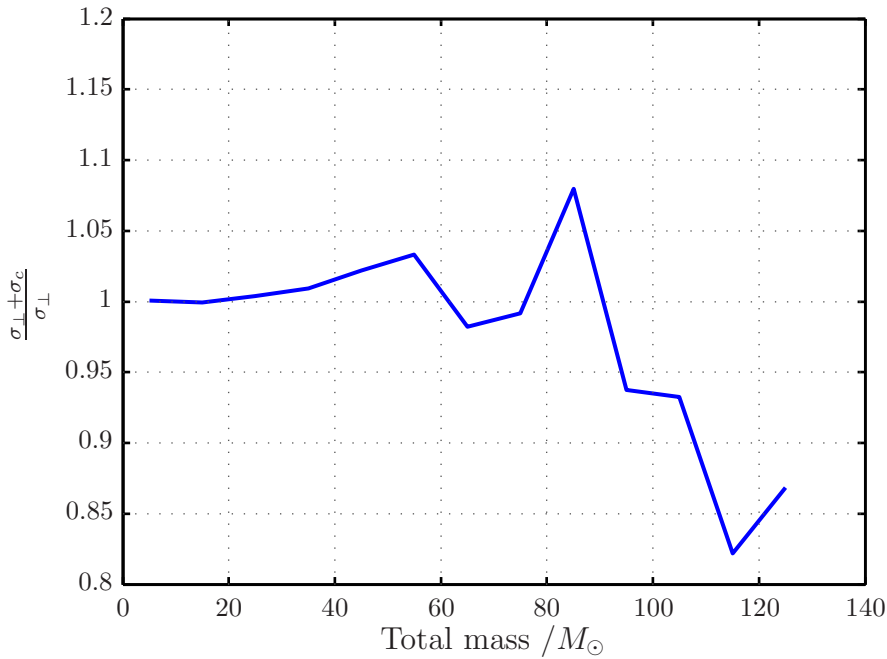}
\caption[Orthogonality check of \acs{FWF} harmonics]
{The ratio of the correct template normalisation, $\sigma_h$, to the
normalisation using only the diagonal terms,
$\sigma_\perp$.
In this example the waveforms are of constant mass ratio 4,
overhead a \acs{LIGO} detector, with an inclination
angle of $45^\circ$.}  
\label{fig:normcheck}
\end{figure}

The effect of assuming that (\ref{eq:north}) is always true, and  
filtering the three 
harmonics separately can be studied by calculating the ratio
$\sigma_h$:$\sigma_\perp$. If the ratio is
always greater than unity, the template would be over-normalised and thus the
\ac{SNR} would be underestimated, which may be acceptable within a certain
tolerance. In fact, if the ratio is always close to unity an overestimation
of \ac{SNR} could also be acceptable. Figure~\ref{fig:normcheck} 
shows $\sigma_h:\sigma_\perp$ 
plotted against the total mass for a single set of parameters,
and reveals that it would most likely not be appropriate to proceed with
the assumption that (\ref{eq:north}) is true - the difference is
$\sim5\%$ at a total mass of $60\Msun$. 
Furthermore, only one choice of mass ratio and inclination angle has been
examined. The relative amplitudes of the first and third harmonics, with 
respect to the
second harmonic, increase with mass ratio 
and also take a maximum at an inclination angle of $90^\circ$. Therefore, one
would expect a greater difference between $\sigma_h$ and $\sigma_\perp$
for different choices of parameters.

It is clear that if the harmonics were orthogonal to each other, i.e.,
\be
\left<\bar{h}_i,\bar{h}_j\right> = \delta^i_j\, ,
\ee
then $\sigma_c=0$ and $\sigma_h :\sigma_\perp=1$. Thus we can 
avoid problems of over/under-normalising the templates by using
any linear transformation that orthogonalises the
components of $h$.
The simplest approach is to use Gram-Schmidt
orthonormalisation. However,  
we will use a matrix to transform the original template $h$ into an 
template $h^\prime$, with orthogonal components.
Such an approach is adopted as the 
transformation matrix becomes useful elsewhere in the filtering
algorithm (see Section~\ref{sec:constraint}).
In later discussions we shall consider this transformation as a coordinate
change from the original template basis to the orthonormal 
basis.

In practice, the template consists of six components as each harmonic
has two polarisations. Hence before orthonormalisation we have
\be
\label{eq:normtmplt}
\left<h,h\right> = 
\sum_{i=1}^6 \sum_{j=1}^6 \alpha_i \alpha_j 
\left<\hat{h}_i,\hat{h}_j\right>\, ,
\ee
where
\be
\hat{h}_k = \frac{h_k}{\sqrt{\left<h_k,h_k\right>}}\, ,
\ee
and $\alpha_i$ are unknown coefficients, which will vary depending upon
the template parameters, such as symmetric mass ratio, sky location and 
the inclination angle.
Let us define a matrix
\be
A_{ij} = \left<\hat{h}_i,\hat{h}_j\right>\, ,
\ee
such that
\be
\label{eq:normmat}
\left<h,h\right> = 
\alpha^{T} A\, \alpha\, . 
\ee
It is obvious that $A_{ij}$ is symmetric and the non-diagonal terms represent
the cross-correlation between the six components of $h$.
One can introduce a matrix, 
$S$, with the properties that it is real and unitary, and that it 
transforms $A_{ij}$ to a diagonal 
matrix:
\be
A^\prime = S^{-1}A\, S\, ,
\ee
allowing us to rewrite (\ref{eq:normmat}) as
\be
\left<h,h\right> = 
\alpha^T\left(SS^{-1}\right) A\, \left(SS^{-1}\right)\alpha\, , 
\ee
or
\be
\label{eq:normat2}
\left<h,h\right> = 
\left(\alpha^TS\right) A^\prime \left(S^{-1}\alpha\right)\, . 
\ee

Let us define a \emph{template vector}
\be
\textup{\textbf{h}} = \left\{
  \hat{h}_1,\hat{h}_2,\hat{h}_3,\hat{h}_4,\hat{h}_5,\hat{h}_6\right\}\, ,
\ee 
allowing us to make the following transformations with the matrix $S$:
\begin{subequations}\begin{align}
\textup{\textbf{h}} \rightarrow
  \textup{\textbf{h}}^\prime & = S^{-1}\textup{\textbf{h}}\, ,\\
\alpha \rightarrow \alpha^\prime & = S^{-1}\alpha\, ,\\
\alpha^T \rightarrow \alpha^{\prime T} & = \alpha^T S\, .
\end{align}\end{subequations}
which gives
\be
\left<h,h\right> = 
\alpha^{\prime T} A^\prime\, \alpha^\prime\, ,
\ee
where we know that $A^\prime_{ij}$ is diagonal and consequently 
there are no \emph{cross} terms on the RHS.
Since the template and its vector are related as
\be
h^\prime = \alpha^{\prime T}\textup{\textbf{h}}^\prime\ , 
\ee
we find
\be
\left<h^\prime,h^\prime\right> = 
\alpha^{\prime T} \left<\textup{\textbf{h}}^\prime,
                      \textup{\textbf{h}}^{\prime T} \right>\alpha^\prime\, .
\ee

All that remains to be done is to find the matrix $S$ used in the 
transformation,
which is straightforward - as $S$ diagonalises $A_{ij}$, it is simply
constructed from the \emph{eigenvectors} of $A_{ij}$.

Thus far, we have a method that orthorgonalises the six components $h_i$ 
that form the template $h$. However, we need to satisfy the normalisation
condition $\left<h^\prime,h^\prime\right>=1$. 
Since
\be
h^\prime = 
\sum_{j=1}^6 \alpha^\prime_j \hat{h^\prime}_j\, ,
\ee
and
\be
\left<h^\prime,h^\prime\right> = 
\sum_{i=1}^6 \sum_{j=1}^6 \alpha^\prime_i \alpha^\prime_j \delta_{ij}\, ,
\ee
it is clear that the templates would be normalised if the coefficients 
satisfied the following
\be
\label{eq:maxcon}
\sum_{j=1}^6 {\alpha^\prime_j}^2 = 1\, .
\ee
Hence (\ref{eq:maxcon}) will be used as a constraint in the maximisation
of the SNR below.

\subsection{Maximisation}
The maximisation of the \ac{SNR} is very different to
that calculated in (\ref{eq:maxfilter}), although it turns out to be
straightforward. The \ac{SNR}, $\rho$, of the orthonormalised template, 
$\bar{h}^\prime$, with some data, $x$, is
\be
\label{eq:snramp}
\rho = \left<x,\bar{h^\prime}\right> = \sum_{j=1}^6 \alpha^\prime_j 
\left<x,\hat{h^\prime}_j\right>\, .
\ee
In order to maximise the \ac{SNR} over the unknown coefficients,
$\alpha^\prime_j$,
of the template, we shall use (\ref{eq:maxcon}) as a constraint. 
It is then convenient to introduce a Lagrange multiplier,
$\lambda$, and maximise the following quantity, $\Lambda$, with respect
to $\alpha^\prime_j$ and $\lambda$,
\be
\label{eq:maximise}
\Lambda = \sum_{j=1}^6 \alpha^\prime_j \left<x,\hat{h^\prime}_j\right>
- \lambda\left[\sum_{k=1}^6 \alpha_k^{\prime2} - 1\right]\, .
\ee
Finding $d\Lambda/d\alpha^\prime_j=0$ and $d\Lambda/d\lambda=0$, yields
\be
\label{eq:lag1}
\left<x,\hat{h^\prime}_j\right> - 2\lambda\alpha^\prime_j = 0\, ,
\ee
and
\be
\label{eq:lag2}
\sum_{k=1}^6 \alpha_k^{\prime2} = 1\, .
\ee
An obvious solution to (\ref{eq:lag1}) and (\ref{eq:lag2}) is
\be
\label{eq:lagsol1}
\alpha^\prime_j = \frac{\left<x,\hat{h^\prime}_j\right>}
{\sqrt{
\sum_{k=1}^6{\left<x,\hat{h^\prime}_k\right>}^2
}}
\ee
and
\be
\label{eq:lagsol2}
\lambda = \frac{1}{2}\sqrt{
\sum_{k=1}^6{\left<x,\hat{h^\prime}_k\right>}^2}\, .
\ee
By substituting (\ref{eq:lagsol1}) in to (\ref{eq:snramp}) we find
\be
\label{eq:maxsnramp}
\rho_\text{max} = 
  \frac{\sum_{j=1}^6\left<x,\hat{h^\prime}_j\right>
        \left<x,\hat{h^\prime}_j\right>} 
       {\sqrt{
        \sum_{k=1}^6{\left<x,\hat{h^\prime}_k\right>}^2}
       }
 = 
  \sqrt{\sum_{l=1}^6{\left<x,\hat{h^\prime}_l\right>}^2}\, .
\ee
The maximisation of the \ac{SNR} is, therefore, simply the sum of squares
of the filtered orthonormal vectors that make up the template.

We will also find it useful to define an \ac{SNR} vector in the 
primed coordinates,
\be
\label{eq:snrvec}
\rho^\prime_i =
  \left<x,\hat{h^\prime}_i\right>\, .
\ee

For proof that the above maximises the SNR, see Section~\ref{sec:maxproof}.

\subsection{Overview of the 0.5PN template filtering algorithm}
~\label{sec:algorithm}
The algorithm that implements the orthonormalisation and filtering is
 described in a stepwise fashion below. N.B.: the 
orthonormalisation transformations are applied to the Fourier transformed
template
components, $\widetilde{h}_{+,\times i}$; 
i.e., all transformations etc. are calculated in
the \ac{FD}, as we are only interested in the \ac{SNR} time series at the
end of the algorithm.
\begin{enumerate}
\item Initially, given the component masses of the template, $m_1$ and $m_2$,
the amplitudes of the three $+$ polarisations (\ref{eq:amplitudes}) are
calculated, along with the phase. The amplitudes of the first and third
harmonics are dependent on the inclination angle, $i$.
However, the amplitudes will all be normalised (it is their evolution that
is important) so one can choose any value, for $i$ other than
$i=0$, so that the first and third harmonic amplitudes are non-zero.
(In the case of equal mass 
templates, the first and third harmonic are correctly set to zero).

\item The $+$ polarisations of the three harmonics, $h_{+k}$, are 
constructed and Fourier transformed, giving $\widetilde{h}_{+k}$.
Before orthogonalising the templates, there exists a
simple relation between the \ac{FT}s of the two polarisations, namely,
\be
\widetilde{h}_{+k} = i\widetilde{h}_{\times k}\, ,
\ee    
allowing all six components of $\widetilde{h}_k$ to be calculated from
the three components $h_{+k}$.
We now have a vector 
$\widetilde{h}=
  \left[\widetilde{h}_{+1},\widetilde{h}_{\times 1},\widetilde{h}_{+2},
  \widetilde{h}_{\times 2},\widetilde{h}_{+3},\widetilde{h}_{\times 3}\right]$
that is to be orthonormalised as described above.

\item As the amplitude of the first and third harmonic may be 
orders of magnitude below the dominant harmonic, one can encounter
problems when computing the transformation matrix. For that reason
the components $\widetilde{h}_i$ are normalised \emph{before} the
transformation matrix is calculated.

\item The matrix $A_{ij}$ is calculated and the transformation matrix, 
$S^{-1}$, is constructed from its eigenvectors. 

\item The transformation 
$\widetilde{\textup{\textbf{h}}}\rightarrow 
  {\widetilde{\textup{\textbf{h}}}}^\prime = 
    S^{-1}\widetilde{\textup{\textbf{h}}}$
yields the orthogonal template
\footnote{The calculations of $A_{ij}, 
A^\prime_{ij}, S, S^1$ and $\bar{\widetilde{h}}^\prime$
are performed using functions from the GNU Scientific Library~\cite{gnu}.}. 

\item Although the template components were normalised \emph{before} the
transformation to alleviate potential numerical issues, the transformed
components 
$\widetilde{h}^\prime_i$ need to be re-normalised to give the
orthonormal template $\bar{\widetilde{h}}^\prime$.
 
\item Finally, the \ac{SNR} is given by (\ref{eq:maxsnramp}).
\end{enumerate}
\clearpage

\section{Initial results}
The filtering algorithm was tested in three ways:
\begin{itemize}
\item Studying the \emph{ambiguity function}, i.e., the overlap of
a single signal with a bank 0.5\ac{PN} templates.
\item Comparing the \emph{overlap} and \emph{faithfulness} (the overlap of
the template with the same parameters as the signal) of the 
0.5\ac{PN} template family with standard \ac{RWF} template families
against a random set of \ac{FWF} (2\ac{PN}) injected signals.
\item Repeating the above study with the signals injected into Gaussian noise
at a fixed \ac{SNR} of 10.
\end{itemize}

To perform the above tests, a template bank was required. In these tests, the
template bank metric was calculated using the \ac{SPA}, as described in
Section~\ref{sec:tmpltbank}, with a minimum match of $0.99$. 
The same metric was used for both the \ac{RWF} and 0.5\ac{PN} templates,
rather than computing a new template bank for the 0.5\ac{PN} templates
\footnote{To compute an optimal metric spacing for the 0.5PN templates
would likely be a complicated task.}.
Using the same metric  provides a good comparison of the
two template families and, in any case, would likely
understate the performance of the 0.5\ac{PN} templates.

We are interested in using the 0.5PN templates as a better, but not an exact,
representation of nature's gravitational waves in comparison to 
RWF templates.
Hence, the injected signals were at a higher order of
2\ac{PN} \emph{in amplitude}.
The TT3 approximant was used for the phase evolution, at 2PN in order, for
the signals 
\emph{and} both the 0.5PN and RWF templates.
A further comparison was also made with RWF templates using
the SPA phase approximant at 2PN.
However, there was negligible difference between the results of the
two \ac{RWF}
models and therefore, on the following pages, only the results
obtained
using the \ac{TT3} approximant are shown.

All \ac{TD} waveforms were tapered using the method to be set out in
Chapter~\ref{winchap} and all of the tests were performed using the
\ac{LIGO} design \ac{PSD}, with a lower cutoff frequency of $40\, \rm Hz$.

\subsection{Ambiguity of the 0.5PN templates}
The ambiguity function measures the overlap of all the templates in
a bank for a given signal, forming a surface in the template parameter
space that should be peaked around
the true value of the signal parameters. The ambiguity function, therefore,
gives an indication of the parameter estimation performance of a template
bank - the sharper the peak, the
more likely the correct parameters will be recovered in the presence of noise.

\begin{figure}[!p]
\hspace{-0.3cm}
\begin{minipage}[h]{7.5cm}
\centering
\psfragfig{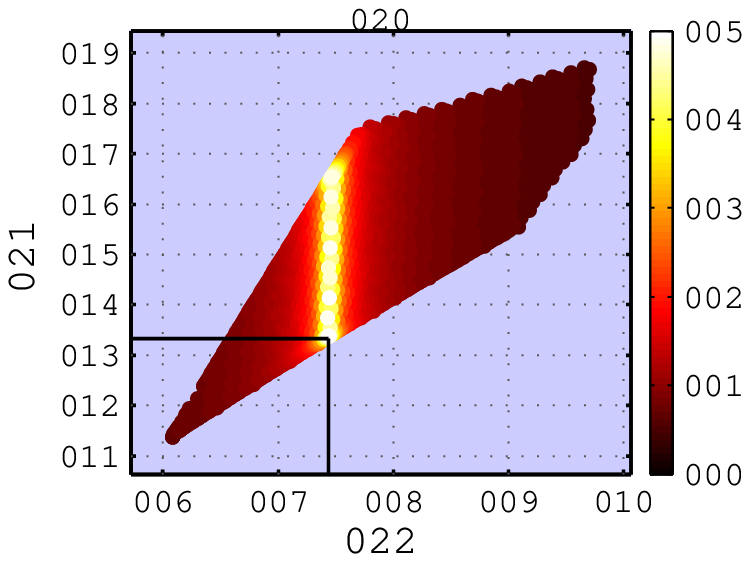}
\end{minipage}
\hspace{0.15cm}
\begin{minipage}[h]{7.5cm}
\centering
\psfragfig{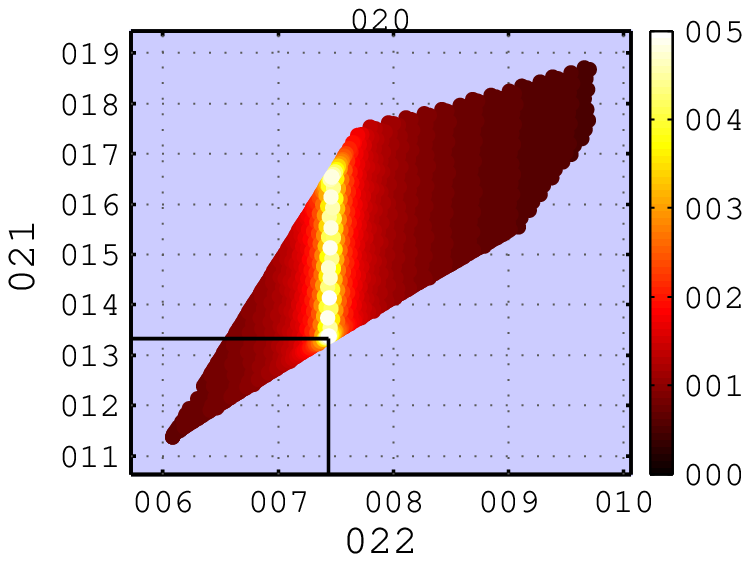}
\end{minipage}
\caption[\acs{BNS} ambiguity function]
{The ambiguity function of a \acs{BNS} signal - $(1.38,1.42)\Msun$ - for the 
surrounding region of the template bank.
The results for the 0.5\ac{PN} 
templates (left) and \ac{RWF} templates (right) are indistinguishable.}
\label{fig:ambbns}
\end{figure}

\begin{figure}[!h]
\hspace{-0.3cm}
\begin{minipage}[h]{7.5cm}
\centering
\psfragfig{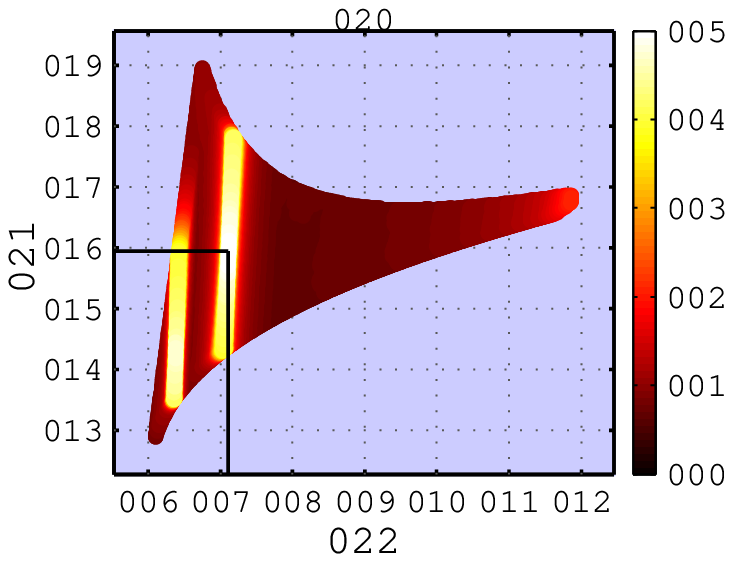}
\end{minipage}
\hspace{0.15cm}
\begin{minipage}[h]{7.5cm}
\centering
\psfragfig{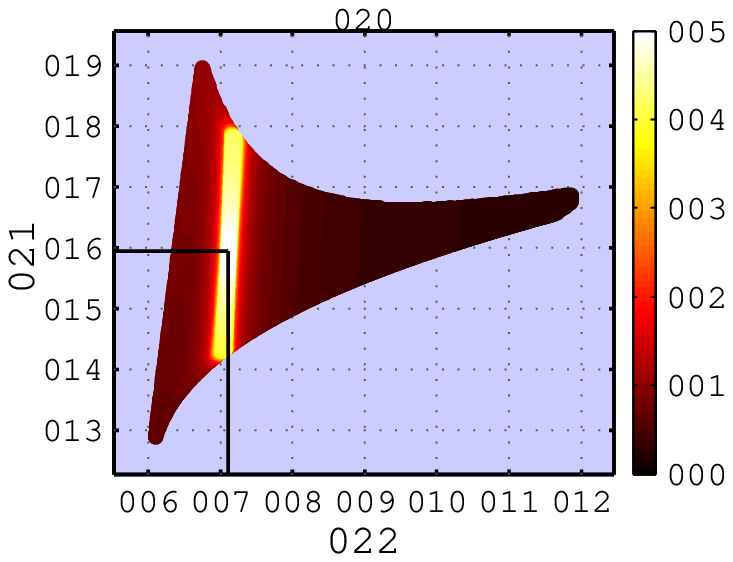}
\end{minipage}
\caption[\acs{NSBH} ambiguity function]
{The ambiguity function of a \acs{NSBH} signal - $(1.4,10)\Msun$. The are
two peaks in the function for the 0.5\ac{PN} templates (left)
as opposed to a single maximum for the \ac{RWF} templates (right).} 
\label{fig:ambnsbh}
\end{figure}

\begin{figure}
\hspace{-0.3cm}
\begin{minipage}[h]{7.5cm}
\centering
\psfragfig{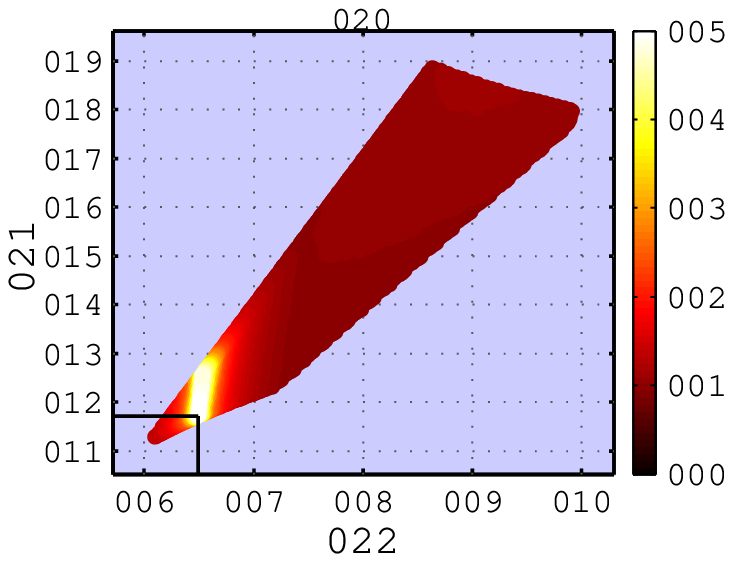}
\end{minipage}
\hspace{0.15cm}
\begin{minipage}[h]{7.5cm}
\centering
\psfragfig{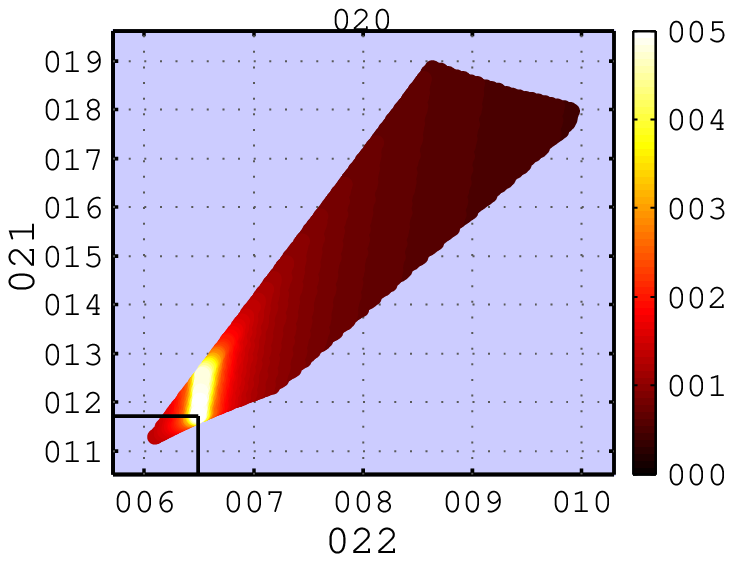}
\end{minipage}
\caption[\acs{BBH} ambiguity function I]
  {The ambiguity function of a \acs{BBH} signal - $(4.8,5.2)\Msun$ - for the 
  surrounding region of the template bank. 
  The 0.5PN templates (left) have a slightly
  larger overlap in the region away from the signal parameters when compared
  to the RWF templates (right).}
\label{fig:ambbbh1}
\end{figure}

\begin{figure}
  \hspace{-0.3cm}
  \begin{minipage}[h]{7.5cm}
    \centering
    \psfragfig{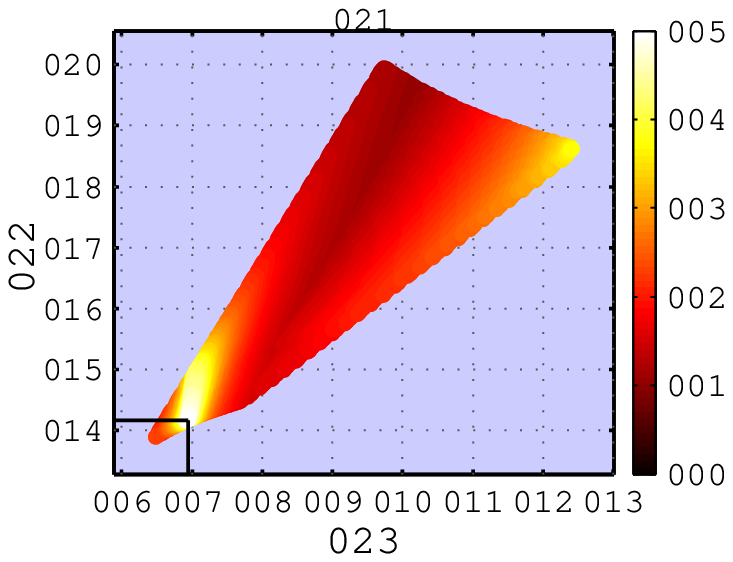}
  \end{minipage}
  \hspace{0.15cm}
  \begin{minipage}[h]{7.5cm}
    \centering
    \psfragfig{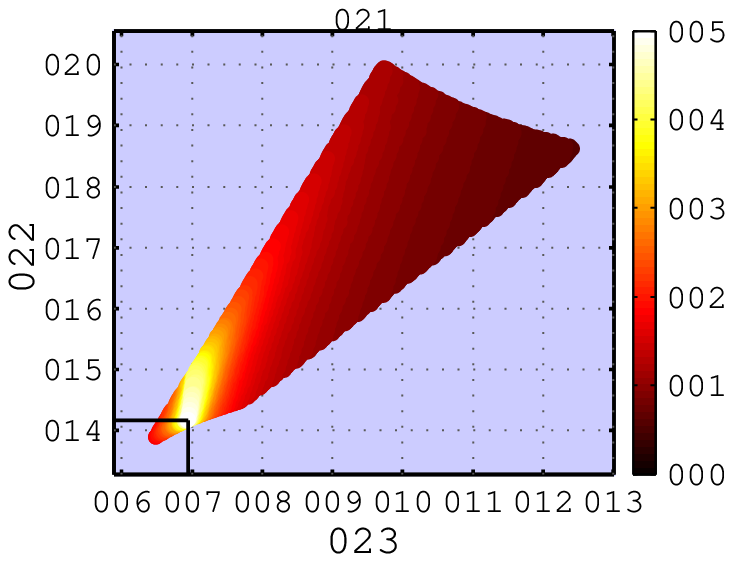}
  \end{minipage}
  \caption[\acs{BBH} ambiguity function II]
  {The ambiguity function of a \acs{BBH} signal - $(9.5,10.5)\Msun$ - for the 
  surrounding region of the template bank. The 0.5PN templates (left) have
  larger overlap in the region away from the signal parameters when compared
  to the RWF templates (right).}
  \label{fig:ambbbh2}
\end{figure}

\begin{figure}
  \hspace{-0.3cm}
  \begin{minipage}[h]{7.5cm}
    \centering
    \psfragfig{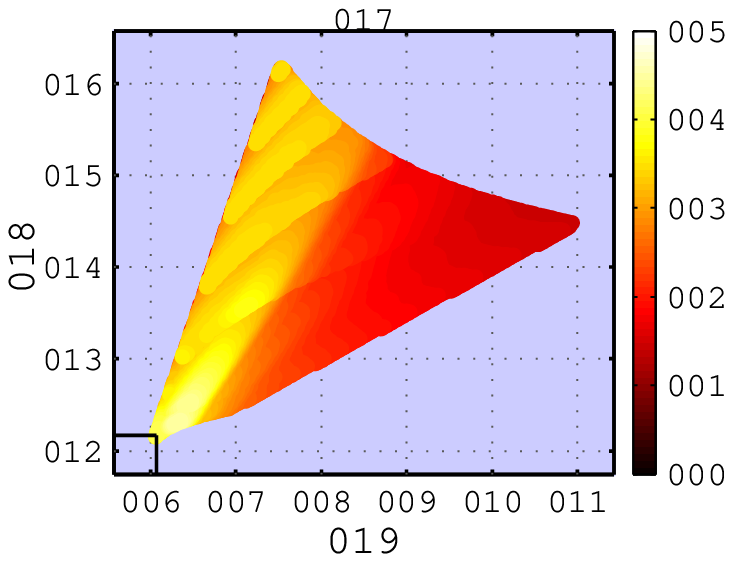}
    \end{minipage}
  \hspace{0.15cm}
  \begin{minipage}[h]{7.5cm}
    \centering
    \psfragfig{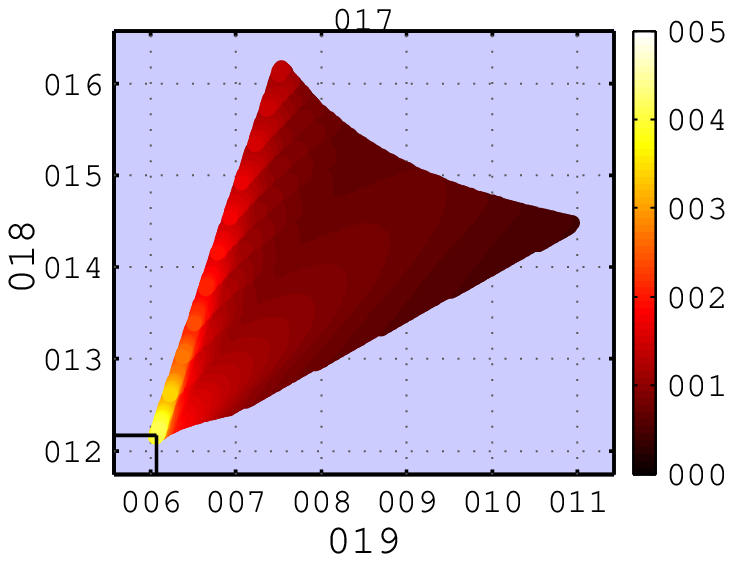}
  \end{minipage}
  \caption[\acs{IMBBH} ambiguity function I]
  {The ambiguity function of a \acs{IMBBH} signal - $(40,60)\Msun$.
  There peak is not well defined
  for the 0.5\ac{PN} templates (left). In contrast, 
  the RWF templates (right), show a sharper peak, although the maximum overlap
  is lower (0.81 compared to 0.90).}
  \label{fig:ambimbbh1}
\end{figure}

\begin{figure}
\hspace{-0.3cm}
\begin{minipage}[h]{7.5cm}
\centering
\psfragfig{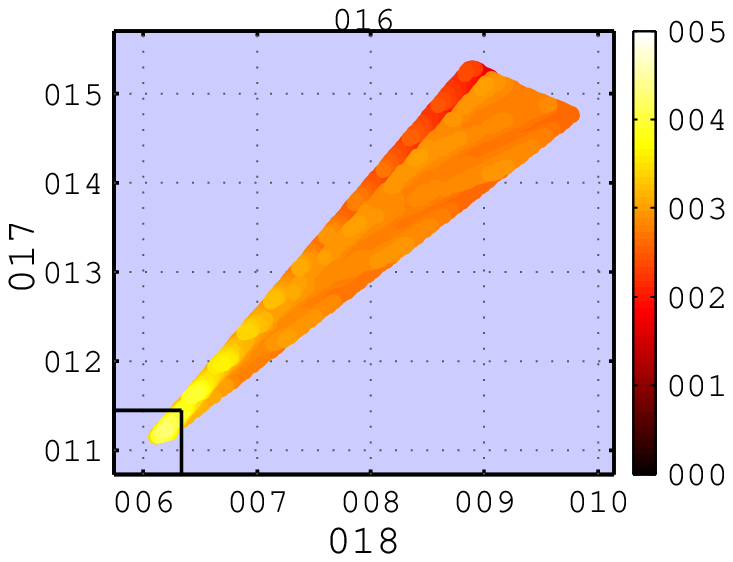}
\end{minipage}
\hspace{0.15cm}
\begin{minipage}[h]{7.5cm}
\centering
\psfragfig{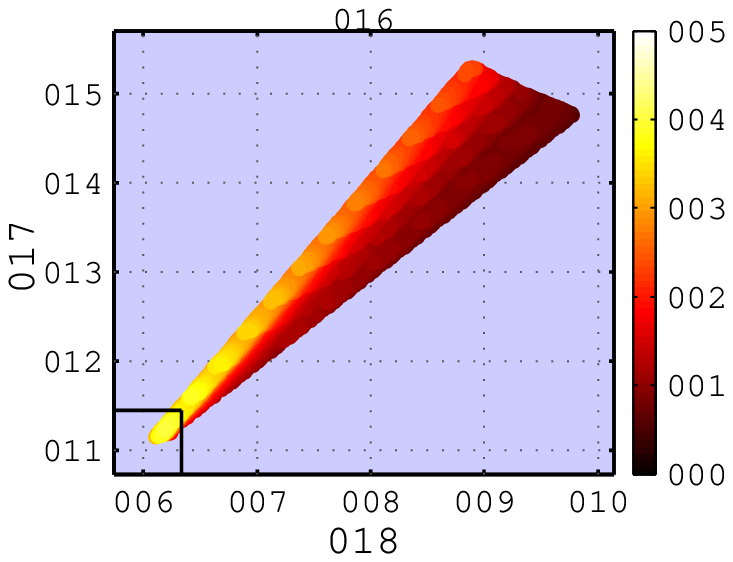}
\end{minipage}
\caption[\acs{IMBBH} ambiguity function II]
  {The ambiguity function of an \acs{IMBBH} of $(10,100)\Msun$.
  There is no well defined maximum
  for the 0.5\ac{PN} templates (left) - the overlap is close to 
  unity for the entire region shown
  As in Figure~\ref{fig:ambimbbh1},
  the RWF templates (right), show a clearer peak,
  but  do not recover the signal as well.}
\label{fig:ambimbbh2}
\end{figure}

Figures~\ref{fig:ambbns} -~\ref{fig:ambimbbh2} show the ambiguity function 
for both the 0.5\ac{PN} and \ac{RWF} (\ac{TT3}) templates for a variety of
signals: \ac{BNS}, \ac{NSBH}, \acp{BBH} and \acp{IMBBH}.
The signal parameters are located where the black lines 
meet on these figures.
 
The ambiguity function of the \ac{NSBH} signal (Figure~\ref{fig:ambnsbh} 
[left])
has \emph{two} peaks for the 0.5\ac{PN} templates, giving an insight
into potential problems with parameter estimation as, in the presence
of noise, it is highly likely that a signal could be detected by a template
at the secondary maximum. A secondary peak is also present to a lesser extent
in the ambiguity function of the $(9.5,10.5)\Msun$ \ac{BBH} signal 
(Figure~\ref{fig:ambbbh2} [left]). 

Figure~\ref{fig:ambimbbh1} and Figure~\ref{fig:ambimbbh2} 
show that for \ac{IMBBH} systems, the ambiguity
functions of the 0.5\ac{PN} templates do not have a well defined peak. 
Indeed, for the $(40,60)\Msun$ system (Figure~\ref{fig:ambimbbh1}), 
the ambiguity function is roughly constant and
conceivably any one of the templates
may recover a signal in the presence of noise.

Issues with the parameter estimation of the 0.5PN templates will
be discussed in far greater detail in Section~\ref{sec:pest}.
\clearpage

\subsection{Overlap and faithfulness}
\begin{figure}
\begin{minipage}[h]{16cm}
\centering
\psfragfig{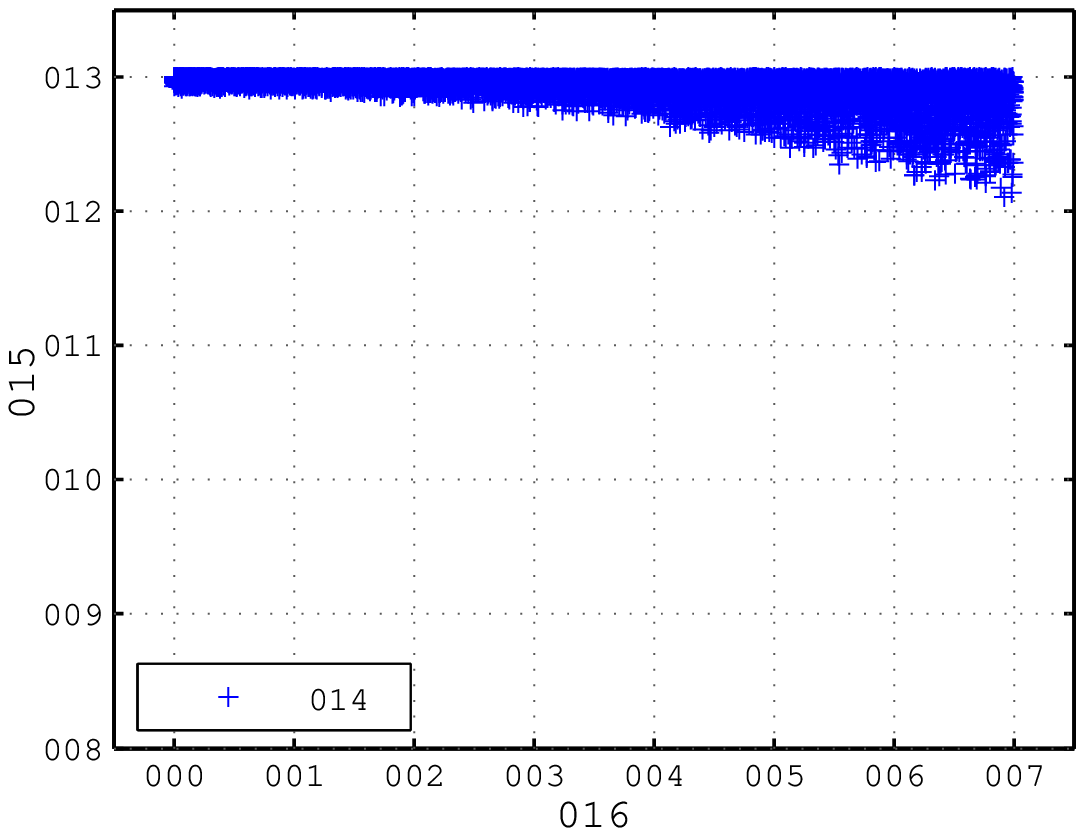}
\end{minipage}
\vspace{1cm}
\begin{minipage}[h]{16cm}
\centering
\psfragfig{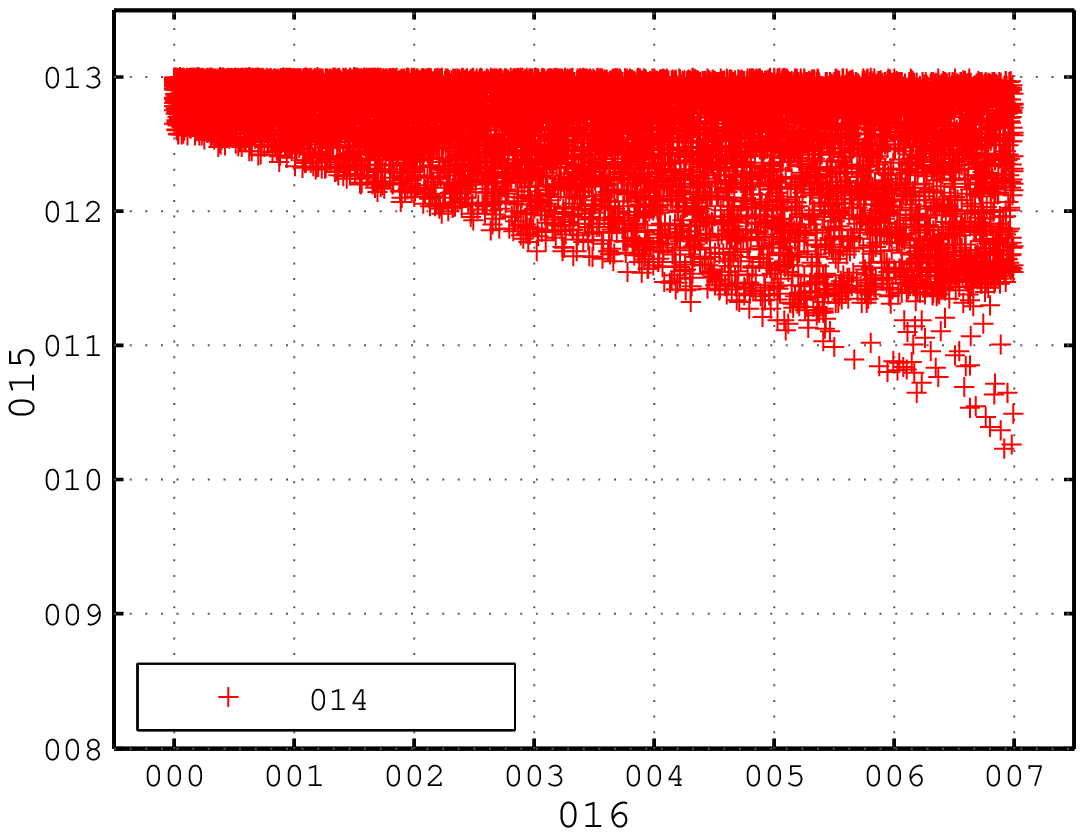}
\end{minipage}
\caption[Overlap with initial 0.5PN templates]
{The maximum overlaps of the 0.5PN and RWF template banks (top and bottom,
 respectively) with \ac{FWF} (2PN) signals.}
\label{fig:ovlpinit}
\end{figure}

\begin{figure}
\begin{minipage}[h]{16cm}
\centering
\psfragfig{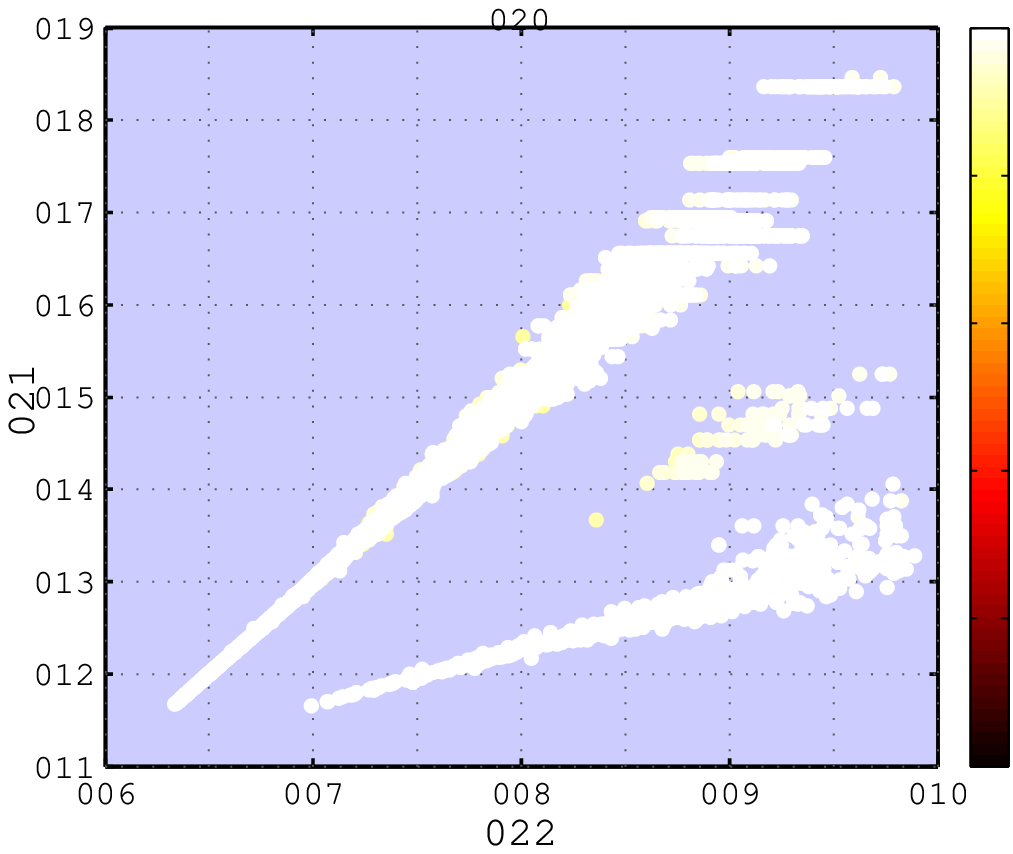}
\end{minipage}
\vspace{1cm}
\begin{minipage}[h]{16cm}
\centering
\psfragfig{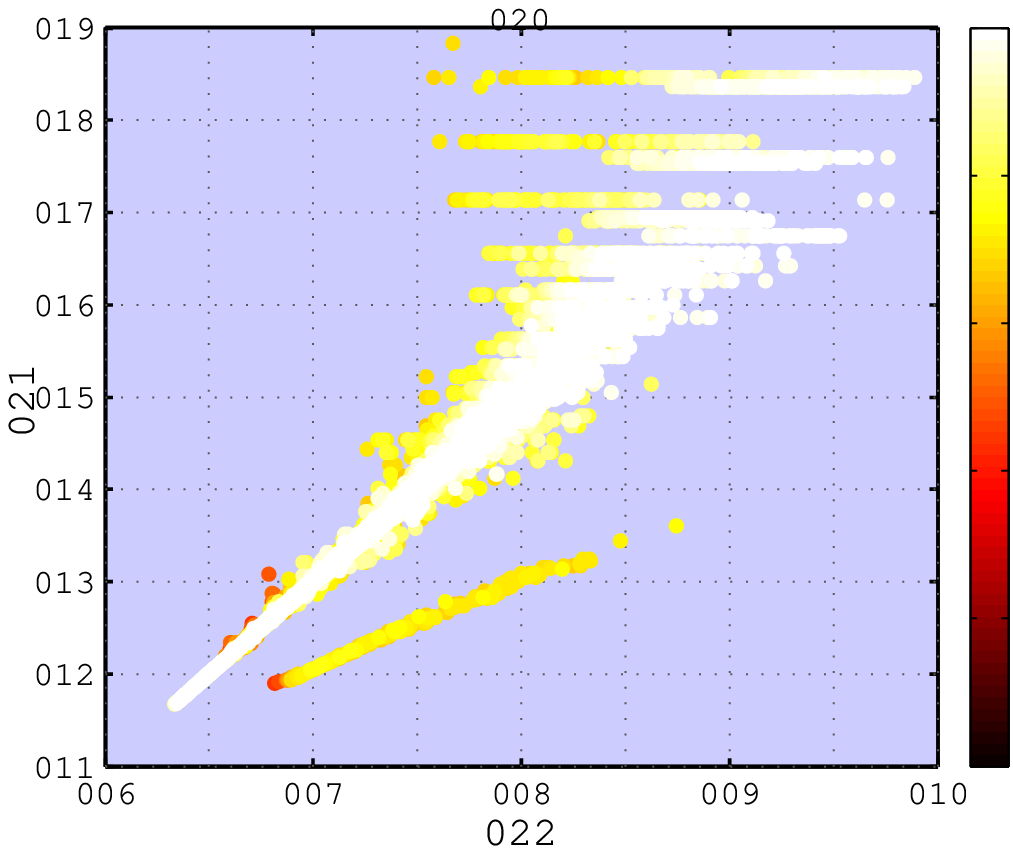}
\end{minipage}
\caption[Recovered chirp mass with initial 0.5PN templates]
{The recovered chirp mass corresponding to the maximum overlaps
(Figure~\ref{fig:ovlpinit}) for the 0.5\acs{PN} templates (top) and the
\acs{RWF} (\acs{TT3}) templates (bottom).}
\label{fig:peinit}
\end{figure}

A Monte-Carlo simulation of 10,000 trials was performed with each trial
calculating the overlap of a random \ac{FWF} (2\ac{PN}) signal
with an entire template bank.
The signals had a total
mass of between 20-90$\Msun$, with a minimum component mass of $1\Msun$;
the template bank was created for the same 
mass range as the signals.
In each trial, the template that had the largest overlap with
the signal was recorded, see Figure~\ref{fig:ovlpinit} (top).
As the template and signal are 
not of the same family, the largest overlap may not occur for the template
of the same parameters. Therefore, the \emph{faithfulness} - the overlap
of the template with the same parameters as the signal
\footnote{Recall that the template parameters are only the component masses;
all other parameters are maximised over.} - was also recorded in each trial,
see Figure~\ref{fig:ovlpinit} (bottom).
\begin{figure}
\begin{minipage}[h]{16cm}
\centering
\psfragfig{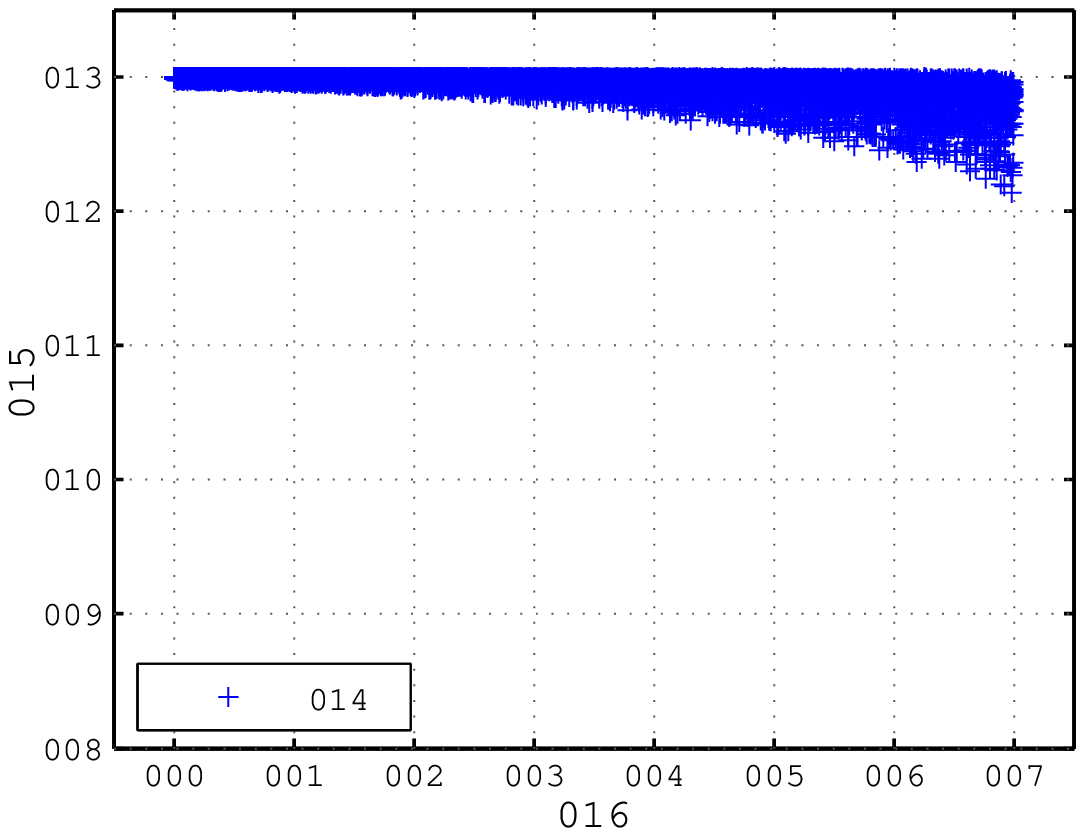}
\end{minipage}
\vspace{1cm}
\begin{minipage}[h]{16cm}
\centering
\psfragfig{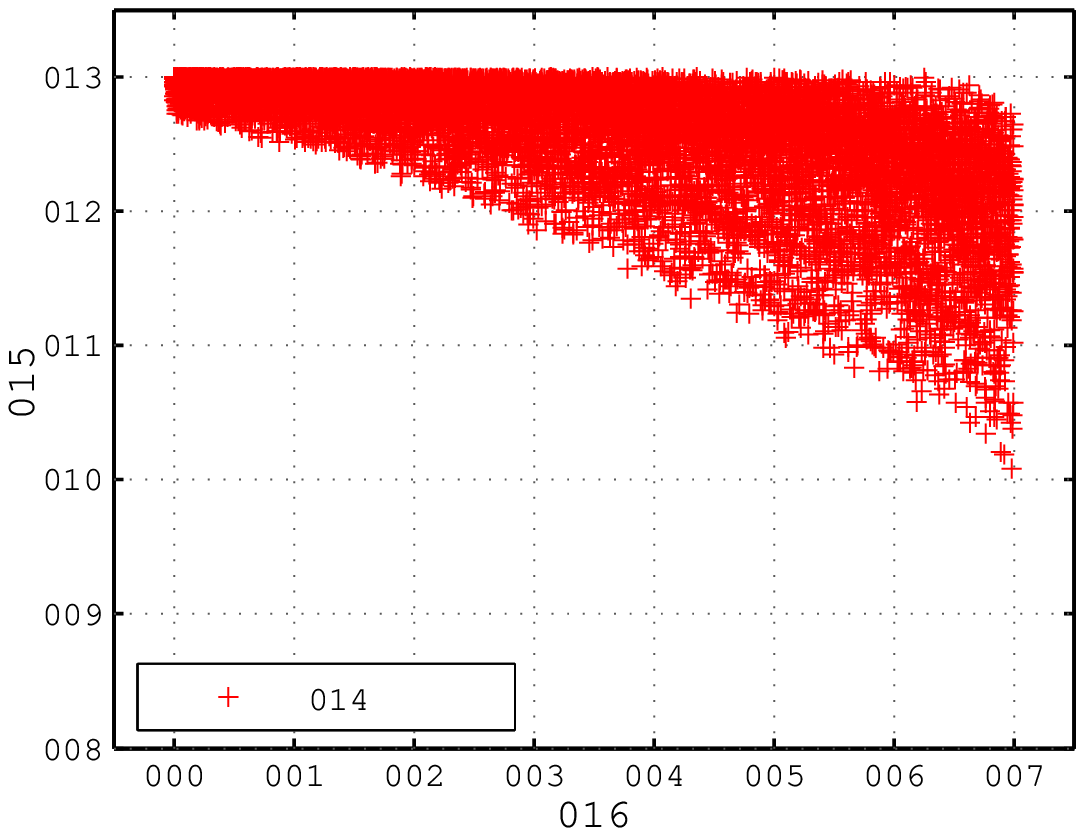}
\end{minipage}
\caption[Faithfulness with initial 0.5PN templates]
{The faithfulness of the 0.5PN and RWF templates (top and bottom,
 respectively) with \ac{FWF} (2PN) signals.}
\label{fig:faithinit}
\end{figure}
The maximum overlap of the 0.5\ac{PN} template bank with the signals 
is consistently higher than that of the \ac{RWF} templates, with the
difference becoming clearer above 40$\Msun$. The faithfulness of the templates
show the same behaviour for both template families.
The results shown in Figure~\ref{fig:ovlpinit} are promising, but 
when the recovered parameters are compared, the 0.5\ac{PN} templates do
not fare so well. Figure~\ref{fig:peinit} shows the recovered chirp mass,
i.e., the chirp mass of the template with the largest overlap, for the
0.5\ac{PN} templates (top) and the \ac{RWF} (\ac{TT3}) templates (bottom).
The parameter estimation is comparable for low values of chirp mass, but for 
values above $1.8\Msun$ three distinguishable
`bands' exist for the 0.5\ac{PN} templates; 
one that recovers the chirp mass well
and two that underestimate the chirp mass. There is also a band that
underestimates the chirp mass when using the RWF templates. 

The parameter estimation problem is discussed investigated further 
in Section~\ref{sec:pest}.
\clearpage

\subsection{Signal and noise simulations}
Signal and noise simulations demonstrate how the 0.5\ac{PN} templates
will perform when analysing `real' detector 
data\footnote{Under the assumption that the
detector noise is Gaussian.}. 
The previous Monte-Carlo simulations were repeated with the same input
parameters,
but with the addition of simulated Gaussian noise, generated as described in
Section~\ref{sec:bhhdev}. The signals were normalised such that they were
injected at a fixed \ac{SNR} of 10
(a reasonable value for detection in \ac{LIGO}).
Fluctuations in the noise affect the power distribution
of the signal (i.e., the observed spectrum) and, therefore, influence
the parameter estimation.

\begin{figure}
  \begin{minipage}[h]{16cm}
    \centering
    \psfragfig{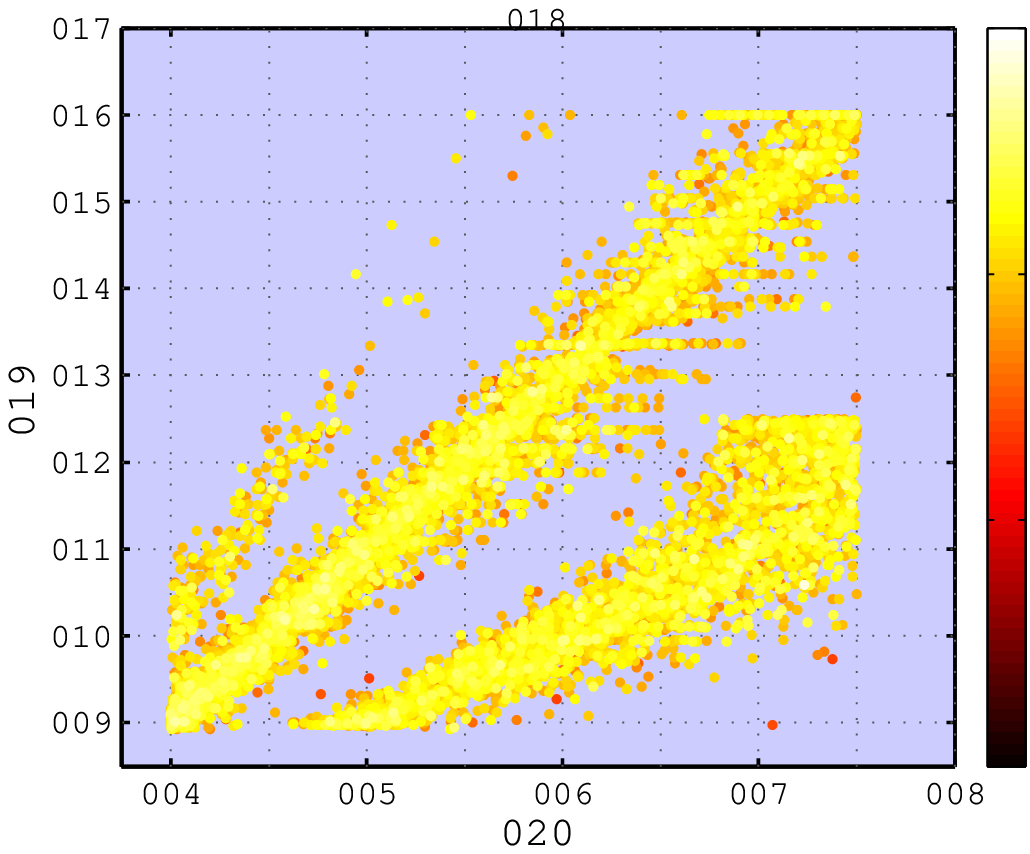}
  \end{minipage}
  \vspace{1cm}
  \begin{minipage}[h]{16cm}
    \centering
    \psfragfig{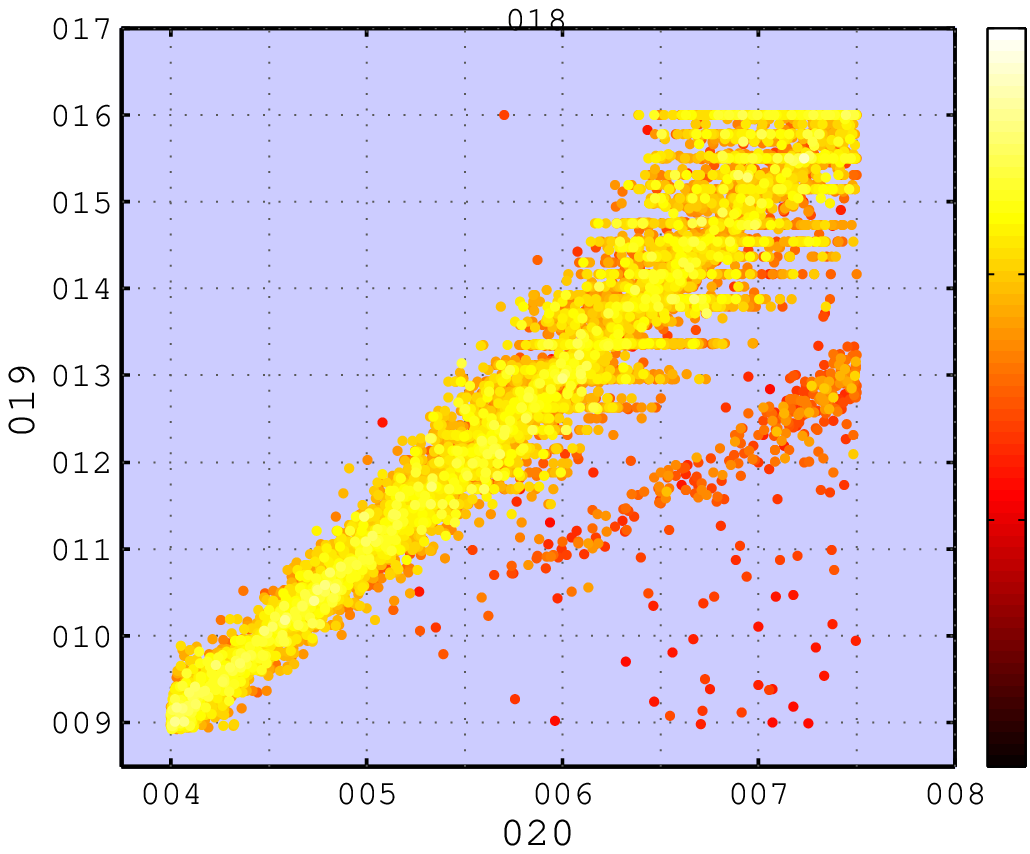}
  \end{minipage}
  \caption[Recovered mass at \acs{SNR} 10 with 0.5\acs{PN} and \acs{RWF}
           templates]
          {The recovered total mass when using the 0.5\ac{PN} templates (top)
           compared to the \acs{RWF} (\acs{TT3}) templates (bottom).
            The colour map 
           shows the value of the recovered \acs{SNR}.}
  \label{fig:snr10_1}
\end{figure}

Figure~\ref{fig:snr10_1} shows the recovered
total mass plotted against injected total mass for both
the 0.5\ac{PN} and \ac{RWF} templates.
The value of the recovered \ac{SNR} is also shown
on the plot as a colour gradient. As expected from the overlap study, 
the 0.5\ac{PN} templates recover the \ac{SNR} well in comparison with
 the \ac{RWF}
templates where the recovered \ac{SNR} is greatly reduced
for the higher mass signals.
However, as seen before, the parameter estimation with the 0.5\ac{PN} templates
is poor, and certainly worse than with the RWF templates.
There are three distinct bands of recovered mass for
the 0.5\ac{PN} templtates with one that overestimates the mass and one that
underestimates it.  Figure~\ref{fig:snr10_1} (top) indicates that a
signal of $40\Msun$ could be recovered by a template of $\sim15$, $40$ or
$60\Msun$ when using the 0.5\ac{PN} templates.

The \ac{RWF} (\ac{TT3}) filter is not useful above 
$\sim50\Msun$ for either recovered \ac{SNR} or recovered mass. N.B.: where
the parameter esitmation is poor with the RWF, it underestimates the mass.
\clearpage

\subsection{The parameter estimation problem}
The band that \emph{underestimates} the recovered mass in 
Figure~\ref{fig:snr10_1} appears to have a clear threshold at template 
total masses
of approximately $50\Msun$, i.e., it ceases to exist above that value, which
corresponds to the mass at which the
first harmonic does not enter the sensitive band and so is not present in the
template. It is, therefore, no great leap of faith to conclude that 
the underestimating band occurs when the dominant harmonic of the 
signal, which is much larger in amplitude than the its first and third
harmonics, is recovered by the first harmonic of the template.
One could then conclude that the band that \emph{overestimates} the 
recovered mass is caused by the template's third harmonic recovering the 
dominant harmonic of the signal. 
Such an effect may be qualitatively understood by comparing the
frequency evolution of the signals. Table~\ref{tab:peflso} shows the
signal \ac{FLSO} for two choices of total mass. In each case the
\ac{FLSO}s of the template harmonics are shown for three different
masses, chosen such that the \ac{FLSO} of one of the harmonics
matches that of signal.
\begin{table}
\centering
\begin{tabular}{|c|c|c|c|c|c|}
\hline
  \multirow{2}{*}{Signal Mass /$\Msun$} &  
  \multirow{2}{*}{Signal \acs{FLSO} /$\rm Hz$} & 
  \multirow{2}{*}{Template Mass /$\Msun$} &  
  \multicolumn{3}{|c|}{Template FLSO /$\rm Hz$} \\
  & & & \multicolumn{3}{|l|}{\ \ 1 \quad\ \ \ 2 \quad\quad\ \ 3} \\
\hline
\hline
  \multirow{3}{*}{ 40 } & 
  \multirow{3}{*}{ $\sim110$ } & 
      20  & $\sim\textbf{110}$ & $\sim220$ & $\sim330$ \\
  & & 40  & $\sim55$ & $\sim\textbf{110}$ & $\sim165$ \\
  & & 60  & $\sim36$ & $\sim73$  & $\sim\textbf{110}$ \\
\hline
  \multirow{3}{*}{60} & 
  \multirow{3}{*}{ $\sim73$ } & 
      30 & $\sim\textbf{73}$ & $\sim147$ & $\sim220$ \\
  & & 60 & $\sim36$ & $\sim\textbf{73}$ & $\sim110$  \\
  & & 90 & $\sim24$ & $\sim48$ & $\sim\textbf{73}$ \\
\hline
\end{tabular}
\caption[Comparison of signal and template \acsp{FLSO}]
{The signal \acs{FLSO} is compared to the \ac{FLSO} of each of the three
harmonics of the 0.5PN template for three different 
choices of template total mass, such that in each case one of the 
template harmonic's FLSO matches that of the signal.}
\label{tab:peflso}
\end{table}
The information in this table goes some  way to explaining why in 
Figure~\ref{fig:snr10_1} it was observed that a signal of $40~\Msun$
could be recovered by a template of approximately $15\Msun$ and $60\Msun$ 

\begin{figure}
  \centering
  \psfragfig{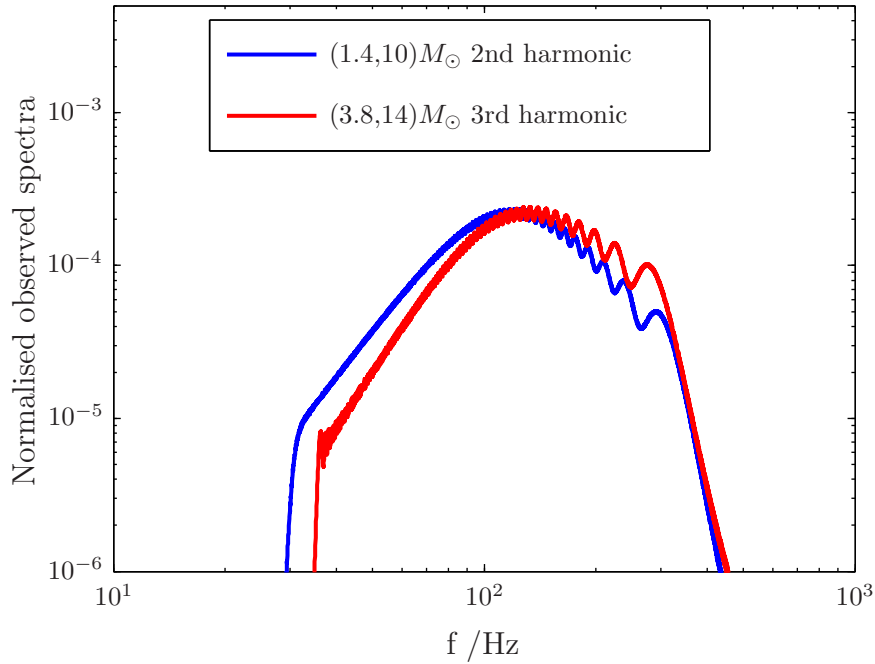}
  \caption[Observed spectra of second and third harmonic]
  {The observed spectra of the second harmonic of a $(1.4,10)\Msun$ waveform
  compared to the third harmonic of a $(3.8,14)\Msun$ waveform. Both are
  normalised to unity.}
  \label{fig:pespec}
\end{figure}

Comparing the \ac{FLSO}s gives some insight into the parameter estimation
problem, but, as we have learned in section~\ref{sec:pe}, the parameter
estimation depends upon the observed spectra. Let us turn our attention
to the ambiguity function. Figure~\ref{fig:ambnsbh} showed that
there are two peaks in the ambiguity function of a $(1.4,10)\Msun$ system
with the 0.5\ac{PN} templates. The second peak occurs at approximately
$(3.8,14)\Msun$, which is an overestimation that we believe is caused by the
third harmonic of the template matching the dominant harmonic of the signal.
Figure~\ref{fig:pespec} shows the observed spectra of the second and third 
harmonic of the aforementioned masses, respectively.

N.B.: the RWF templates also underestimate
the mass for large signal masses, which
is likely to be due to the second (and only) harmonic of the template
matching the third harmonic of the signal.  Therefore, the underestimating
band seen with the 0.5PN templates may also caused by the same effect.

\section{Implementing a constraint on the template harmonics}
\label{sec:constraint}
It should be of no great surprise that the parameter estimation problem exists.
The first and third harmonics are an order $(v/c)$ smaller in amplitude
than the dominant harmonic, yet we place no constraint on the \ac{SNR} 
contributions from each harmonic, allowing the dominant harmonic of the
signal to have a greater correlation with one of the sub-dominant template
harmonics. 
The simplest constraint one can place on the template is to check that the
contributions to the \ac{SNR} from the first and third template
harmonics do not
exceed their expected values when compared with the contribution of the 
dominant harmonic.
In order to implement
such a constraint there are several things to consider.
Firstly, the  prior information regarding the amplitudes of the
harmonics applies only to the non-transformed templates. Once the 
templates are orthonormalised, it no longer makes physical sense to discuss
the components. We will, therefore,
need to implement a constraint based on the physical template,
before the transformation. 
Secondly, the contributions to the template
of the first and third harmonics will
vary, not only with the parameters of the template bank,
but also with the sky angles and inclination angle
that are maximised over when filtering.

\subsection{Relative amplitudes of the harmonics}
Returning to the original 
three component template of (\ref{eq:tmplt}), we wish to find
the maximum possible contribution to the \ac{SNR} from the first and third
harmonics relative to the dominant second harmonic.
Let us define the amplitude of each harmonic by its expected $\text{SNR}^2$,
i.e., $\left<h_i,h_i\right>$.
The relative amplitudes of the harmonics
varies across the template bank parameter space and also 
depends upon the location of
the source and its polarisation.
However, the
first and third harmonic take a maximum
amplitude when the inclination angle is $90^\circ$ (\ref{eq:amplitudes}),
which means the
the emitted graviational waves are 
linearly polarised and the dependence on sky location becomes
irrelevant. For linearly polarised waves, $a_{\times k}=0$ and
(\ref{eq:tmplt}) simplifies to
\be
h(t) = \sum_{k=1}^3 F_+ a_{+k} V_{k} \cos k\varphi(t) \ ,
\ee
with the dependence on the sky location and polarisation angle
contained in a common factor $F_+$. 
The maximum relative amplitudes are, therefore,
given by
\be
R_{12} = 
  \frac{\left<h_1,h_1\right>}{\left<h_2,h_2\right>}
  =
  \frac
  {\left<a_{+1} V_{1} \cos \varphi(t), a_{+1} V_{1} \cos \varphi(t)\right>}
  {\left<a_{+2} V_{2} \cos 2\varphi(t), a_{+2} V_{2} \cos 2\varphi(t)\right>}
\ee
and
\be
R_{32} = 
  \frac{\left<h_3,h_3\right>}{\left<h_2,h_2\right>}
  =
  \frac
  {\left<a_{+3} V_{3} \cos 3\varphi(t), a_{+3} V_{3} \cos 3\varphi(t)\right>}
  {\left<a_{+2} V_{2} \cos 2\varphi(t), a_{+2} V_{2} \cos 2\varphi(t)\right>}
  \, .
\ee

\subsection{Constraining the \acs{SNR}}
As we wish to implement a constraint in the original template basis,
using the relative amplitudes of the
harmonics, we will transform the \ac{SNR} vector, $\rho^\prime_i$, to $\rho_i$
using the inverse of the transformation matrix, which is simply $S$,
\be
\rho^\prime_i \rightarrow \rho_i = S\rho^\prime_i\, .
\ee

Let us consider a \emph{three-component} \ac{SNR} vector, 
$\overset{(3)}{\rho_\mu}$, 
where the first component consists of contributions from both phases
of the first harmonic etc.
Essentially we wish to use the maximum ratio of the relative amplitudes 
of the harmonics as a constraint on the ratio 
$\overset{(3)}{\rho_1}:\overset{(3)}{\rho_2}$.
However, as we are in the
original basis, we must also consider
the cross correlation between the harmonics.
The \ac{SNR} of the first harmonic, $\overset{(3)}{\rho_1}$, is defined as
\be
\overset{(3)}{\rho_1} = 
  \left<h_1,h_1\right> + \beta_{12}\left<h_2,h_2\right> 
  + \beta_{13}\left<h_3,h_3\right>\, ,
\ee
where $\beta_{12}$ and $\beta_{13}$ are unknown correlations of the
first harmonic with the second and third harmonic 
respectively.
The ratio $\overset{(3)}{\rho_1}:\overset{(3)}{\rho_2}$ is then
\be
\label{eq:snrrat}
\frac{\overset{(3)}{\rho_1}}{\overset{(3)}{\rho_2}} = 
  \frac{
        \left<h_1,h_1\right> + \beta_{12}\left<h_2,h_2\right>
        + \beta_{13}\left<h_3,h_3\right>
       }
       {
        \left<h_2,h_2\right> + \beta_{12}\left<h_1,h_1\right>
        + \beta_{23}\left<h_3,h_3\right>
       }\, .
\ee
Let us now divide the top and bottom of the RHS of (\ref{eq:snrrat}) by 
$\left<h_2,h_2\right>$,
\be
\label{eq:snrrat2}
\frac{\overset{(3)}{\rho_1}}{\overset{(3)}{\rho_2}} = 
\frac{
      \frac{\left<h_1,h_1\right>}{\left<h_2,h_2\right>} 
      + \beta_{12}
      + \beta_{13}\frac{\left<h_3,h_3\right>}{\left<h_2,h_2\right>}
     }
     {1   
      + \beta_{12}\frac{\left<h_1,h_1\right>}{\left<h_2,h_2\right>}
      + \beta_{23}\frac{\left<h_3,h_3\right>}{\left<h_2,h_2\right>}
     }\, .
\ee
We wish to place a constraint on the maximum allowed ratio (\ref{eq:snrrat2}).
It is clear that the numerator, $\overset{(3)}{\rho_1}$, has a maximum
value when the
relative amplitudes of the first and third harmonics are at a maximum
and the correlations between the harmonics are also at a maximum.
For simplicity, we will heuristically assume that this also 
gives the maximum value of the ratio 
(\ref{eq:snrrat2}). 
Our constraint on the first harmonic, $C_1$, is, therefore,
\be
C_1 = \frac{\overset{(3)}{\rho_1}}{\overset{(3)}{\rho_2}} = 
\frac{R_{12} + c_{12}  + c_{13}R_{32}}
     {1 + c_{12}R_{12} + c_{23}R_{32}}\, ,
\ee
where $c_{12}$, $c_{13}$ and $c_{23}$ are the maximum correlations between
the harmonics over both phases (i.e., the overlap), calculated using the method
of Damour et al.~\cite{Damour:1998zb} (see \ref{sec:maxcor} for details).
Following the same reasoning we set the constraint on the
third harmonic, $C_3$,
\be
C_3 = \frac{\overset{(3)}{\rho_3}}{\overset{(3)}{\rho_2}} = 
\frac{R_{13} + c_{13}R_{13} + c_{12}}
     {1 + c_{12}R_{12} + c_{23}R_{32}}\, .
\ee

\subsection{Implementation}
For a given template, 
the \emph{six-component} \ac{SNR} vector, $\rho^\prime_i$,
is calculated and transformed to the original coordinates, giving $\rho_i$.
Recall that the first two components of $\rho_i$ relate to the two phases
of the first harmonic etc. The ratio of the \emph{three-component}
\ac{SNR}, $\overset{(3)}{\rho_1}:\overset{(3)}{\rho_2}$, 
is then calculated and the following inequality is
evaluated:
\be
\sqrt{
      \frac{\rho_1^2 + \rho_2^2}
           {\rho_3^2 + \rho_4^2}
    } \leq C_1\, .
\ee
If the inequality is violated, the \ac{SNR} for the template is set to $0$, 
i.e., it is discarded at that point in the \ac{SNR} time series.
Likewise, for the third harmonic
the inequality is
\be
\sqrt{
      \frac{\rho_5^2 + \rho_6^2}
           {\rho_3^2 + \rho_4^2}
    } \leq C_3\, ,
\ee
and the template will be discarded if the inequality is violated.

N.B.: before the inverse transformation can be performed, the components must
be \textit{un-normalised}, 
i.e., the inverse of Step $6$ in Section~\ref{sec:algorithm} must be applied.

\section{Results}
\begin{figure}
  \centering
  \psfragfig{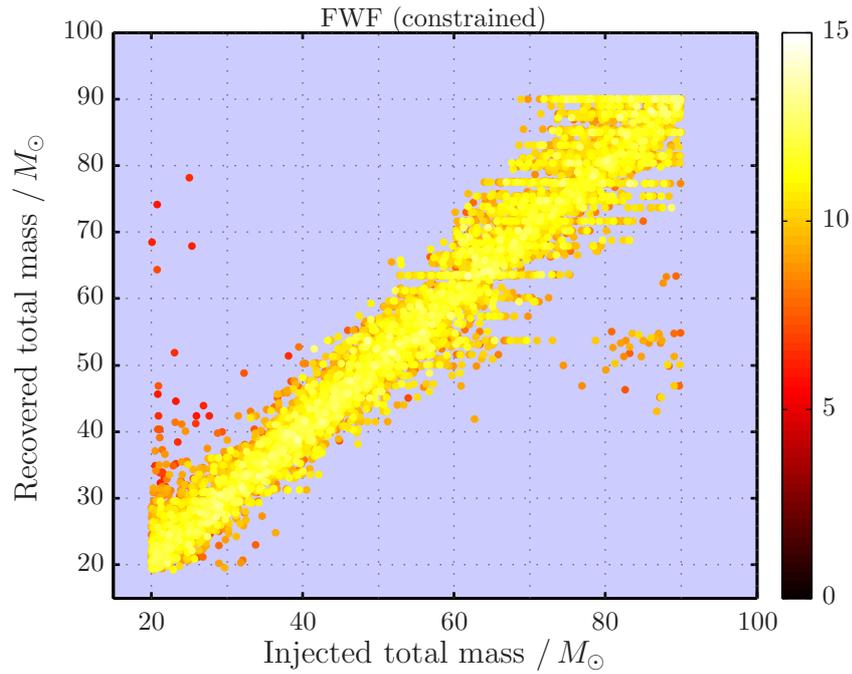}
  \caption[Recovered mass at \acs{SNR} 10 with constrained 0.5\acs{PN} 
           templates]
  {The recovered total mass when using the constrained  0.5\ac{PN} templates.
   The colour map shows the value of the \acs{SNR}.}
  \label{fig:conssnr}
\end{figure}

\begin{figure}
  \centering
  \psfragfig{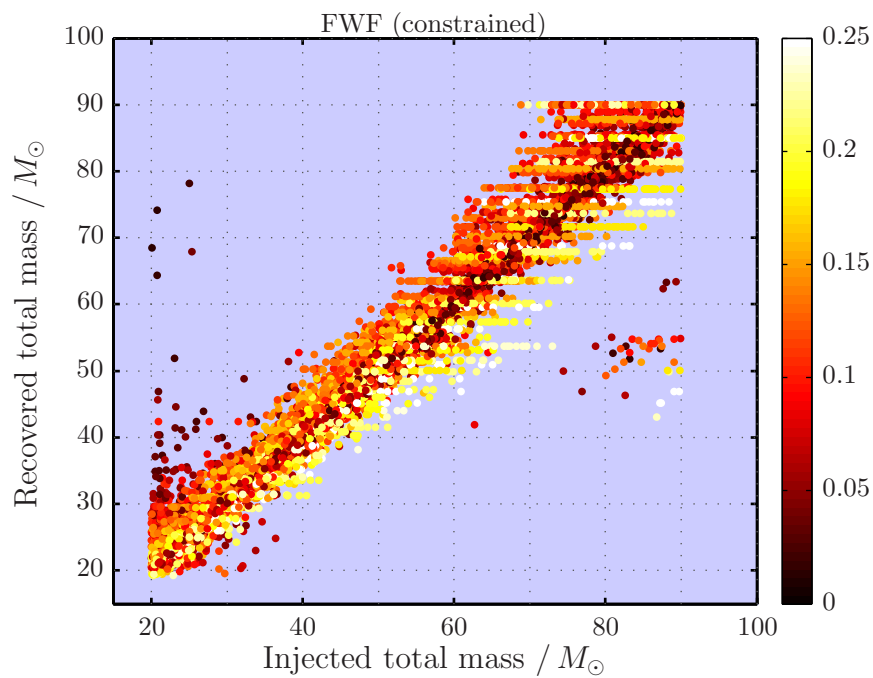}
  \caption[Recovered mass at \acs{SNR} 10 with the 0.5\acs{PN} templates, 
           showing the symmetric mass ratio of the templates]
  {The recovered total mass when using the constrained 0.5\ac{PN} templates
  The colour map shows the value of the template symmetric mass
  ratio. The poor parameter estimation appears where the templates are the 
  least constrained.}
  \label{fig:conseta}
\end{figure}

Figure~\ref{fig:conssnr} shows the recovered mass using the constrained
0.5\ac{PN} templates, repeating the signal and noise simulation as before,
with a fixed \ac{SNR} of $10$.
When compared to the original results (Figure~\ref{fig:snr10_1} [top]), it is
clear that the constraint has improved the 0.5\ac{PN} filtering algorithm.
The bands that overestimate and underestimate the total mass
no longer exist, with the exception of a few templates at low mass that 
overestimate and a small band at high mass that
underestimate. Figure~\ref{fig:conseta} shows the recovered
mass and the symmetric mass ratio of the templates.
Where the parameter estimation is poor the templates are of a small symmetric
mass ratio, which is understandable as these templates have the least 
constrained values of the first and third harmonics. With the constraint 
implemented, the 0.5PN templates outperform the RWF templates at masses above
$\sim40\Msun$.


\section{Parameter estimation study}
\label{sec:pest}
Figure~\ref{fig:conssnr} indicates that parameter estimation is improved with
use of the
constrained
0.5\ac{PN} and is better than that of the RWF templates.
Here a deeper study of the parameter estimation is
presented. Further Monte-Carlo simulations were performed, this time of
1,000 trials for a range of signals masses at different values of
\ac{SNR} and symmetric mass ratio, namely
$M_T = \left[30,45,67.5,80\right]$, $\rho = \left[8,16,64\right]$ 
and $\eta = \left[0.050,0.075,0.111,0.167,0.25\right]$. 
We examine the error in recovered chirp mass as this depends upon both
the total mass and the symmetric mass ratio and present examples of the
most and least improved results\footnote{I.e., the most improved by
qualitatively studying histograms of the errors.}
when using the 0.5\ac{PN} templates in comparison with the RWF templates.

\subsection{Low \acs{SNR}}
At an \ac{SNR} of $\rho=8$ there is little difference between the
constrained 0.5PN templates and the RWF templates for signals of total mass
$30\Msun$ and $45\Msun$, but some improvements are seen for the non-symmetric,
high mass binaries.

\begin{figure}[p]
  \centering
  \psfragfig{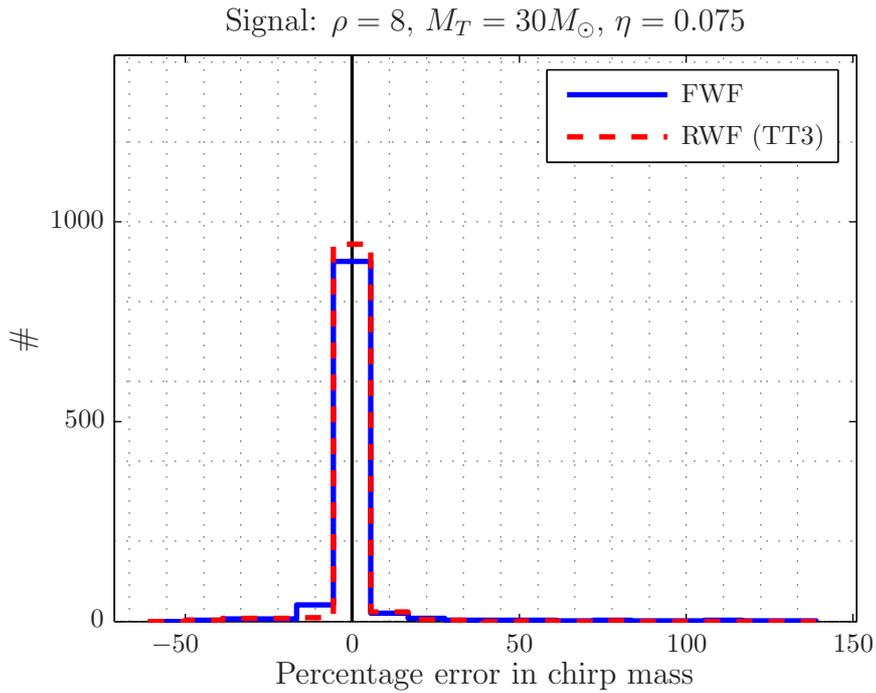}
  \caption[Percentage error in chirp mass. $\rho=8$, $M_T=30\Msun$,
           $\eta=0.075$]
  {Histogram of the percentage error in chirp mass for a low \acs{SNR},
   low mass signal of symmetric mass ratio $\eta=0.075$. There is 
   negligible
   difference between the performance of the two different
   template families.}
  \label{fig:pem30snr8}
\end{figure}

\begin{figure}[h]
  \centering
  \psfragfig{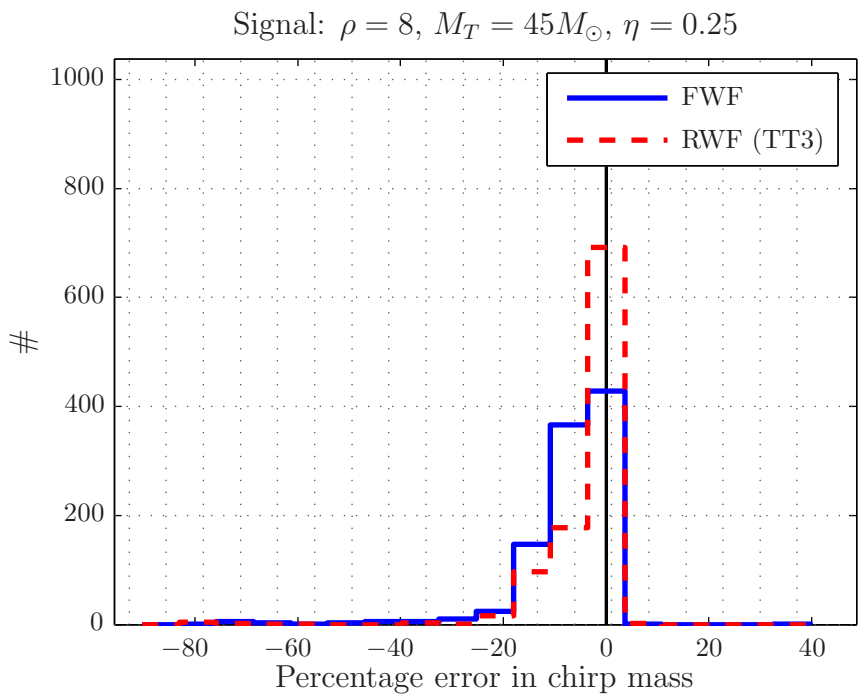}
  \caption[Percentage error in chirp mass. $\rho=8$, $M_T=45\Msun$,
           $\eta=0.25$]
  {Histogram of the percentage error in chirp mass for a low \acs{SNR},
   medium mass signal of equal mass ($\eta=0.25$). The RWF templates perfomr
  better in this case.}
  \label{fig:pem45snr8}
\end{figure}

Figure~\ref{fig:pem30snr8} shows an example where there is little difference
between the 0.5\ac{PN} and \ac{RWF} templates, whereas
Figure~\ref{fig:pem45snr8}, shows an
example where the \ac{RWF} templates outperform the 0.5\ac{PN} templates -
an equal mass signal of total mass $45\Msun$. N.B.: for the other 
choices of symmetric mass ratio using signals of total mass
$45\Msun$, there is little
difference in results between the two template families.

\begin{figure}
  \centering
  \psfragfig{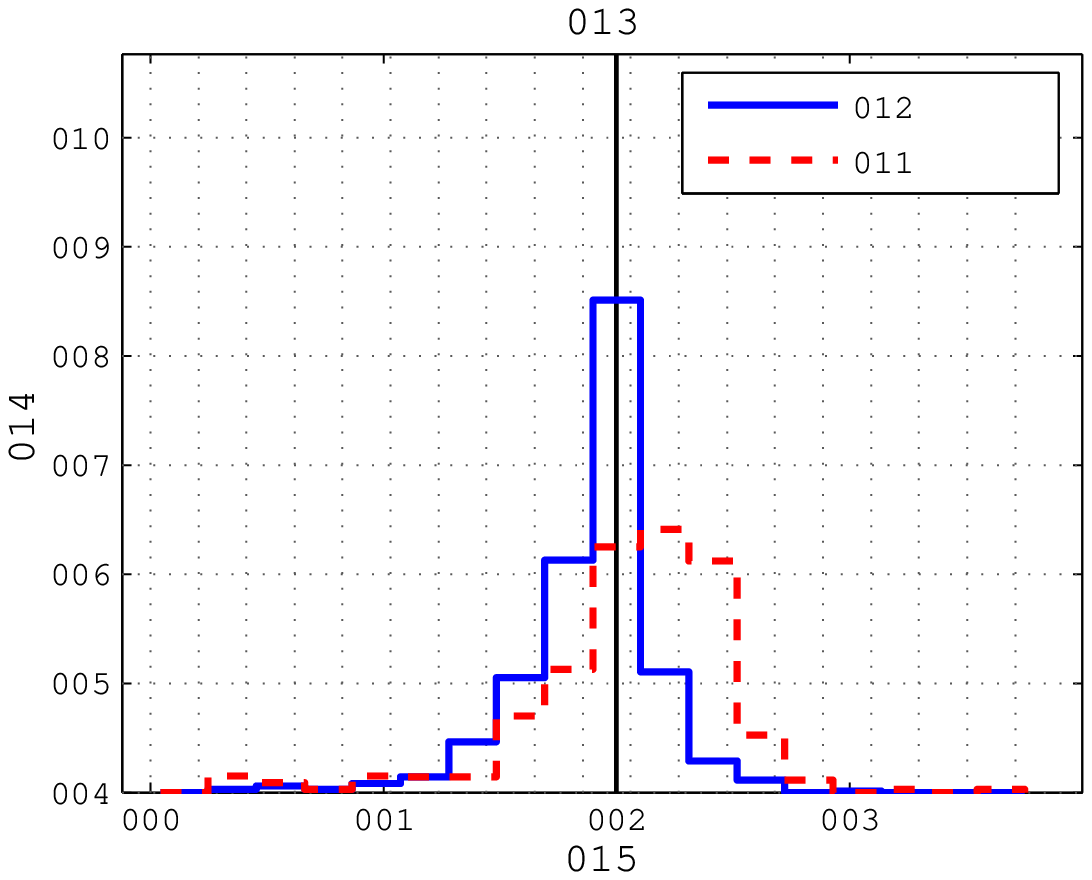}
  \caption[Percentage error in chirp mass. $\rho=8$, $M_T=67.5\Msun$,
           $\eta=0.167$]
  {Histogram of the pecentage error in chirp mass for a low \acs{SNR},
   high mass signal of symmetric mass ratio $\eta=0.167$.}
  \label{fig:pem67.5snr8}
\end{figure}

\begin{figure}
  \centering
  \psfragfig{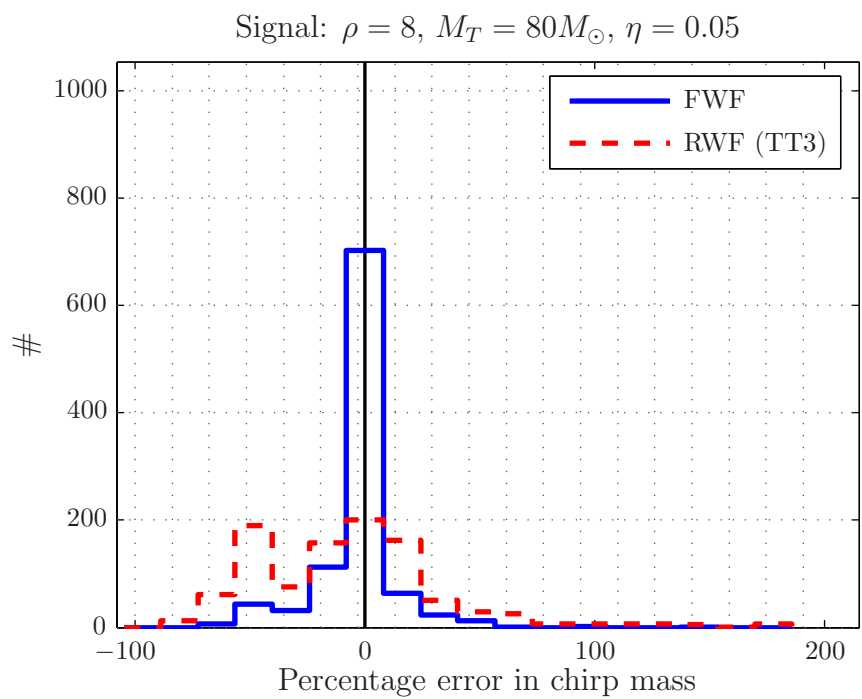}
  \caption[Percentage error in chirp mass. $\rho=8$, $M_T=80\Msun$,
           $\eta=0.05$]
  {Histogram of the percentage error in chirp mass for a low \acs{SNR},
   high mass signal of symmetric mass ratio ($\eta=0.05$).}
  \label{fig:pem80snr8}
\end{figure}

As one would expect from the results shown in Figure~\ref{fig:conssnr}, the
greatest improvements in parameter estimation occur for the high mass
signals where the 0.5\ac{PN} templates clearly 
outperform the \ac{RWF} templates - see Figure~\ref{fig:pem67.5snr8} and 
Figure~\ref{fig:pem80snr8}. The parameter estimation is improvemed 
for all but the equal mass binaries where both where the paremeter
estimation is similar.
\clearpage

\subsection{Increased SNR}
At an \ac{SNR} of $\rho=16$ there is little difference between the two
template families for signals of total masses $30\Msun$ and $45\Msun$. Indeed,
the errors are much smaller in general, due to the increase in \ac{SNR}, 
which is noticeable by comparing the $x$-axis limits of the figures in
this Section with that of the previous Section. 

\begin{figure}[p]
  \centering
  \psfragfig{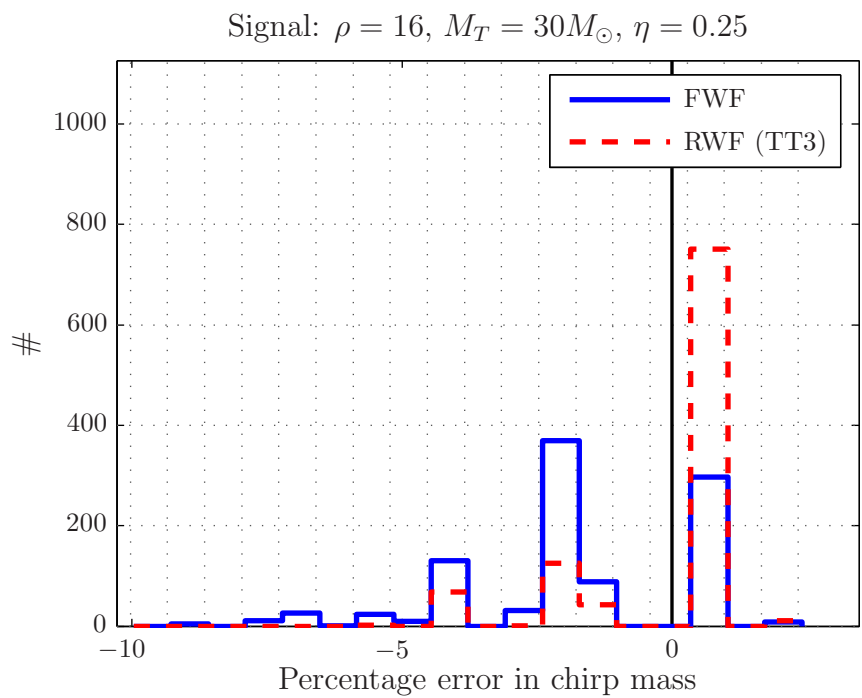}
  \caption[Percentage error in chirp mass. $\rho=16$, $M_T=30\Msun$,
           $\eta=0.25$]
  {Histogram of the percentage error in chirp mass for an intermediate
   \acs{SNR},
   low mass signal of equal mass ($\eta=0.25)$. All
   errors are within $\sim10\%$.}
  \label{fig:pem30snr16}
\end{figure}

\begin{figure}[!h]
  \centering
  \psfragfig{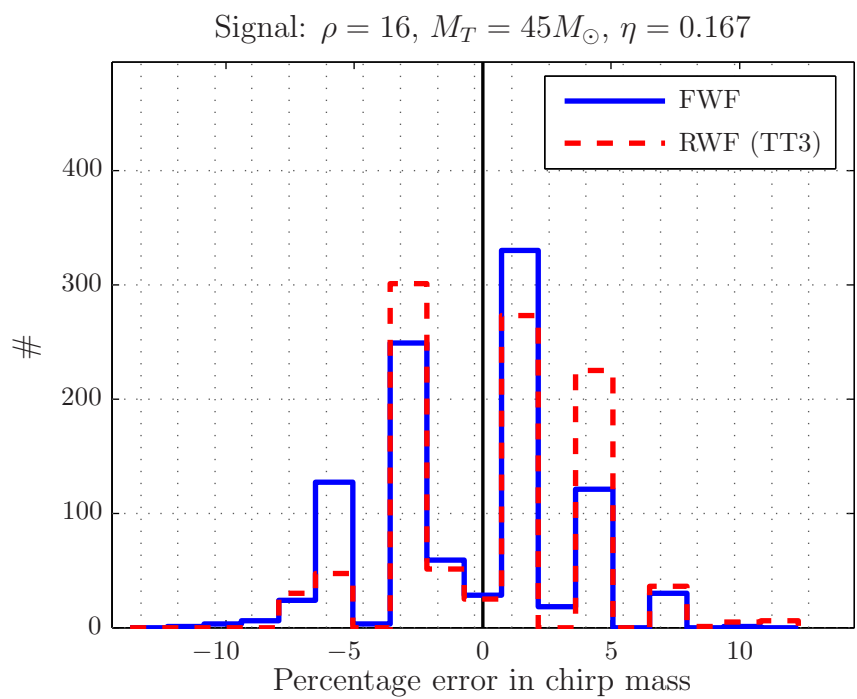}
  \caption[Percentage error in chirp mass. $\rho=16$, $M_T=45\Msun$,
           $\eta=0.167$]
  {Histogram of the percentage error in chirp mass for an intermediate
   \acs{SNR},
   medium mass signal of symmetric mass ratio $\eta=0.167$.
   All errors are within $\sim10\%$.}
  \label{fig:pem45snr16}
\end{figure}

Figure~\ref{fig:pem30snr16} and Figure~\ref{fig:pem45snr16} 
show casses where
there is little difference between the two template families. 
As with the low SNR case, the 
0.5PN templates offers little improvement in the parameter estimation
of equal mass signals.

\begin{figure}
  \centering
  \psfragfig{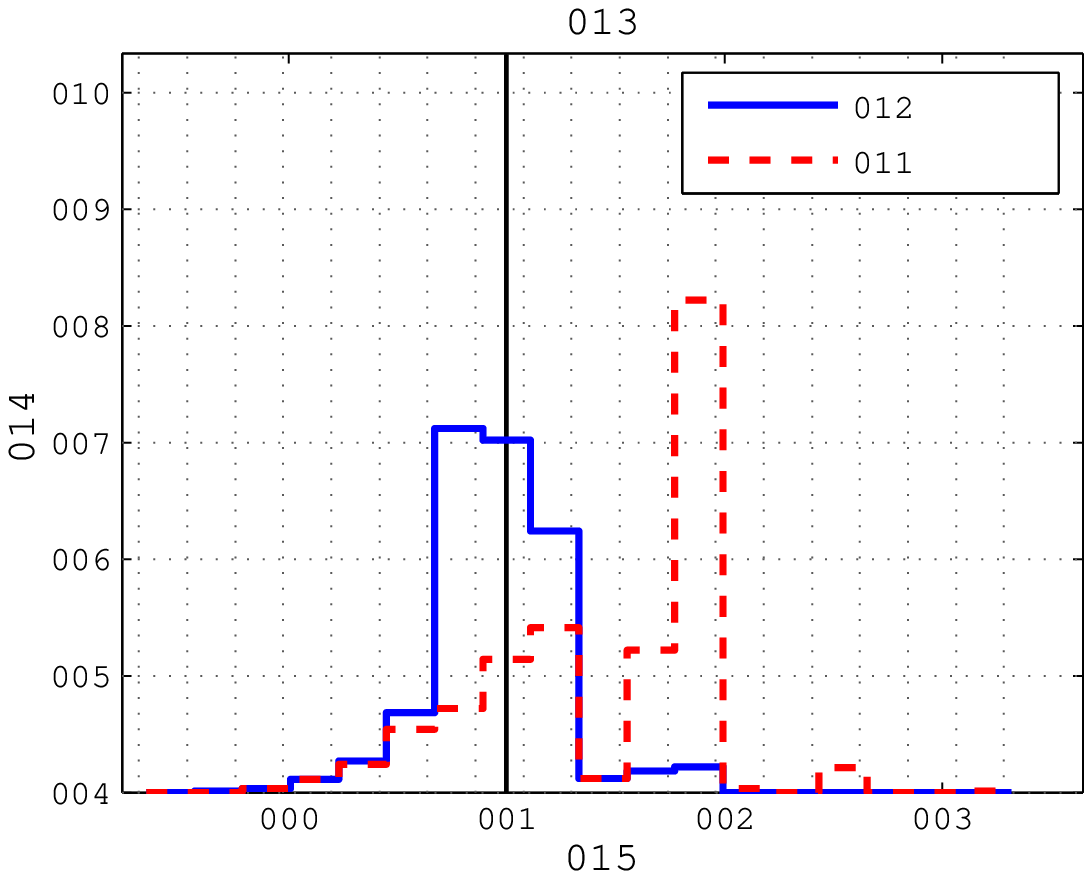}
  \caption[Percentage error in chirp mass. $\rho=16$, $M_T=67.5\Msun$,
           $\eta=0.167$]
  {Histogram of the percentage error in chirp mass for an intermediate
   \acs{SNR},
   high mass signal of symmetric mass ratio $\eta=0.167$.
   The 0.5\ac{PN} templates peak at zero error, whereas the
   the RWF templates show a bias.}
  \label{fig:pem67.5snr16}
\end{figure}

\begin{figure}
  \centering
  \psfragfig{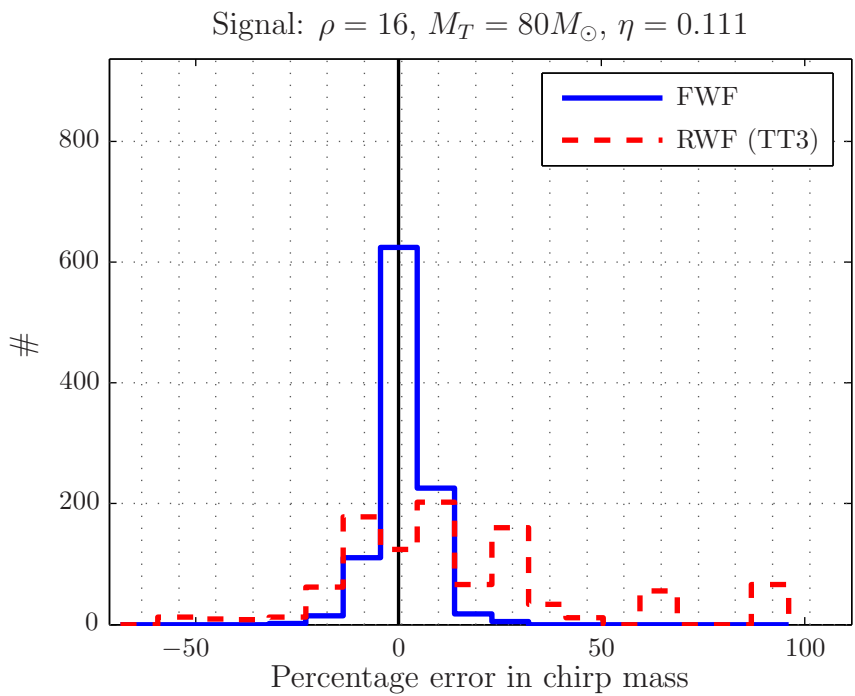}
  \caption[Percentage error in chirp mass. $\rho=16$, $M_T=80\Msun$,
           $\eta=0.111$]
  {Histogram of the percentage error in chirp mass for a high
   \acs{SNR},
   high mass signal of symmetric mass ratio $\eta=0.111$.
   The RWF templates do not exhibit good parameter estimation.}
  \label{fig:pem80snr16}
\end{figure}

At high mass, the parameter estimation improvements with the 
0.5PN templates are quite 
significant, often peaked around zero error, whereas the RWF templates may 
exhibit a bias or have no central peak at all, 
see Figure~\ref{fig:pem67.5snr16} and
Figure~\ref{fig:pem80snr16}. 
\clearpage

\subsection{Large SNR}
At a large SNR of $\rho=64$, the errors are significantly reduced for the low mass systems ($\sim1\%$ - we are approaching the limit where we are measuring  
overlap) and there is very little difference between the results obtained
using either template family. 
However, the parameter estimation for the larger mass systems is  
significantly improved.

\begin{figure}
  \centering
  \psfragfig{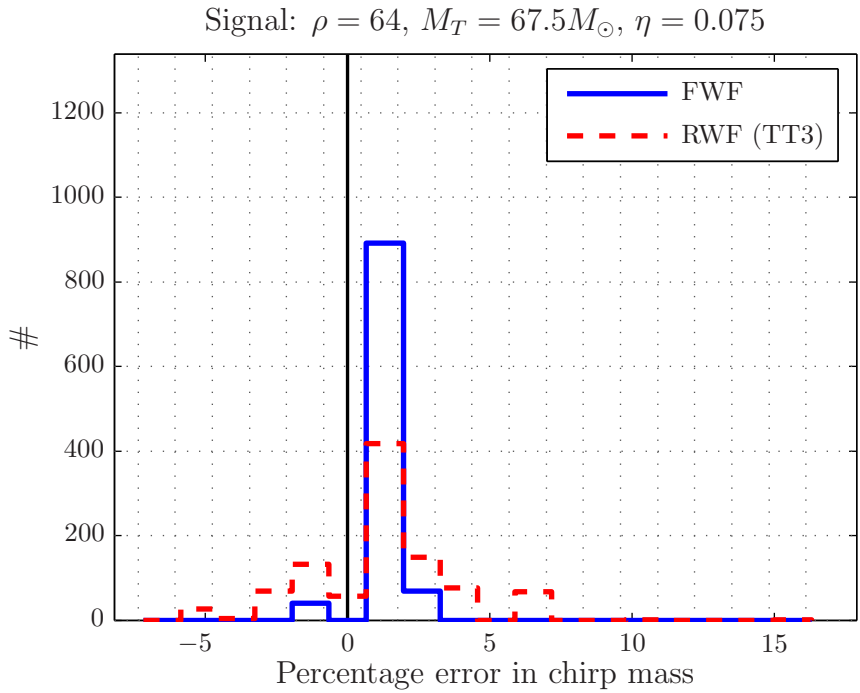}
  \caption[Percentage error in chirp mass. $\rho=64$, $M_T=67.5\Msun$,
           $\eta=0.075$]
  {Histogram of the percentage error in chirp mass for a large
   \acs{SNR},
   high mass signal of symmetric mass ratio $\eta=0.075$.
   The 0.5PN templates outperform the RWF templates.}
  \label{fig:pem67.5snr64}
\end{figure}

\begin{figure}
  \centering
  \psfragfig{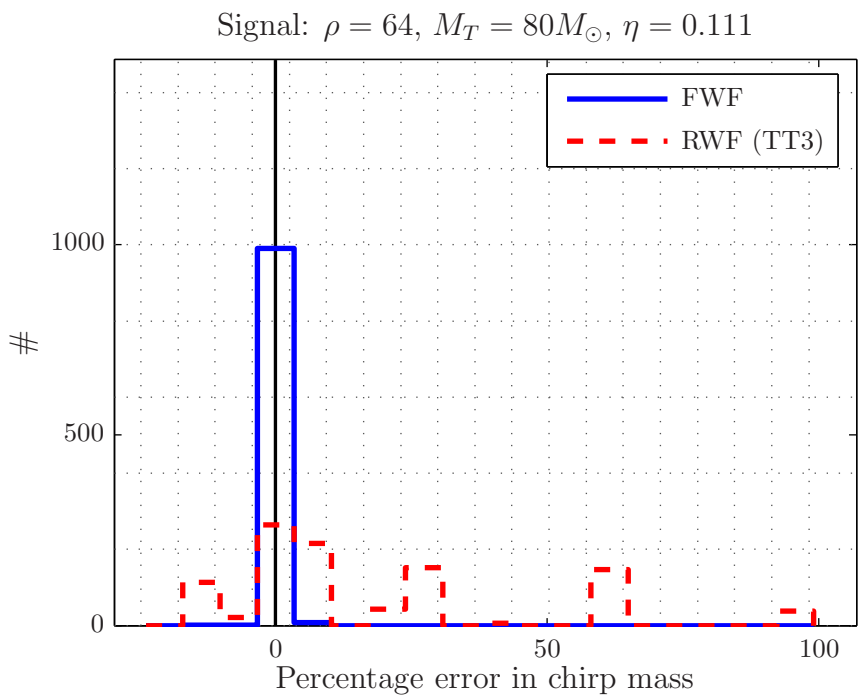}
  \caption[Percentage error in chirp mass. $\rho=64$, $M_T=80\Msun$,
           $\eta=0.111$]
  {Histogram of the percentage error in chirp mass for a large
   \acs{SNR},
   high mass signal of symmetric mass ratio $\eta=0.111$.
   As in Figure~\ref{fig:pem67.5snr64}, 
   the 0.5PN templates outperform the RWF templates.}
  \label{fig:pem80snr64-2}
\end{figure}

Figure~\ref{fig:pem67.5snr64} and Figure~\ref{fig:pem80snr64-2} show two cases
where the 0.5PN templates outperform the RWF templates at large SNR.
\clearpage
\clearpage

\section{Other considerations - The $\chi^2$ distribution}
The LIGO pipeline, described in Section~\ref{sec:pipeline}, is a two-stage
pipeline, where the second stage makes use of the $\chi^2$ veto, which
is a computationally expensive test.
Here, we will not consider how to construct a $\chi^2$ test for the
0.5PN templates. It is clear, however, 
that with six components filtered separately, that any such test 
would be considerably more expensive than the RWF $\chi^2$ test.

\subsection{The $\chi^2$ test with RWF templates and FWF injections}
We have seen in Figure~\ref{fig:spectra} that the spectra of the FWF
and the RWF can differ greatly and since the FWF is a better representation
of nature's gravitational waves, one might expect a significant
impact on a search that uses the effective SNR and $\chi^2$
veto based on the RWF templates.

Table~\ref{tab:chisq} shows the $\chi^2$ excess, i.e,
the $\chi^2$ value in the absence of noise,
for several systems computed
using FWF signals and RWF templates of the same parameters. 
N.B.: if the templates and signals matched exactly, the
$\chi^2$ excess would equal zero.
\begin{table}
\centering
\begin{tabular}{|c|c|c|c|}
\hline
Inclination angle & $\chi^2$ $(3,10)\Msun$ & $\chi^2$ $(3,15)\Msun$ & $\chi^2$ $(3,30)\Msun$\\
\hline
\hline
 $0^0$ & 0.25 & 0.33 & 0.33 \\  
 $45^0$ & 0.21 & 0.31 & 0.39 \\  
 $90^0$ & 0.15 & 0.26 & 0.38 \\  
\hline
\end{tabular}
\caption[The $\chi^2$ excess for FWF signals detected by RWF templates]
{The $\chi^2$ excess for a several systems modelled by the FWF and filtered
with a RWF template in the absence of noise, using the LIGO design PSD.
N.B.: for RWF templates there are 2 degrees of freedom.}
\label{tab:chisq}
\end{table}
However, in the presence of stationary Gaussian noise,
the $\chi^2$ distribution is known~\cite{Allen:2004}, which can be integrated
to give the cumulative probability that a measured $\chi^2$ is consistent
with the template given the presence of the noise. The probability of 
obtaining the $\chi^2$ excesses, or greater, for all the values in
Table~\ref{tab:chisq}, with a matching signal in Gaussian noise, 
is $\sim100\%$.
This result indicates
that a $\chi^2$ test based on the RWF template does not adversely affect
gravitational wave searches. 

\subsection{Degrees of freedom}
\begin{figure}[p]
  \centering
  \psfragfig{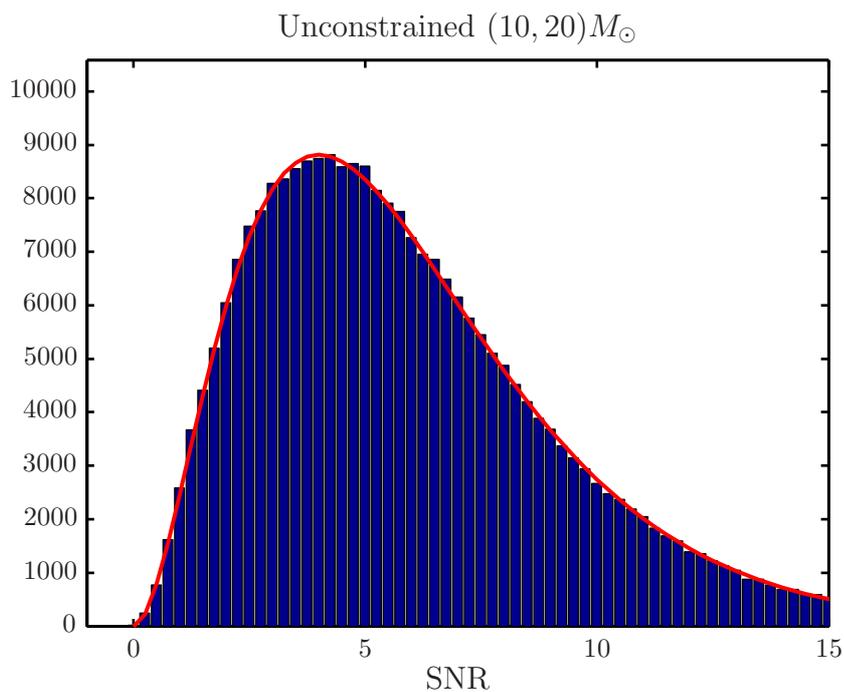}
  \caption[Distribution of SNR time series with the unconstrained 0.5PN 
          filter]
  {Histogram of the SNR time series with the unconstrained 0.5PN filter. As
   expected the distribution follows a classic $\chi^2$ distribution with
   six degrees of freedom (red). The template has component masses
   $(10,20)\Msun$}
  \label{fig:chi10-20}
\end{figure}

\begin{figure}
  \centering
  \psfragfig{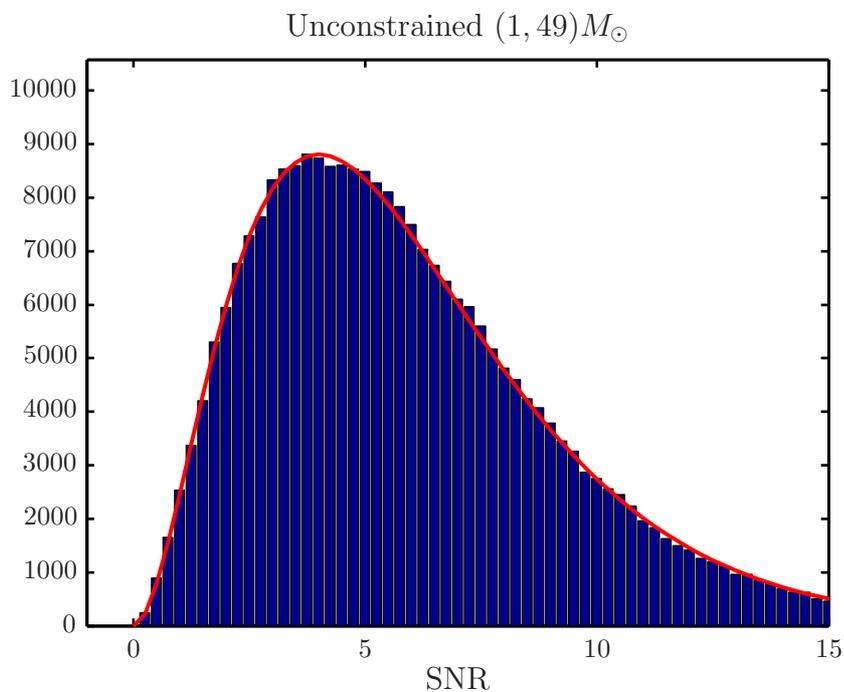}
  \caption[Distribution of SNR time series with the unconstrained 0.5PN filter
           (large mass ratio)]
  {Histogram of the SNR time series with the unconstrained 0.5PN filter. As
   expected the distribution follows a classic $\chi^2$ distribution with
   six degrees of freedom (red). The template has component masses
   $(1,49)\Msun$}
  \label{fig:chi1-49}
\end{figure}

The RWF filtering algorithm has two degrees of freedom, one for each 
phase of the filter. The 0.5PN templates has six components and therefore
six degrees of freedom, although the constraint placed on the templates 
is likely
to have a large effect on the 
$\chi^2$
distribution. Figure~\ref{fig:chi10-20} and
Figure~\ref{fig:chi1-49}  show the distribution of the 
SNR time-series for 
Gaussian noise filtered using the unconstrained 0.5PN templates with 
templates of mass-ratio $1$:$2$ and $1$:$49$ .
The histogram, as expected,
follows a classic $\chi^2$ distribution
with six degrees of freedom.

\subsection{A signal-based veto included in the filter?}
\begin{figure}
  \centering
  \psfragfig{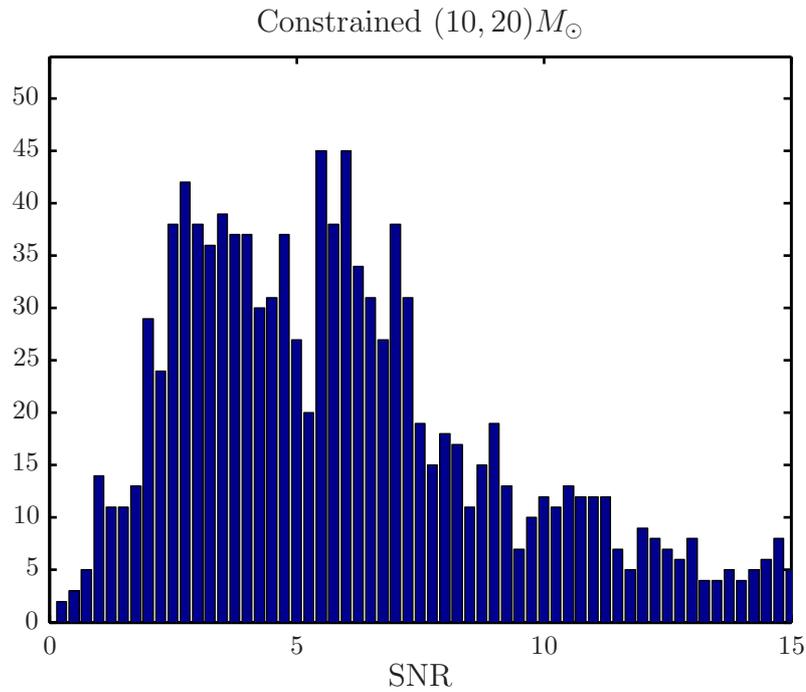}
  \caption[Distribution of SNR time series with the 0.5PN filter II]
  {Histogram of the SNR time series with the 0.5PN filter with all null
   values removed. The template has component masses $(10,20)\Msun$.
   N.B.: the $y$-axis limits are markedly different to those in 
   Figure~\ref{fig:chi10-20}.}
  \label{fig:con1chi10-20}
\end{figure}

\begin{figure}
  \centering
  \psfragfig{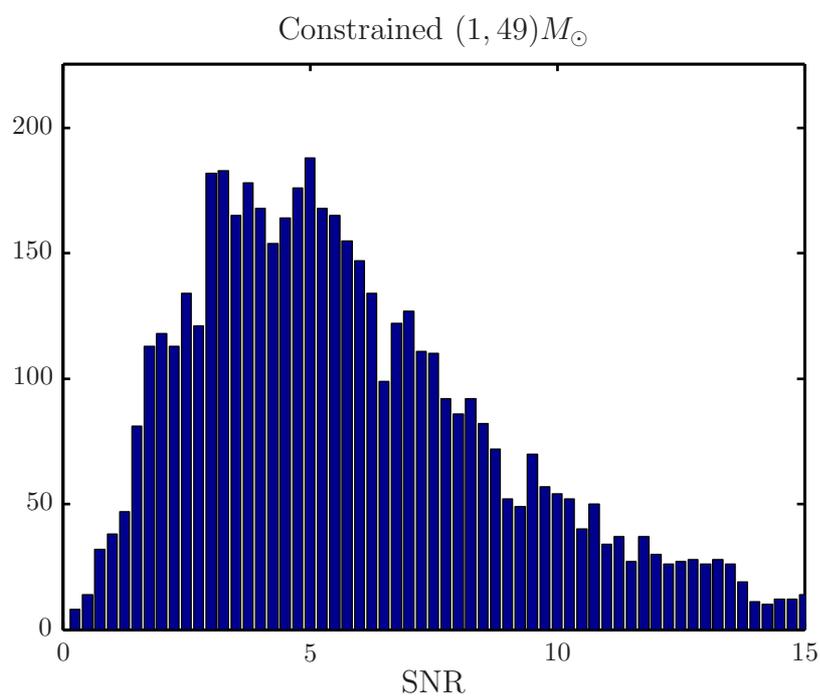}
  \caption[Distribution of SNR time series with the 0.5PN filter
           (large mass ratio) II]
   {Histogram of the SNR time series with the 0.5PN filter with all null
    values removed. The template has component masses $(1,49)\Msun$.
    N.B.: the $y$-axis limits are markedly different to those in
    Figure~\ref{fig:chi1-49}.}
  \label{fig:con1chi1-49}
\end{figure}

The constraint on the 0.5PN filter is, effectively, a signal-based veto. 
A value in the SNR time series is rejected if it does not pass the constraint,
i.e. if it does not look like a signal. We therefore expect the $\chi^2$
distribution to be very different with the constraint implemented.
Figure~\ref{fig:con1chi10-20} and Figure~\ref{fig:con1chi1-49} show the
$\chi^2$ distribution for the same templates as above, but with the constraint
applied, which has a dramatic effect (note the change in the $y$-axis
from Figure~\ref{fig:chi10-20} and Figure~`\ref{fig:chi1-49}). There
were $262144$ points\footnote{Sampled at $4096\, \rm Hz$, a total duration of
$64\, \rm s$.} in the time-series and all but $1133$ and $5044$
were discarded for the $(10,20)\Msun$ and $(1,49)\Msun$ 
constrained templates respectively. 

This result is highly significant as it gives rise to the possibility that
the FAR could be dramatically 
reduced in a gravitational wave search using the constrained 0.5PN 
templates. It is also
promising to see that the same effect is also impressive for the large mass
ratio system where the constraint is less restrictive.

\clearpage
\section{Concluding remarks}
In this chapter, we have seen how to matched filter using templates that
are 0.5\ac{PN} in amplitude and have shown that the mass reach
and parameter estimation are improved with the constrained 
0.5\ac{PN} templates, as opposed to the \ac{RWF} templates, which
fulfils the motivations presented at the start of the chapter.

The constraint implemented in this chapter is primitive and there are many
potential areas for improvement. For one thing, 
the constraint was set using the relative amplitudes of the
individual harmonics, but it would in fact be more appropriate
to set the constraint on the maximum allowed ratio of
(\ref{eq:snrrat2}) and the equivalent for
the third harmonic. Moreover, it is not yet known how the constraint will
perform in real detector data that is non-Gaussian.

Although not discussed in this chapter, the additional complexity of the 
0.5\ac{PN} filtering algorithm leads to longer processing times,
which may be a practical consideration when performing a search for
gravitational waves.
Despite the above concerns, the results presented here
are promising and already proffer improvements on the existing \ac{RWF}
templates. 

In principle, one could extend the three harmonic filter to include higher
harmonics, but at 1\ac{PN} and above amplitude corrections are introduced,
(\ref{eq:tmplt}) and a new approach would be required. It may turn
out that neglecting the amplitude corrections and using only the
higher harmonics is effective. In any case, \emph{even}
higher order waveforms will place additional strains on computational
resources and it may be that they are 
best used not as a detection device, but for following
up interesting triggers from a gravitational wave search.

However, by examining the $\chi^2$ distribution of the constrained 0.5PN
filter we have seen that the constraint is 
potentially a very effective signal-based veto
and could reduce the FAR in a gravitational wave search
significantly. If used in a two-stage pipeline,
with a RWF $\chi^2$ test at the second stage, 
a reduction in triggers due to the decrease in FAR
could mitigate the extra computational expense of the 0.5PN filter making
it a viable algorithm to be used in future gravitational wave searches.

There are many other ways in which this work can be extended. For one thing
the tests in this chapter consider a single \ac{IFO}, yet it would be
interesting to study the effects on coincidence and the \ac{FAR} in the
context of a complete gravitational wave search, similar to that in
Chapter~\ref{howchap}. It would also be interesting to apply the
0.5PN filtering algorithm to IMR waveforms,
which should be used in gravitational wave searches for CBCs of 
the mass ranges studied in this chapter.



\chapter{A tapering window for time-domain templates and simulated signals}
\label{winchap}
\lhead{Chapter~\ref{winchap}
\emph{A tapering window function for TD templates and signals}}
\rule{15.7cm}{0.05cm}

Inspiral signals from binary black holes, in particular those with
masses in the range $10\Msun \lsim M \lsim 1000 \Msun,$ may last for 
only a few cycles within a detector's most sensitive frequency band. The 
spectrum of a square-windowed time-domain signal could contain unwanted 
power that can cause problems in gravitational wave data analysis, 
particularly when the waveforms are of
short duration. There may be leakage of power into 
frequency bins where no such power is expected, causing an excess of false
alarms. 

In this chapter a method of tapering \ac{TD} waveforms is presented that 
significantly reduces unwanted leakage of power, leading to a spectrum 
that agrees very well with that of a long duration signal. The tapered 
window also decreases the false alarms caused by instrumental and 
environmental transients that are picked up by templates with spurious 
signal power. The suppression of background is an important goal in 
noise-dominated searches and can lead to an improvement in the detection 
efficiency of the search algorithms.

The tapering method has proved very useful and has been used in all of the
studies in Chapter~\ref{ampchap}.

\section{Motivations}
We have seen in Chapter~\ref{howchap} that gravitational wave searches
are noise dominated and must use techniques to extract the signal from the
noise. In Chapter~\ref{howchap} the matched filter was used; other
examples include: wavelet transforms for transient signals of unknown 
shape~\cite{Klimenko:2004a,Klimenko:2004b}, coherent search methods for burst 
signals~\cite{ShourovLazzariniSteinSuttonSearleTinto2006}, etc. 
Moreover, we have also discussed 
vetoes based on the expected signal evolution~\cite{Allen:2004} and
instrumental and environmental monitors~\cite{Hanna:2006} that
have been developed 
over the past decade to improve detection probability and mitigate 
false alarms.  Detecting a signal buried in non-stationary noise 
is a challenging problem as some types of non-stationary noise
artefacts can partially mimic the signal.

Many of these techniques involve the computation of a correlation 
integral in which band-passed data are multiplied by the 
\ac{FD} model waveform or the \ac{DFT} 
of the \ac{TD} model (see, e.g.,~\cite{LIGOS1iul}).  
Here we will again consider a matched filtering search for inspiral signals 
where the \ac{DFT} of a \ac{TD} waveform is used to construct 
the correlation. A problem that has not been adequately addressed 
(see, however,~\cite{Arnaud:2007a}) in this context is the 
effect of the window that is used in chopping a
\ac{TD} signal before computing its \ac{DFT}.

Inevitably, all signal analysis algorithms use, implicitly or 
explicitly,  some form of window function.  An 
inspiral waveform sampled from a time when the signal's instantaneous 
frequency enters a detector's sensitive band until the time when it
reaches the \ac{FLSO} implicitly makes use of a square window.  
Signal analysis literature is full of examples of artefacts caused 
by the use of such window functions. Examples are: leakage of power
from the main frequency bins where the signal is expected to lie into
neighbouring bins, loss of frequency resolution and corruption of
parameter estimation~\cite{Windows:2009}.  In this chapter
we will explore the problems caused by using a 
square window and suggest an alternative that cures some of the 
problems.

There is no unique, or favoured, windowing method. One is often
guided by the requirements of a particular analysis at hand.
In this case, a square window is especially bad since the leakage
of power outside the frequency range of interest can lead to
increased \ac{FAR} and poorer estimation of parameters.
One reason for increased \ac{FAR} could be that the
noise glitches in the detector look more like the 
untapered/square-windowed 
waveform and less like a tapered one.
Here we will explore the effect of a smoother window function, presented
in Section~\ref{sec:tapmethod}, which has a far steeper fall-off of 
power outside the frequency range of interest.
Use of this window has cured several problems we had with a square
window. The effect of the new method on waveform spectra is shown in
in Section~\ref{sec:effect}.

Section~\ref{sec:tapsnr} shows how tapering
helps in a more reliable signal spectral estimation and hence a
proper determination of the expected signal-to-noise ratio. Spectral
contamination is worse for larger mass black hole binaries as they
are in the detector's sensitive band for a shorter time and
the window function can only extend over a short time. 
It is for such signals that the tapered window presented here
offers the most improvement.
In Section~\ref{sec:taptriggers} we will see how the rate of triggers
from a matched filtered search can vary depending on the kind of
window function used. Finally, in Section~\ref{sec:tappe}
we see
what effect the window function has on parameter estimation,
before drawing the conclusions of the study in Section~\ref{sec:tapconc}.

\section{Window functions and their temporal and spectral characteristics}
\label{sec:tapmethod}
Let $h(t)$ denote a continuous differentiable function, for example
a gravitational wave signal emitted by a \ac{CBC}, and 
let $\widetilde{h}(f)$ denote the \ac{FT} of $h(t)$ defined by
\be
\widetilde{h}(f) = 
  \int_{-\infty}^\infty h(t)\, \exp(2\pi ift)\, {\rm d}t.
\ee
In reality the signal does not really last for an infinite time. The
\ac{FT} of a signal of finite duration lasting, say, 
from $-{T}/{2}$ to ${T}/{2},$ can be represented either by setting the 
limits of the integral to go from $-{T}/{2}$ to ${T}/{2}$ or by using 
a window function. The latter is preferred so as to preserve the 
definition of the \ac{FT}. 

A window function is a function that has either a finite 
support or falls off sufficiently rapidly as $t\rightarrow \pm \infty.$ 
Two simple windows that have finite support are the square window 
$s_T(t)$ defined by 
\be
s_T(t) = 
  \begin{cases}
    1 &\text{for} \quad -\frac{T}{2} \le t \le \frac{T}{2}\\
    \hfill 0 &\text{otherwise}\, .\\
  \end{cases}
\ee
and the triangular window $b_T(t)$ defined by 
\be
b_T(t) =
  \begin{cases}
    (1-2|t|/T) &\text{for} \quad -\frac{T}{2} \le t \le \frac{T}{2}\\
    0          &\text{otherwise}\, .\\
  \end{cases}
\ee
Neither the square nor the triangular window are 
differentiable everywhere. 
As a result, they are not functions of finite bandwidth. 
In other words, their \ac{FT}s, $\widetilde{s}_T(f)$ 
and $\widetilde{b}_T(f)$, do not have finite 
support in the \ac{FD}: $|\widetilde{s}(f)| > 0$ 
for $-\infty \le f \le \infty.$ 
In the case of a square window the \ac{FT} $\widetilde{s}(f)$ is a 
sinc function, $|\widetilde{s}_T(f)| = T {\rm sinc}(\pi f T),$ which is
peaked at $f=0,$ with a width $\pi/T$ and falls off as $f^{-1}$ as 
$f \rightarrow \pm \infty.$ The lack of finite support in the Fourier
domain could sometimes cause problems, especially when the width, $T$,
of 
the window in the \acf{TD} is too small. For functions
that have infinite bandwidth the sampling theorem does not hold but
this is not a serious drawback if the \ac{FT} falls off sufficiently
fast above the Nyquist frequency. However, there could be other
issues when the window leads to leakage of power outside a region
of interest as we shall see below.

\subsection{The Planck-taper window function}
A signal $h(t)$ with the window $w_T(t)$ applied to it, in other words
the windowed signal $h_w(t),$ is defined by
\begin{equation}
\label{eq:hw}
h_w(t) = h(t) w_T(t).
\end{equation}
The convolution theorem states that the \ac{FT} of the product of
two functions $h(t)$ and $w_T(t)$ is the convolution of individual
\ac{FT}s:
\begin{subequations}
\begin{align}
\label{eq:conv}
\widetilde{h}_w(f) & = 
  \int_{-\infty}^\infty h(t) w_T(t) \exp(2\pi i f t) {\rm d}t\, ,\\
\label{eq:conv2}
 & = 
  \int_{-\infty}^\infty \widetilde{h}(f') \widetilde{w}_T(f-f') {\rm d}f'\, ,
\end{align}
\end{subequations}
We can now see why a window whose power in the \ac{FD} does not 
fall off sufficiently rapidly
might be problematic. The convolution integral will have contributions
from all frequencies. Suppose we are interested in matched filtering the
data with an inspiral signal from a compact coalescing binary whose 
instantaneous frequency varies from $f_a$, at time $t_a$, to $f_b$, at
time $t_b$. One would normally achieve this by using a square window 
$s_T(t)$ that is centred at $(t_a+t_b)/2$ with width $T=t_b-t_a.$ However,
we can see from (\ref{eq:conv2}) that the convolution integral will
have contributions from outside the frequency range of interest.

\begin{figure}
  \begin{minipage}[h]{16cm}
    \centering
    \psfragfig{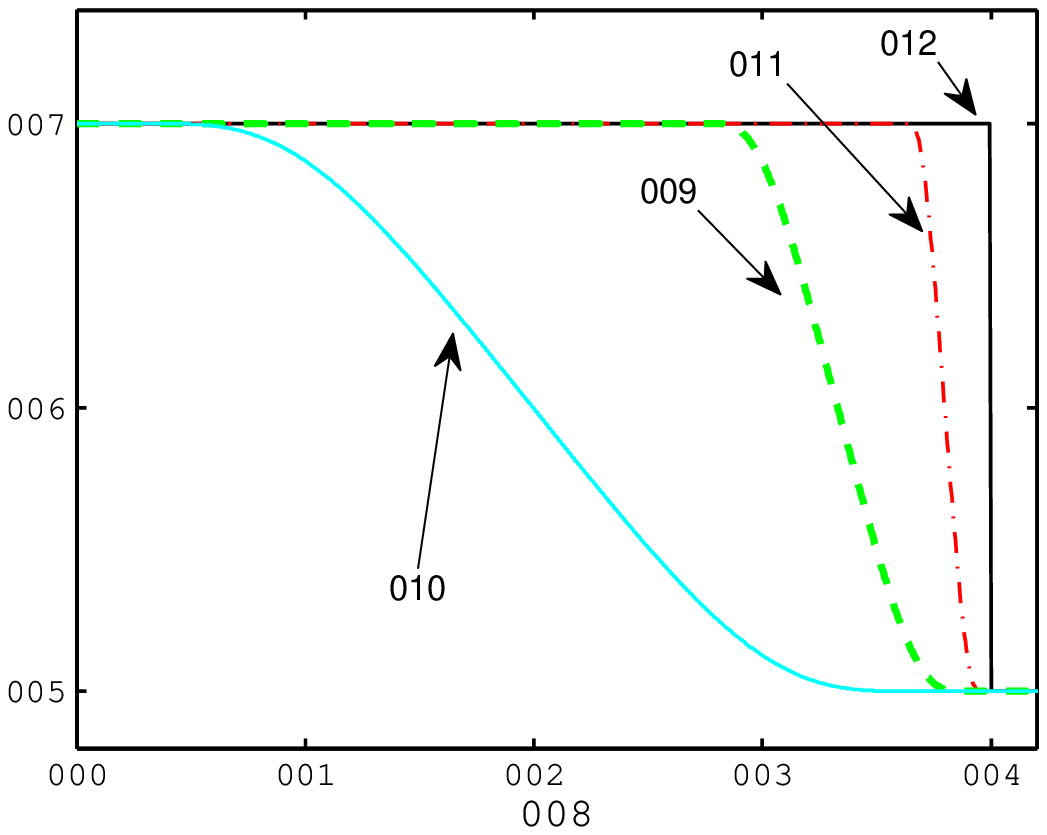}
  \end{minipage}
  \vspace{1cm}
  \begin{minipage}[h]{16cm}
    \centering
    \psfragfig{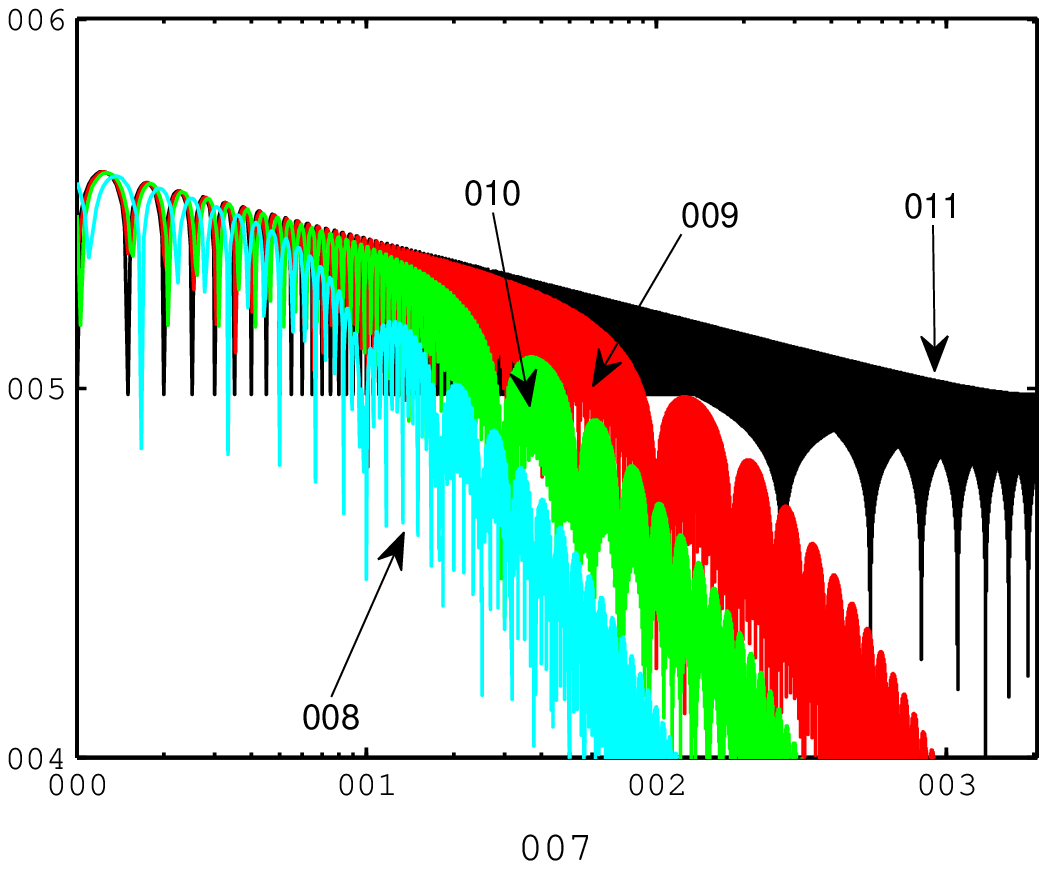}
  \end{minipage}
  \caption[The Planck-taper window in the \acs{TD} and the \acs{FD}]
  {The Planck-taper window in the \ac{TD} for three  
  different choices of the parameter $\epsilon=0.01,\,0.033,\,0.1,$
  (top).  For reference the square window with the same effective width as the
  Planck-taper window has also been plotted.
  The bottom plot shows their corresponding spectra.}  
  \label{fig:windows}
\end{figure}

To circumvent this problem we define a new window function 
that falls off rapidly outside the frequency range of interest.
Inspired by the tapering function used in Damour et 
al.~\cite{Damour:2000gg}, we define the new function $\sigma(t)$ by
\begin{eqnarray}
\label{eq:taper}
\sigma_T(t; \epsilon) = 
\left \{
\begin{array}{llll}
 &  & \hfill 0, & \text{for} \quad t \le t_1\, ,\\[0.5cm]
 \dfrac{1}{\exp(z(t))+1}, &  & z(t)=\dfrac{t_2-t_1}{t-t_1} +
    \dfrac{t_2-t_1}{t-t_2}, & \text{for} \quad t_1  < t < t_2\, ,\\[0.5cm]
 & \quad & \hfill 1, & \text{for} \hfill t_2 \le t \le t_3\, ,\\[0.5cm]
 \dfrac{1}{\exp(z(t))+1}, & & z(t)=\dfrac{t_3-t_4}{t-t_3}+
  \dfrac{t_3-t_4}{t-t_4}, & \text{for} \hfill t_3  < t < t_4\, ,\\[0.5cm]
 & & \hfill 0, & \text{for} \quad t_4 \le t\, ,
\end{array}
\right .
\end{eqnarray}
where
\begin{align}
t_1 &= -\dfrac{T}{2}\, ,\\
t_2 &= -\dfrac{T}{2}\left(1-2\epsilon\right)\, ,\\
t_3 &= \dfrac{T}{2}\left(1-2\epsilon\right)\, ,\\
t_4 &= \dfrac{T}{2}\, .
\end{align}
Here $T$ is the width of the window and $\epsilon$ is the fraction of the 
window width over which the window function smoothly rises from 0 at $t=t_1$ 
to 1 at $t=t_2$ or falls from 1 at $t=t_3$ to 0 at $t=t_4.$
We shall call $\sigma(t)$ the \textit{Planck-taper window} as 
the basic functional
form is that of the Planck distribution. The motivation for choosing this 
window function is to reduce the leakage of power in the \ac{FD} 
but at the same time not to lose too much of the length of the signal in the 
\ac{TD}. The choice of $\epsilon$ will affect both aspects significantly.
Figure~\ref{fig:windows}  shows the window function for several 
choices of the parameter $\epsilon=0.01, 0.033$ and $0.1$ with their
 corresponding spectra.
At lower frequencies the spectrum of the Planck-taper window falls 
off at the same rate (i.e., $1/f$) as a square window. But beyond a
certain frequency $f_0 \sim (\epsilon T)^{-1},$ the spectrum falls off 
far faster.

A key feature the Planck-taper window is the
fraction of the window width that is flat, i.e., the choice of $\epsilon$,
which we will automate to be waveform-dependent, see 
section~\ref{sec:tapimp} below.

\subsection{Implementation of the window}
\label{sec:tapimp}
\begin{figure}[htp]
  \centering
  \psfragfig{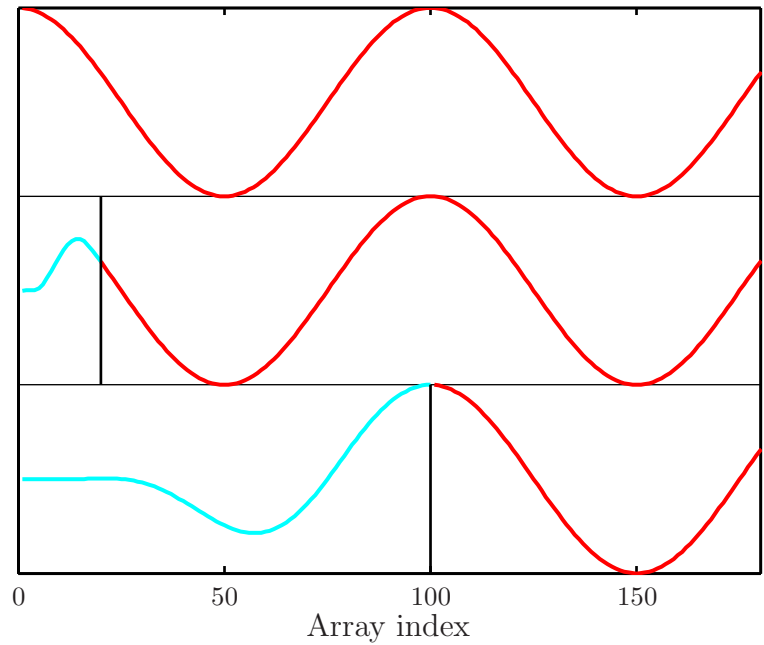}
  \caption[Application of the Planck-taper windows]
{The window function has been applied to the start of a cosine wave (top curve)
using two methods. In the first case it is applied from $j=1$ up to an
arbitrary choice of $j=20$ (middle), whereas in the second case it is applied
up to the second maximum at $j=100$ (bottom).
The lighter coloured parts of the middle and bottom curves (to the left of the
black vertical lines) show where the taper has been applied.}
\label{fig:taperapplication}
\end{figure}

We may discretise  (\ref{eq:taper}) by replacing $t,t_1,t_2, t_3, t_4$
with the array indices $j,j_1,j_2, j_3, j_4.$
In this notation the parameter epsilon is approximated by
$\epsilon \simeq (j_2-j_1)/N,$
where $N$ is the number of data points in the waveform.
The start and end of the
waveform are denoted by $j_1$ and $j_4$, respectively.
The values of $j_2$
and $j_3$ have to be chosen judiciously to avoid leakage of power.
We shall choose $j_2$ and $j_3$ to be the array indices corresponding
to the second stationary point after $j_1$ and before $j_4$
(see Figure~\ref{fig:taperapplication}). Applying the transition stage
of $\sigma$ from a crest/trough ensures that the window does not
have a sudden impact
on the behaviour of the waveform. The first stationary point would not be an
appropriate choice as it may occur within only a few array points of
$j_1$ or $j_4,$ causing $\epsilon$ to be too small. One could
choose the 3rd, 4th or 5th, but using such later maxima would reduce
the genuine power of the waveform more than what might be acceptable.

\subsection{Comparison with other windows}
Here we shall not compare the performance of Planck-taper with other
commonly used windows, e.g., Bartlet, Hann or Welch. Such windows
transition between $0$ and $1$
over $j=1,\ldots,N/2$, where the window is of length $N$,
producing significant differences between $h(t)$ and $h_w(t)$ in (\ref{eq:hw}).
The power is therefore suppressed at the beginning and end of $h(t)$.
This is acceptable when computing the \ac{PSD} of a data segment,
but would cause a problem if applied to a template waveform as the phase 
and amplitude of $h(t)$ are both instantaneous functions of $t$,
with the most power at the end of the waveform. More generally, the noise
tends to be stationary (see Chapter~\ref{howchap}) whereas the signal is 
not.

Windows with properties similar to Planck-taper, such as having a central flat
region, do exist. For example, the Tukey window~\cite{1978IEEE}, which has been
used in gravitational-wave data analysis recently~\cite{2007PhRvD}, may offer
a good comparison. However, a key feature in our study of the Planck-taper
window is the waveform-dependent adjustment of $j_2$ and $j_3$. Whilst this
automation could be considered separately from the Planck-taper window and
used on other windows defined by the points $j_{1,2,3,4}$, we have not
done so here. Given the shared features of the Tukey window with
Planck-Taper one might expect similar results.

\section{Effect of the window function on the signal spectrum}
\label{sec:effect}
In this section we will examine the power spectrum of
the waveform of a coalescing binary emitted during the inspiral phase.
As we have seen 
the waveforms are modelled using the \ac{PN} approximation. However,
even within the \ac{PN} approximation, there are several different
ways in which one might construct the waveform
\cite{Damour:2000zb,Buonanno:2009zt}.
Two such models widely used in the search for compact binary
coalescences are \ac{TT3} and the \ac{SPA}.  \ac{TT3} is a \ac{TD}
signal model in which the amplitude and phase of the signal are both
explicit functions of time. In the so-called restricted \ac{PN} approximation
the signal consists of the dominant harmonic at twice the orbital
frequency, but not higher order \ac{PN} corrections consisting of other
harmonics, and the phase is a \ac{PN} expansion that is currently known
to ${\cal O}(v^7)$ in the expansion parameter $v$ -- the relative velocity
of the two stars. The \ac{SPA} is the Fourier transform of the \ac{TT3}
model obtained
by using the stationary phase approximation to the Fourier integral
\cite{SathyaDhurandhar:1991}.  A template belonging to the \ac{TT3} model
is defined for times when the gravitational wave frequency is within the
detector's sensitivity band until it reaches \ac{FLSO}. This means one is
in effect multiplying a square window with a continuous function.

\begin{figure}
  \begin{minipage}[h]{16cm}
    \centering
    \psfragfig{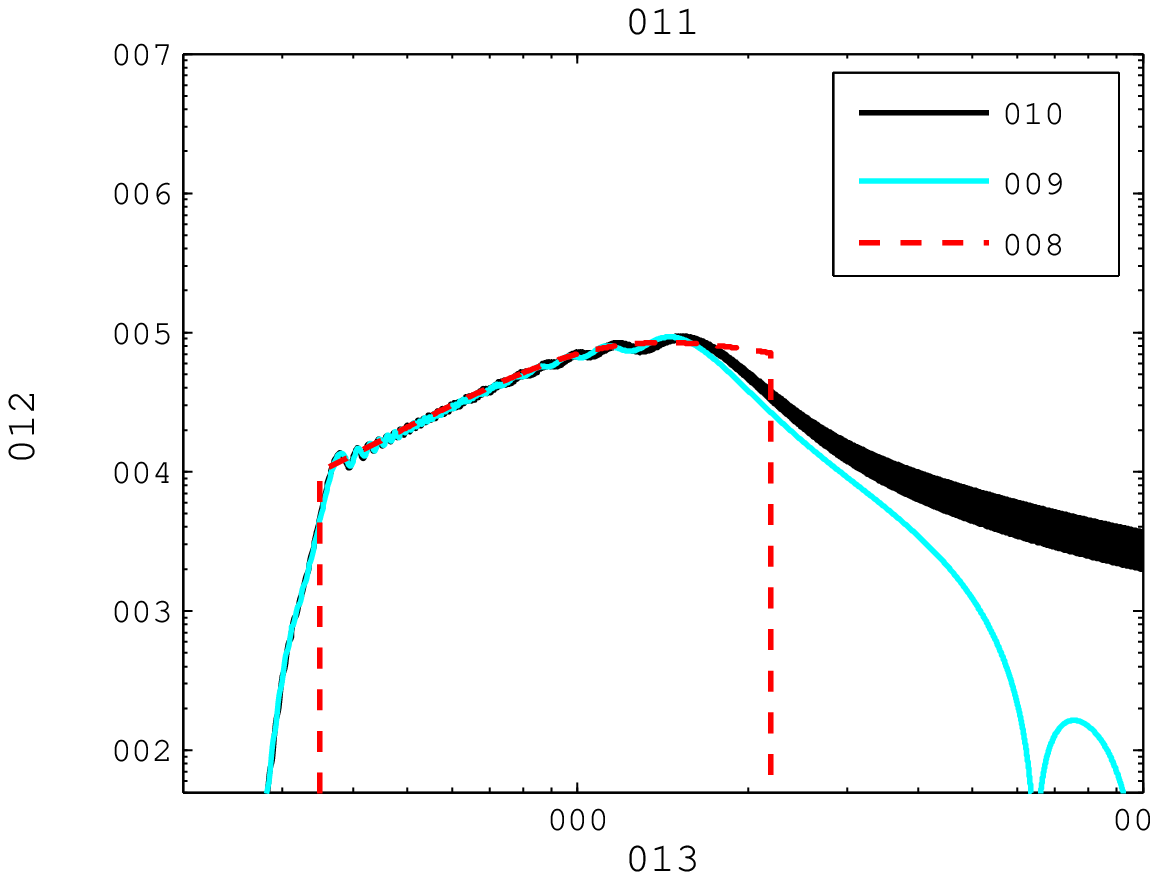}
  \end{minipage}
  \vspace{1cm}
  \begin{minipage}[h]{16cm}
    \centering
    \psfragfig{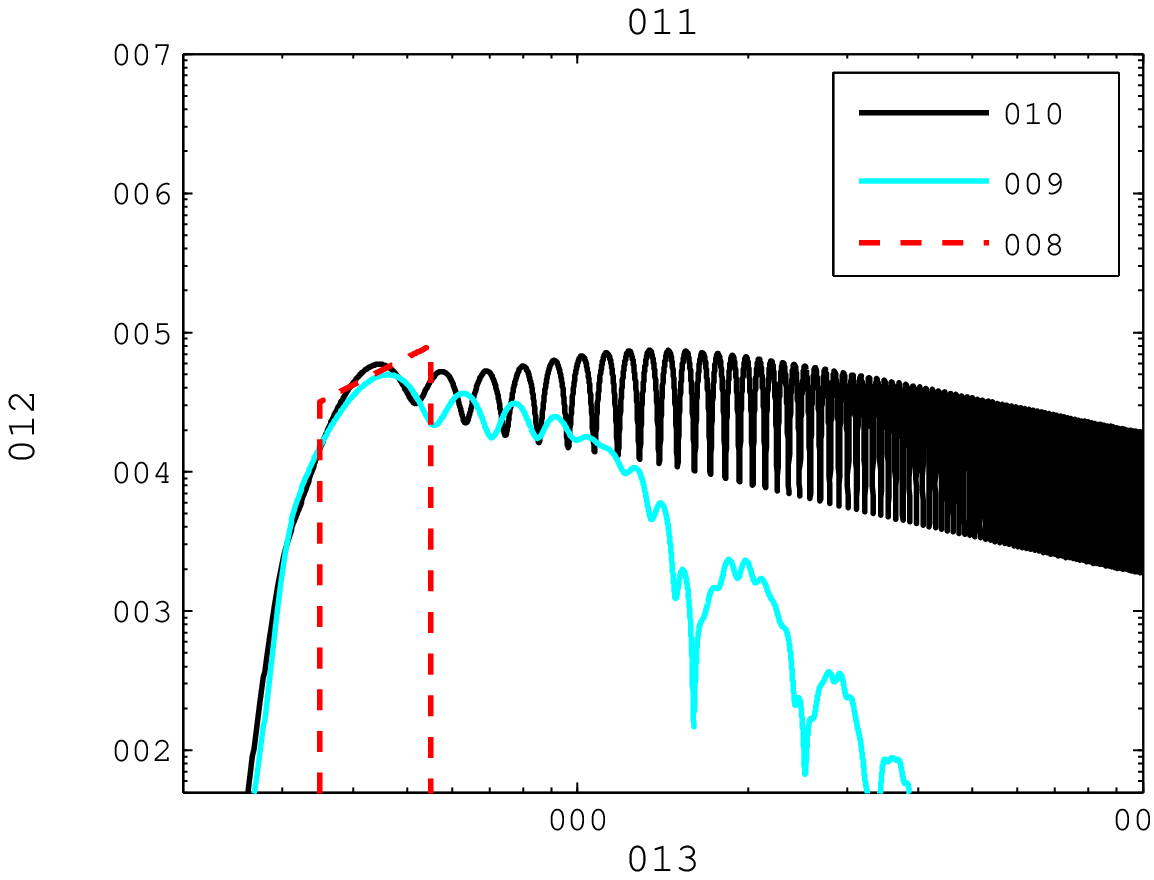}
  \end{minipage}
  \caption[Spectra of a $20\Msun$ and a $80\Msun$ inspiral using both 
           square and Planck-taper window]
          {The \ac{SNR} integrand produced with both the square and Plank-taper
           window, where the waveform is generated from a
           frequency of $35$ Hz to the \ac{FLSO} of the source,
           computed using the LIGO design \acs{PSD} for sources of total mass 
           $20\Msun$ (top) and $80\Msun$ (bottom).
           In both cases, the \ac{SNR} integrand falls off far faster
           with the use of the Planck-taper window compared to the
           square window.}
  \label{fig:spectra2}
\end{figure}

Figure~\ref{fig:spectra2} shows the \ac{SNR} integrand of the
\ac{SPA}, computed using the initial \ac{LIGO} design
\ac{PSD}~\cite{Damour:2000zb}. The inspiral waveform
is defined from a lower cut-off frequency of $35\,\rm Hz$ up to its \ac{FLSO},
for $20\, \Msun$ and $80\,\Msun$ equal-mass binaries.
The \ac{DFT} of the \ac{TT3}, generated between the same frequencies,
with a square window (or rather no window), labelled H$_S$, and
with the Planck-taper window, labelled H$_\sigma$, are also plotted.
Where the Planck-taper window is used the excess power
(that above \ac{FLSO}) decreases rapidly and the spectrum is closer
to that of the \ac{SPA}.

\section{Effect of the window function on the estimation of the
signal-to-noise ratio}
\label{sec:tapsnr}

Gravitational wave searches for known signals, such as those emitted by
\acp{CBC} \cite{Collaboration:2009tt,Abbott:2009qj},
rely upon signal models for two primary reasons. Firstly, they are used as
templates to matched filter the data. Secondly, they are injected into the data
as simulated signals to estimate the efficiency of the detector to detect
such signals.  If the signal/template models are
generated in the \ac{TD} then they must undergo a \ac{DFT} if the data
are analysed in the \ac{FD} as is the case for the current \ac{LIGO}
matched filtering code.

The expectation value for the $\text{SNR}^2$ of a signal in 
stationary Gaussian
noise, when the signal and template match exactly (\ref{eq:snrest}), 
may be expressed discretely
\begin{equation}
\label{eq:tapsnr}
\left<h,h\right>
  \simeq 4\Delta f \sum_{k=1}^{N/2-1}\frac{{|\widetilde{h}_k|}^2}{S_{nk}}\, ,
\end{equation}
The discretised evaluation of the \ac{SNR} is
often used in numerical calculations. 
Here $\widetilde{h}_k,$ $k=0,\ldots,N/2,$
is the \ac{DFT} of the signal defined for positive frequencies
and $S_{nk}$ is the discretised one-sided \ac{PSD}.

As we have seen 
the amplitude of an inspiral signal increases with the total mass of
the system; conversely, the \ac{FLSO} of the signal is inversely
proportional to the total mass.  Therefore,
as the total mass of a system increases, the amplitude of the signal and
the \ac{FLSO} will have opposing effects.
For lower mass systems, the increasing amplitude causes the \ac{SNR}
to increase as a function of the total mass. However, for higher mass
systems, the reduction in the \ac{FLSO} causes the signal to have less
power in band. As a result, the \ac{SNR} will decrease as a function
of the total mass. The relatively low \ac{FLSO} of the higher mass templates,
coupled with their short duration, lead them to be particularly
susceptible to artefacts of spectral leakage in the \ac{DFT}.

\begin{figure}[htp]
  \centering
  \psfragfig{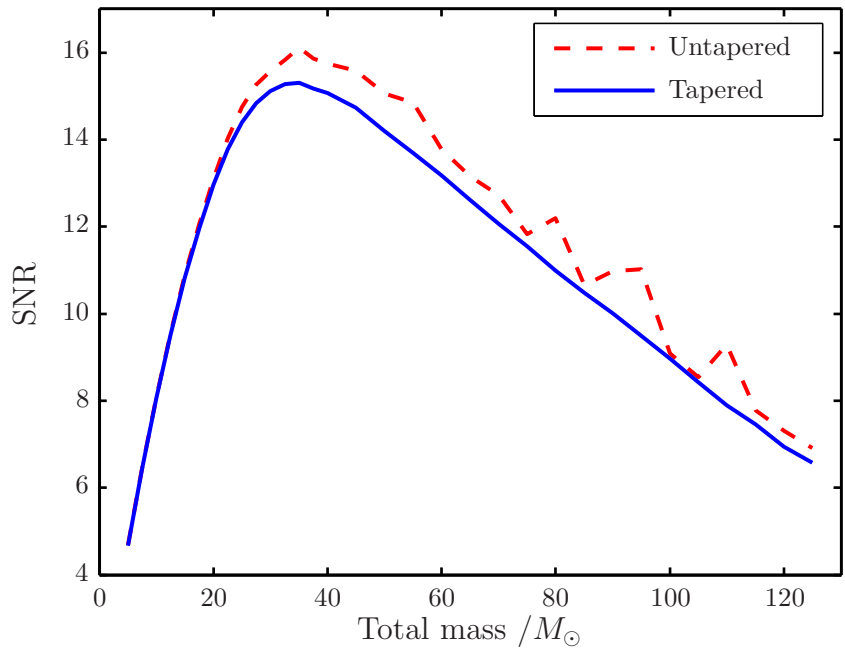}
\caption[SNR vs.~mass with and without tapering]
{The \ac{SNR} vs.~the total mass of the source for signals corresponding to
compact binary systems directly overhead a detector of initial LIGO design
\ac{PSD}. The \ac{SNR} is obtained using the \ac{DFT} of \ac{TD}
waveforms with a square window (dashed curve) and with the Planck-taper window
(solid curve). Here the systems are overhead the detector at an effective
distance of $65\, \rm Mpc$, using a fixed mass ratio of $5:1$ and a fixed
inclination angle of $45^\circ$.}
\label{fig:snrvsmass}
\end{figure}

Figure~\ref{fig:snrvsmass} shows the \ac{SNR} for \ac{TT3} inspiral
waveforms that are $2$\ac{PN} in amplitude and phase, plotted as a
function of the total mass for two choices
of the window function: the dashed curve corresponds to the
square window and the solid curve to the Planck-taper window. All
other parameters are the same in both cases.  When the Planck-taper window
is used, the curve exhibits the expected behaviour, whereas in the
case of a square window , the \ac{SNR} curve is `jagged' which is unexpected
given that stationary Gaussian noise was used in the estimation of the
\ac{SNR}. This behaviour is most likely explained by the excess power from the
\ac{DFT} of the waveform.

It should be noted that integrating to \ac{FLSO} rather than Nyquist in Eq.\
(\ref{eq:tapsnr}), is not considered appropriate here. Firstly, the higher
harmonics in the amplitude corrected waveforms contain power above \ac{FLSO}
(which becomes more significant for high mass systems).
Secondly, cutting off the integration at \ac{FLSO} is essentially
the application of a square window to the template waveform in the
frequency domain. This will lead to leakage of power in the time domain
which is not a desirable feature. The problem
of using a square-windowed \ac{TD} template as our matched filter is not
that there is power above \ac{FLSO}; it is that the excess power in this
region, present due to windowing, but not present in a genuine signal will
lead to unnecessary false alarms in a search.

\section{Effect of window functions on trigger rates}\label{sec:taptriggers}
To assess the effect that tapering of templates has on trigger rates, we have
applied the \ac{LSC} \ac{CBC} pipeline~\cite{LIGOS3S4Tuning,LIGOS3S4all,
Collaboration:2009tt,Abbott:2009qj,Allen:2005fk}
to data taken during the \ac{S4},
which took place from February 22 - March 23, 2005. The basic topology of
the pipeline is similar to that used in many previous
searches~\cite{LIGOS3S4all,Collaboration:2009tt}, with
the pipeline used in~\cite{Abbott:2009qj}
described in detail in Chapter~\ref{howchap}, from which we recall the
trigger generation:
\begin{itemize}
\item The template bank is chosen such that the loss of SNR due to having a
finite number of templates is no more than $3\%$ for any signal belonging
to a given family of waveforms~\cite{hexabank,BBCCS:2006}.
\item Matched filter the data with the generated templates. A trigger is
generated at times when the SNR is larger than a given threshold. The output
of this stage is a list of \textit{first-stage single-detector triggers}.
\item Check for coincident events between different detectors. For an event
to be deemed coincident, the parameters seen in at least two detectors (for
instance, the masses of the system, the time of coalescence, \ldots)
should agree to within a certain tolerance \cite{Robinson:2008}.
The output of this stage is a list of \textit{first-stage coincident triggers}.
\item Re-filter the data using only templates associated with coincident
triggers. This time, the triggers are subjected to further
signal-based vetoes, some of which are computationally costly, such as the
chi-squared veto~\cite{Allen:2004}. This produces a list of
\textit{second-stage single-detector triggers}.
\item Check for coincident events between detectors using the 
second-stage single-detector triggers. This produces a list of 
\textit{second-stage coincident triggers}.
\end{itemize}

In this study the data were filtered using the \ac{EOB} templates
\cite{BuonannoDamour:1999,BuonannoDamour:2000,Damour:2000zb}, tuned to recent
results in numerical relativity~\cite{Buonanno:2007pf,Damour:2007yf},
with a total mass in the range $25-100M_\odot$. This choice agrees with
the templates used to search for signals from high-mass \ac{CBC}s in
data from \acf{S5}. Because the \ac{EOB} waveforms used as templates
contain the inspiral, merger and ringdown phases,
there was no need to taper the end of the waveform.
Therefore, in this case, the taper specified in (\ref{eq:taper}) was only
applied to the start of the waveform. Although this may reduce the effect
the taper has in comparison to tapering both ends of an inspiral-only template,
it is of more interest to evaluate the performance in a realistic search case.
N.B.: the tapering window is
\textit{explicitly applied to the
template waveform} where the length of the waveform is less than the
length of the data segment that is matched filtered. In this study no window
has been applied to the data segment.

\begin{figure}
  \begin{minipage}[h]{16cm}
    \centering
    \psfragfig{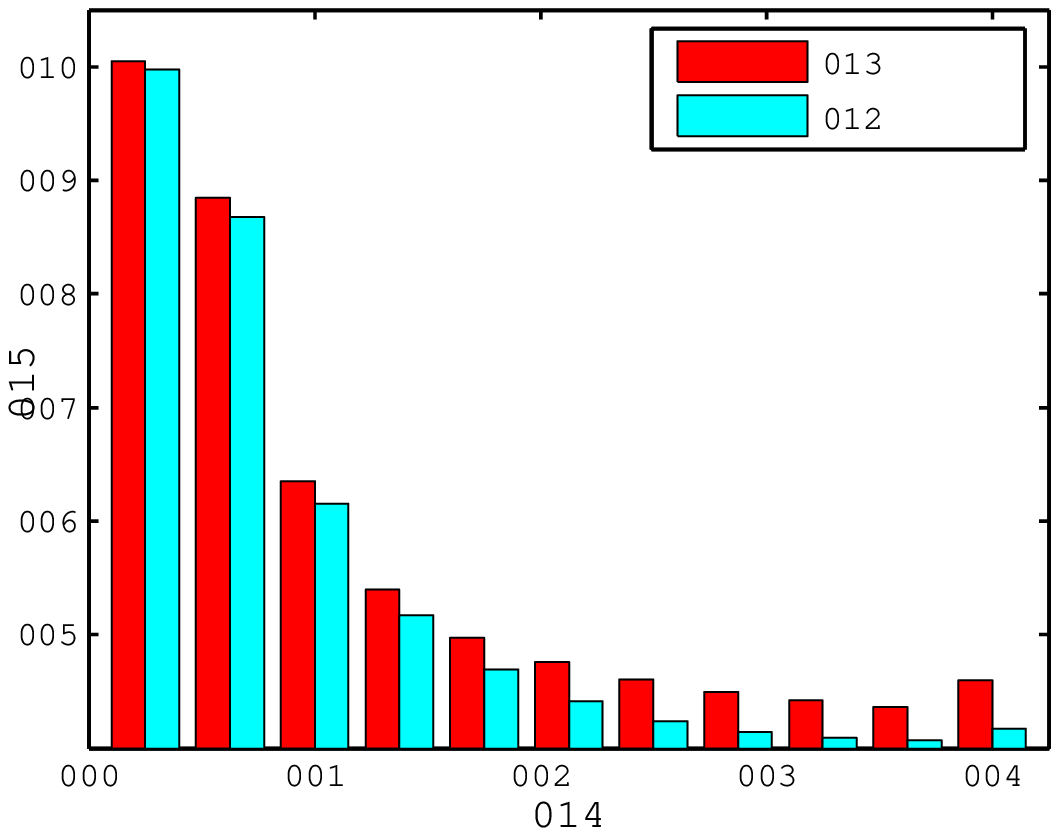}
  \end{minipage}
  \vspace{1cm}
  \begin{minipage}[h]{16cm}
    \centering
    \psfragfig{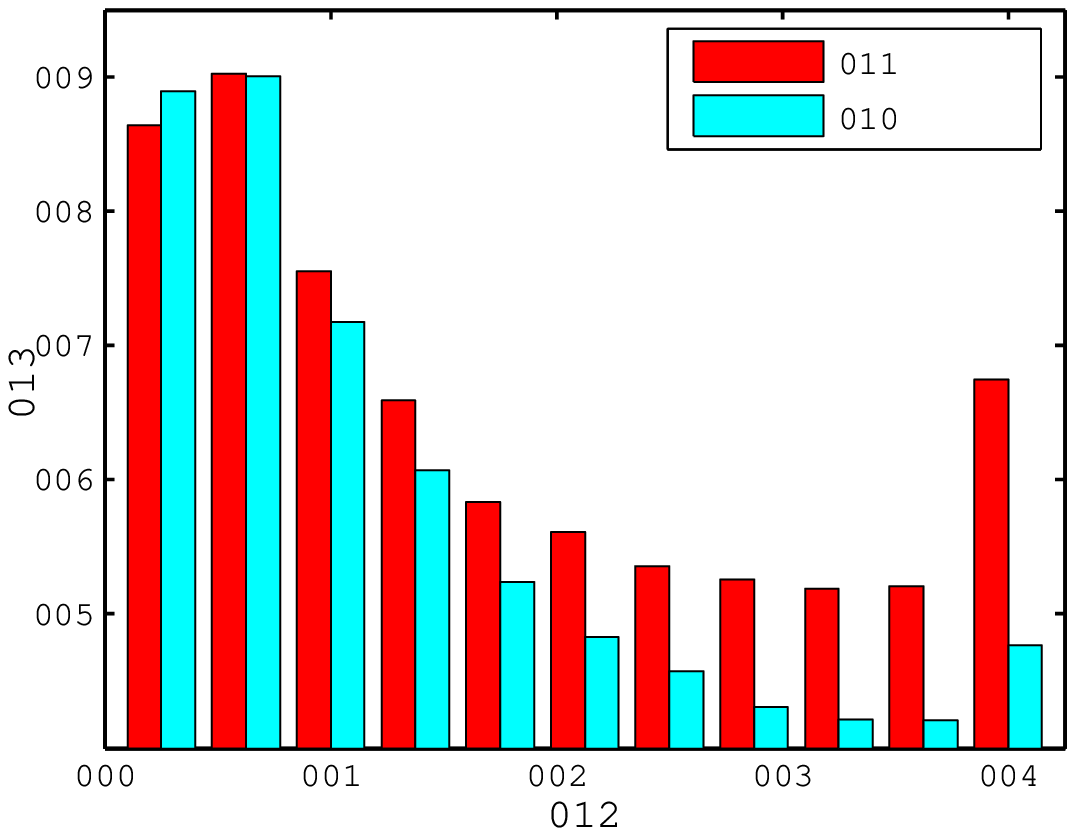}
  \end{minipage}
  \caption[First stage triggers with and without tapering]
   {Number of triggers recovered by match filtering the S4 data
    with and without tapering applied to the templates for the first stage
    (top) and the second stage (bottom) 
    where consistency checks and coincidence tests \cite{Robinson:2008}
    in the time-of-arrival and masses of the component stars have been
    applied.}
\label{fig:triggers}
\end{figure}

Figure~\ref{fig:triggers}
shows the number of triggers as a function of
total mass with and without tapering for the first and second stages of the
pipeline. It can be seen that the number of
triggers is generally higher when the templates are not tapered. The only
exception seems to be the lowest mass bin in the second-stage coincident
triggers, where the opposite is true.
However, the difference in the number of triggers in this bin is not large,
and is likely just a statistical anomaly.
For first-stage single-detector
triggers, the number of triggers using tapered templates is $84\%$
of that obtained using un-tapered templates. The number of second-stage
coincident triggers when using tapered templates is $71\%$ of
that obtained for un-tapered templates.
The difference in trigger rates is more significant
at higher masses. This is because the template waveforms for these systems
terminate at a frequency within or below the most sensitive frequency band
of the detector, making any leakage of power to higher frequencies more
significant (see Figure \ref{fig:spectra2}). The reduced
trigger rate indicates that applying the taper
function to the templates could aid in reducing the false alarm rate in a
search for high mass \ac{CBC}s.

\section{Effect of windowing on detection efficiency and parameter estimation}
\label{sec:tappe}
The same data used in section \ref{sec:taptriggers} were re-analysed, but with
simulated gravitational wave signals (injections) added.
The injections were of the same family as the templates used in section
~\ref{sec:taptriggers}, allowing the detection
efficiencies and accuracy of parameter estimation using tapered vs.~untapered
templates to be compared. There was negligible difference in 
the error in recovered chirp mass and arrival time at both
single detector first stage triggers and coincident second stage triggers.
Although the detection efficiency was not explicitly measured as a function
of distance, the number of injections recovered was found to be
nearly identical
in the two cases, with less than $1\%$
fewer injections found when using tapered 
templates.  Given the vast reduction in the trigger rates shown in
Section~\ref{sec:taptriggers}, this indicates that an improvement in detection
efficiency \textit{can be expected} when using tapered templates.

The above studies were performed first with tapering applied to the
injections and then repeated without - the difference between the
results was negligible.

\section{Concluding remarks}\label{sec:tapconc}
The Planck-taper window leads to spectra for \ac{TD}
waveforms that more closely match their \ac{FD} analogs, containing
significantly less power at unexpected frequencies when compared with the use
of a square window. This is achieved by automating the implementation of the
window.

If tapering is applied to templates in a
gravitational wave search the trigger rates are reduced, especially for
high mass templates, without any significant change in detection efficiency.
In a search, foreground triggers can be ranked by their probability of
occurring as a background trigger; thus if background triggers are reduced,
a given foreground trigger may appear more significant.
Another benefit of reduced trigger rates is that the computational cost of a
search will decrease. Indeed the studies here demonstrate that the 
Planck-taper windowing method would be
beneficial when used in a high mass search.

The tapering method could also be useful in low latency data analysis
techniques where \ac{TD} templates
are divided into sub-templates of different frequency ranges, and
match filtered individually~\cite{Acernese:2006uu}. The relative shortness of
some templates in the higher frequency bands potentially compounds
the problem of using a square window, and tapering the templates may go
some way to alleviating this issue.


\chapter{Black Hole Hunter: The game that lets YOU search for 
gravitational waves} 
\label{hunter}
\lhead{Chapter~\ref{hunter}.
\emph{BHH: The game that lets YOU search for 
gravitational waves}}
\rule{15.7cm}{0.05cm}

\label{sec:hunter}
A collaboration of gravitational wave physics groups from 
the United Kingdom and Germany presented the exhibit `Can you hear black
holes collide?' at the Royal Society Summer Science Exhibition 2008 
in London.  The exhibit gave the public insight into how
gravitational waves are generated, how gravitational wave detectors
function, and how searches for gravitational waves are performed.  The `Black
Hole Hunter' computer game was developed to illustrate the
challenges of searching for a gravitational wave signal in noisy data.
The game was popular with attendees at the exhibition and has
subsequently been used in many other outreach projects.  The game's 
website, \emph{ www.blackholehunter.org}, currently receives approximately 
one-thousand unique visitors each month.  

\section{Searching for gravitational waves}
\label{sec:bhhintro}
  
Gravitational wave experiments are  in an exciting era. A
global network of first generation \ac{IFO}s have been used to search
for gravitational waves and have already made statements about our
Universe, e.g., ~\cite{Abbott:2008fx,Collaboration:2009tt,
Abbott:2009qj,GRB070201}.
Furthermore, the detectors are currently undergoing 
upgrades to reach ever more impressive levels of 
sensitivity~\cite{Smith:2009bx}. This
provides an ideal opportunity to inspire public interest
and excitement in science.  There is a large outreach effort 
in the gravitational wave community,
including public education centres~\cite{Cavaglia:2008},
teaching projects in schools, and a travelling gravitational waves
exhibit~\cite{Cavaglia:2009}.  
 
\section{`Can you hear black holes collide?'}
\label{sec:blackholehunter}
 
The Royal Society annually hosts a summer science exhibition at its
offices in central London.  This exhibition, which is open to the
general public, aims to inform visitors of the latest developments and
discoveries in all fields of science and inspire young people's interest
and involvement in science. The Royal Society Summer Exhibition 2008 
\cite{roysoc2008} consisted of twenty-three exhibits and two additional
art and history of science exhibits each. These exhibits covered a vast range of
scientific fields from bioscience to astrophysics, and the exhibition was
attended by several thousand visitors over four days.
 
 
Among the exhibits selected for the summer exhibition in 2008 was `Can
you hear black holes collide?' presented by a collaboration of British
and German gravitational wave researchers.  Detectors 
such as LIGO and GEO are sensitive to gravitational waves
with frequencies between approximately $50\, \rm Hz$ 
and a few thousand $\rm Hz$.  This range is
comparable to frequency range of the human ear, motivating the choice of
title.
 
The goals of this exhibit were two-fold: to give the public an idea of
what gravitational waves are; and how we go about searching for them.
The exhibition featured a short, looping video to attract
visitors. A `rubber sheet universe' was
used to illustrate Einstein's concept of space-time and curvature and to
demonstrate heuristically how orbiting bodies might emit gravitational radiation.
A fully-functional table-top interferometer was used to
explain and demonstrate to visitors the basic principles of 
laser interferometric detectors. In order to illustrate the
methods and challenges involved in searching for gravitational waves,
the `Black Hole Hunter' game was available to play on multiple
computers.
 
Additionally a group of researchers 
actively involved in gravitational wave science were stationed at the 
exhibit to talk to visitors and to answer their questions and a variety 
of handouts were distributed which provided visitors
with website addresses and further information on the exhibit allowing
them to continue learning more on gravitational waves after the
exhibition. 
 
\section{The Black Hole Hunter game}
\label{sec:bbhgame}

 
The aim of the Black Hole Hunter game is to give the player 
insight into the various techniques used, and challenges faced, in the
search for gravitational waves.  There are many potential sources of
gravitational waves, but the game focused on those emitted
during the merger of binaries consisting of 
black holes and/or neutron stars.  These systems
produce a characteristic `chirp' waveform which sweeps upwards in both
frequency and amplitude as the stars draw closer to merger.
 
The game begins by showing the player a graph 
of the
gravitational wave signal 
from a binary merger, as a \ac{TD} waveform, 
and playing a short audio clip of the 
waveform~
\footnote{Although the signal frequencies are within human hearing range
they were in fact shifted to higher pitches because
typical laptop speakers and headphones were not deemed adequate at low
frequencies.}. 
The player is then told that he/she must `detect' this gravitational
wave signal.  Once the player has listened to the signal he/she is
presented with four graphs, and their corresponding audio clips, of simulated
data output from a gravitational wave detector, one of which contains the
signal.  The \ac{SNR}, which determines the relative
amplitudes of the signal and the simulated detector noise, varies
depending on the difficulty level. The idea is that the player must work
in a similar way to real search algorithms and match the 
gravitational wave signal to what he/she can see or hear in the noisy
data.  Interestingly, it is much easier to pick out a signal by
listening to the audio clips than by looking at the plots.
 
Once the player has decided which of the four data streams
contains the signal, he/she selects an answer and the game reveals whether
it is correct by showing which of the data streams contained the
signal and the position of the signal in the noise.  If the chosen
answer is correct the player will proceed on to the next level where the
SNR will be lower, and thus the signal is harder to find. 
If the wrong answer is selected
the player will be able to try again with a different signal at the same
SNR.  This repeats until the player runs out of `lives' or reaches the
furthest level. The player can choose
between {\it beginner}, {\it intermediate} or {\it advanced} at 
the start of the game,
which adjusts the SNR of the first and hardest levels accordingly.
 
To demonstrate some of the problems faced in real gravitational wave
data analysis (and to make the game more fun), the hardest levels also
contain `glitches' in some of the simulated data. The glitches are designed
to confuse the player.  They are either short sine waves of random frequency 
with Gaussian envelopes or \emph{other} simulated gravitational waves that 
are similar to the signal, but shorter in duration. 
The hardest levels contain simulated data with several glitches of both
kinds!
 
As well as giving a basic demonstration of the problems data analysts
face in searching for gravitational wave sources, the Black Hole Hunter
game aims to teach the player more about gravitational physics in
general.  This is achieved in two ways during the game. Firstly, the
home page and the `Game Over' pages of Black Hole Hunter both have an
information bar on the right hand side, which contains links to a
variety of pages where the player can find out more about gravitational
wave physics, and even actively participate in real gravitational wave
research through the \emph{einstein@home} project~\cite{einsteinathome}.
Secondly, when the player has given their answer he/she is presented
with a prominent `Did you know?' box. The box contains a snippet of
information about gravitational physics and an associated internet link
leading to more information.
There are nearly one hundred different pieces of information, 
so it is unlikely that a player will encounter the
same `Did you know?' twice.  
\begin{figure}[htbp]
\centering
\includegraphics[width=14cm]{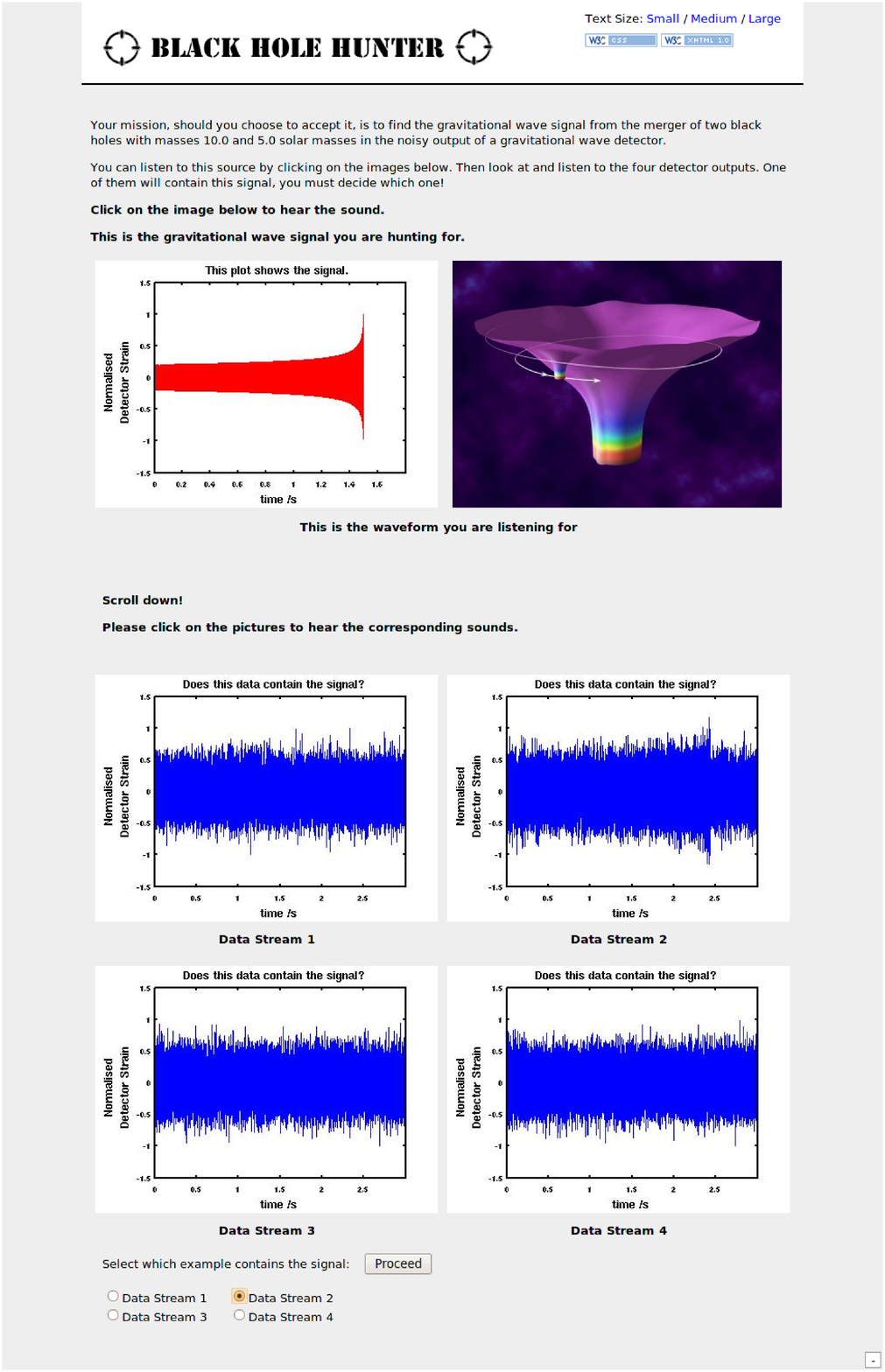}
\caption[Black hole hunter game]{This is the web page that the player sees when
playing Black Hole hunter. The four data sets
are plotted, one of which contains the 
signal. The player listens to the data by clicking on the plots before
selecting their answer at the bottom of the page.}
\label{fig:bhhpage}
\end{figure}
 
In addition to the website the Black Hole Hunter game has been
modified to run on a local machine without requiring access to the
internet. This version is available in German as well as English.
 
\section{Downloadable ringtones}
\label{sec:bbhringtones}
 
In addition to the game itself, the Black Hole Hunter website also gave
players the opportunity to download gravitational wave ringtones. These
consisted of short snippets of sound or music in WAV and MP3 format
which are suitable for use as a ringtone on a mobile phone. The
ringtones themselves were produced by manipulating sound files generated
from the expected gravitational wave signals of a variety of sources.
The manipulations included significant editing, pitch shifting, layering
signals on top of each other, and applying a number of audio effects.
These processes were performed using audio editing software such as
Cubase\cite{Cubase}, LMMS~\cite{LMMS} and Audacity~\cite{Audacity}.
 
\section{Response to the Black Hole Hunter game}
\label{sec:bbhresponse}
 
Black Hole Hunter has been
used in exhibitions in the UK and Germany, as a teaching aid in
Australia and is forming a major part of a travelling exhibition in the
USA~\cite{Cavaglia:2009}.  Visitors to these exhibitions typically
include school teachers, schoolchildren and their parents.

Following its launch at the 2008 exhibition, 
Black Hole Hunter was featured in a New Scientist blog
\cite{new_scientist_blog} and linked from the Einstein@Home web
site~\cite{einsteinathome}.  With this publicity, in the first month the
site received 3123 unique visitors (IP addresses) from at least 66
different countries. In 2009 the website recorded nearly $1000$ 
unique visitors each month.


\section{Development}
\label{sec:bhhdev}
The development of the game was broken in to two parts, firstly 
the media files that contain the simulated signals and data as both audio and 
images and secondly the development of the web pages that keep the player's
score and presents the correct media. The author was responsible for the 
former, with some help from Patrick Sutton who generated 
the simulated \ac{LIGO} noise and Ian Harry who calculated the \ac{SNR}s
of the different difficulty levels.

The core of the code required to generate the media is contained in a single
MATLAB function \emph{mp3BlackHoleMusic.m}. 
This function requires four 
input parameters, the first three relating to the signal, namely the component 
masses of the binary signal and the inclination 
angle of the source; the final input parameter is the duration of the data
in seconds. The function generates the simulated signal, 
the \ac{LIGO} noise, `glitches' and adds the signal to the noise for five
 different values of \ac{SNR}. The output of the function is a variety of audio
and image files, everything needed for a particular set of signal parameters 
to be used in the game. The function can be run multiple times over with 
different input choices to create enough variation for the online game.

\subsection{Simulating the signals}
The \ac{TT3} \ac{PN} inspiral waveform was coded in 
MATLAB~\cite{Blanchet:2004ek}.
The waveforms are evolved according to a dimensionless
time parameter $\tau$, which decreases from an initial value until it
reaches $81/16$, the value it has when the orbital separation of
the two objects is $r=6\Msun$, i.e. at \ac{FLSO}. Before the waveform
is generated we know the required duration in seconds and the sampling rate
of the output audio file. A simple calculation then reveals the number of 
discrete steps, the step size, $\Delta\tau$, and finally the initial value
$\tau_0$. Once all values of $\tau$ are known a 2\ac{PN} waveform with
the chosen parameters is generated.

\subsection{Simulating the noise}
The coloured noise is created in the \ac{FD}, by multiplying
a frequency array of Gaussian random amplitudes with the \ac{LIGO}
design \ac{PSD}. The noise then undergoes an \ac{IFT}, which means
the length of the array and the frequency resolution must be set
correctly so that the noise is of the correct length and sample rate in the
\ac{TD}. 

\subsection{Adding the signal to the noise}
The first step of the process is to divide the signal and data 
by their maximum amplitudes plus ``epsilon'' respectively so that both
have a maximum value of just under 1, as the function that produces the
audio files from the arrays clips any data with amplitudes greater than 1.
The signal and nose  are then saved as audio files so that the player can
hear the signal before playing and the noise can be used as one of the three
data without the signal.
Before adding the signal to the noise it is scaled by a chosen factor 
that sets the difficulty. The duration of the noise is twice that of the signal.
Therefore, if the noise is of length $T$, the last point of the signal is placed
at random between $T/2$ and $T$. The simulated data that contains the signal
is again divided by its maximum amplitude plus epsilon. This process is
iterated over with different scaling factors for the difficulty levels. At
the time of the development, the levels were set `by ear' with the 
\ac{SNR}s estimated retrospectively.

\subsection{Simulating glitches}
There a several types of glitches introduced at random in the harder levels.
Firstly, inspiral signals of different parameters are added. The other glitches
are sine-Gaussians of a random frequency, in some cases several different
glitches are added at the same time. The glitches were not modelled on
real causes of data noise, but were engineered to make the game more 
interesting. The duration of each glitch was set to $1/5$ of the noise
and normalised to have a maximum amplitude half that of the noise.



\chapter*{Concluding remarks}
\addtotoc{Concluding remarks}
\lhead{\textit{Concluding remarks}}
\rule{15.7cm}{0.05cm}

It is currently an exciting time in gravitational wave research. The
LIGO and Virgo detectors have recently collected the 
most sensitive gravitational wave strain data ever measured;
as a result, analyses have produced upper limits on the
rates of various astrophysical sources in the nearby Universe. The detectors
are currently undergoing further 
commissioning that will increase their sensitivity,
and hence their horizon distance,  
by a factor of $\sim10$. This improved sensitivity equates to a
factor of $\sim1000$ increase in the volume of the observable 
Universe. The expected rate of \acp{CBC} detectable by 
advanced LIGO-Virgo networks
may be as high as $400$ per year or, more realistically, $40$ per 
year~\cite{:2010cf}. It is not an implausible suggestion that gravitational 
waves will be directly detected by ground-based 
interferometric detectors before the
centenary of Einstein's completed theory of general relativity, in 2016.

In Chapter 1 and Chapter 2 we learned the nature of gravitational waves
and how they may be detected. Gravitational waves are generated by 
acceleration of the mass quadrupole moment, they are
transverse and propagate through vacua at the speed of light. Gravitational
waves from CBCs may be modelled using the PN approximation. We saw 
that the gravitational wave strain upon the Earth from a 
coalescing binary source
at a distance of $100\ \rm Mpc$ would produce a strain of the right 
amplitude and frequency to be detectable by ground-based 
interferometric detectors such as LIGO.  

In Chapter~\ref{howchap} we covered the derivation of the matched filter
and saw how it was used in a search pipeline on a subset of 
LIGO's S5 data, that placed the following upper limits on the rates with 
$90\%$ confidence: \ac{BNS} - $1.4\times10^{-2}\ \rm yr^{-1}\rm L^{-1}_{10}$;
\ac{BBH} - $7.3\times10^{-4}\ \rm yr^{-1}\rm L^{-1}_{10}$ and
NSBH - $3.3\times10^{-3}\ \rm yr^{-1}\rm L^{-1}_{10}$. Although these upper
limits are 1-2 orders of magnitude above the optimistic predicted rates
they are a significantly lower than those obtained from the S51YR search
alone. 

In Chapter 4 we set out the motivations behind using higher order waveforms
in gravitational wave data analysis and then developed a filtering algorithm
that used templates of 0.5PN in amplitude. The algorithm required
significant development with key changes in normalisation
and maximisation in comparison to the RWF algorithm.
A matrix was used to transform the templates from their
original basis to an orthonormal basis before computing the SNR.
A constraint was set on the SNR of the relative harmonics by
transforming it back to the original basis and comparing with the
expected maximum values. The final results were promising with 
improvements in both
the detection (SNR value) and the parameter estimation observed, which
matched the original motivations. Furthermore, by studying the 
SNR time-series we observed that the constraint appears to be a very 
effective signal-based veto in terms of eliminating noise, which could lead to
a reduced FAR. 
There is great potential for the FWF filtering algorithm, even at 0.5PN,
and perhaps a further developed version
will play a part in the analysis pipeline of the next
generation detectors.

In Chapter 5 we examined a new method of windowing that tapers
the start and/or end of a waveform using an algorithm that finds the
near-optimal place to apply the taper, ensuring that the
transitions are smooth. The new method resulted in a better estimation for
the SNR, a more `realistic' representation of the signal in the FD and
reduced trigger rates when tested with a LIGO high mass pipeline
in LIGO's S4 data. Furthermore, the method did not significantly affect
the number of detected injections, indicating that detection efficiency 
would be improved with use of the window due to the reduction in background.

Finally, we ended with a description of Black Hole Hunter, an exciting
outreach project that aims to teach the public about gravitational 
waves and the efforts to detect them.



\addtocontents{toc}{\vspace{2em}} 

\appendix 


\chapter{Introduction}
\label{AppIntro}
\lhead{Appendix . \emph{Introduction}}

\section{The energy-momentum tensor}
\label{sec:tmunu}
The energy-momentum tensor contains information on the matter 
and energy that
causes the curvature of spacetime. Its components represent the following:
\begin{itemize}
\item $T^{00}$ is the relativistic mass density;
\item $T^{0i}$ is the flux of momentum in the $i$ direction;
\item $T^{ij}$ is the rate of flow of the $i$ component
of momentum in the $j$ direction. These components are often referred to
as the \emph{stress} components for $i\neq j$ and the \emph{pressure}
components for $i=j$.
\end{itemize}
N.B.: $T^{\mu\nu}=T^{\nu\mu}$.

\section{The amplitude matrix $A_{\mu\nu}$}
The Lorentz gauge condition (\ref{eq:gauge1}) is only satisfied if
\be
A_{\mu\nu}k^\nu = 0\, ,
\ee
which implies that the amplitude matrix is orthogonal to $k^\nu$.

\chapter{Gravitational waves radiated from binary systems}
\label{AppendixB}
\lhead{Appendix . \emph{Gravitational waves radiated from binary systems}}
\rule{15.7cm}{0.05cm}

\section{The Lambda tensor}
\label{sec:lambda}
The Lambda tensor, $\Lambda_{ij,kl}$, upon contraction with any symmetric
tensor, $B_{ij}$, yields the transverse and traceless part, i.e.,
\be
B^{TT}_{ij} = \Lambda_{ij,kl}B_{kl}\, .
\ee
The Lambda tensor is defined as
\be
\Lambda_{ij,kl}(\hat{\textup{\textbf{n}}}) 
  = \delta_{ik}\delta_{jl} - \frac{1}{2}\delta_{ij}\delta_{kl}
  - n_jn_l\delta_{ik} - n_in_k\delta_{jl}
  + \frac{1}{2}n_kn_l\delta_{ij} + \frac{1}{2}n_in_j\delta_{kl}
  + \frac{1}{2}n_in_jn_kn_l\, .
\ee

\section{Centre-of-mass, single body representation}
\label{sec:eobsimp}
For a point particle, following a trajectory $x_0(t)$
in flat spacetime the energy momentum tensor is
\be
T^{\mu\nu}(t,\textup{\textbf{x}})
  = \dfrac{p^\mu p^\nu}{\gamma m} \delta^{(3)}
      \left(\textup{\textbf{x}} - \textup{\textbf{x}}_0(t)\right)\, ,
\ee
where
\be
p^\mu = \gamma m(dx_0^\mu/dt)\, ,
\ee
is the four-momentum and
\be
\gamma = {\left(1 - v^2\right)}^{-\frac{1}{2}}\, ,
\ee
where
\be
v^2 := \frac{dx^i}{dt}\frac{dx_i}{dt}\, . 
\ee

\section{Moments}
\subsection{Taylor expansion of the energy-momentum tensor}
In Section~\ref{sec:lvexp} the polarisations (\ref{eq:lvexp}) are
written as a Taylor expansion of the energy-momentum tensor.
Firstly, (\ref{eq:efesol2}) is written in terms of the \acf{FT} of 
$T_{\mu\nu}$, which only consists of frequencies $\omega\leq\omega_s$ where
$\omega_sa\ll1$. Under these conditions it is clear that the 
exponent in the \ac{FT} can be expanded, which is equivalent to the Taylor
expansion in the \acf{TD}:
\be
T_{kl}\left(t - D + \textup{\textbf{x}}^\prime\cdot\hat{\textup{\textbf{n}}}, 
                \textup{\textbf{x}}^\prime\right)
 = T_{kl}\left(t-D,\textup{\textbf{x}}^\prime\right)
 + \textup{\textbf{x}}^{\prime}_i\textup{\textbf{n}}^i\delta T_{kl}
 + \frac{1}{2}
  \textup{\textbf{x}}^{\prime}_i\textup{\textbf{x}}^{\prime}_j
  \textup{\textbf{n}}^i\textup{\textbf{n}}^j
  \delta^2T_{kl} + ...\, .
\ee

\subsection{Moments of the source}
\label{sec:mos}
In Section~\ref{sec:lvexp} the expansion of the energy-momentum tensor
is expressed as the moments, $S^{ij}$, of the the stress components of
$T^{ij}$,
which have the following definitions:
\begin{subequations}
\label{eq:momident}
\begin{align}
S^{ij}(t) & = \int d^3xT^{ij}(t,\textup{\textbf{x}})\, ,\\ 
\label{eq:momidentS2}
S^{ij,k}(t) & = \int d^3xT^{ij}(t,\textup{\textbf{x}})x^k\, ,\\ 
S^{ij,kl}(t) & = \int d^3xT^{ij}(t,\textup{\textbf{x}})x^kx^l\, . 
\end{align}
\end{subequations}
We also introduced the moments of the energy density, which
are defined as
\begin{subequations}
\label{eq:momident2}
\begin{align}
M & = \int d^3x T^{00}(t,\textup{\textbf{x}})\, ,\\
M^{i} & = \int d^3x T^{00}(t,\textup{\textbf{x}})x^i\, ,\\
M^{ij} & = \int d^3x T^{00}(t,\textup{\textbf{x}})x^ix^j\, ,\\
M^{ijk} & = \int d^3x T^{00}(t,\textup{\textbf{x}})x^ix^jx^k\, .
\end{align}
\end{subequations}
Similarly the moments of the momentum density are defined as
\begin{subequations}
\label{eq:momident3}
\begin{align}
P^{i} & = \int d^3x T^{0i}(t,\textup{\textbf{x}})\, ,\\
P^{i,j} & = \int d^3x T^{0i}(t,\textup{\textbf{x}})x^j\, ,\\
P^{i,jk} & = \int d^3x T^{0i}(t,\textup{\textbf{x}})x^jx^k\, .
\end{align}
\end{subequations}

\subsection{Identities}
\label{sec:momident}
In linearised theory there are a number of identities that exist between
the moments. These are obtained by defining a volume $V$ that is larger than
the source, such that $T^{\mu\nu}=0$ outside $V$, and applying the
conservation law $\partial_\mu T^{\mu\nu}=0$. To first order
the identities are
\begin{subequations}
\begin{align}
\label{eq:conmom1}\dot{M} & = 0\, ,\\
\dot{M}^i & = P^i\, ,\\
\dot{M}^{ij} & = P^{i,j} + P^{j,i}\, ,\\
\dot{M}^{ijk} & = P^{i,jk} + P^{j,ki} + P^{k,ij}\, ,
\end{align}
\end{subequations}
and
\begin{subequations}
\begin{align}
\label{eq:conmom2}\dot{P}^i & = 0\, ,\\
\dot{P}^{i,j} & = S^{ij}\, ,\\
\dot{P}^{i,jk} & = S^{ij,k} + S^{ik,j}\, .
\end{align}
\end{subequations}
It is from these identities that we find (\ref{eq:sij}). N.B.:
(\ref{eq:conmom1}) and (\ref{eq:conmom2}) are the conservation of mass and
momentum respectively, whilst it can also be shown that $S^{ij} - S^{ji}=0$,
which corresponds to the conservation of angular momentum.

\section{The \acl{TT3} phase approximant}
\label{sec:tt3phase}
The \ac{TT3} approximant up to 2PN in order~\cite{Buonanno:2009zt}:
\be
\begin{split}
\varphi_\text{TT3}(t) = 
&\  \varphi_0 - \frac{1}{\eta\theta^5}
    \biggl[ 1 + \left(\frac{3715}{8064} + \frac{55}{96}\eta\right)\theta^2 
    - \frac{3\pi}{4}\theta^3 \\
&\  + \left( \frac{9275495}{14450688} + \frac{284875}{258048}\eta 
     + \frac{1855}{2048}\eta^2\right)\theta^4 \\
&\  + \left(\frac{38645}{21504} - \frac{65}{256}\eta\right)\ln
    \left(\frac{\theta}{\theta_{LSO}}\right)\pi\theta^5\biggr]\, ,
\end{split}
\ee
where
\be
\theta = {\left[\frac{\eta(t_0 -t)}{5M}\right]}^{-\frac{1}{8}}\, ,
\ee
$\varphi_0$ is a constant and $\theta_{LSO}$ is the value of $\theta$ at the
time of \ac{ISCO}.

\section{The inspiral gravitational wave polarisations up to 2PN}
\label{sec:hphc}
The gravitational wave polarisations from inspiralling compact binaries up to
2PN are~\cite{Blanchet:2008je}:
\begin{subequations}
\begin{align}
H^{(0)}_+ & = \left(1+\cos^2i\right)\cos2\varphi\, ,\\
H^{(0)}_\times & = 2\cos i\sin2\varphi\, ,
\end{align}
\end{subequations}
\begin{subequations}
\begin{align}
H^{(0.5)}_+ & = 
  -\Delta \sin i\left[
    \left(\dfrac{5}{8} + \dfrac{1}{8}\cos^2i\right)\cos\varphi
    -\left(\dfrac{9}{8} + \dfrac{9}{8}\cos^2i\right)\cos\left(3\varphi\right)
  \right]\, ,\\
H^{(0.5)}_\times & = 
  -\Delta \sin i\cos i\left[-\dfrac{3}{4}\sin\varphi 
    + \dfrac{9}{4}\sin\left(3\varphi\right)
  \right]\, ,
\end{align}
\end{subequations}
Continued on following page.
\clearpage

\begin{subequations}
\begin{align}
\begin{split}
H^{(1)}_+ &  =
  -\cos2\varphi\biggl[\frac{19}{6} + \frac{3}{2}\cos^2i 
    - \frac{1}{3}\cos^4i 
    + \eta\left(-\frac{19}{6}+\frac{11}{6}\cos^2i +\cos^4i\right)\biggr]\\
&\ \ \   
  +\cos4\varphi\biggl[\frac{4}{3}\sin^2i\left(1+cos^2i\right)
    \left(1-3\eta\right)\biggr]\, ,
\end{split}\\
\begin{split}
H^{(1)}_\times &  =
  -\cos i\sin2\varphi\biggl[\frac{17}{3} - \frac{4}{3}\cos^2i 
    + \eta\left(-\frac{13}{3} + 4\cos^2i\right)\biggr]\\
&\ \ \ 
  -\cos i\sin^2i\sin4\varphi\biggl[-\frac{8}{3}\left(1-3\eta\right)
    \biggr]\, ,
\end{split}
\end{align}
\end{subequations}
\begin{subequations}
\begin{align}
\begin{split}
H^{(1.5)}_+ &  =
  - \sin i\Delta\cos\varphi\biggl[\frac{19}{64} + \frac{5}{16}\cos^2i
    - \frac{1}{192}\cos^4i\\
&\ \ \ 
   + \eta\left(-\frac{49}{96} + \frac{1}{8}\cos^2i
   + \frac{1}{96}\cos^4i\right)\biggr]\\
&\ \ \ 
  - \cos2\varphi\left[-2\pi\left(1+\cos^2i\right)\right]\\
&\ \ \ 
  - \sin i\Delta\cos3\varphi\biggl[-\frac{657}{128} - \frac{45}{16}\cos^2i
    + \frac{81}{128}\cos^4i\\
&\ \ \ 
    + \eta\left(\frac{225}{64} - \frac{9}{8}\cos^2i - \frac{81}{64}\cos^4i
    \right)\biggr]\\
&\ \ \ 
  - \sin i\Delta\cos5\varphi\biggl[\frac{625}{384}\sin^2i
    \left(1+\cos^2i\right)\left(1-2\eta\right)\biggr]\, , 
\end{split}\\
\begin{split}
H^{(1.5)}_\times & = 
  - \sin i\cos i\Delta\sin\varphi\biggl[
    \frac{21}{32} - \frac{5}{96}\cos^2i + \eta\left(-\frac{23}{48}
    + \frac{5}{48}\cos^2i\right)\biggr]\\
&\ \ \ 
  + 4\pi\cos i\sin2\varphi\\
&\ \ \ 
  - \sin i\cos i\Delta\sin3\varphi\biggl[-\frac{603}{64} 
    + \frac{135}{64}\cos^2i + \eta\left(\frac{171}{32} - \frac{135}{32}
    \cos^2i\right)\biggr]\\
&\ \ \
  -\sin i\cos i\Delta\sin5\varphi\biggl[\frac{625}{192}
    \left(1-2\eta\right)\sin^2i\biggr]\, ,
\end{split}
\end{align}
\end{subequations}
\begin{subequations}
\begin{align}
\begin{split}
H^{(2)}_+ &  =
  -\pi\sin i\cos\varphi\biggl[-\frac{5}{8}-\frac{1}{8}\cos^2i\biggr]\\
&\ \ \ 
  -\cos2\varphi\biggl[\frac{11}{60} + \frac{33}{10}\cos^2i 
    + \frac{29}{24}\cos^4i-\frac{1}{24}\cos^6i\\
&\ \ \
    +\eta\left(\frac{353}{36} - 3\cos^2i-\frac{251}{72}\cos^4i +
    \frac{5}{24}\cos^6i\right)\\
&\ \ \ 
    +\eta^2\left(-\frac{49}{12} + \frac{9}{2}\cos^2i - \frac{7}{24}\cos^4i
    -\frac{5}{24}\cos^6i\right)\biggr]\\
&\ \ \ 
  -\pi\sin i\Delta\cos3\varphi\biggl[\frac{27}{8}
    \left(1+\cos^2i\right)\biggr]\\
&\ \ \ 
  -\frac{2}{15}\sin^2i\cos4\varphi\biggl[59 + 35\cos^2i - 8\cos^4i
    -\frac{5}{3}\eta\left(131+59\cos^2i-24\cos^4i\right)\\
&\ \ \ 
    +5\eta^2\left(21-3\cos^2i-8\cos^4i\right)\biggr]\\
&\ \ \ 
  -\cos6\varphi\biggl[-\frac{81}{40}\sin^4i\left(1+\cos^2i\right)
    \left(1-5\eta+5\eta^2\right)\biggr]\\
&\ \ \ 
  -\sin i\Delta\sin\varphi\biggl[\frac{11}{40}+\frac{5\ln2}{4}
    +\cos^2i\left(\frac{7}{40} +\frac{\ln2}{4}\right)\biggr]\\
&\ \ \ 
  -\sin i\Delta\sin3\varphi\biggl[\left(-\frac{189}{40} 
    + \frac{27}{4}\ln\left(\frac{3}{2}\right)\right)
    \left(1+\cos^2i\right)\biggr]
\end{split}\\
\begin{split}
H^{(2)}_\times & = 
  -\sin i\cos i\Delta\cos\varphi\biggl[-\frac{9}{20}-\frac{3}{2}\ln2\biggr]\\
&\ \ \ 
  -\sin i\cos i\Delta\cos3\varphi\biggl[\frac{189}{20}-\frac{27}{2}\ln
    \left(\frac{3}{2}\right)\biggr]\\
&\ \ \
  +\sin i\cos i\Delta\frac{3\pi}{4}\sin\varphi\\
&\ \ \ 
  -\cos i\sin2\varphi\biggl[\frac{17}{15} + \frac{113}{30}\cos^2i -
    \frac{1}{4}\cos^4i\\
&\ \ \ 
    + \eta\left(\frac{143}{9} -\frac{245}{18}\cos^2i + \frac{5}{4}\cos^4i
    \right)\\
&\ \ \ 
    +\eta^2\left(-\frac{14}{3}+\frac{35}{6}\cos^2i-\frac{5}{4}\cos^4i
    \right)\biggr]\\
&\ \ \ 
  -\sin i\cos i\Delta\sin3\varphi\biggl[\frac{27\pi}{4}\biggr]\\
&\ \ \ 
  -\frac{4}{15}\cos i\sin^2 i\sin4\varphi\biggl[
    55-12\cos^2i-\frac{5}{3}\eta\left(119-36\cos^2i\right)
    +5\eta^2\left(17-12\cos^2i\right)\biggr]\\
&\ \ \ 
  -\cos i\sin6\varphi\biggl[-\frac{81}{20}\sin^4i\left(1-5\eta+5\eta^2
    \right)\biggr]\, .
\end{split}\\
\end{align}
\end{subequations}

\chapter{Higher order waveforms in data analysis}
\label{AppendixD}
\lhead{Appendix . \emph{Higher order waveforms in data analysis}}
\rule{15.7cm}{0.05cm}

\section{Maximisation proof}
\label{sec:maxproof}

\paragraph*{Lemma:}
In (\ref{eq:maximise}) maximising $\Lambda$ over $\alpha_i$ and $\lambda$
yields the maximum of $\rho$ with the constraint 
\be
\sum_{i=1}^6\alpha^{\prime2}_i=1\, .
\ee

\paragraph*{Proof:}
Suppose that another quantity $\gamma_{i=1,\ldots,6}$ exists such that
\be
\label{eq:gamma}
\sum_{i=1}^6\gamma_i = 1\, ,
\ee
and
\be
\label{eq:maxproof}
\rho[\gamma_i] > \rho[\alpha^\prime_i]\, .
\ee
However, (\ref{eq:gamma}) means that
\be
\rho[\gamma_i] = \Lambda[\gamma_i]\, .
\ee
Yet $\alpha^\prime_i$ maximise $\Lambda$ which would give
\be
\Lambda[\gamma_i] \le \Lambda[\alpha^\prime_i] = \rho[\alpha^\prime_i]\, ,
\ee 
but that contradicts (\ref{eq:maxproof}).

\section{The log-normal distribution}
\label{sec:logn}
The log-normal distribution, $f(x;\mu,\sigma)$,
is the probability distribution of a random 
variable whose logarithm is normally distributed:
\be
f(x;\mu,\sigma) = 
  \frac{1}{
    x \sigma \sqrt{2 \pi}}\, e^{-\frac{(\ln x - \mu)^2}{2\sigma^2}},
    \ \ x>0\, ,
\ee
where $\mu$ is the mean and $\sigma$ is the standard deviation of the 
distribution, respectively.

\section{Maximum correlation between two templates}
\label{sec:maxcor}
In~\cite{Damour:1998zb} it is shown how to find the minimum and maximum
correlation between two \emph{two-phase} templates, for the case of
any time lag between the templates arrival time. We are interested in the
maximum correlation. Given two templates, or two harmonics of a 0.5PN 
template, $a$ and $b$, the process is as follows:
\begin{enumerate}
  \item Compute the following:
    \begin{align}
      A & = {\left<a_+,b_+\right>}^2 + {\left<a_+,b_\times\right>}^2\, ,\\
      B & = {\left<a_\times,b_+\right>}^2 
                 + {\left<a_\times,b_\times\right>}^2\, ,\\
      C & = \left<a_+,b_+\right>\left<a_\times,b_+\right> 
               + \left<a_+,b_\times\right>\left<a_\times,b_\times\right>\, .\\
      \end{align}
  \item The maximum overlap, $p$,  between $a$ and $b$ is then given by
    \be
      p = {\left[\frac{A+B}{2} + 
            {\left[{\left(\frac{A-B}{2}\right)}^2 + C^2\right]}^{\frac{1}{2}}
          \right]}^2 
    \ee
  \item Compute $p$ over all values of time and record the maximum value.
\end{enumerate}

\chapter{Miscellany}
\label{AppHigher}
\lhead{Appendix . \emph{Miscellany}}


\section{ACTD logo}
\begin{figure}[h!]
  \centering
  \includegraphics[width=14cm]{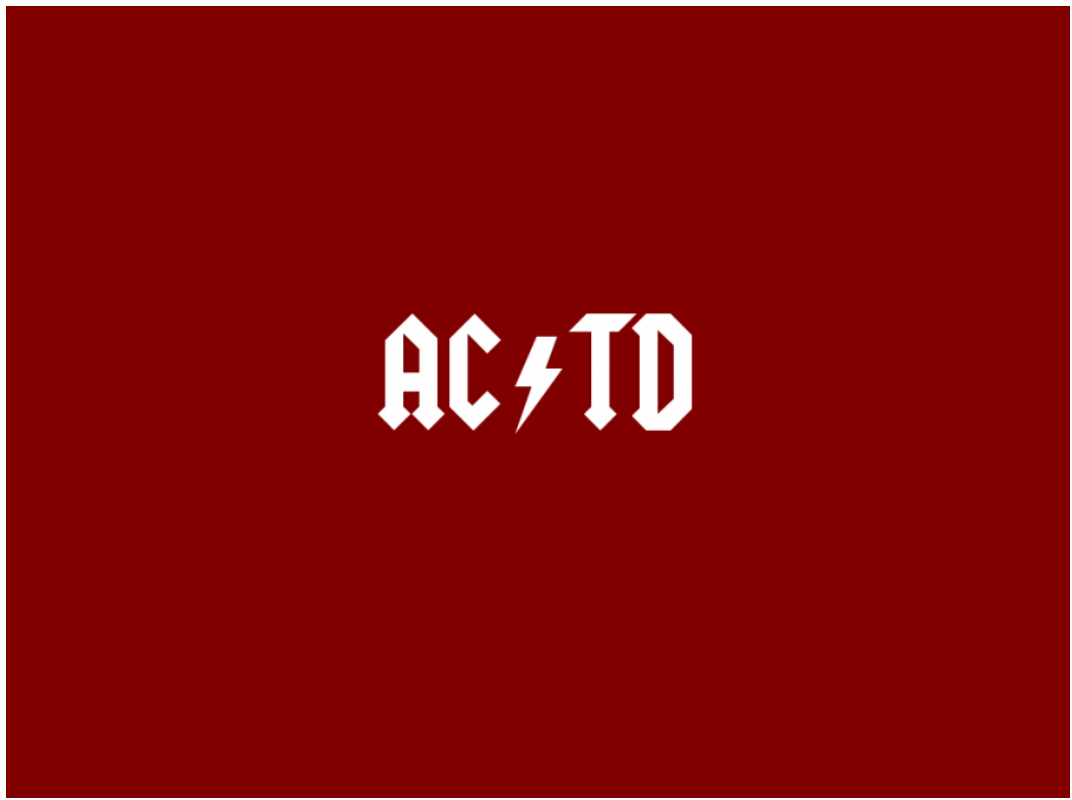}
  \caption[ACTD logo]
  {The only appropriate logo for the \acs{ACTD} codes.}
  \label{fig:actd}
\end{figure}

No self-respecting student can dare to develop new code without designing
an appropriate logo. The 0.5\ac{PN} filtering algorithms are written in
codes named with the acronym \emph{\ac{ACTD}}. Thus there was only one
logo suitable... cf.~Figure~\ref{fig:actd}.
\clearpage

\section{SVN commit history}
\begin{figure}[h]
  \centering
  \psfragfig{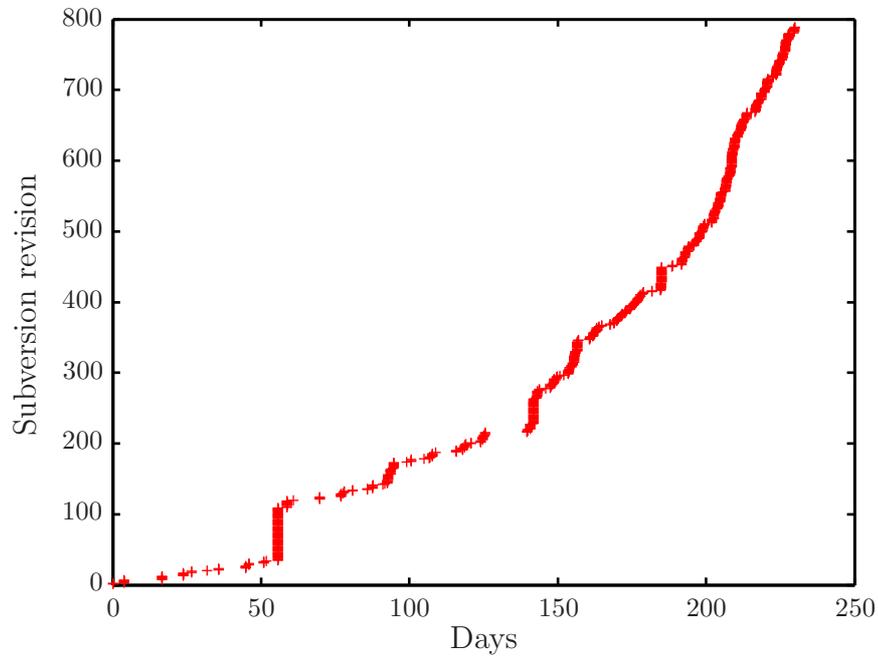}
  \caption[SVN commit history]
  {SVN commit history of this thesis.}
  \label{fig:svn}
\end{figure}

The author's progress in writing this thesis is shown in Figure~\ref{fig:svn}.
N.B.: at the outset the author committed files individually before
realising that several file changes could be covered in one commit. Therefore,
the actual increase in work rate is slightly under exaggerated.

\onehalfspacing

\addtocontents{toc}{\vspace{2em}}  
\backmatter

\label{Bibliography}
\lhead{\emph{Bibliography}}  
\bibliographystyle{unsrtnat}  
\bibliography{Bibliography}  

\end{document}